\documentclass[sigplan,screen,authorversion,nonacm]{acmart}

\renewcommand\footnotetextcopyrightpermission[1]{} 
\copyrightyear{2026}
\acmYear{2026}
\setcopyright{cc}
\acmConference[]{}{}

\author{Gaetano Coccimiglio}
\email{gccoccim@uwaterloo.ca}
\affiliation{%
  \institution{University of Waterloo}
  \city{Waterloo}
  \state{Ontario}
  \country{Canada}
}

\author{Trevor Brown}
\email{trevor.brown@uwaterloo.ca}
\affiliation{%
  \institution{University of Waterloo}
  \city{Waterloo}
  \state{Ontario}
  \country{Canada}
}

\author{Srivatsan Ravi}
\email{srivatsr@usc.edu}
\affiliation{%
  \institution{University of Southern California}
  \city{Los Angeles}
  \state{California}
  \country{USA}
}

\usepackage{xcolor}
\usepackage{wrapfig}
\usepackage{subcaption}
\usepackage{listings}
\usepackage{graphicx} 
\usepackage{url}
\usepackage{hyperref}
\usepackage{cleveref}
\usepackage{multirow}
\usepackage{ulem}

\usepackage{soul} 

\definecolor{drawioorange}{HTML}{D79B00}
\definecolor{codegreen}{rgb}{0,0.6,0}
\definecolor{codegray}{rgb}{0.5,0.5,0.5}
\definecolor{codepurple}{rgb}{0.58,0,0.82}
\definecolor{backcolour}{rgb}{0.95,0.95,0.92}
\definecolor{anti-flashwhite}{rgb}{0.95, 0.95, 0.96}
\definecolor{highlight}{rgb}{1.0, 0.0, 0.0} 

\lstdefinestyle{mystyle}{
  float=t,
  mathescape=true,
  backgroundcolor=\color{anti-flashwhite},
  commentstyle=\color{codegray},
  numberstyle=\tiny\color{codegray},
  stringstyle=\color{codepurple},
  basicstyle=\ttfamily\footnotesize,
  breakatwhitespace=false,         
  breaklines=true,
  numberblanklines=false,
  captionpos=b,                    
  keepspaces=false,                 
  numbers=left,                    
  numbersep=5pt,                  
  showspaces=false,                
  showstringspaces=false,
  escapeinside={<@}{@>},
  showtabs=false,                  
  tabsize=1,
  keywordstyle=\color{codegreen},
  otherkeywords={thread_local,then,for,each,true,false, class, struct, def, goto, uint64_t}, 
  morekeywords={type,subtype,break,continue,if,else,elif,end,loop,while,do,done,exit, when,then,return,read,and,or,not,boolean,procedure,invoke,iteration,until,wait}
}
\newcommand{\tmname}{Multiverse}

\title{Multiverse: Transactional Memory with Dynamic Multiversioning$^{*}$}



\begin{document}

\begin{abstract}
Software transactional memory (STM) allows programmers to easily implement concurrent data structures.
STMs simplify atomicity.
Recent STMs can achieve good performance for some workloads but they have some limitations.
In particular, STMs typically cannot support long-running reads which access a large number of addresses that are frequently updated.
Multiversioning is a common approach used to support this type of workload.
However, multiversioning is often expensive and can reduce the performance of transactions where versioning is not necessary.

In this work we present \tmname, a new STM that combines the best of both unversioned TM and multiversioning.
\tmname~ features versioned and unversioned transactions which can execute concurrently.
A main goal of \tmname~ is to ensure that unversioned transactions achieve performance comparable to the state of the art unversioned STM while still supporting fast versioned transactions needed to enable long running reads.

We implement \tmname~ and compare it against several STMs.
Our experiments demonstrate that \tmname~ achieves comparable or better performance for common case workloads where there are no long running reads.
For workloads with long running reads and frequent updates \tmname~ significantly outperforms existing STMS.
In several cases for these workloads the throughput of \tmname~ is several orders of magnitude faster than other STMs.
\end{abstract}

\maketitle

\begingroup
\renewcommand{\thefootnote}{\fnsymbol{footnote}}
\footnotetext[1]{This version of the paper corrects two minor issues.
In \Cref{erratum} with regards to version list traversals, a suitable version has timestamp strictly less than the transaction's versioned timestamp (previously this was incorrectly \textit{less than or equal}).
In \Cref{alg:txn-interface} the commit clock is obtained before read set validation (previously those lines of the pseudocode were flipped in the opposite order).}
\endgroup

\section{Introduction}
Software transactional memory (STM) is a synchronization mechanism that allows users to execute sequences of memory accesses as atomic \textit{transactions}.
STMs make atomicity easy but not necessarily fast.
STMs typically perform read-only transactions using optimistic synchronization.
In such STMs~\cite{herlihy1993transactional, dice2006transactional, lev2009anatomy, ramalhete2024scaling, felber2008dynamic, fraser2007concurrent, memoryprinciples, ennals2006software, shavit1995software}, memory addresses are typically associated with version numbers that are used to determine whether a transaction's reads are consistent. If not, a transaction aborts and retries.

For workloads where transactions are small (accessing few addresses) and contention is low, TMs can typically achieve good performance. 
On the other hand, STMs often struggle to handle transactions that read a large number of addresses that are frequently updated.
This type of transaction is very likely to repeatedly abort.

A classical solution in data structures and databases for supporting large read-only operations is to utilize \textit{multiversioned concurrency control (MVCC)}, which allows one to take atomic \textit{snapshots} of memory \cite{afek1993atomic, wu2017empirical}.
Much of the early work on MVCC focused on maintaining consistent versions and avoiding an unbounded number of versions per address.
Many of these MVCC designs which focus only on providing atomic snapshots are typically expensive.
Recent work has made significant practical improvements to MVCC \cite{blelloch2024verlib, wei2021constant}.

There have been some attempts to combine multiversioning with STM \cite{lu2013generic, perelman2011smv, riegel2006snapshot}.
These existing approaches typically guarantee the weaker correctness conditions of snapshot isolation \cite{berenson1995critique} or serializability \cite{bernstein1987concurrency}, as opposed to (the stronger) opacity \cite{tm-book}.
Without opacity, aborted transactions are allowed to observe inconsistent state which can lead to various problems \cite{dice2014pitfalls}.
Furthermore, maintaining multiple versions can incur substantial overhead, reducing performance in the case where versioning is not required.
In such cases, traditional \textit{unversioned} 
TMs are preferable.

\begin{figure}[t]
    \begin{subfigure}{1.0\linewidth}
        \centering
        \includegraphics[width=0.85\linewidth]{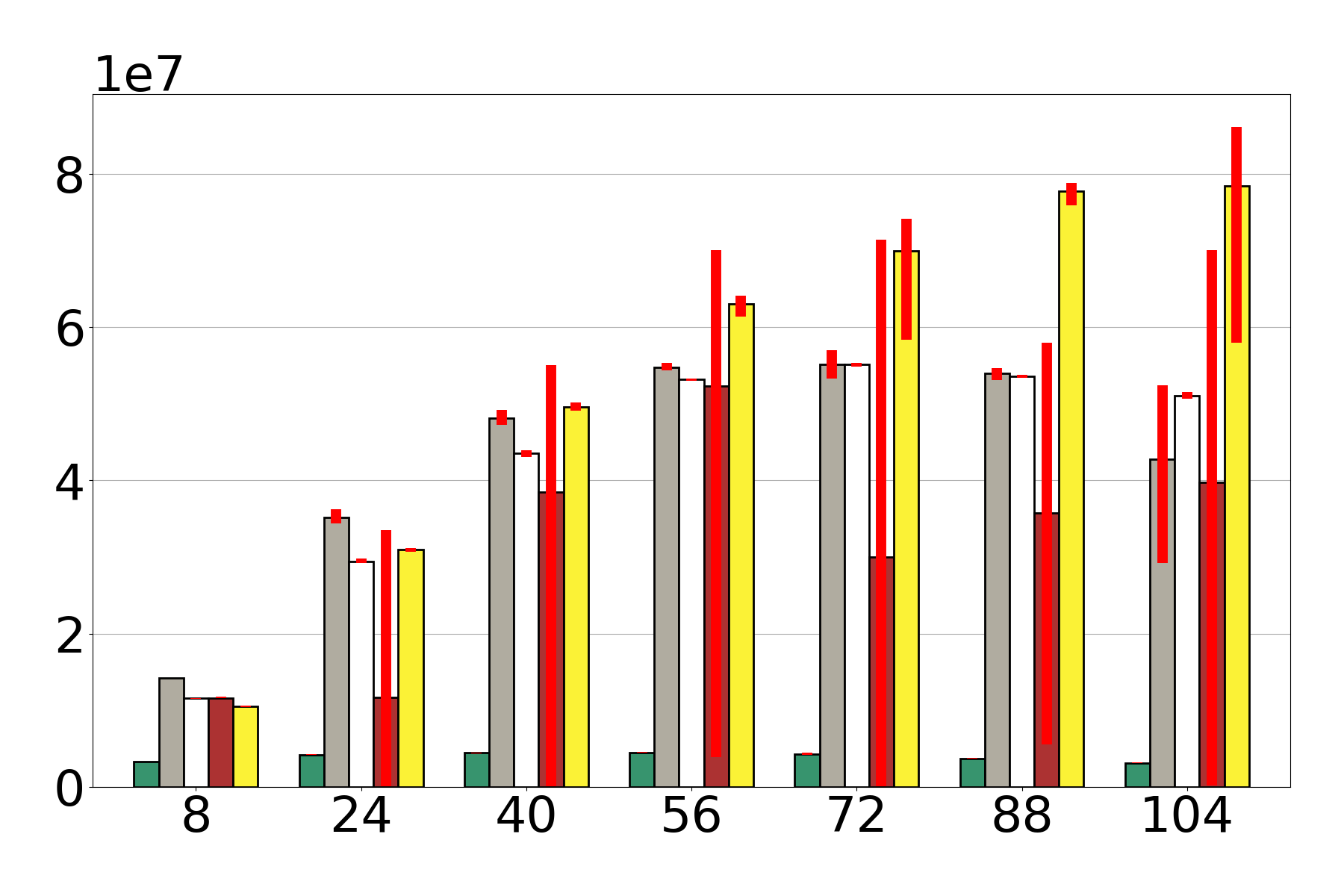}
    \end{subfigure}  
    \begin{subfigure}{1.0\linewidth}
        \centering
        \includegraphics[width=1.0\linewidth]{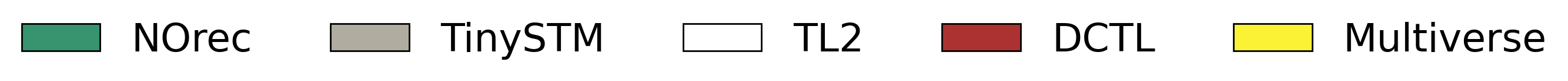}
    \end{subfigure} 
    \vspace{-4mm}
    \caption{
    (a,b)-tree benchmark with an 89.99\% search, 0.01\% RQ, 5\% insert, 5\% delete workload using a uniform key access pattern. RQ size is 10k (1\% of prefill size). Y-axis is ops/sec. X-axis is number of threads.
    }
    \Description{}
    \label{fig:intro-fig}
\end{figure}

In this work we present \tmname, a new opaque STM that leverages recent advancements in MVCC~\cite{blelloch2024verlib} to enable read-only transactions that access a large number of addresses to commit even in the presence of many concurrent updates that would otherwise cause potentially unbounded aborts.
Unlike prior multiversioned STMs, our primary design goal is to ensure that \textit{our unversioned transactions} are competitive with the \textit{fastest unversioned STM}, Deferred Clock Transactional Locking (DCTL) at present \cite{ramalhete2024scaling}, while also ensuring that our \textit{versioned} transactions are fast as possible subject to the constraint that they do not add substantial overhead to unversioned transactions.
As \Cref{fig:intro-fig} shows, \tmname\ can outperform the fastest unversioned STM even for workloads with very few range queries (RQs), where versioning is not always necessary.

Our algorithm features unversioned and versioned transactional code paths. 
Transactions begin as unversioned, and transition to the versioned code path based on a heuristic function that considers the number of times the transaction has aborted, the thread's recent transactions' behavior, and the recent behavior of other transactions in the TM system.
Our design associates versions with individual addresses at the word level of granularity.
We dynamically switch addresses between versioned and unversioned states.
Switching addresses between unversioned and versioned states is subtle and deeply connected to both correctness and performance.

A key insight in our work is that versioning of addresses should be done quite differently in different workloads. 
More specifically, if a versioned transaction only needs to access a small number of addresses under low contention, it is most efficient to leave most addresses unversioned, and allow concurrent non-versioned update transactions to proceed with as little overhead as possible.
On the other hand, if a versioned transaction needs to access a very large number of addresses under high contention, it is more efficient to preemptively (and globally) force all concurrent updating transactions to preserve old versions of all addresses.

Our TM thus features two global modes that we switch between based on a heuristic. 
In \textit{Mode~Q}, transactions on the \textit{versioned code path} are responsible for \textit{marking addresses as versioned}, and transactions on the unversioned code path can be largely oblivious to versioned transactions. 
A versioned transaction in Mode~Q that encounters an unversioned address will abort if the address has changed since the transaction began.
So, this mode is suitable when transactions that require multiversioning only access relatively few memory addresses, and/or are infrequent.

If aborts due to encountering unversioned addresses are frequent, then Mode~U can help substantially. 
In \textit{Mode~U}, transactions on the \textit{unversioned} code path \textit{that write to memory} are responsible for marking addresses as versioned, and transactions on the versioned code path can simply behave as if all relevant addresses are already versioned.
Mode~U is crucial for enabling high performance when versioned transactions perform a large number of accesses that are likely to be aborted by concurrent updates.



The complexities of versioning are completely hidden from the user, and \tmname\ does not require any modifications to a program's memory layout---only replacement of variable types with analogous transactional types.
This is the gold standard in TM. 
By avoiding changes to the memory layout, we preserve the cache behavior of the underlying program as much as possible, limit intrusive changes to the program, and allow standard object serialization techniques for disk storage or network transfer of transactional objects.
%
We utilize separate parallel lock- and version-tables so that when the versioning mechanism is not actively engaged, the cache behavior of unversioned transactions is as similar as possible to the state of the art in unversioned STM.

\textbf{Contributions:}
\begin{itemize}
    \item We introduce \tmname, a novel opaque STM that combines unversioned STM and MVCC.
    It features a distinct usage of dynamic multiversioning and multiple TM modes that adapt the behavior of the TM to fit the needs of the workload \S\ref{sec:alg}.
    \item To our knowledge, \tmname\ is the first full featured opaque multiversioned STM implemented in C++ that has proper memory management which avoids crashes that would occur in other STMs like TL2 or DCTL \S\ref{sec:impl}.
    \item We implement \tmname\ and experimentally evaluate it on a real system comparing it against several existing opaque STMs \S\ref{sec:eval}.
\end{itemize}

\section{Background}

\subsection{Transactional Memory}
A transaction is a sequence of transactional accesses, (reads and writes), performed on a set of \textit{transactional addresses}.
A transaction either \textit{commits} and appears as a single indivisible step, or \textit{aborts} and has no visible effect.

A TM implementation provides operations to start a transaction, read and write transactional addresses, commit a transaction, and voluntarily abort a transaction.
If two transactions are concurrent and one or both of the transactions writes to an address that the other has already accessed then we say that these transactions conflict.
Conflicts cause transactions to abort.
A transaction that aborts due to a conflict will typically be retried until it either succeeds and commits, or until it is voluntarily aborted.
We call the set of all addresses read by a transaction its \textit{read set} and the set of all addresses written by a transaction its \textit{write set}.


\begin{figure*}
    \centering
    \includegraphics[width=0.9\linewidth]{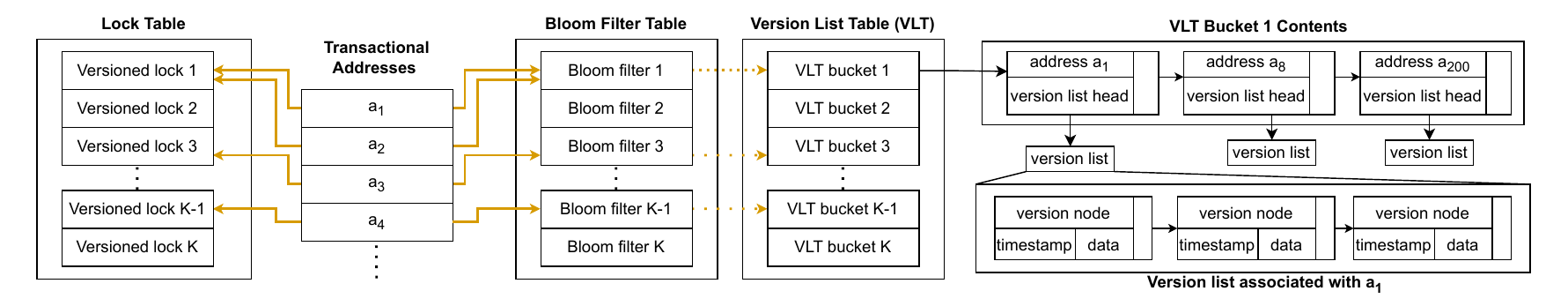}
    \vspace{-4mm}
    \caption{
    Data structures used in \tmname.
    In this example addresses $a_1$ and $a_2$ map to the first VLT bucket but only $a_1$ is versioned.
    The \textcolor{drawioorange}{orange arrows} indicate a mapping while the black arrows indicate a memory pointer.
    The dotted arrow from the bloom filter to the VLT is indicating that we access the bloom filter first before the VLT.
    }
    \label{fig:vlt}
    \Description{}
\end{figure*}

\subsection{Correctness}
Snapshot isolation (SI) \cite{berenson1995critique}, serializability, strict serializability \cite{bernstein1987concurrency} and the stronger opacity \cite{tm-book} are common correctness conditions used in TM.
A history of \textit{committed} transactions is serializable if it is equivalent to some serial history.
Strict serializability further requires that the history maintains the real-time order of transactions.
Opacity requires that the history of all transactions (including aborted ones) be equivalent to some sequential history.
In other words, opacity requires that \textit{all} transactions must observe consistent state.
As discussed in \cite{dice2014pitfalls}, one should care about opacity since it prevents various problems.
In particular, without opacity, one loses even single-threaded invariants since transactions could observe inconsistent state and continue running.
The weaker SI, intuitively, allows transactions to perform (consistent) reads in the past, but write in the present, which can be difficult to use correctly. 
Opacity is most common in TM.

\section{Algorithm}
\label{sec:alg}


\tmname\ is an opaque word based STM in which \textbf{both addresses and transactions} can either be \textit{unversioned} or \textit{versioned}.
Transactions always begin as unversioned.
Transactions that write remain unversioned.
Unversioned read-only transactions will switch to versioned after some number of attempts or under certain conditions discussed in \Cref{sec:impl}.

A core goal of \tmname\ is that our unversioned transactions should match the performance of the fastest unversioned STM (DCTL), and our versioned transactions should be as fast as possible subject to the former. 
This goal motivated many of our design choices.
By default, \tmname\ prioritizes the performance of unversioned transactions.
Similar to DCTL, the leading STM, we use a global clock and transactional addresses are protected by \textit{versioned locks}.
The data structures used in \tmname\ are illustrated in \Cref{fig:vlt} and described below.

\subsection{Word Based Dynamic Versioning}
\label{sec:dynamic-versioning}
Addresses in \tmname\ are initially unversioned, meaning they do not have associated version lists.
An address can \textit{become versioned} if we determine that maintaining additional versions is likely to reduce aborts.
Likewise, if we later determine that we do not need additional versions of a particular address then we can \textit{unversion} the address by removing and freeing (via epoch-based reclamation) its version list.

\subsubsection{Versioning Addresses}
Versioning an address requires creating and associating a version list with the address.
Versioning and unversioning of an address is done while holding the associated address lock.
This ensures no concurrent updates can modify the address. 

To avoid changing a program's memory layout, we store all versioned locks and version lists in hash tables which we refer to as the \textit{lock table} and \textit{Version List Table (VLT)}.
Each bucket in the VLT is a linked list.
Within a bucket, each node contains (1) a pointer to the head of a version list, (2) the address for which the version list is tracking changes, and (3) a pointer to the next bucket node.
The VLT and lock table are identical in size, which allows us to use the same mapping function from addresses to entries for both, and enables a convention that an address' lock also protects its version list.
%

When we create a new version list, we insert an initial version.
This requires a timestamp and the data.
For the data, we take the last consistent value of the address.
Since we must hold the lock before versioning an address, the last consistent value is simply the current value of the address.
Choosing the timestamp is more nuanced.
A timestamp will correspond to some value of the global clock.
We attempt to take the earliest possible timestamp (details in \Cref{sec:txns}).

\subsubsection{Checking if an Address is Versioned}
Determining whether or not a particular address is versioned requires traversing the associated bucket in the VLT.
We know that the address is versioned if we find a node in the bucket that has the address. 
We use bloom filters to make this efficient. 
Each address is associated with a bloom filter.
When an address becomes versioned we add it to the bloom filter.
To determine if an address is versioned we first check the bloom filter.
If we do not find the address in the bloom filter we know the address is unversioned.
Similar to the VLT, we store these bloom filters in a separate table of identical size.


\subsubsection{Unversioning Addresses}
\label{sec:unversioning}
We unversion entire VLT buckets rather than individual addresses.
There are several reasons for this approach.
First, one cannot remove items from a bloom filter---one can only reset it.
Resetting the filter means all addresses that map to that VLT bucket are now unversioned.
Doing this periodically is worthwhile since otherwise bloom filters will slowly fill up, and produce many false positives.
Second, there is value in keeping the length of a VLT bucket small to reduce the overhead of traversals.
Any ``collateral damage'' in unversioning can affect performance but not correctness.

Unversioning is performed by a background thread. 
Unversioning a VLT bucket requires removing and freeing the linked list in the bucket along with all of the version lists in the bucket.
Before unversioning, the background thread must claim the associated lock.
We determine when to unversion a bucket using a heuristic which considers the timestamps of version lists in the bucket as well as the current global clock version.
We discuss this further in \Cref{sec:when-to-unversion}.

\subsection{Transaction Paths Basic Overview}

\subsubsection{Unversioned Transactions}
The basic execution of an unversioned transaction follows an approach similar to DCTL.
At the start of each attempt, the transaction will read and record a local copy of the global clock to acquire its \textit{read clock}.
For each TM access of an address, the version of the lock protecting the address is validated against the transaction's read clock.
TM writes use encounter time locking and writing. 
If the address being written to is versioned, the write \textit{also} updates its version list (keeping \textit{both} the version list, and the location where an unversioned transaction would read, up to date).
Any modifications to version lists are marked to-be-determined (TBD) until the transaction commits, which prevents versioned transactions from seeing inconsistent states.
Unversioned transactions keep track of a read set and two write sets.
The \textit{standard write set} tracks \textit{all} addresses written by the transaction and the \textit{versioned write set} tracks only the \textit{versioned} ones (so \textit{versioned} $\subseteq$ \textit{standard}). 
Depending on the TM mode, even unversioned transactions may cause addresses to become versioned (Mode~U).

At commit time, the read set is revalidated.
If this succeeds, the transaction rereads the global clock to obtain a \textit{commit clock}.
Then, any versioned addresses that were written have their TBD marks removed, and all locks are released (and their versions updated to the commit clock).
Any validation failures cause the transaction to abort and retry, and after sufficiently many aborts, it may become versioned.
See \Cref{sec:txns} for optimizations to unversioned transactions.


\subsubsection{Versioned Transactions}
Only read-only transactions can be versioned. 
We briefly discuss the possibility of versioned \textit{writing} transactions in \Cref{sec:snapshot-iso}.
A versioned transaction begins in the same way as an unversioned transaction, except that its read clock also serves as a \textit{versioned timestamp}.
For each address read, the transaction will determine if the address is versioned.
Depending on the TM mode, versioned transactions will either cause these addresses to become versioned (Mode~Q), or rely on relevant addresses being versioned already by updaters (Mode~U).

Accesses to versioned addresses involve version list traversals to find a suitable version. 
Such traversals are blocked by TBD markers forcing the executing thread to wait.
The traversal continues once TBD markers are removed.
If a suitable version is found, the data at that version is returned.
Otherwise the transaction aborts and retries.

Versioned transactions additionally save some information in shared memory for use by the background thread in determining when to unversion addresses.
Specifically, on the first attempt of a versioned transaction the thread will save its \textit{initial versioned timestamp}, and when a versioned transaction commits,
it computes the difference between the current global clock, and its initial versioned timestamp, and saves the resulting \textit{commit timestamp delta} 
(see \Cref{sec:when-to-unversion}).


\begin{table*}[t]
\resizebox{0.85\textwidth}{!}{%
\begin{tabular}{c|c|c|c|c|}
\cline{2-5}
                                        & Mode~Q          & Mode QtoU (Transient)  & Mode~U           & Mode UtoQ (Transient)   \\ \hline
\multicolumn{1}{|c|}{Unversioned} &
  \begin{tabular}[c]{@{}c@{}}Writes add versions iff\\ address is already versioned\end{tabular} &
  \begin{tabular}[c]{@{}c@{}}Writes forced to \\ version\end{tabular} &
  \begin{tabular}[c]{@{}c@{}}Writes forced to\\ version\end{tabular} &
  \begin{tabular}[c]{@{}c@{}}Writes forced to \\ version\end{tabular} \\ \hline
\multicolumn{1}{|c|}{Versioned} &
  Reads version &
  Reads version &
  \begin{tabular}[c]{@{}c@{}}Reads assume all \\ addresses are versioned\end{tabular} &
  \begin{tabular}[c]{@{}c@{}}Versioned txns forced\\ back to Mode~Q\end{tabular} \\ \hline
\multicolumn{1}{|c|}{Background Thread} & Unversioning enabled & Unversioning disabled & Unversioning disabled & Unversioning disabled \\ \hline
\end{tabular}%
}
\caption{Differences between TM modes for versioned and unversioned transactions along with the background thread.}
\label{tab:tm-modes}
\end{table*}

\subsection{TM Modes}

\tmname\ dynamically switches between four modes: two \textbf{stable} (Mode Q \& U) and two \textbf{transient} (Mode QtoU \& UtoQ).
Each mode changes the behavior of both unversioned and versioned transactions.
A summary of the modes appears in \cref{tab:tm-modes}.
In \textit{Mode~Q}, versioned transactions are responsible for versioning addresses, while unversioned transactions are largely oblivious to ongoing versioned transactions.
Mode~Q optimizes for unversioned transactions by ensuring new versions are added on demand only by versioned transactions that rely on them.
This mode is suitable when there is low contention, or transactions are small enough that the relevant addresses can be versioned before a conflict occurs.

\begin{figure*}[t]
    \centering
    \includegraphics[width=0.95\linewidth]{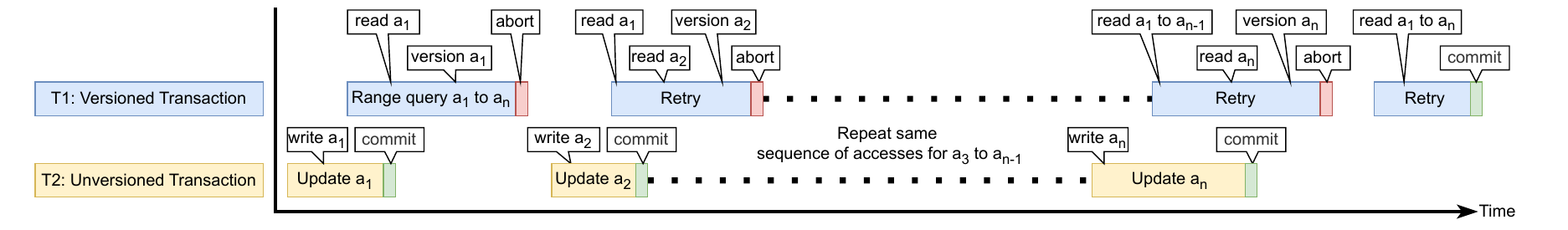}
    \vspace{-4mm}
    \caption{Example execution where Mode~Q is not suitable. All addresses are initially unversioned. The versioned transaction T1 needs to read addresses $a_1$ to $a_n$ but it must perform $O(n^2)$ accesses to commit as a result of aborts caused by conflicts with the concurrent unversioned transaction. T1 would perform only $n$ accesses if the addresses were already versioned.}
    \label{fig:mode1-bad-exec}
    \Description{}
\end{figure*}

\begin{figure*}[t]
    \centering
    \includegraphics[width=0.95\linewidth]{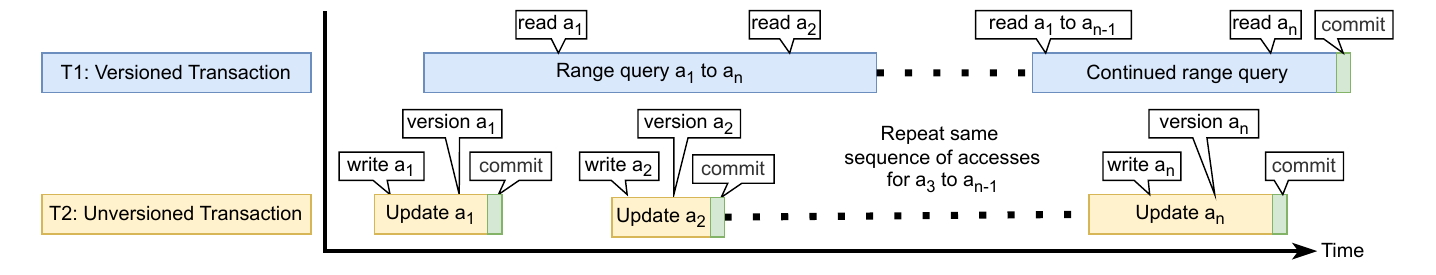}
    \vspace{-4mm}
    \caption{Example execution with the same transactions from \Cref{fig:mode1-bad-exec} but now the TM is in Mode~U forcing the unversioned transaction to version each address it updates. The versioned transaction commits without any aborts.}
    \label{fig:mode2-good-exec}
    \Description{}
\end{figure*}

However, Mode~Q is not suitable under high contention, or when there are long running transactions.
\Cref{fig:mode1-bad-exec} shows an example execution where the TM is in Mode~Q and a versioned transaction requires $n^2$ accesses to commit a transaction over $n$ addresses. 
In this case, it would be better to use 
\textit{Mode~U}, which optimizes for versioned transactions.
\Cref{fig:mode2-good-exec} shows an example execution with the same transactions and accesses from \Cref{fig:mode1-bad-exec}, but in this case the TM is in Mode~U and the versioned transaction can commit without aborting.

In Mode~U, unversioned transactions that write are forced to version addresses while versioned transactions operate as if every address is already versioned.
Mode~U essentially enables \textit{global versioning for writes}.
%
As a result, if an address is not versioned, then it has not been written since the TM entered Mode~U.
So, versioned transactions can safely read unversioned addresses without versioning them. 
\Cref{sec:mode2-reads} discusses the subtleties that arise if an address becomes versioned \textit{while} it is being read. 
%

The current (global) TM mode is visible to all transactions, and each transaction has a \textit{local mode} it operates in. 
Before each attempt, a transaction records the current TM mode and uses it as its local mode.
It is possible for the local mode to differ from the TM mode, since the TM mode can change after the transaction decides its local mode. 
For this reason, we cannot immediately change (globally) from TM Mode~Q to Mode~U since there might still be ongoing transactions operating \textit{locally} in Mode~Q. 
To ensure correct transitions between (global) modes, we use two intermediate modes. 
To transition the TM from Mode~Q to Mode~U, we first transition to \textit{Mode QtoU}.
The purpose of this mode is to allow ongoing local Mode~Q writers to commit or abort before the TM enters Mode~U.
Similarly, when we transition from Mode~U back to Mode~Q, we first transition into \textit{Mode UtoQ}, allowing ongoing versioned transactions in local Mode~U to commit or abort before the TM enters Mode~Q.
Together, these transient modes ensure that Mode~U transactions can always rely on writing transactions to version all written addresses---a property that concurrent Mode~Q writing transactions would violate.

\subsubsection{Transitioning between TM Modes}
The TM mode can only change in a fixed order: Mode~Q, Mode QtoU, Mode~U, Mode UtoQ, Mode~Q, and so on.
The TM begins in Mode~Q, and while in Mode~Q, any transaction can attempt transitioning the TM to Mode QtoU.
All other mode transitions are performed by the same background thread that handles unversioning.
The specifics of how we decide when to transition between modes is discussed in \Cref{sec:mode-switching}.

\subsection{Correctness and Progress}

\begin{theorem}
    \tmname\ guarantees weak progressiveness and opacity.
\end{theorem}

\paragraph{Weak Progressiveness}
In \tmname, a conflict will occur if a transaction attempts to validate a lock that is already locked.
A conflict will also occur if a transaction attempts to validate a lock with a version that is greater than or equal to the transaction's read clock.
Assuming there is a one-to-one mapping of locks to addresses, both scenarios can only arise as a result of a concurrent (update) transaction.
When two transactions conflict \tmname\ does not guarantee that one of the transactions will commit, meaning it is not strongly progressive.
However, since aborts only occur due to conflicts, \tmname\ does guarantee weak progressiveness. 

\paragraph{Opacity}
A full proof of opacity can be found in \cite{coccimiglio2026thesis}\footnote{The full proof appears in the PhD dissertation of first author Gaetano Coccimiglio. The dissertation is currently under submission at the University of Waterloo. It will be publicly available at \url{https://uwspace.uwaterloo.ca/}.}.
The proof follows a standard opacity pattern: take an arbitrary finite execution, complete it, and construct a sequential witness that explains every observed read using proof-only source tags.  
The key device is a serialization graph whose edges record real-time order, read sources, and conflicts with later writes; 
proving this graph acyclic gives an order of transactions that preserves real time.  
In that order, each successful read has a well-defined latest preceding source from a committed write, and the temporary proof tags can be erased to obtain a legal application-level history.

\subsection{Weakening Update Correctness to Snapshot Isolation to Allow Versioned Writes}
\label{sec:snapshot-iso}
Intuitively, snapshot isolation allows transactions to read in some prior version and then write into the current version.
This is somewhat simple to support in \tmname.
To do so, we would provide a special snapshot isolation code path which the user could explicitly invoke (only in cases where an application can tolerate the diminished correctness/atomicity guarantee).
On this code path, a transaction would follow the usual approach taken on \tmname 's versioned code path for all reads and \tmname 's unversioned code path for writes.
In essence, a read-only snapshot isolation transaction is the same as a standard \tmname\ snapshot transaction, and an updating snapshot isolation transaction is a snapshot in the past followed by an atomic DCTL-style update in the present.

\section{Implementation Details}
\label{sec:impl}

\label{sec:txns}
In this section we describe the full details of the versioned and unversioned code paths for each of the TM modes.
We include full pseudocode in \Cref{alg:txn-interface} to \Cref{alg:bgthread}.


\begin{figure}[tp]
\begin{lstlisting}[label={alg:txn-interface}, frame=single, language=c++, caption={Pseudocode for TM interface functions}]
thread locals: localModeCounter, localMode, tid,
  rClock, attempts, stickyModeU, readOnly, 
  readCnt, versioned, commitTSDelta,  
  consecSmallTxns, 

beginTxn():
  setjmp()
  localModeCounter = globalModeCounter
  localMode = getMode(globalModeCounter)
  rClock = globalClock
  reset txn data // logs and heuristic stats
  announce stickyModeU <@\color{black}{and}@> localModeCounter 
    
tryCommit():
  if readOnly 
    if versioned
        announce commitTSDelta
        update consecSmallTxns 
        stickyModeU = heuristic(readCnt)
    return
  commitClock = globalClock
  validateReadSet(rClock)  
  versionedWriteSet.unsetTBDs(commitClock)
  writeSet.releaseLocks(commitClock)
  consecSmallTxns++

abort():
  writeSet.rollback()
  snapshotWriteSet.rollback()
  clear eventual frees 
  nextClock = gClock.increment()  
  writeSet.unlock(nextClock);
  if readOnly         
    if heuristic(localMode, readCnt, attempts)
      globalMode.transitionToMode(QtoU)
    versioned = heuristic(readCnt, attempts)      
  attempts++
  longjmp() // retry at start of beginTxn
\end{lstlisting}
\Description{}
\end{figure}

\begin{figure}[tp]
\begin{lstlisting}[label={alg:utils}, frame=single, language=c++, caption={Pseudocode for utility functions along with versioned lock version list node types}]
type VersionedLock: [locked, version, tid, flag]
type VListNode: [olderNode, timestamp, data, tbd]

traverse(vlist):
  vNode = vlist.head
  if vNode.timestamp == myd->rClock 
    abort    
  while vNode.tbd and vNode.timestamp < rclock
    vNode = vlist.head //reread head
  while vNode.timestamp >= rclock or vNode.timestamp == deletedTs
    if vNode.timestamp == rclock 
        abort()
    vNode = vNode.olderNode
    if vNode == null abort
  return vNode.data
  
validateLock(lockState, rClock):
  if lockState.tid == tid return true
  if lockState.locked return false
  return lockState.version < rClock
\end{lstlisting}
\Description{}
\end{figure}

\subsection{Mode Q Code Paths}
\label{sec:mode1}
\paragraph{Update Transactions}
Transactions that write (update), are always unversioned.
A TM write (\texttt{TMWrite} in \Cref{alg:tm-write}) to an address begins by computing the mapping of the address to an index in the lock table.
This same index is used for accessing the bloom filter table and the VLT.
Next the associated lock is read.
The lock may be marked to indicate that it is held by a concurrent transaction solely for the purpose of versioning, in which case we wait for the lock to be released.
The lock version is then validated against the transaction's read clock (\texttt{validateLock} in \Cref{alg:utils}).

The transaction then attempts to claim the lock.
After claiming the lock, the transaction checks if the address is versioned.
If the address is versioned then the transaction will perform a \textit{versioned write} (\texttt{tryWriteToVersionList} in \Cref{alg:tm-write}).
If the transaction has already written to the address then the head of the list will be marked TBD and the transaction will just update the data of the TBD version.
If this is the first write to the address by this transaction then it adds a new version to the version list. 
For the new version's timestamp, we use the transaction's versioned timestamp and mark it TBD.
The address is then added to the transaction's versioned write set.
Finally, the transaction then performs the \textit{in-place write} to update the location read by unversioned transactions before adding the address to its standard write set.

When an update transaction commits (\texttt{tryCommit} in \Cref{alg:txn-interface}), the read set is revalidated.
If the validation succeeds, the transaction will read the global clock to get its commit timestamp.
The transaction then removes all of the TBD markers from addresses in its versioned write set before releasing its write set locks.

Any validation failures cause the transaction to abort (\texttt{abort} in \Cref{alg:txn-interface}).
If the transactions aborts, all of its writes will be rolled back.
This includes rolling back any versioned writes by replacing the TBD marked timestamps with a \textit{deleted timestamp} and retiring the added version.
The deleted timestamp ensures that concurrent versioned transactions are not permanently blocked waiting for a TBD marked timestamp to be resolved.

\paragraph{Read-only Transactions}
\label{erratum}
We track the number of TM reads performed during a read-only transaction.
On abort, this read count is used to determine if the transaction is more likely to commit in Mode U.
When a read-only transaction first begins it will always start as an unversioned transaction (\texttt{TMRead} in \Cref{alg:tm-read}).
Unversioned reads follow an approach similar to (unversioned) writes.
We find associated lock, wait while the lock state indicates that the address is currently being versioned then validating the lock state against our read clock.
An unversioned transaction which fails to commit after $\mathcal{K}_{1}$ attempts will switch to the versioned path.
$\mathcal{K}_{1}$ is a tunable parameter.

Versioned readers in local Mode Q (\texttt{modeQ\_versionedRead} in \Cref{alg:tm-read}) begin by checking if the address is versioned.
If the address is versioned then the transaction will traverse the version list beginning from the newest version to find a suitable version with a timestamp less its versioned timestamp.
The traversal can safely skip over versioned nodes if the node has a timestamp greater than the transaction's versioned timestamp or 
if the node is marked with deleted timestamp.
The transaction must abort if the traversal reaches node that has a timestamp equal to the transaction's versioned timestamp.
This traversal is blocked if the newest version is suitable but marked TBD.
If a suitable version is found the data is returned otherwise the transaction is aborted.

If the address is unversioned then the transaction will version it.
This requires updating the associated lock version to both claim the lock and mark it to indicate versioning is in progress.
The transaction will repeatedly try to claim the lock until it is successful after which it will validate the lock version by comparing against its read clock.
If this validation fails then, after versioning the address, the transaction must abort.
The transaction must then re-check if the address is versioned since it is possible that a concurrent transaction versioned it while this transaction was waiting for the lock.
If the address is still unversioned, the transaction will version it.
This requires allocating a new version list, a new versioned node to serve as the initial version and a new VLT bucket node.
For the initial version, we take the current value of the address for the data and, when the TM is in Mode Q, we use the lock version as the timestamp.
The new VLT bucket containing the address and new version list is inserted at the front of the VLT bucket.
The versioning process completes by inserting the address into the associated bloom filter after which the lock is released.

\begin{figure}[tp]
\begin{lstlisting}[label={alg:tm-write}, frame=single, language=c++, caption={Pseudocode for TM Write}]
TMWrite(addr, value):
  lockState = reread lock until flag is false
  if not validateLock(lockState, rClock) abort()
  if not tryLock(lockState, rClock) abort()
  standardWriteSet.add(addr, *addr)    // undo log
  oldVal = *addr
  *addr = value
  if localMode == ModeQ
    tryWriteToVersionList(addr, value)
    return
  // Modes QtoU or U or UtoQ
  vlist = tryGetVList(addr)
  if vlist == null vlist = new VersionList
    ts = firstObsModeUTs
    if ts invalid then ts = lockstate.version
    vNode0 = [olderNode = null, timestamp = ts,
             data = oldVal, tbd = false]
    vlist.head = vNode0
    bloomFltr.tryAdd(addr) 
  if vlist.head.tbd then vlist.head.data = value
  else
    vNode = [olderNode = head, timestamp = rClock,
             data = value, tbd = true]    
    vlist.head = vNode 
    versionedWriteSet.add(addr)
    eventualFree(vNode.next) 
    
tryWriteToVersionList(addr, value)
  if not bloomFltr.contains(addr) return
  if vlist = tryGetVList(addr) is null return
  if vlist.head.tbd then vlist.head.data = value
  else
    vNode = [olderNode = head, timestamp = rClock,
             data = value, tbd = true]    
    vlist.head = vNode   // protected by addr lock
    versionedWriteSet.add(addr)
    eventualFree(vNode.next) // not freed if abort
\end{lstlisting}
\Description{}
\end{figure}

\begin{figure}[tp]
\begin{lstlisting}[label={alg:tm-read}, frame=single, language=c++, caption={Pseudocode for TM Read and ModeQ TM Read}]
TMRead(addr):
  readCnt++
  if versioned and localMode == modeQ
    return modeQ_versionedRead(addr)
  if versioned and localMode == modeU
    return modeU_versionedRead(addr)
  data = *addr
  lockState = reread lock until flag is false
  if not validateLock(lockState, rClock) abort()
  readSet.add(addr)
  return data

modeQ_versionedRead(addr):  
  if not bloomFltr.tryAdd(addr)  // exists already
    vlist = tryGetVList(addr)
    if vlist return traverse(vlist)
  return versionThenRead(addr)

versionThenRead(addr):
  lockState = lockAndFlag(addr)
  data = *addr
  ts = firstObsModeUTs
  if ts invalid then ts = lockstate.version
  add new version(ts, data)
  unlock(addr)
  if not validateLock(lockState, rClock) abort()
  return data
\end{lstlisting}
\Description{}
\end{figure}

\begin{figure}[tp]
\begin{lstlisting}[label={alg:tm-read2}, frame=single, language=c++, caption={Pseudocode for ModeU TM Read}]
modeU_versionedRead(addr):
  lastVer = -1, lastVal = null, didRetry = false
  retry:
  if bloomFltr.contains(addr) 
    vlist = tryGetVList(addr)
  if vlist != null return traverse(vlist)
  // this addr is not versioned  
  val = *addr  
  lockState = read lock  
  validVer = lockState.version < rClock or 
   (firstObsModeUTs != -1 and firstObsModeUTs < rClock)
  if didRetry
    verChanged = lockState.version != lastVer
    valChanged = val != lastVal    
    if verChanged return lastVal
    if lockState.locked and validVer and not verChanged and not valChanged
      return lastVal
    if lockState.unlocked and validVer return lastVal
    abort()
  else 
   if lockState.locked
     lastVer = lockState.version
     lastVal = val 
     didRetry = true
     goto retry
    else if validVer return val
    else abort()   
\end{lstlisting}
\Description{}
\end{figure}

\subsection{Mode U Code Paths}
\paragraph{Update Transactions}
In Mode U, and any mode other than Mode Q, updaters perform the same steps as in Mode Q up to and including checking if the address is versioned.
If the address is not versioned, updaters in local Mode U must version it by following the same steps as versioned readers in Mode Q (except the updater will already hold the lock).

In Mode U, when versioning an address, we can apply an optimization when choosing the timestamp of the initial version.
Specifically, rather than using the lock version, we can use the timestamp that existed immediately after the TM entered into Mode U.
This \textit{first observed Mode U timestamp} is recorded by the background thread and stored in shared memory.
Using an earlier timestamp will reduce the likelihood of aborting versioned readers.
In Mode U, if a transaction needs to version an address, then it must be the first transaction to write to the address (otherwise it would already have been versioned).
Since no writes occurred since the TM transitioned to Mode U, it is safe to use the earlier timestamp.
%
(This optimization is also applied by read-only transitions in local Mode Q if the TM concurrently transitioned into Mode U after the reader obtained its local mode).
The first observed Mode U timestamp is invalidated by the background thread before it transitions back to Mode Q ensuring that we only apply this optimization in Mode U.
After versioning, the updater must also perform the versioned write and in-place write following the same steps as in Mode Q.
Likewise, committing or aborting also follows the same steps as in Mode Q.

\begin{figure}[t]
    \centering
    \includegraphics[width=0.95\linewidth]{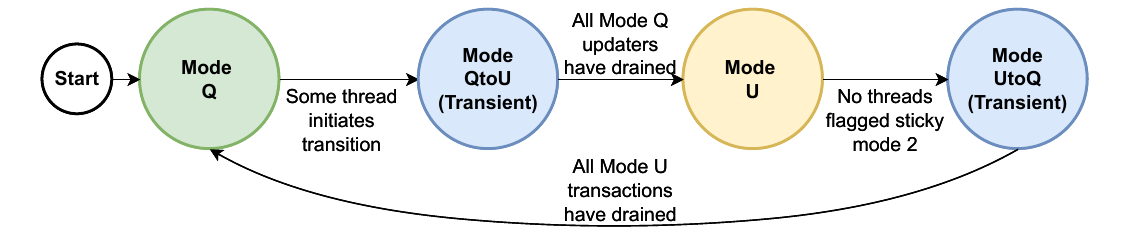}
    \vspace{-4mm}
    \caption{State transition diagram of the TM mode.}    
    \Description{TM Modes and the associated requirements for mode transitions.}
    \label{fig:tm-modes}
\end{figure}

\paragraph{Read-only Transactions}
\label{sec:mode2-reads}
When a read-only transaction is in local Mode U (\texttt{modeU\_versionedRead} in \Cref{alg:tm-read2}), it still begins as unversioned and the execution of unversioned reads is the same as in Mode Q.
When reading an address, versioned read-only transactions in local Mode U still need to check if the address is versioned.
If the address is already versioned then we perform the same traversal of the version list as in Mode Q.

If a versioned transaction in Mode U encounters an unversioned address then we know that there has not been any writes to this address since the transition to Mode U, otherwise, some other writer would have already versioned the address.
However, we check if an address is versioned before reading the data.
Since these steps are not done atomically together, it is possible for a concurrent transaction to update the address after we observe that it is unversioned, which could result in the versioned transaction observing inconsistent data.
To prevent this, if a versioned transaction encounters an unversioned address it will read and make a local copy of the data at the address and the associated lock state.
If the address is unlocked, then it is safe to return the data that we read so long when the address becomes versioned its initial version will have a timestamp that is less than the transactions versioned timestamp.
This is the case if the lock version is less than the transaction's versioned timestamp or if the the first observed Mode U timestamp is valid and is less than the transaction's versioned timestamp.

In the case that the (unversioned) address is locked then the versioned transaction re-checks if the address is versioned.
If the address is still unversioned, the transaction must redo the reads of the lock and data.
An address in Mode U can be locked iff a writer is concurrently updating the address or due to a lock table collision.
If the address is still unversioned but we observe a change in the lock version, then the address must have been locked as a result of a lock table collision, since otherwise, the writer holding the lock would have versioned the address.
Alternatively, if the address is still unversioned and we do not observe a change in the lock version or the data, then the lock could be held by a transaction seeking to update this address.
However, in this case, our first read of the data must have occurred before any such update, since again, otherwise, the address would have been versioned.
In this case it is safe to return the first value that the transaction read so long as when the address is versioned it will be versioned at a timestamp that is less than the transactions versioned timestamp.
The requirements for this to be the case are the same as in the case when the address as unlocked.

We record in shared memory, a global \textit{minimum Mode U read count}, which is the minimum number of reads performed by versioned transactions that commit in Mode U.
At commit, versioned transactions in local Mode U will update the minimum Mode U read count if they performed fewer TM reads.
When a transaction in local Mode Q aborts, this value is used to predict it is more likely to commit in Mode U.

\begin{figure}[tp]
\begin{lstlisting}[label={alg:bgthread}, frame=single, language=c++, caption={Pseudocode for Background Thread}]
bgThread() {
  while (!stopBgThread) {
    currModeCounter = globalModeCounter
    if getMode(currModeCounter) != ModeQ
      //we are in ModeQtoU
      waitForWorkers(currModeCounter);
      currModeCounter = transitionMode(currModeCounter, ModeU);
      //we are in ModeU
      firstObsModeUTs = currModeCounter
      waitForWorkers(currModeCounter);
      currModeCounter = transitionMode(currModeCounter, modeUtoQ);
      //we are in ModeUtoQ
      waitForWorkers(currModeCounter);
      firstObsModeUTs = -1
      currModeCounter = transitionMode(currModeCounter, modeQ);      
    //we are in ModeQ  
    foreach bucket in VLT
        latestVer = findLatestVersionInBucket(b);
        if globalClock - latestVer >= threshold
            unversion(bucket)
    
waitForWorkers(modeCounter)
  foundThreadAtOldMode = false
  while (true)
    foreach thread in activeThreads
      if thread.localModeCounter < modeCounter
        foundThreadAtOldMode = true
          break
    if !foundThreadAtOldMode
      return
\end{lstlisting}
\Description{}
\end{figure}

\subsection{How to Switch Between Modes}
\label{sec:mode-switching}

We utilize a monotonically increasing integer for the TM mode.
Transitions require incrementing the TM mode.
\Cref{fig:tm-modes} shows a state transition diagram which summarizes when \tmname\ transitions between TM modes.
A transaction in local Mode Q, can attempt a \textit{compare-and-swap (CAS)} operation to transition the TM mode from Mode Q to Mode QtoU.
After $\mathcal{K}_{2}$ attempts, an unversioned or versioned read-only transaction will attempt the CAS iff its read count is greater than or equal to the minimum Mode U read count.
A versioned transaction will always attempt the CAS after $\mathcal{K}_{3}$ attempts. Both $\mathcal{K}_{2}$ and  $\mathcal{K}_{3}$ are tunable parameters.

Any thread that attempts this CAS sets a thread-local \textit{sticky bit} to indicate that it wants to operate in Mode U. 
The background thread (\texttt{bgThread} in \Cref{alg:bgthread}) will inspect this bit to decide whether to remain in Mode U.
This flag bit is removed after the thread completes $\mathcal{S}$ consecutive \textit{small} transactions, where $\mathcal{S}$ is a tunable parameter.
The size of a transaction refers to the number of TM reads performed by the transaction.
Any unversioned transaction is considered small.
Each thread dynamically computes its own \textit{small transaction read count} to be $\frac{1}{\mathcal{S}}$ times the size of the transaction that the thread first committed after its last attempt of the CAS.

The background thread handles all TM mode transitions whenever the TM is not in Mode Q.
The background thread will determine when to transition by examining each transaction's local mode along with the per-thread sticky bit of each thread.
The background thread will iterate over the relevant data of all active threads to determine when it can transition out of the current TM mode.
Mode transitions by the background thread are performed via atomic writes.

The transition from Mode QtoU to Mode U will occur once the background thread completes an iteration of the relevant data without observing any update transactions with a local mode whose value is less than the value of current TM mode.
Immediately after this transition, the background thread will read and save in shared memory, the first observed Mode U timestamp (to be used in the optimized path when versioning).
The transition from Mode U to Mode UtoQ occurs once the background thread completes an iteration of the relevant data without observing any threads with the sticky Mode U flag.
Finally, the transition from Mode UtoQ back to Mode Q occurs once the background thread completes an iteration of the relevant data without observing any versioned transactions with a local mode whose value is less than the value of the current TM mode.
Immediately prior to performing this transition, the background thread invalidates the first observed Mode U timestamp.

\subsection{How to Unversion}
\label{sec:when-to-unversion}
Unversioning is only enabled in Mode Q.
We unversion any VLT bucket in which there is a sufficiently large difference between the most recent timestamp of any version in the bucket and the current global clock version.
We utilize a heuristic approach to determine a suitable difference which is computed as follows:
First the background thread will compute the average of all transactions' commit timestamp deltas.
It will then add the average to a list.
The background thread will repeat this process until the list reaches a size of $\mathcal{L}$ where $\mathcal{L}$ is given as a tunable parameter.
The background thread then sorts the list into descending order, and then computes the average of a prefix of the list.
The length of the prefix, $\mathcal{P}$, is also a tunable parameter.
Next the background thread iterates over each VLT bucket and unversions the bucket if the difference between the current global clock version and most recent timestamp in the bucket is larger than the average of the prefix.
To unversion, the background thread will acquire the associated lock.
It then removes and retires every node in the bucket along with every version list in those nodes after which the lock is released.

There is a subtle scenario in which a single address should be unversioned (instead of unversioning the entire VLT bucket).
More specifically, the version list of a single address must be invalidated or removed if the address is reclaimed while it is still versioned.
This is necessary since otherwise, the address could be returned from a future allocation and would appear to still be versioned but the associated version list could contain versions that point to freed memory. 

\subsection{Memory Management}
\label{sec:memory-management}
During a transaction, all allocations are buffered such that they can be rolled back if the transaction aborts.
We utilize a simple epoch based reclamation (EBR) mechanism to enable safe memory reclamation within transactions.
EBR pairs naturally with TM since we can tie the epoch management into transaction commits and aborts.
Immediately after an update transaction adds a new version to a version list, the previous version is retired.
However, if the transaction aborts then the previous version should not be reclaimed.
Thus, when we rollback the effects of an update transaction we also revoke any of its retires.
Any of the new versions added by an aborted update transaction will also be retired (these retires will not be revoked).

Algorithms like TL2 and DCTL benefit from the fact that they do not require read-only transactions to revalidate read sets.
While this tends to lead to better performance, it permits a crucial memory reclamation race condition.
Consider a concurrent singly linked list where the synchronization is handled by DCTL.
Suppose our list contains four nodes: $A, B, C$ and $D$ (linked in that order).
Furthermore, consider two threads $t_1$ and $t_2$.
$t_1$ will execute a transaction to read the entire list from $A$ to $D$ and $t_2$ will concurrently execute a transaction to remove the latter half of the list by removing $C$ and $D$ via a single write to change $B$'s next pointer to \texttt{null}.
Let $t_1$ begin and progress until it reaches $C$.
During its traversal, $t_1$ will pass all validation.
Now we let $t_2$ begin.
$t_2$ will traverse the list until it reaches $B$, successfully validating each read and adding the addresses to its read-set.
$t_2$ will then successfully claim the lock protecting $B$'s next pointer before performing the write.
At commit time, $t_2$ will successfully revalidate its read-set and commit.
Now, $C$ and $D$ have been unlinked from the list but since $t_1$ will not revalidate any of its reads, it will not be aborted.
If $t_2$ then, deallocates $C$ and $D$, $t_1$ could experience a segmentation fault when it continues its traversal.
TL2 also permits this scenario.

In section 2.3 of the TL2 paper, the authors mention the problem of memory reclamation but they only discuss an issue where we want to free objects that exist within the write-set of a transaction.
Their proposed reclamation scheme only works if the issue can be caught during read-set validation, which will never occur in our example.
This problem is also briefly discussed in \cite{blelloch2025tlf} and the authors similarly avoid it via EBR.
Alternatively, we could require read-only transactions perform revaluation or we could utilize an allocator where the internal state is accessed via transactions.
Both alternatives have significant performance drawbacks.
\section{Evaluation}
\label{sec:eval}

We implemented \tmname\ in C++.
The implementation is publicly available \cite{coccimiglio_2025_18099743}\footnote{The code can be found at: \url{https://zenodo.org/records/18099743}}.
The code was compiled with GCC 10.3.0 with an optimization level of -O2.
We compare against several existing STMs which also guarantee opacity.
Specifically TL2 \cite{dice2006transactional}, DCTL \cite{ramalhete2024scaling}, NOrec \cite{dalessandro2010norec} and TinySTM \cite{felber2008dynamic}.
These STMs are described in \Cref{sec:related}.
For TL2 and TinySTM we use the author's public implementations.
For NOrec we used same implementation as in \cite{brown2019cost}.
While DCTL has no public implementation, the authors have a public implementation of a similar algorithm \cite{ramalhete2021efficient} which we could easily adapt into DCTL.
We used the same benchmark as \cite{brown2020non} 
for our evaluation.
We ran all experiments on a single AMD EPYC 7662 processor which has 64 cores and 128 hardware threads. 
All results report the average of 5 trials.
The measurement period of each trial is 20 seconds. 
In this section we focus on a representative sample of our results using an (a,b)-tree.
In \Cref{sec:additional-experiments} we show other data structures including an internal AVL tree, an external binary search tree and a hashmap.

\paragraph{Tunable Parameters}
For TL2 we use the GV4 global clock implementation.
DCTL requires specifying the number of aborts before falling back to a starvation free mode for which we use 100 (the maximum used in \cite{ramalhete2024scaling}).
For \tmname\ we use $\mathcal{K}_{1}$=100, $\mathcal{K}_{2}$=16, $\mathcal{K}_{3}$=28, $\mathcal{S}$=10, $\mathcal{L}$=10 and $\mathcal{P}$=10\%.
For both \tmname\ and DCTL we use the same linear backoff as in \cite{ramalhete2021efficient}.
Unmentioned parameters use default values.

\begin{figure*}[t!]
    \begin{subfigure}{0.02\linewidth}        
        \raisebox{0.5\height}{\rotatebox{90}{0 Updaters}}
    \end{subfigure}
    \begin{subfigure}{0.97\linewidth}
        \begin{subfigure}{0.24\linewidth}
            \centering
            90\% Search, 0\% RQ
            \includegraphics[width=1\linewidth]{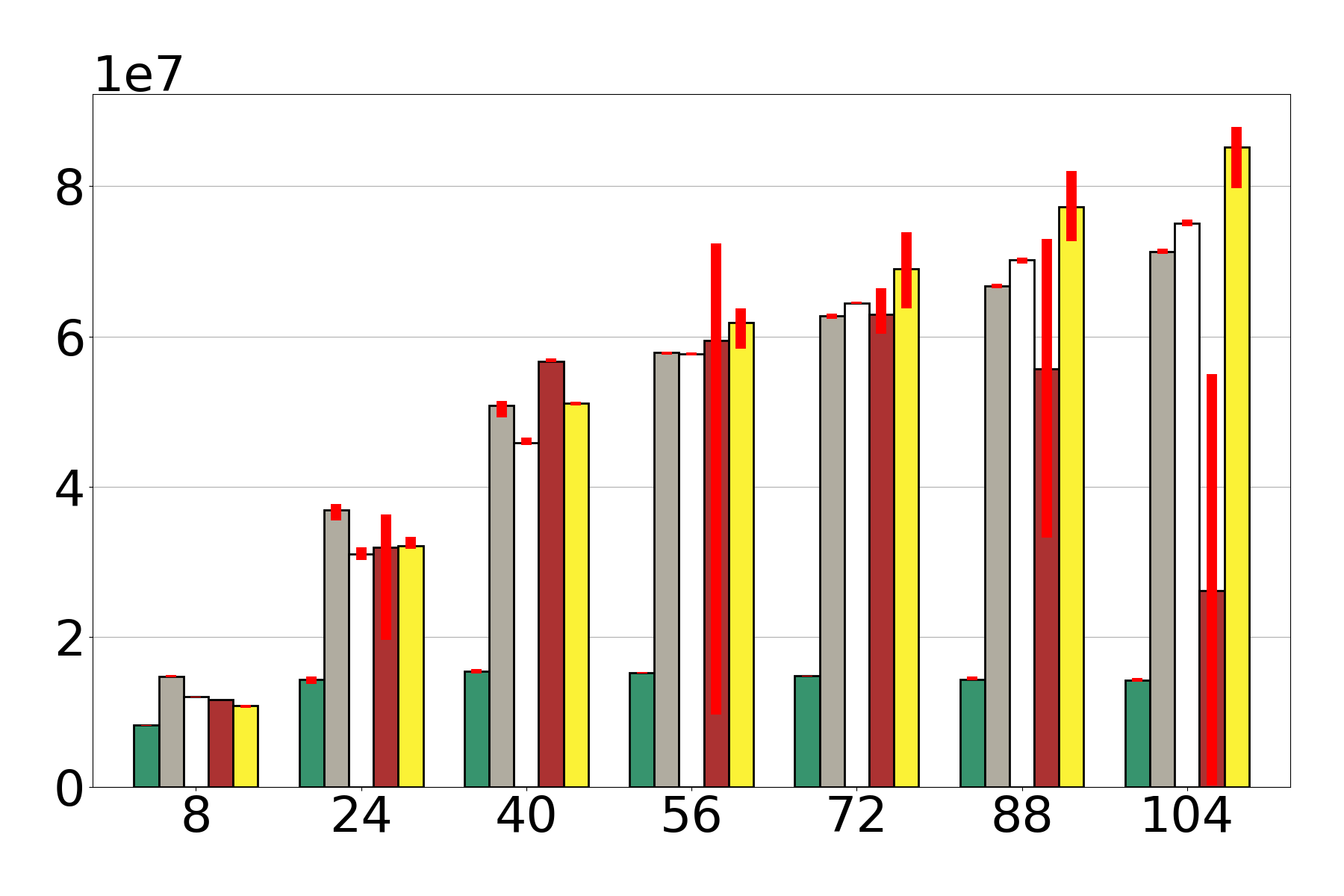}
        \end{subfigure} 
        \begin{subfigure}{0.24\linewidth}
            \centering        
            89.99\% Search, 0.01\% RQ
            \includegraphics[width=1\linewidth]{plots/abtree/node0/10krq/Throughput-IN_DEL=5.0_5.0-RQ=0.01-RQ_SIZE=10000-RQ_T=0-UPDT_T=0-mkey=2000000.png}
        \end{subfigure}
        \begin{subfigure}{0.24\linewidth}
            \centering
            90\% Search, 0\% RQ
            \includegraphics[width=1\linewidth]
            {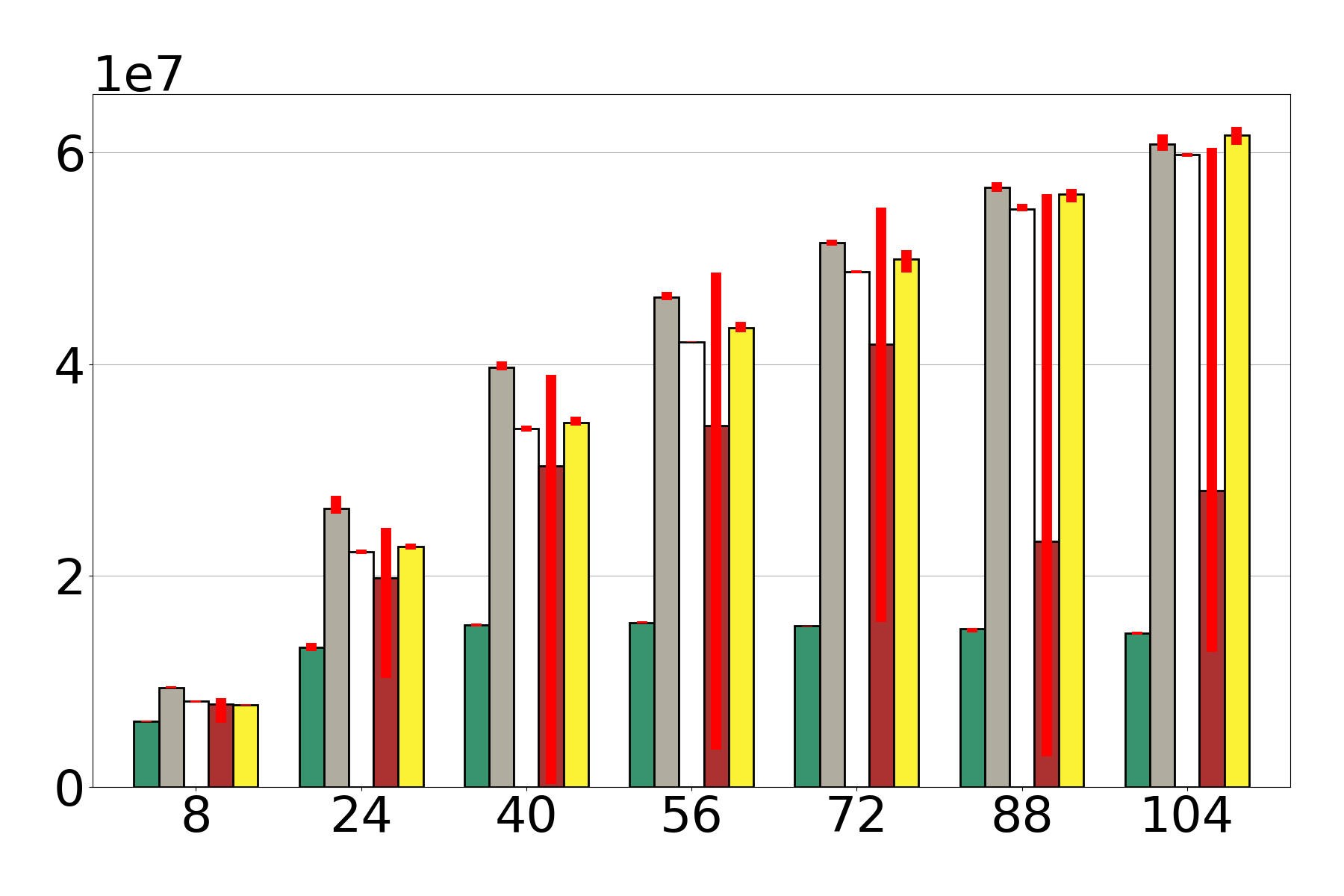}
        \end{subfigure}
        \begin{subfigure}{0.24\linewidth}
            \centering    
            89.99\% Search, 0.01\% RQ
            \includegraphics[width=1\linewidth]
            {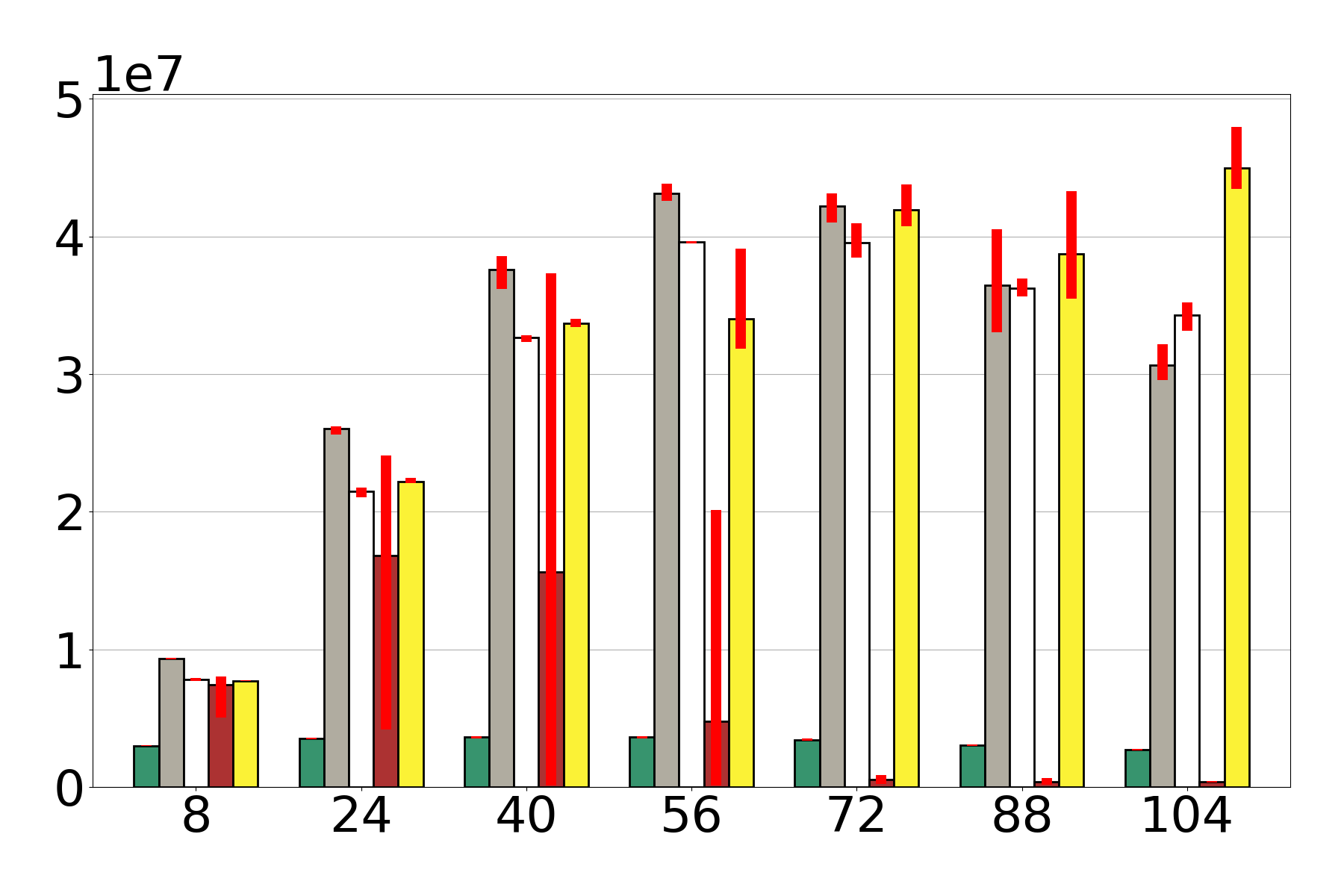}
        \end{subfigure}
    \end{subfigure}
    \begin{subfigure}{0.02\linewidth}
        \raisebox{0.3\height}{\rotatebox{90}{16 Updaters}}
    \end{subfigure}
    \begin{subfigure}{0.97\linewidth}
        \begin{subfigure}{0.24\linewidth}
            \centering
            \includegraphics[width=1\linewidth]{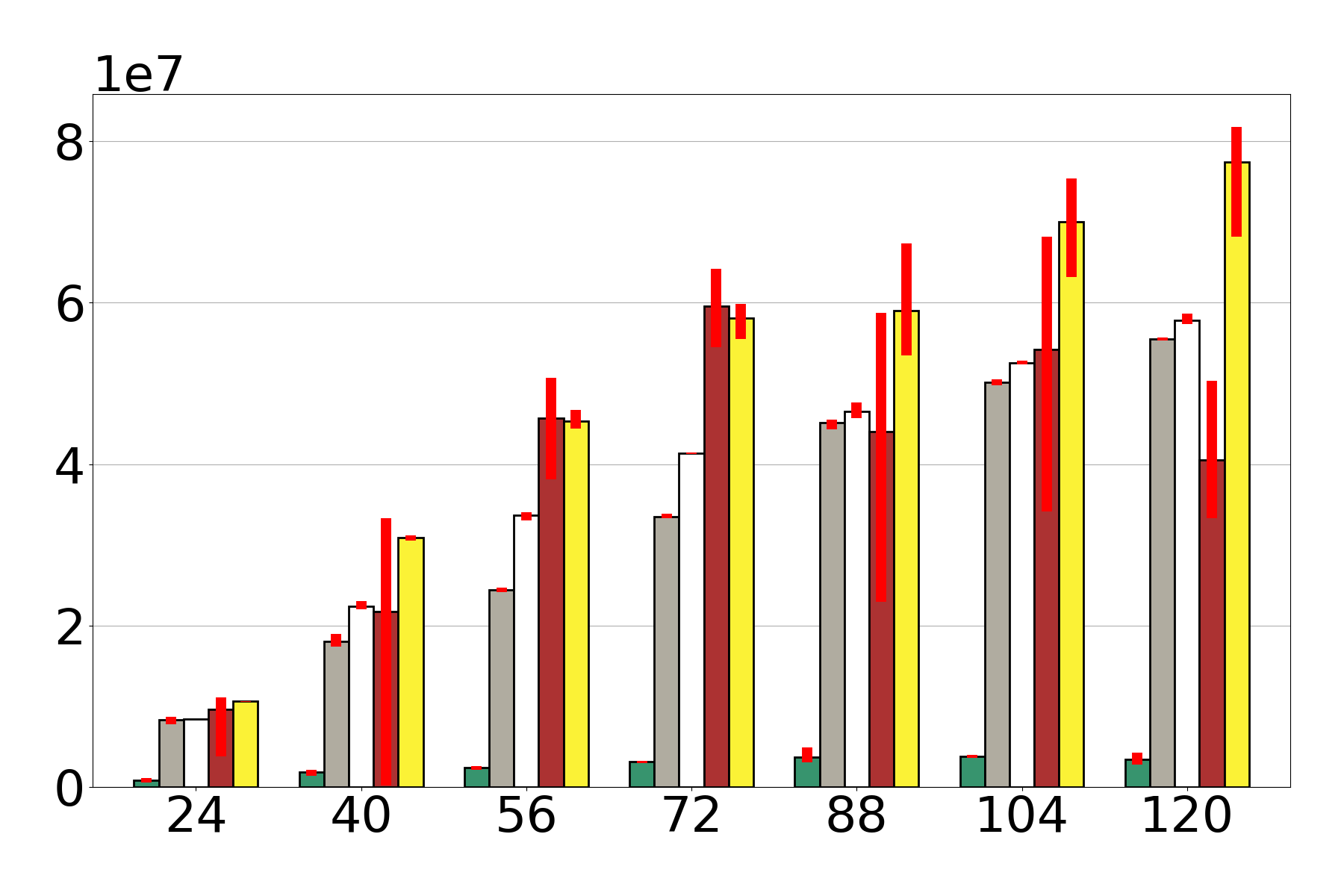}
        \end{subfigure} 
        \begin{subfigure}{0.24\linewidth}
            \centering        
            \includegraphics[width=1\linewidth]{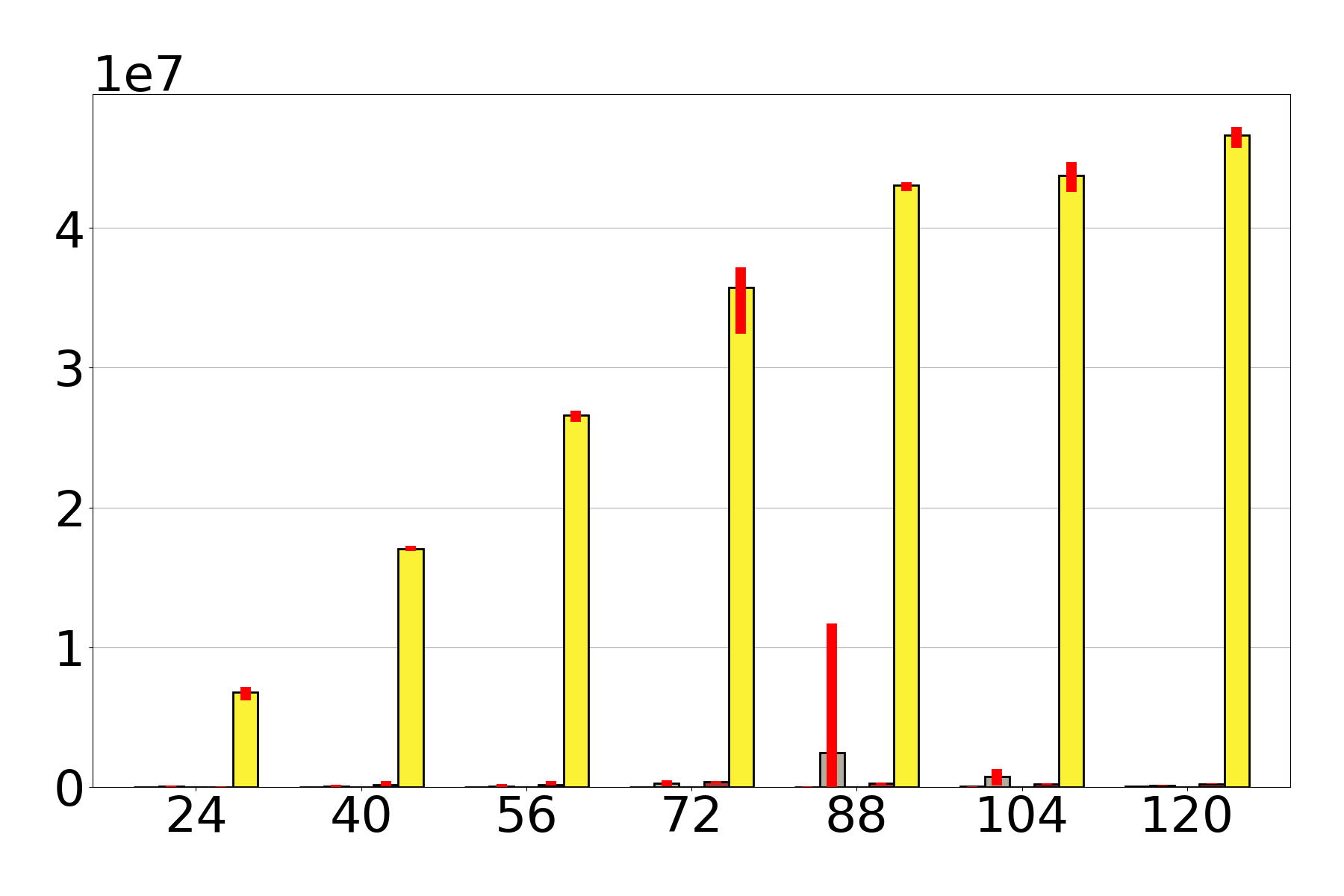}
        \end{subfigure}
        \begin{subfigure}{0.24\linewidth}
            \centering    
            \includegraphics[width=1\linewidth]
            {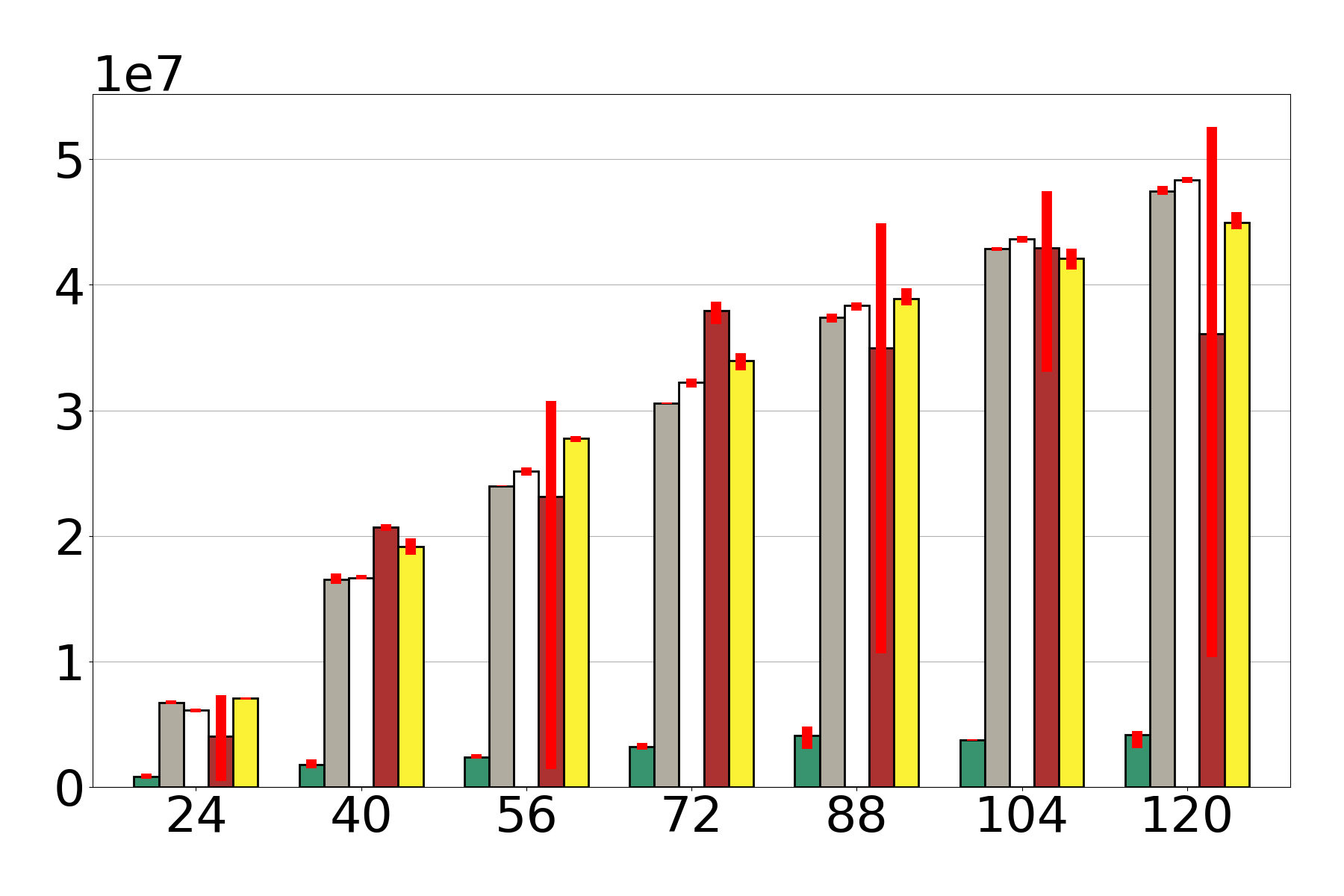}
        \end{subfigure}
        \begin{subfigure}{0.24\linewidth}
            \centering    
            \includegraphics[width=1\linewidth]{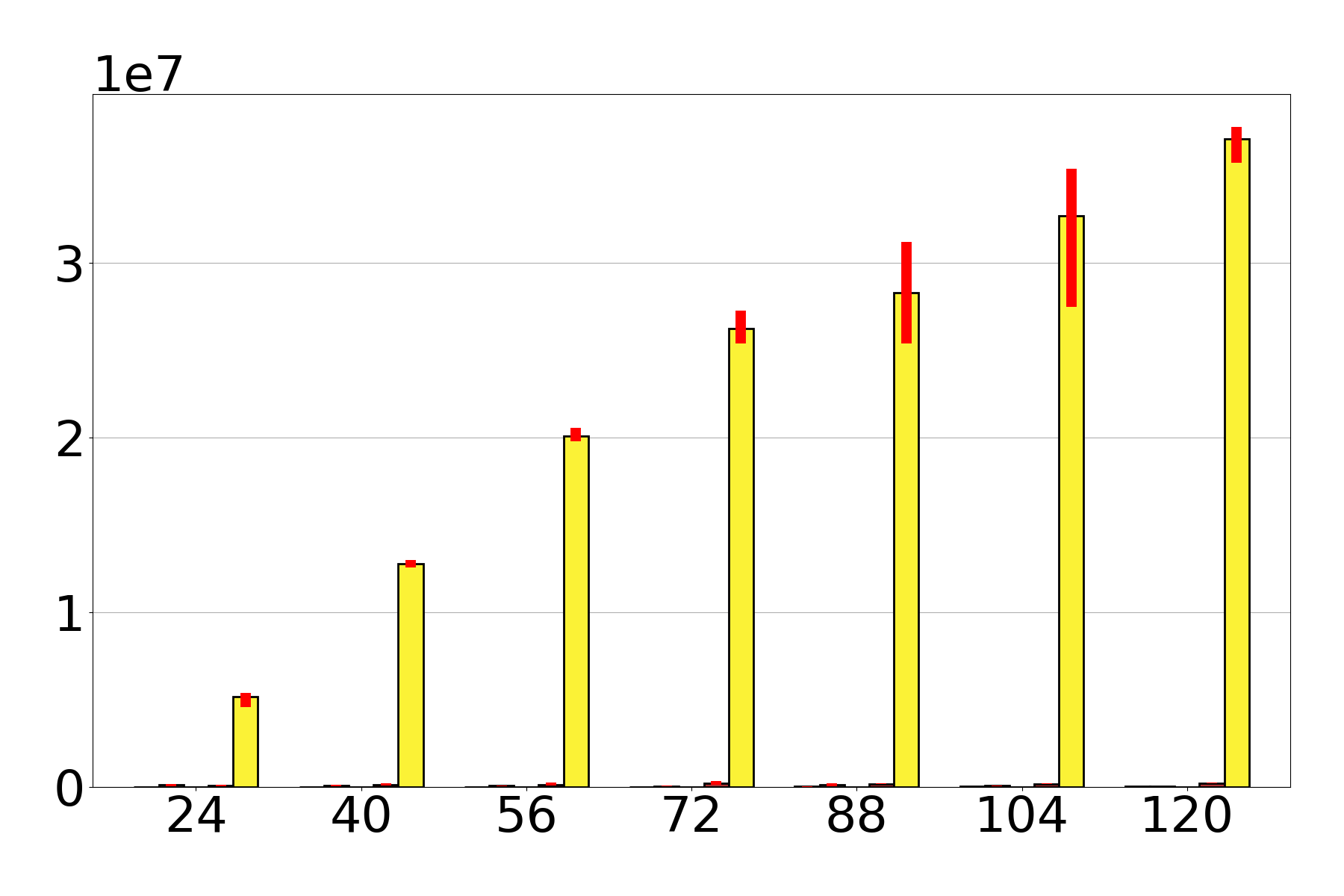}
        \end{subfigure}
    \end{subfigure}

    \begin{subfigure}{0.02\linewidth}        
        \raisebox{0.5\height}{\rotatebox{90}{1 Updater}}
    \end{subfigure}
    \begin{subfigure}{0.97\linewidth}
        \begin{subfigure}{0.24\linewidth}
            \centering
            80\% Search, 0\% RQ
            \includegraphics[width=1\linewidth]{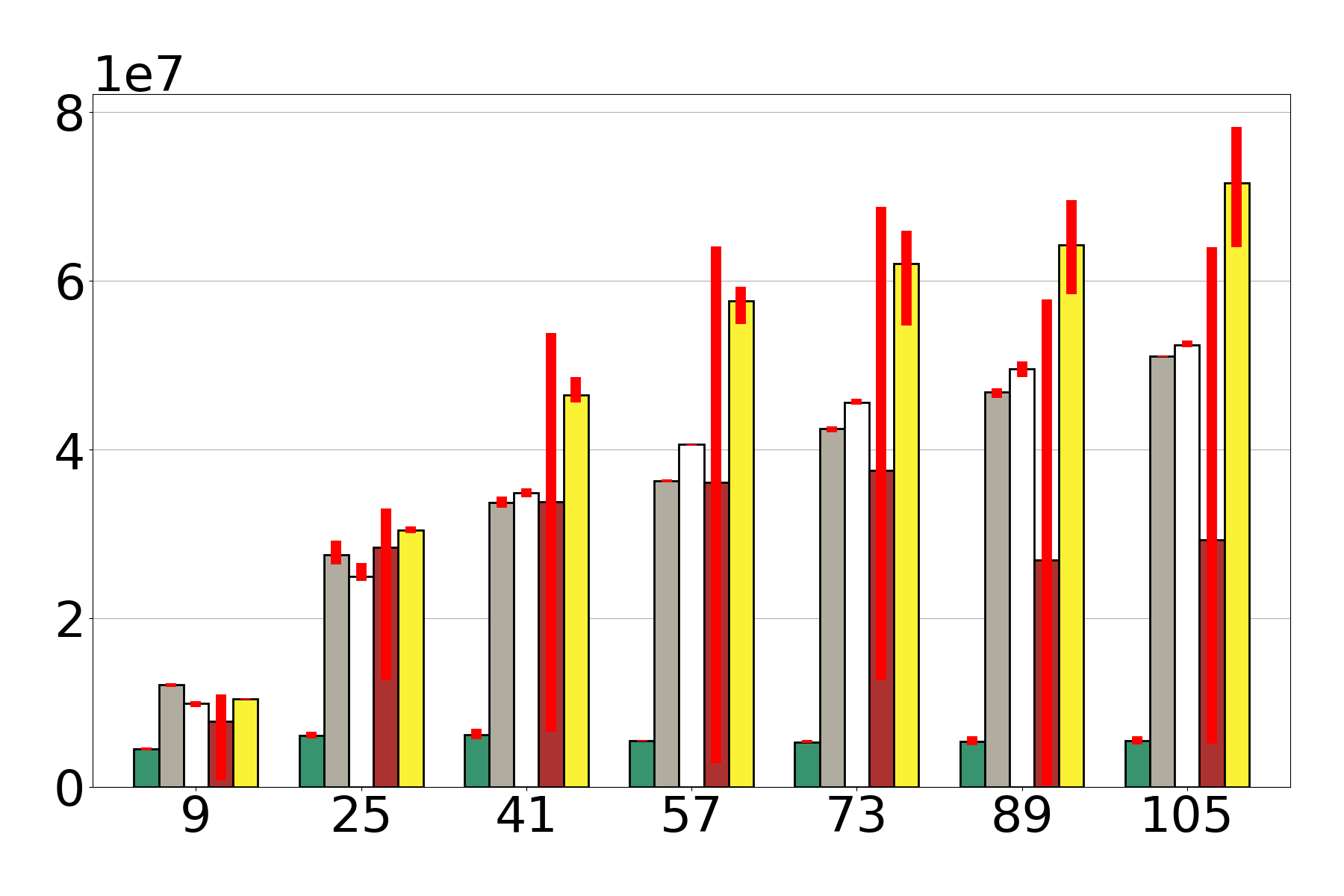}
        \end{subfigure} 
        \begin{subfigure}{0.24\linewidth}
            \centering        
            79.99\% Search, 0.01\% RQ
            \includegraphics[width=1\linewidth]{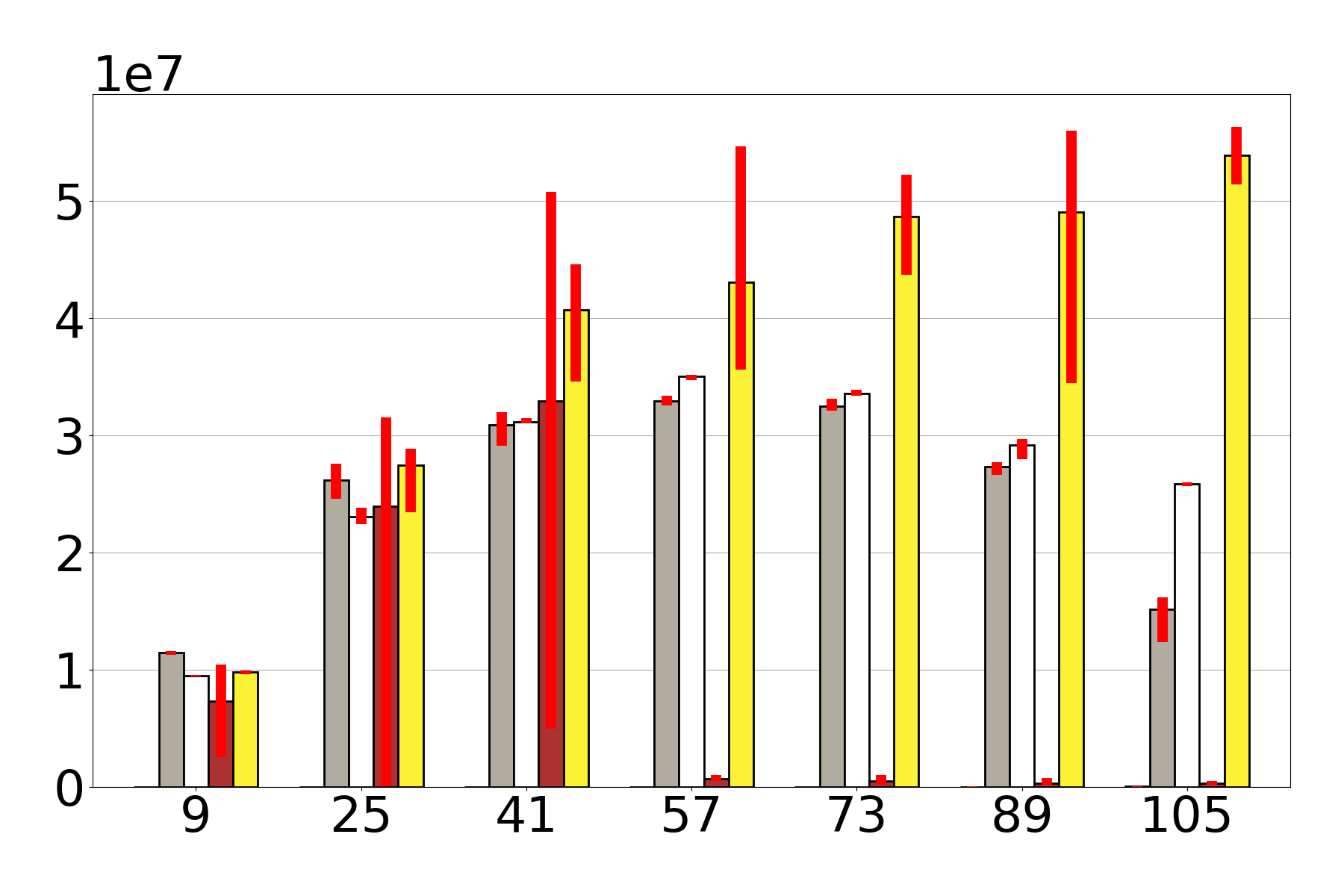}
        \end{subfigure}
        \begin{subfigure}{0.24\linewidth}
            \centering
            80\% Search, 0\% RQ
            \includegraphics[width=1\linewidth]
            {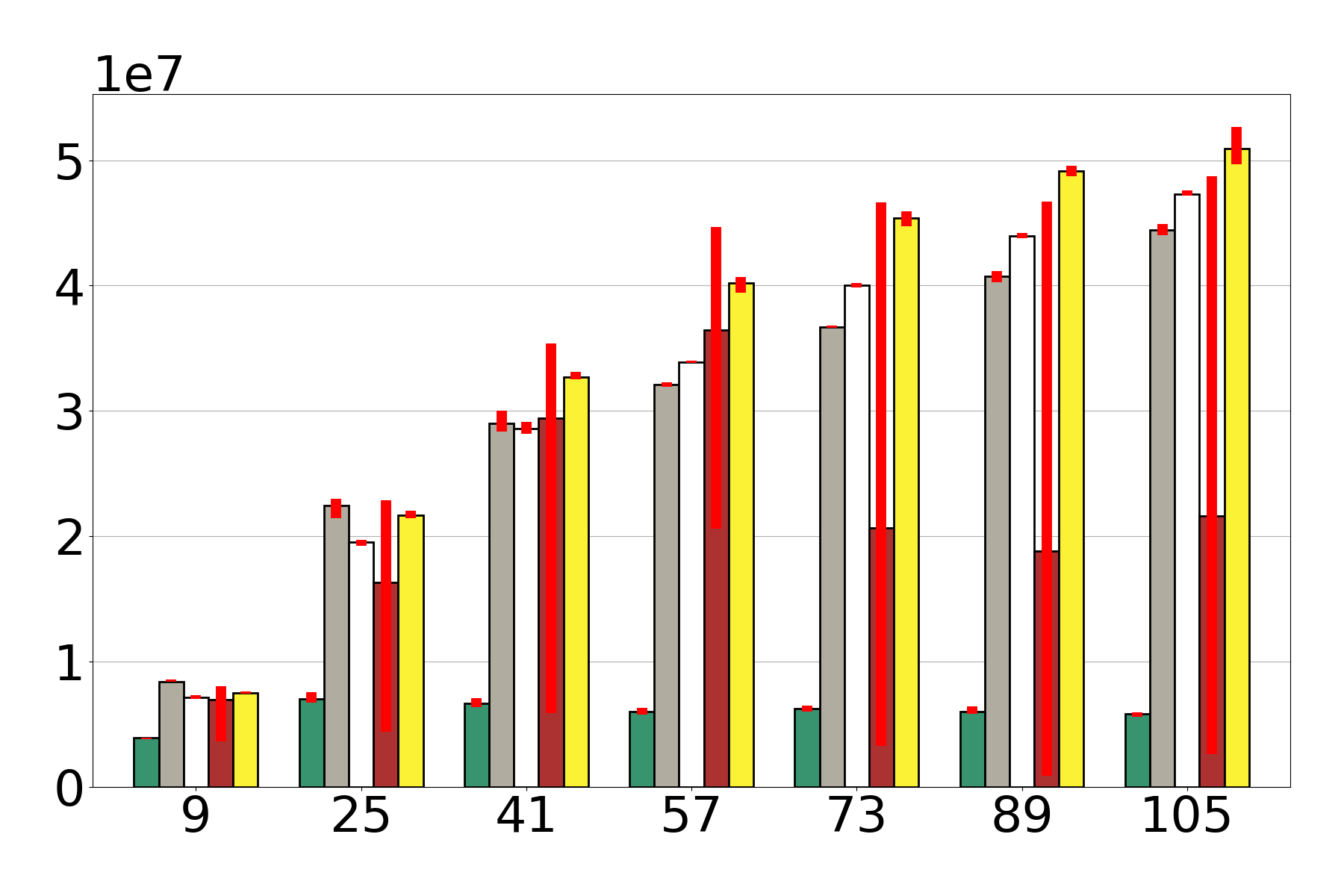}
        \end{subfigure}
        \begin{subfigure}{0.24\linewidth}
            \centering    
            79.99\% Search, 0.01\% RQ
            \includegraphics[width=1\linewidth]
            {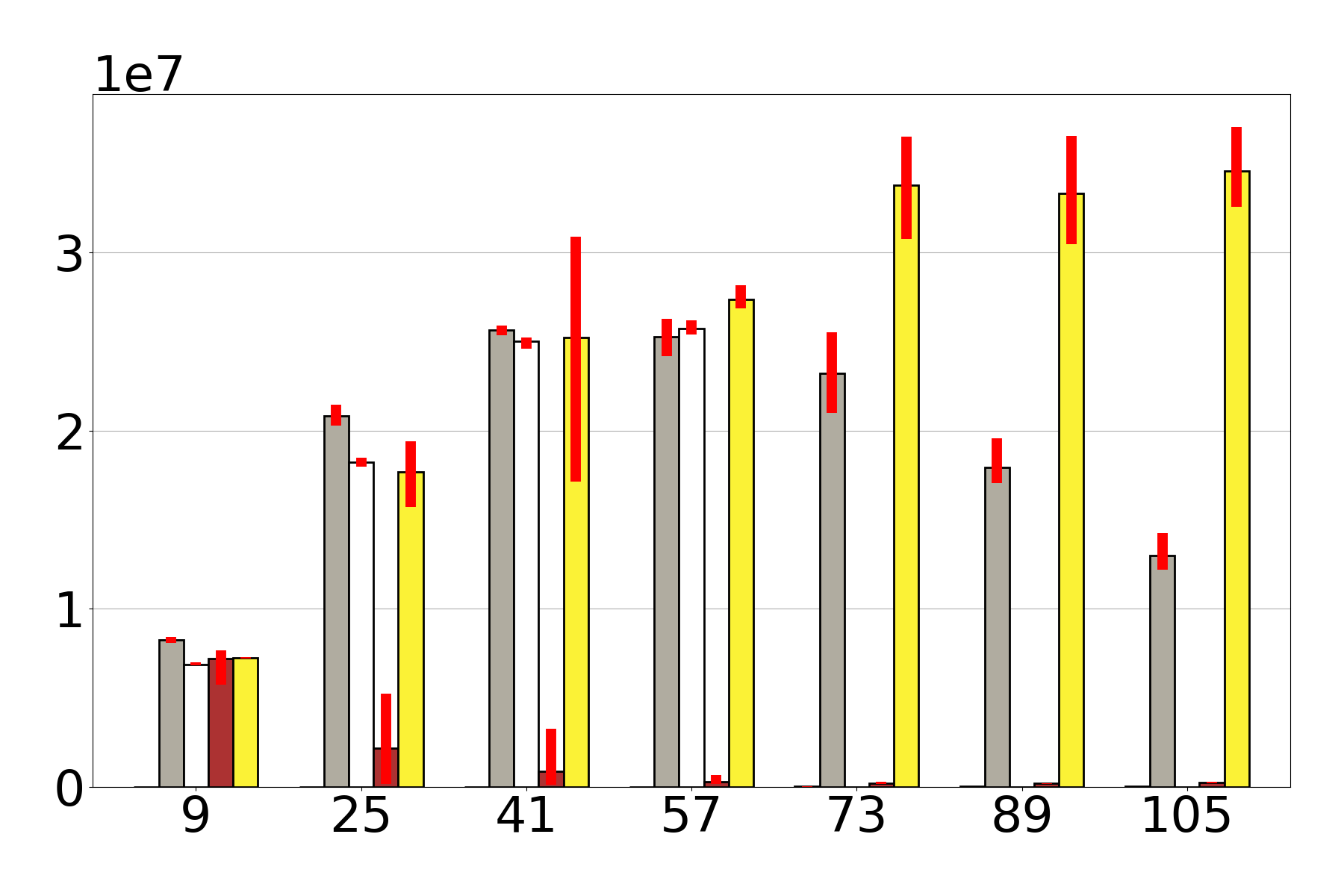}
        \end{subfigure}
    \end{subfigure}
    
    \begin{subfigure}{1.0\linewidth}
        \centering
        \begin{subfigure}{0.45\linewidth}
            \caption{Uniform Key Access Pattern}
        \end{subfigure}
        \hfill
        \begin{subfigure}{0.45\linewidth}
            \caption{Zipfian Key Access Pattern. Exponent = 0.9}
        \end{subfigure}
    \end{subfigure} 
    \begin{subfigure}{1.0\linewidth}
        \centering
        \includegraphics[width=0.4\linewidth]{plots/legend.png}
    \end{subfigure}     
    \vspace{-8mm}
    \caption{\centering Throughput for (a,b)-tree, a=4 b=16, prefilled to 1 million keys. Y-axis: average ops/sec. X-axis: number of threads. 
    In all workloads the remaining percentage of work is equal parts insert and delete.    
    RQ size is 10k (1\% of prefill size).
    }
    \Description{}
    \label{fig:throughput-abtree2}
\end{figure*}

\begin{figure*}[t]
    \includegraphics[width=1\linewidth]{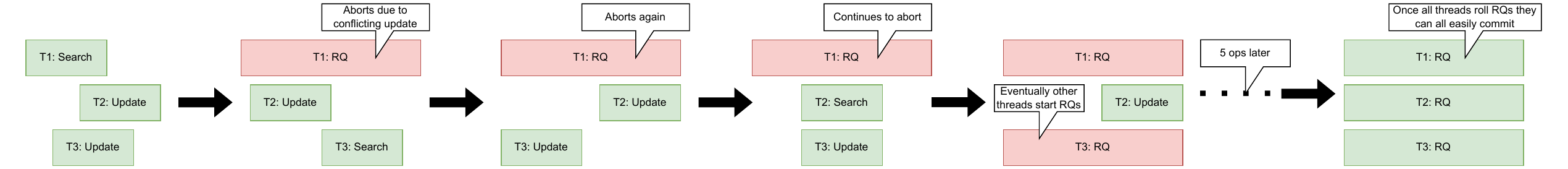}
    \caption{Example of how an algorithm that does not have proper support for range queries can still successfully perform range queries when the tested workload is flawed. 
    This workload is 10\% range queries and 90\% other short operations (search and update).
    In expectation, 1 in 10 operations by any thread will be a range query. 
    No single thread is able to commit a range query (because of concurrent updates) until all of the threads are executing range queries.
    }
    \Description{}
    \label{fig:bad-experimental-setup}
\end{figure*}

\paragraph{Experimental Setup}
We use our own benchmark because prior TM benchmarks are not compelling if we are interested in supporting long queries (see Supplementary A).
The goal of our experimental setup is to test a plausibly realistic workload where an algorithm must be able to perform large range queries (RQs) reliably in the presence of updates in order to obtain high performance overall.
One might think that this can be trivially achieved simply by testing workloads where the distribution of work contains some amount of RQs.
However, if a workload has all threads perform some amount of searches, inserts, deletes and RQs then the performance results can incorrectly prop-up algorithms that do not actually support RQs.
In that scenario, RQs would always abort forever while other threads continue to perform other operations.
A thread attempting a RQ that is repeatedly aborting can effectively wait until other threads that were completing conflicting operations happen to roll the dice and determines that they must now also perform a RQ.
Eventually all (or most) threads will be performing RQs at the same time at which point they can all easily succeed.
To further illustrate the problem consider the example shown in \Cref{fig:bad-experimental-setup}.
Here, the workload is 10\% RQs.
This means that, in expectation, 1 in 10 operations by any thread will be a range query.
In this example, the algorithm does not have proper support for RQs.
As a result, when a thread attempts a RQ, due to conflicts, it aborts forever until all threads happen to also start executing RQs at which point all operations are read-only so the RQs easily commit.

To avoid this issue, we add dedicated \textit{updater} threads.
All operations performed by dedicated updaters will never commit as read-only (unlike regular inserts or deletes which, in expectation, will be read-only half of the time).
Since the updaters never perform read-only operations they can continually interfere with RQs.
This means an algorithm needs to be able to deal with contention in order to successfully perform RQs consistently!
Since these dedicated updater threads can continue to perform operations successfully even if an algorithm has no ability to perform RQs, we do not count the throughput of the dedicated updaters towards the overall throughput (otherwise we would reward those algorithms with no proper RQ support).
Sadly the above still requires careful thought. 
Having some percentage of RQs in our workload may slow down other operations dramatically. 
For example, if an algorithm can perform 1000 RQs/sec, and 10$\%$ of all operations are RQs, then, in expectation, for every RQ done, you can only do 9 other operations.
In our example, this means that the total throughput will be at most 10x the speed of RQs or 10k ops/sec. 
Thus, if we evaluate workloads with even a modest percentage of RQs, we would be measuring almost entirely the cost of performing the RQ. 
For this reason, we evaluate workloads with less than 1\% RQs.

There are some application benchmarks for TM such as STAMP \cite{minh2008stamp}, TPC-C \cite{tpcc} and YCSB \cite{cooper2010benchmarking} however these benchmarks have various problems which make them not interesting if we care about performance for workloads with long running queries in the presence of frequent updates.
STAMP was designed to demonstrate the advantages of first generation TM which had no practical support for large queries.
TPC-C is almost an append-only workload despite defining many transaction types, more than 90\% of transactions are simple, append-only new-order or payment transactions.
In popular research implementations of TPC-C, like DBx1000's, the default index type is a hash table, which can not reasonably be used to implement RQs, and ad-hoc solutions are hard-coded instead \cite{yu2014staring}.
YCSB-A is a rather boring nearly read-only workload.
There are no row insertions or deletes.
One can interpret our benchmark as a generalization of YCSB-A in which we not only update rows, but also insert and delete rows.
Issues with TPC-C and YCSB are discussed in detail in section 7 of \cite{avni2016persistent}.

One can interpret our benchmark as a generalization of YCSB-A in which we not only update rows, but also insert and delete rows.
We also started developing our own more sophisticated TPC-C style application benchmark but we chose to leave that to future work.

\paragraph{Preserving Short Query Performance} 
For workloads that do not feature RQs, versioning is typically not necessary.
In these workloads, we want \tmname\ to maintain the performance of the state of the art STM.
As demonstrated in columns 1 and 3 of \Cref{fig:throughput-abtree2}, \tmname\ achieves this goal achieving throughput comparable or better to all of the STMs that we test.
This is possible because our versioned transactions are only executed on-demand.

\paragraph{Supporting Long Running Queries}
For workloads with some percentage of RQs, versioning is useful.
As seen in the first row of \Cref{fig:throughput-abtree2}, when there are no dedicated updaters to cause conflicts with the RQs all of the TMs can achieve reasonable throughput but \tmname\ still performs better.
On the other hand, in the second row of \Cref{fig:throughput-abtree2} we show that when there are some dedicated updaters and RQs, \tmname\ significantly outperforms the other TMs.
In many cases the throughput of the other TMs is too low to display.
Even worse, it is common for the other TMs to have transactions reach their maximum allowed aborts and quit.

\paragraph{DCTL Starvation Freedom}
One will notice that the variance in DCTL is very high.
In many cases for workloads without RQs, DCTL's maximum throughput is similar or better compared to the other TMs.
However, its minimum performance is often near zero.
This is a result of its irrevocable transactions which must claim locks on reads (which can abort other transactions).
Since only a single transaction can be irrevocable at any time, this can lead to many or all transactions waiting to execute on the irrevocable path. 


\paragraph{Time Varying Workloads}
To understand the benefit of our different TM modes, we experiment with time varying workloads.
Specifically, we split each trial into 4 intervals of 5 seconds and we change the workload in each interval.
In these experiments we measure throughput over time using an additional background thread which captures the throughput every 200ms.
We also separately show implementations of \tmname\ where we disable mode switching and force the initial mode to Mode~Q or Mode~U respectively.

\begin{figure}
    \centering
    \begin{subfigure}{1.0\linewidth}
        \centering
        \includegraphics[width=1.0\linewidth]{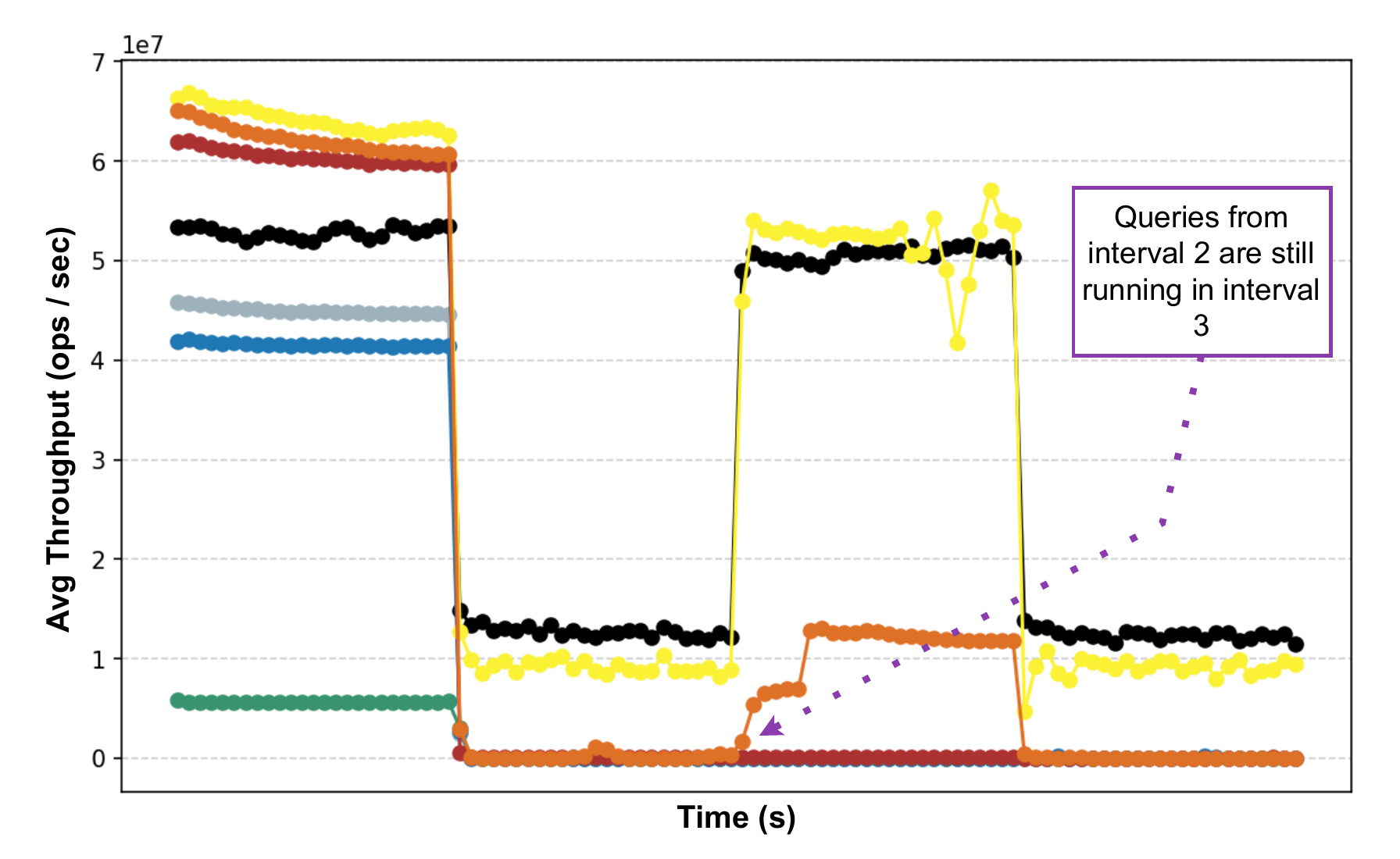}
    \end{subfigure} 
    \begin{subfigure}{1.0\linewidth}
        \centering
        \includegraphics[width=1.0\linewidth]{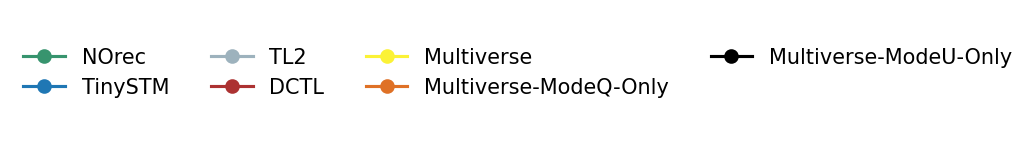}
    \end{subfigure}     
    \vspace{-8mm}
    \caption{
    Throughput over time for an (a,b)-tree using 64 worker threads for a time-varying workload with 4 intervals. Intervals 1 and 3 have no RQs and no dedicated updaters. Interval 2 and 4 have 0.01\% RQs with a RQ size of 100k and 4 dedicated updaters. All intervals have 10\% insert, 10\% deletes and the remaining work is searches (point queries). 
    }
    \Description{}
    \label{fig:intervals}
\end{figure}

We use a workload with 2 repeated intervals where interval 1 and 3 have no RQs and no dedicated updaters and interval 2 and 4 have 0.01\% RQs with a large RQ size of 100k and 4 dedicated updaters. 
All 4 intervals have 10\% insert and 10\% deletes with the remaining work being searches (point queries).
\Cref{fig:intervals} shows the results of this experiment.
In interval 1 our Mode~Q only implementation performs noticeably better than our Mode~U only implementation and the opposite is true once we introduce RQs.
Our implementation with mode switching enabled achieves performance comparable to the better of the mode restricted implementations in each interval despite us not investing much effort into finely tuning our parameters for mode switching.

One might expect all TMs to increase in throughput when RQs are removed from the workload but this does not occur.
When we change intervals, newly generated work conforms to the new workload however, any queries that started in the previous interval must finish before the thread can continue under the new workload (this matches the reality of varying workloads in the real world).
In other words, a thread running a large RQ will continue to attempt the RQ until it succeeds.
This means that algorithms that do not support the larger RQs are likely to be stuck retrying those RQs forever even after an interval change.

\begin{figure}[t!]
    \begin{subfigure}{0.02\linewidth}        
        \raisebox{0.5\height}{\rotatebox{90}{0 Updaters}}
    \end{subfigure}
    \begin{subfigure}{0.97\linewidth}
        \begin{subfigure}{0.49\linewidth}
            \centering
            90\% Search, 0\% RQ
            \includegraphics[width=1\linewidth]{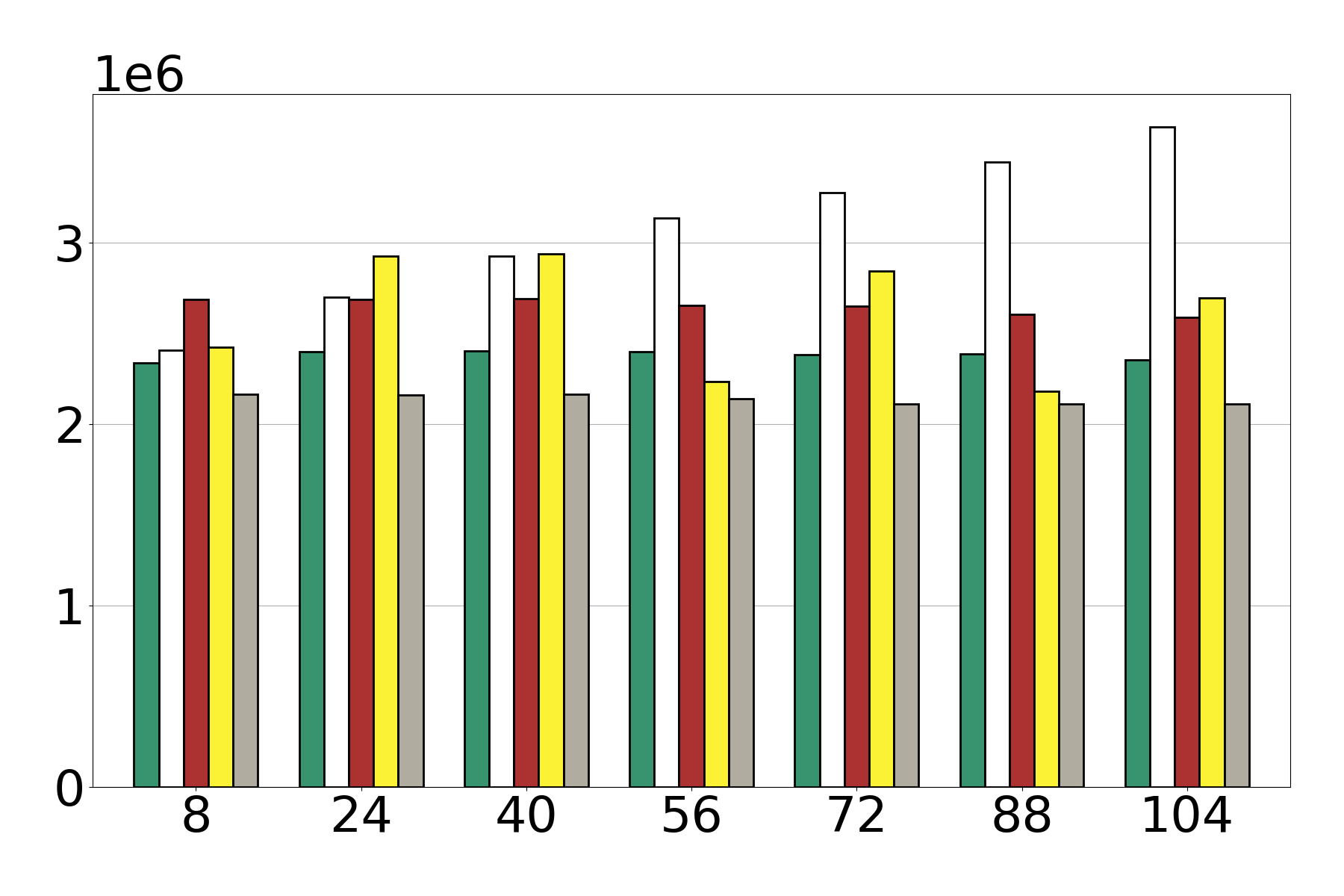}
        \end{subfigure} 
        \begin{subfigure}{0.49\linewidth}
            \centering        
            89.99\% Search, 0.01\% RQ
            \includegraphics[width=1\linewidth]{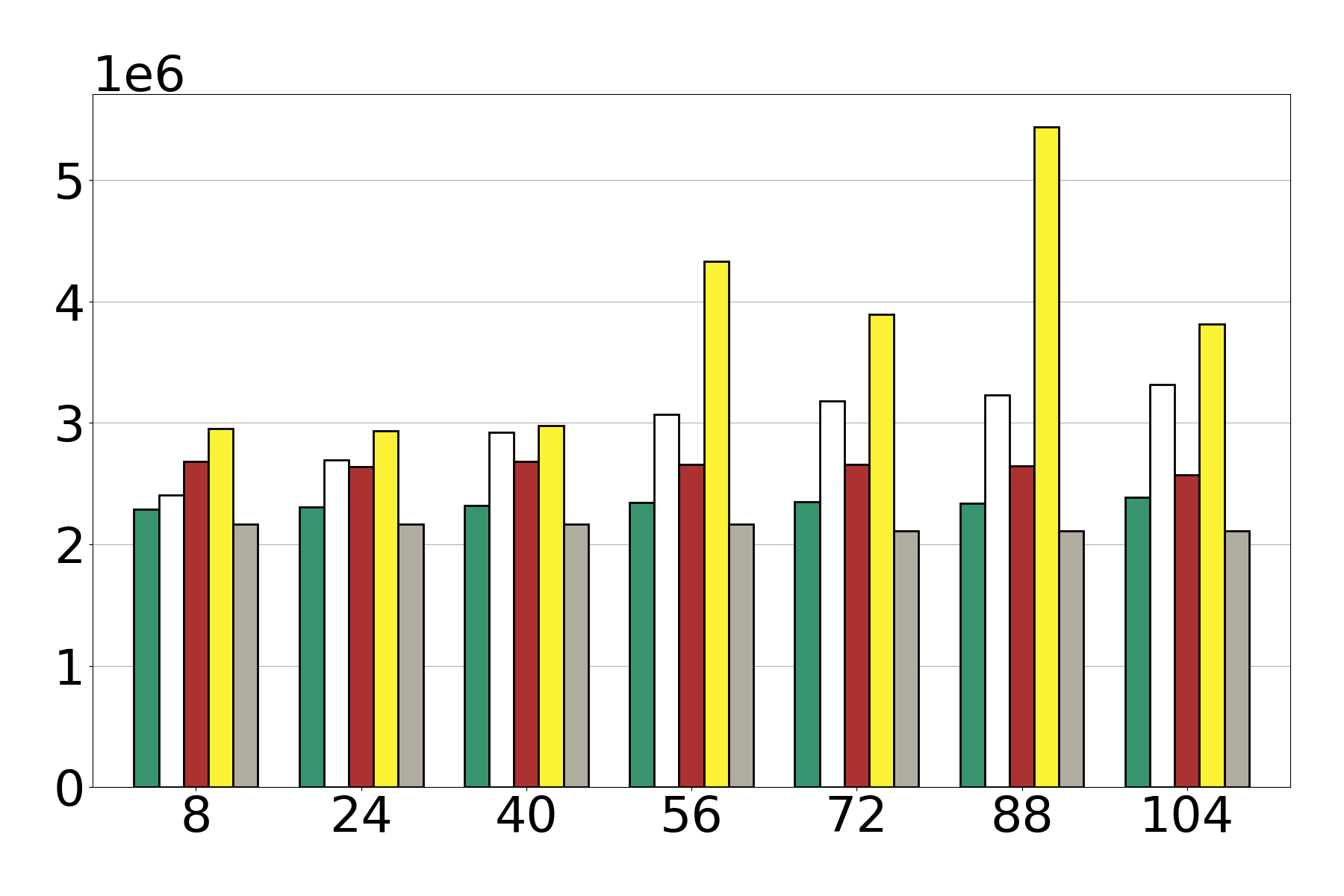}
        \end{subfigure}
    \end{subfigure}
    \begin{subfigure}{1.0\linewidth}
        \centering
        \includegraphics[width=1.0\linewidth]{plots/legend.png}
    \end{subfigure}     
    \vspace{-8mm}
    \caption{
    Maximum memory usage for the same (a,b)-tree from \Cref{fig:throughput-abtree2} using a uniform key access pattern. Y-axis: max resident memory in KB. X-axis: number of threads.
    }
    \Description{}
    \label{fig:memory-abtree}
\end{figure}

\paragraph{Memory Usage}
\Cref{fig:memory-abtree} shows the maximum memory usage for the same (a,b)-tree from row one of \Cref{fig:throughput-abtree2}.
In general, it is expected that multiversioned algorithms will require more memory compared to unversioned algorithms.
However, as a result of our dynamic multiversioning approach, we only pay the cost of additional memory requirements when multiple versions are actually needed.
For workloads without RQs the maximum memory used by \tmname\ is comparable to (and sometimes lower than) DCTL.

\begin{figure}[t!]
    \begin{subfigure}{0.02\linewidth}        
        \raisebox{0.5\height}{\rotatebox{90}{16 Updaters}}
    \end{subfigure}
    \begin{subfigure}{0.97\linewidth}
        \begin{subfigure}{0.49\linewidth}
            \centering
            90\% Search, 0\% RQ
            \includegraphics[width=1\linewidth]{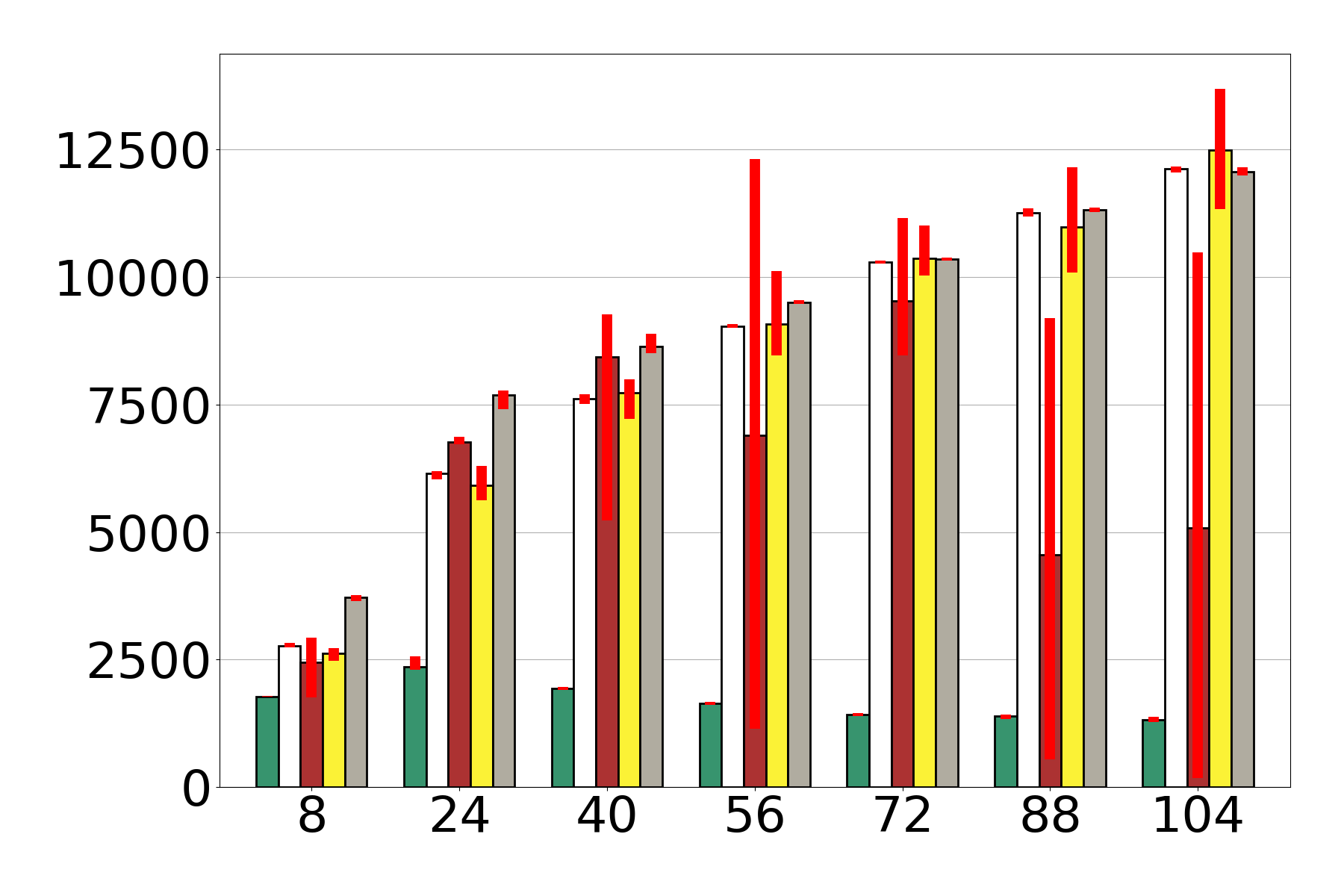}
        \end{subfigure} 
        \begin{subfigure}{0.49\linewidth}
            \centering        
            89.99\% Search, 0.01\% RQ
            \includegraphics[width=1\linewidth]{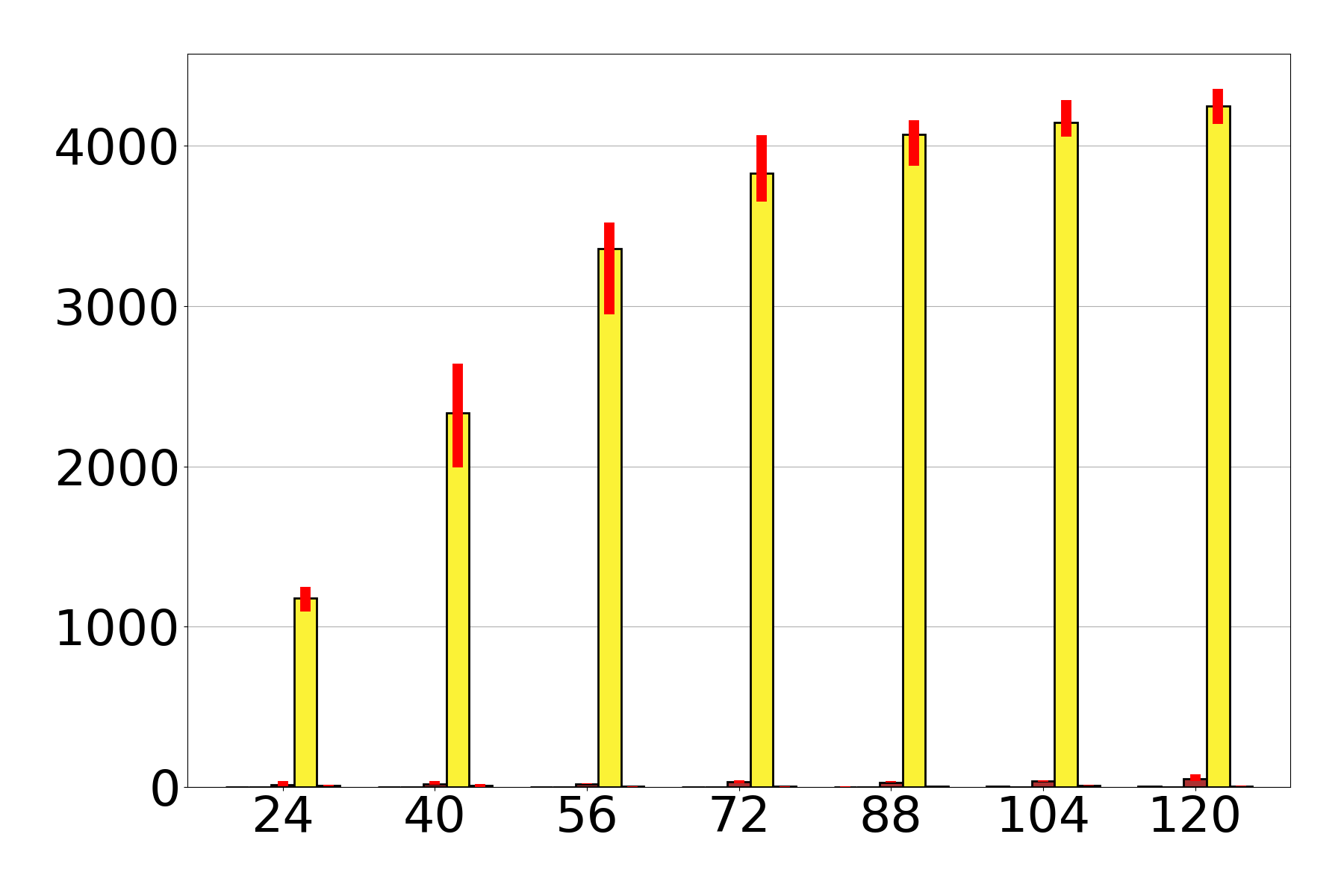}
        \end{subfigure}
    \end{subfigure}
    \begin{subfigure}{1.0\linewidth}
        \centering
        \includegraphics[width=1.0\linewidth]{plots/legend.png}
    \end{subfigure}     
    \vspace{-8mm}
    \caption{
    Throughput per joule consumed by the CPU package for the same (a,b)-tree workload from \Cref{fig:throughput-abtree2} using a uniform key access pattern. Y-axis: average ops/sec per joule. X-axis: number of threads.
    }
    \Description{}
    \label{fig:power-pkg-abtree}
\end{figure}

\paragraph{Power Consumption}
Power Consumption is another interesting metric to consider.
We compare the energy efficiency of the TMs via \texttt{perf} by measuring the power consumption of the CPU package in joules (specifically measuring the \texttt{energy-pkg} hardware event).
Note that it is not possible to measure the energy consumption of a specific program or core \cite{khan2018rapl, raffin2024dissecting}.
\Cref{fig:power-pkg-abtree} shows average throughput per joule of energy for the same (a,b)-tree from row two of \Cref{fig:throughput-abtree2}.
\tmname\ is able to leverage increased memory usage to efficiently support RQs resulting in up to 50x improved energy efficiency compared to the next best TM.

\section{Related Work}
\label{sec:related}
%
Transactional Locking II (TL2) \cite{dice2006transactional} is one of the most well known STMs. 
TL2 is an opaque unversioned STM that relies on a global clock and per-address versioned locks.
By default, TL2 uses \textit{buffered writes} and deferred \textit{commit time locking}.
Transactions that write increment the global clock at commit time after locking the write set.
Numerous optimizations of TL2 exist, especially new approaches for managing the global clock.
%
The most recent of these optimizations was presented in a new STM, Deferred Clock Transactional Locking (DCTL), \cite{ramalhete2024scaling}.
DCTL uses encounter time locking and increments the global clock only on aborts.
It has a starvation free mode where a single transaction can become irrevocable.
Multiple non-irrevocable transactions can execute concurrently with a single irrevocable transaction.
%
%
TLRW \cite{dice2010tlrw} is an STM that relies on byte-locks which are a form of read-write locks.
TLRW is unversioned but it does support irrevocable transactions.
It has been shown to be competitive with TL2 for single chip machines in some workloads.
Unfortunately TLRW has no public implementation.
%
%
TinySTM \cite{felber2008dynamic} is another well known opaque unversioned STM.
Like TL2, it relies on a global clock and per-address versioned locks.
TinySTM uses encounter time locking.
%
No Ownership Records (NORec) \cite{dalessandro2010norec} is an unversioned STM that does not use per-address versioned locks.
It uses a single global sequence lock, commit time locking and value based validation.
These STMs do not have proper support for long running read-only transactions.

Verlib \cite{blelloch2024verlib} is the state of the art MVCC mechanism.
Notably, Verlib can sometimes avoid adding indirection when updating addresses.
Verlib was incorporated into a multiversion STM optimized for optimistic data structures in \cite{blelloch2025tlf}.
This STM, which the authors simply refer to as \textit{MV-STM}, was not the main focus of the work, and the authors did discuss whether or not MV-STM satisfies opacity.
After some discussions with the authors, we believe that MV-STM would satisfy opacity.
Unlike \tmname, MV-STM collocates locks with the underlying user data.
This makes MV-STM more similar to object based STMs.
Furthermore, when the transactional variables are pointers, MV-STM steals some bits from the pointer.
It is known that for some data structures, collocating locks with user data can improve performance \cite{coccimiglio2025persistent}.
In this work we explicitly designed \tmname\ to avoid changing the memory layout of the underlying user data. 
We did experiment with storing the bloom filters alongside the locks in the lock table.
In some cases this improved the performance of versioned transactions, however, the resulting cache effects typically had a significant negative impact on the performance of unversioned transactions.

Selective Multi-Versioning (SMV) \cite{perelman2011smv} is an opaque STM that uses an approach somewhat similar to Verlib.
We would have liked to compare against SMV \cite{perelman2011smv}, but it is implemented in Java and heavily relies on garbage collection. 
To our knowledge there is no public non-Java implementation and it is not straightforward to implement it in C++, since we would need to solve the memory management problem.
SMV uses commit time locking.
In SMV updates always add new versions but old versions can be quickly removed. 
Active transactions in SMV append a descriptor to a global list.
A descriptor has a timestamp and keeps references to the old versions of objects that the transaction modified to prevent their reclamation by the garbage collector.
Unlike Verlib which uses a numeric global timestamp, SMV uses the list of descriptors to serve as a global clock which requires different synchronization.
Other non-opaque multiversion STMs have also been proposed \cite{riegel2006snapshot, lu2013generic}.

\section{Conclusion}
In this work we presented \tmname, a new opaque STM that combines the best of both unversioned STM and MVCC.
\tmname\ features dynamic multiversioning and uses multiple TM modes which adapt the behavior of the TM to fit the needs of the workload.
Our experimental evaluation of \tmname\ demonstrated that our TM can match or exceed the performance of existing unversioned STMs even in common case workloads that do not feature long running read-only transactions while still significantly outperforming them for workloads that do feature long running reads.

\begin{acks}
We thank Naama Ben-David for her involvement in numerous discussions during the early development of the algorithm.
Naama's ideas inspired various initial versions of \tmname.
Her insights and feedback were instrumental in overcoming several issues and helped to guide the evolution of the algorithm.

This work was supported by: the Natural Sciences and Engineering Research Council of Canada (NSERC) Collaborative Research and Development grant: CRDPJ 539431-19, the Canada Foundation for Innovation John R. Evans Leaders Fund with equal support from the Ontario Research Fund CFI Leaders Opportunity Fund: 38512, NSERC Discovery Launch Supplement: DGECR-2019-00048, NSERC Discovery Program grant: RGPIN-2019-04227, and the University of Waterloo.
\end{acks}

\balance
\bibliographystyle{ACM-Reference-Format}
\bibliography{refs}

\appendix
\clearpage
\setcounter{section}{0}
\setcounter{page}{1}


{ \noindent\LARGE\bfseries Appendix }

\section{Additional Experiments}
\label{sec:additional-experiments}
Here we include some additional experimental results.
Recall that we do not include the throughput of the dedicated updater threads in any of our results.
We include additional results for the same (a,b)-tree as shown in the main paper, as well as results for an internal AVL tree, an external binary search tree and a hashmap.
For the (a,b)-tree and external binary search tree we use the same implementation as \cite{brown2019cost}.
For the interval AVL tree, we utilize the implementation from \cite{brown2022pathcas}. 
For the hashmap we utilize the implementation from \cite{ramalhete2021efficient}.
We do not use an order-preserving hash function for the hashmap.
This makes typical RQs not meaningful.
For this reason, in the hashmap experiments, instead of RQs we instead execute size queries (SQs) which are atomic size operations that count the number of keys in the map.
The hashmap has a fixed 1 million buckets, where each bucket is a linked list and we prefill to only 100k keys.
For the hashmap we always use at least 1 dedicated updater since the update operations in the hashmap are much simpler compared to updates for the tree data structures and we still want to have some contention via updates.

\begin{figure*}[t!]
    \begin{subfigure}{0.02\linewidth}        
        \raisebox{0.5\height}{\rotatebox{90}{0 Updaters}}
    \end{subfigure}
    \begin{subfigure}{0.97\linewidth}
        \begin{subfigure}{0.32\linewidth}
            \centering
            90\% Search, 0\% RQ
            \includegraphics[width=1\linewidth]{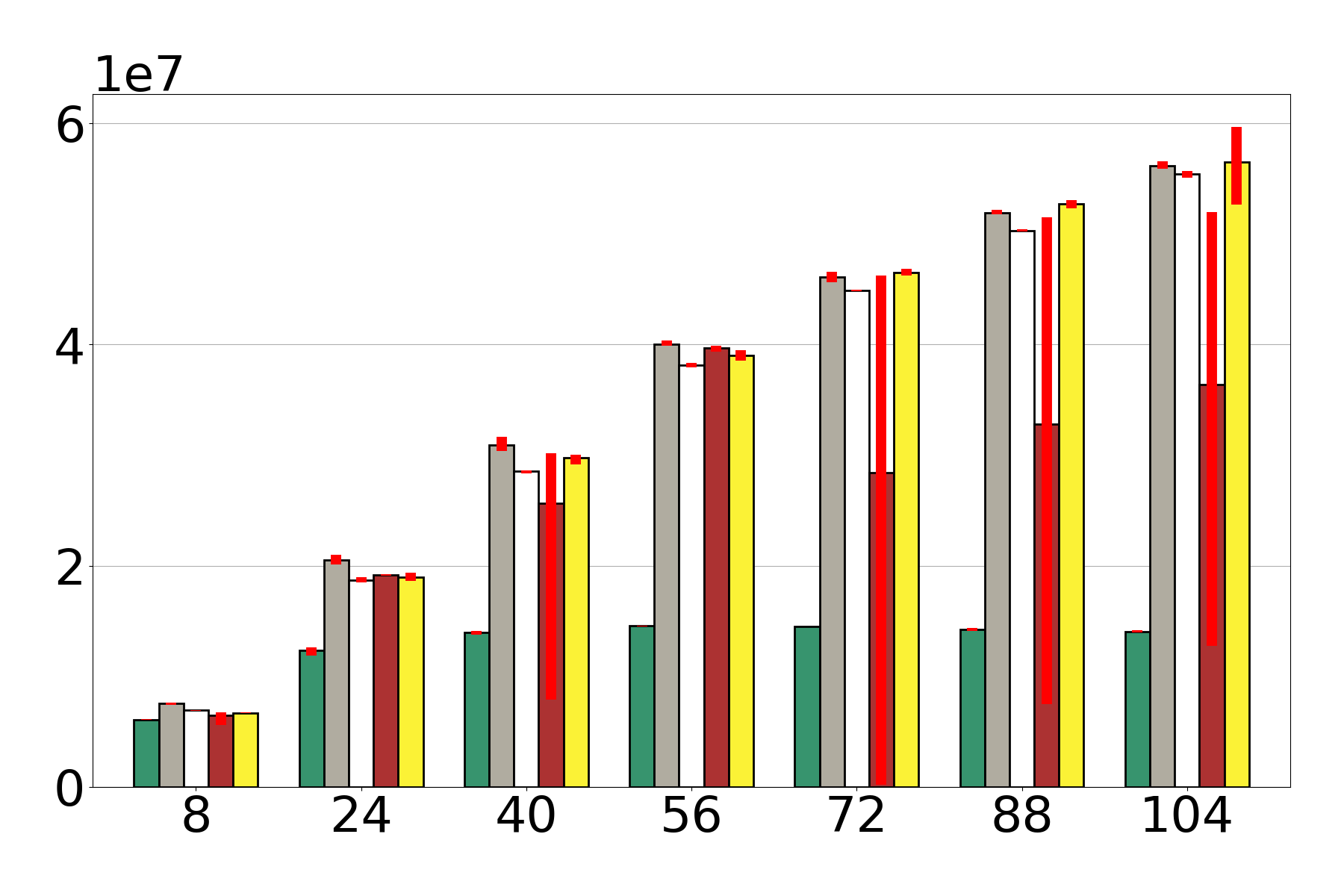}
        \end{subfigure} 
        \begin{subfigure}{0.32\linewidth}
            \centering        
            89.9\% Search, 0.1\% RQ
            \includegraphics[width=1\linewidth]{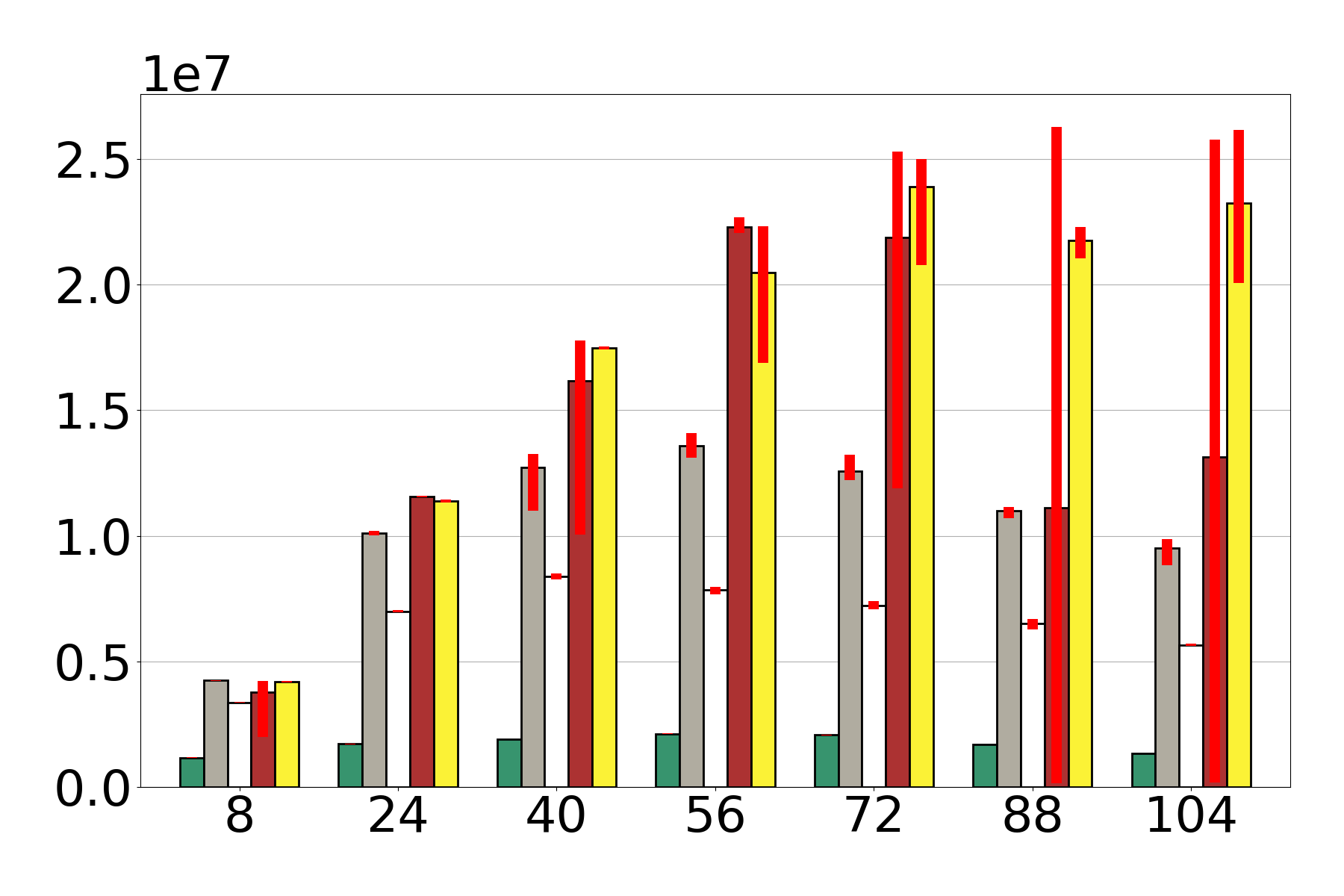}
        \end{subfigure}
        \begin{subfigure}{0.32\linewidth}
            \centering    
            89.99\% Search, 0.01\% RQ
            \includegraphics[width=1\linewidth]{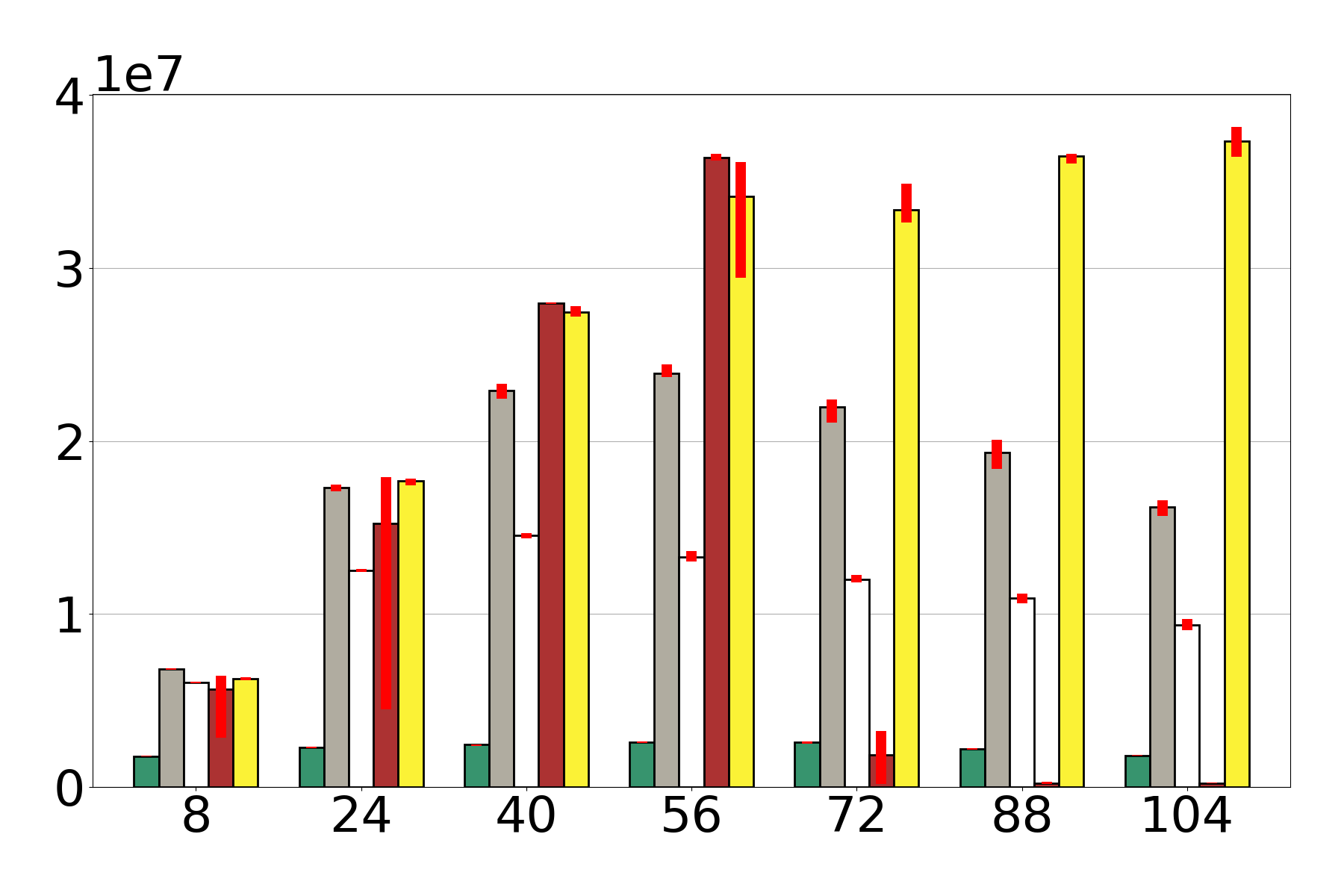}
        \end{subfigure}
    \end{subfigure}
    \begin{subfigure}{0.02\linewidth}
        \raisebox{0.5\height}{\rotatebox{90}{16 Updaters}}
    \end{subfigure}
    \begin{subfigure}{0.97\linewidth}
        \begin{subfigure}{0.32\linewidth}
            \centering
            \includegraphics[width=1\linewidth]{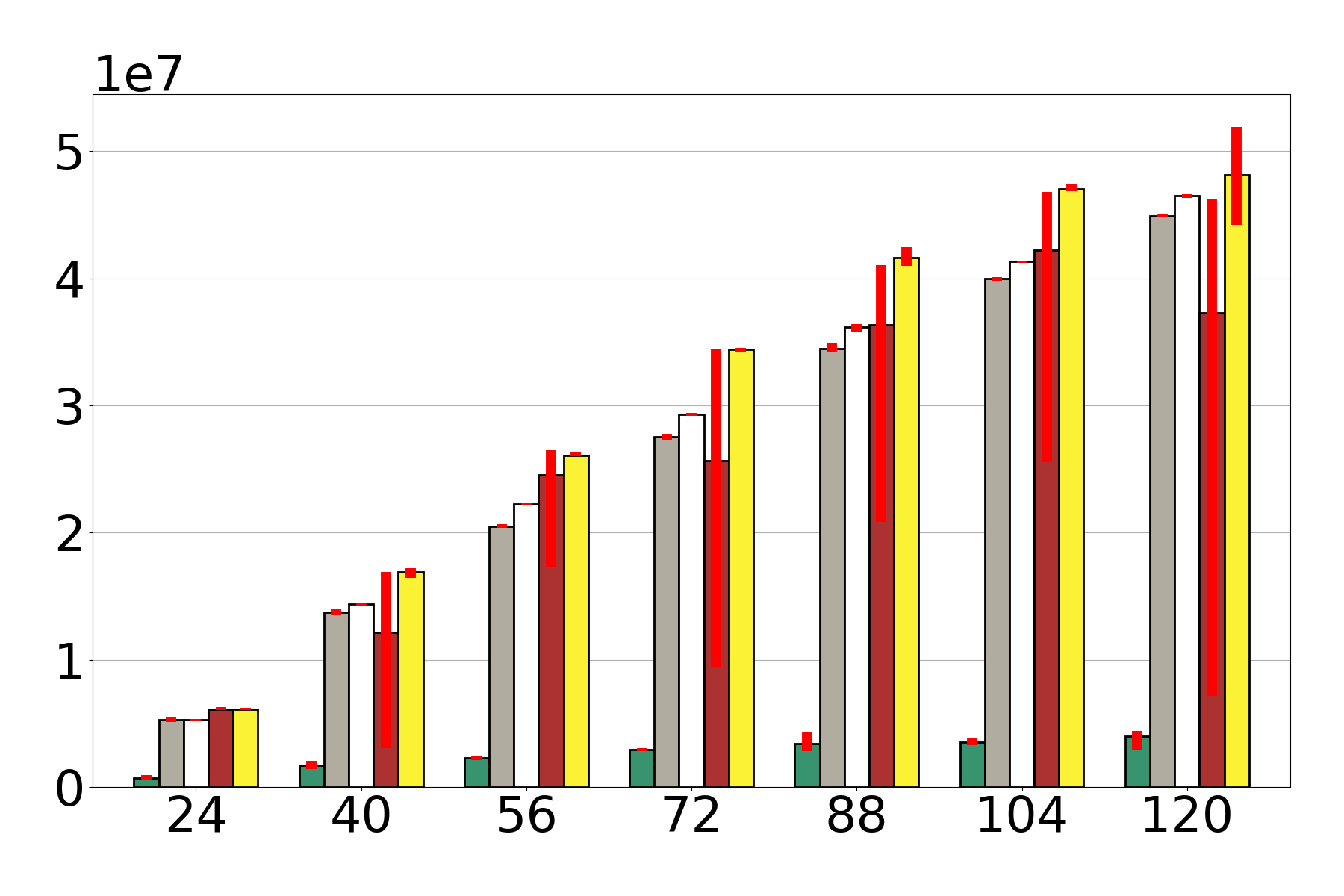}
        \end{subfigure} 
        \begin{subfigure}{0.32\linewidth}
            \centering        
            \includegraphics[width=1\linewidth]{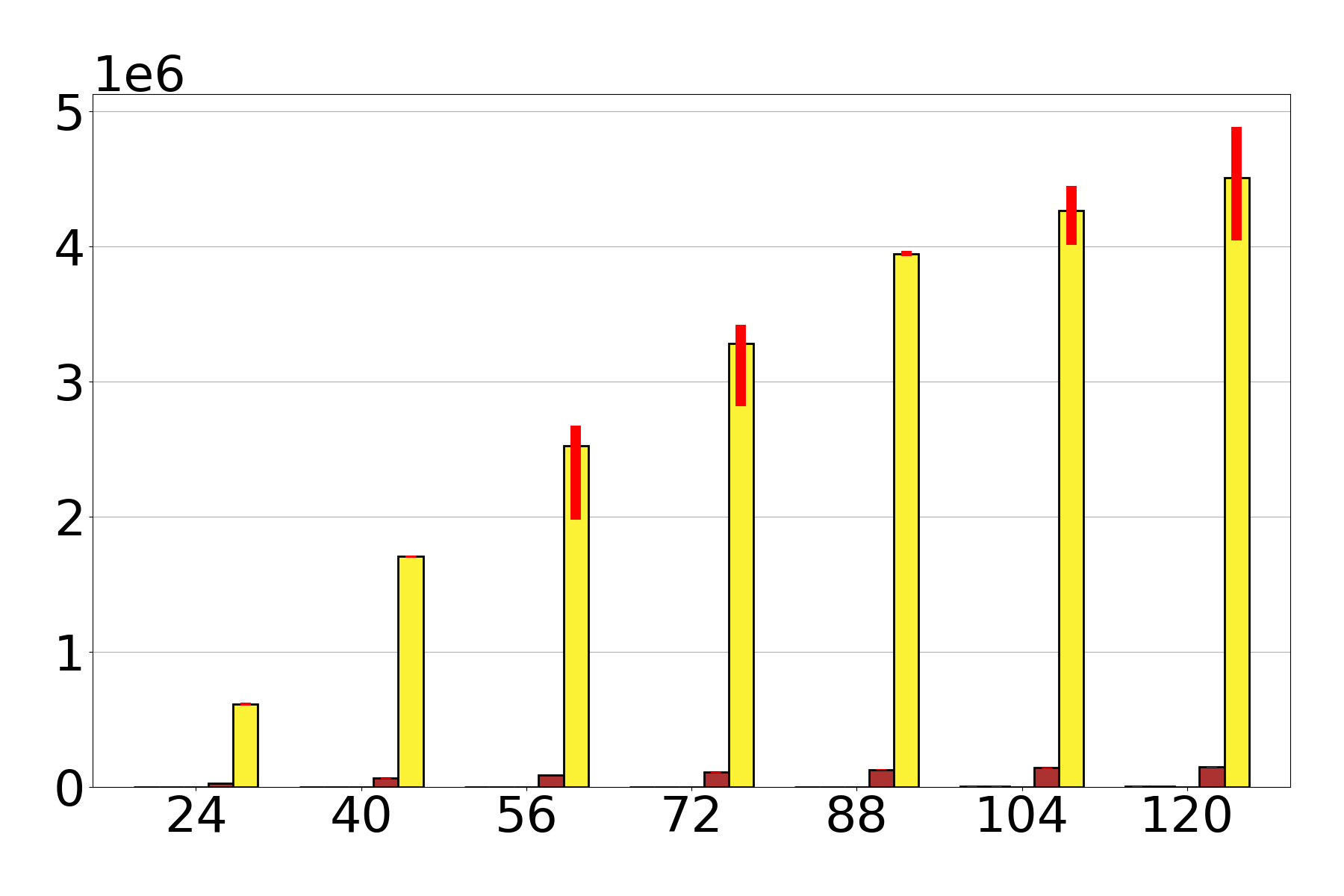}
        \end{subfigure}
        \begin{subfigure}{0.32\linewidth}
            \centering    
            \includegraphics[width=1\linewidth]{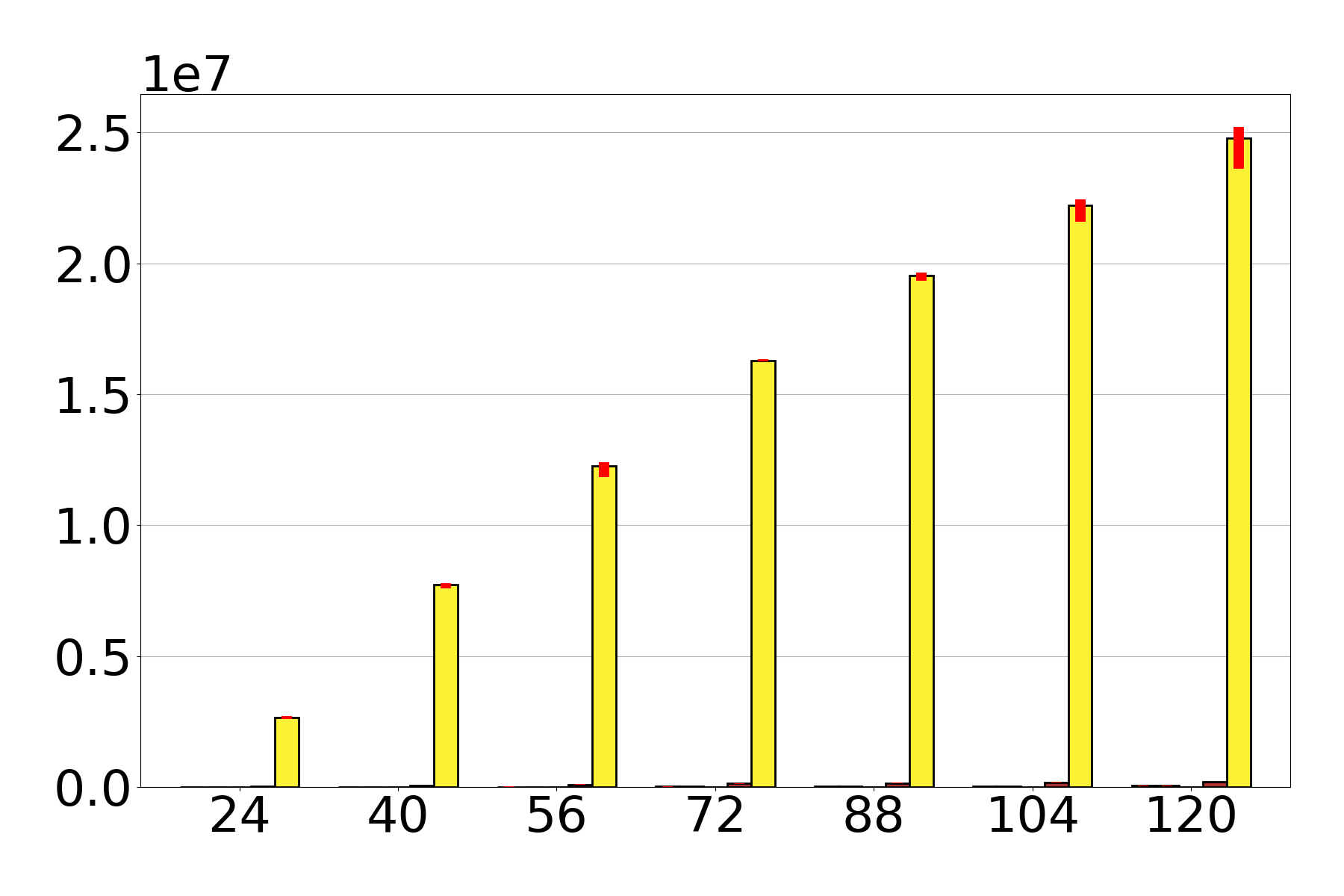}
        \end{subfigure}
    \end{subfigure}
    \begin{subfigure}{1.0\linewidth}
        \centering
        \includegraphics[width=0.4\linewidth]{plots/legend.png}
    \end{subfigure}     
    \vspace{-8mm}
    \caption{
    \centering Throughput for AVL-tree prefilled to 1 million keys using a uniform key access pattern. Y-axis is ops/sec. X-axis is number of threads. All workloads include 5\% insert and 5\% delete. RQ size is 10k (1\% of prefill size). Experiment ran on one AMD EPYC 7662.
    }
    \Description{}
    \label{fig:throughput-avl2}
\end{figure*}

\begin{figure*}[t!]
    \begin{subfigure}{0.02\linewidth}        
        \raisebox{0.5\height}{\rotatebox{90}{0 Updaters}}
    \end{subfigure}
    \begin{subfigure}{0.97\linewidth}
        \begin{subfigure}{0.32\linewidth}
            \centering
            90\% Search, 0\% RQ
            \includegraphics[width=1\linewidth]{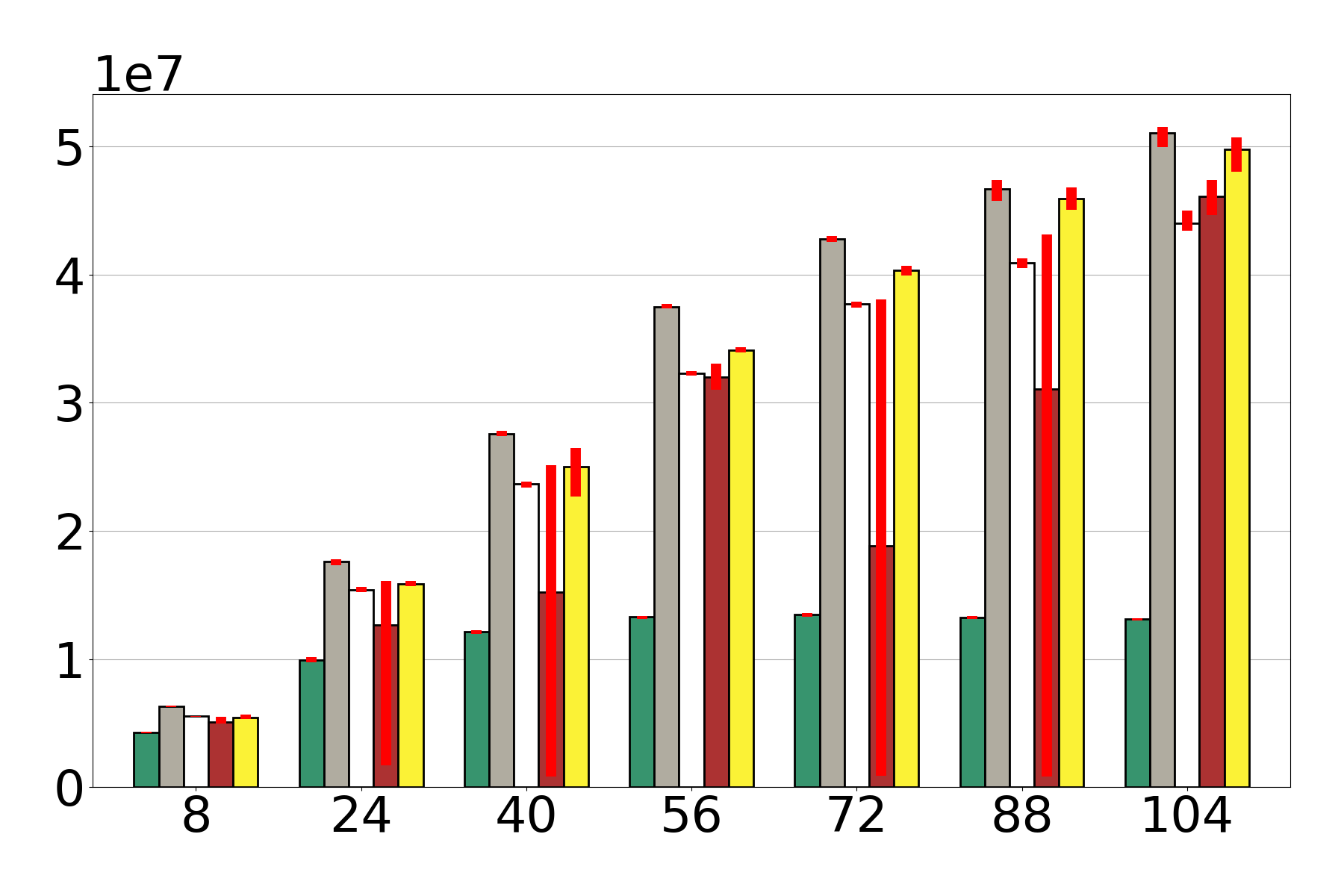}
        \end{subfigure} 
        \begin{subfigure}{0.32\linewidth}
            \centering        
            89.9\% Search, 0.1\% RQ
            \includegraphics[width=1\linewidth]{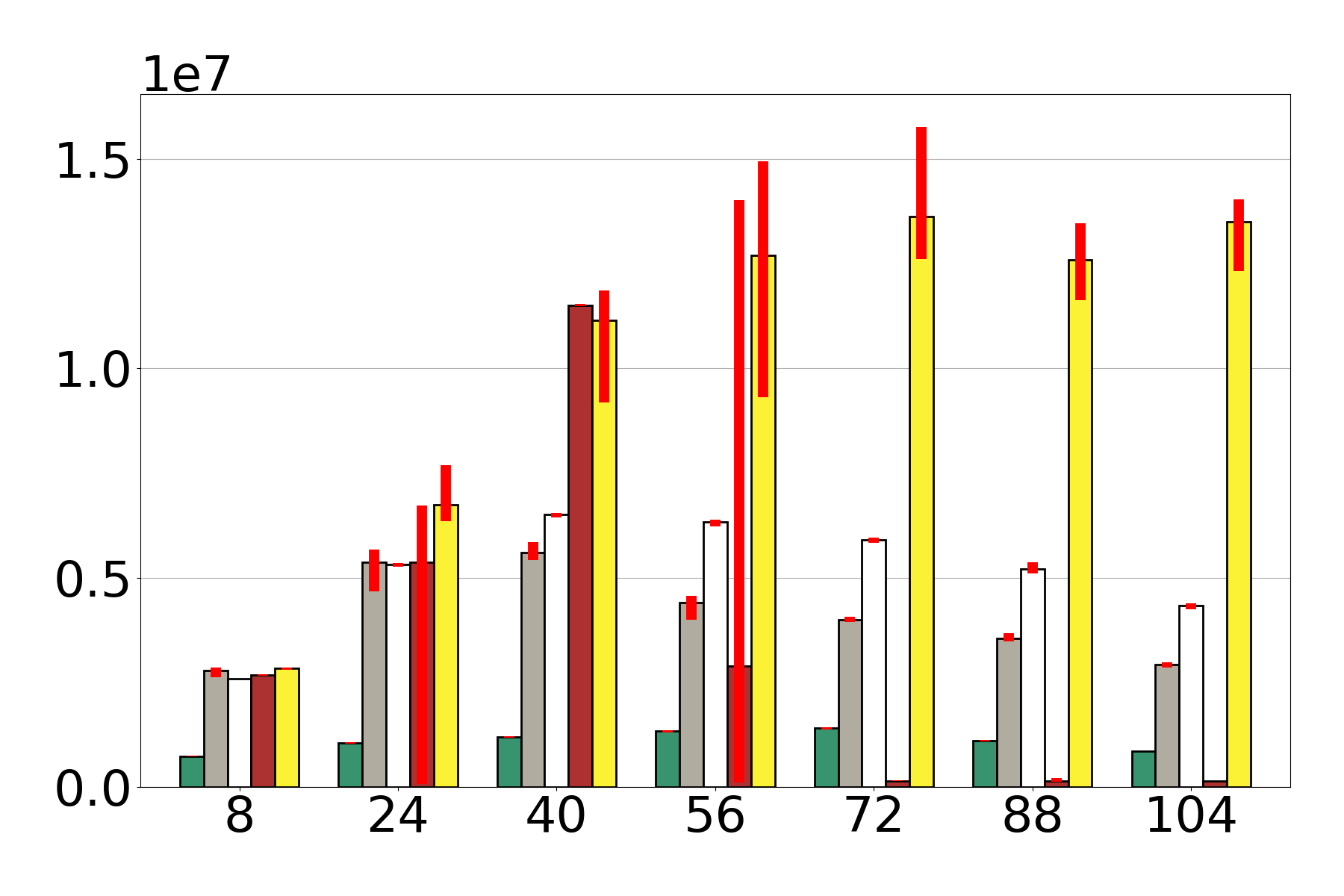}
        \end{subfigure}
        \begin{subfigure}{0.32\linewidth}
            \centering    
            89.99\% Search, 0.01\% RQ
            \includegraphics[width=1\linewidth]{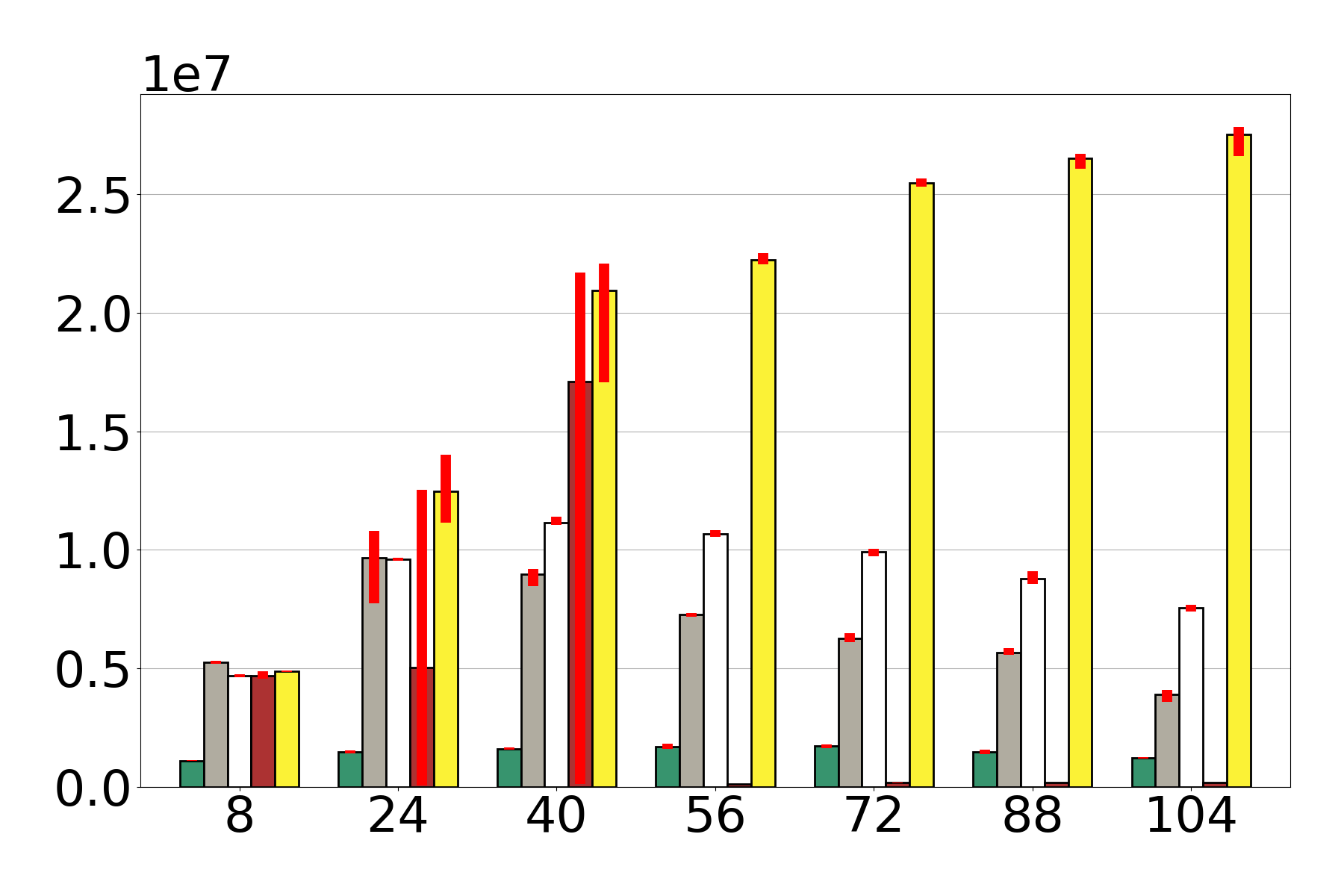}
        \end{subfigure}
    \end{subfigure}
    \begin{subfigure}{0.02\linewidth}
        \raisebox{0.5\height}{\rotatebox{90}{16 Updaters}}
    \end{subfigure}
    \begin{subfigure}{0.97\linewidth}
        \begin{subfigure}{0.32\linewidth}
            \centering
            \includegraphics[width=1\linewidth]{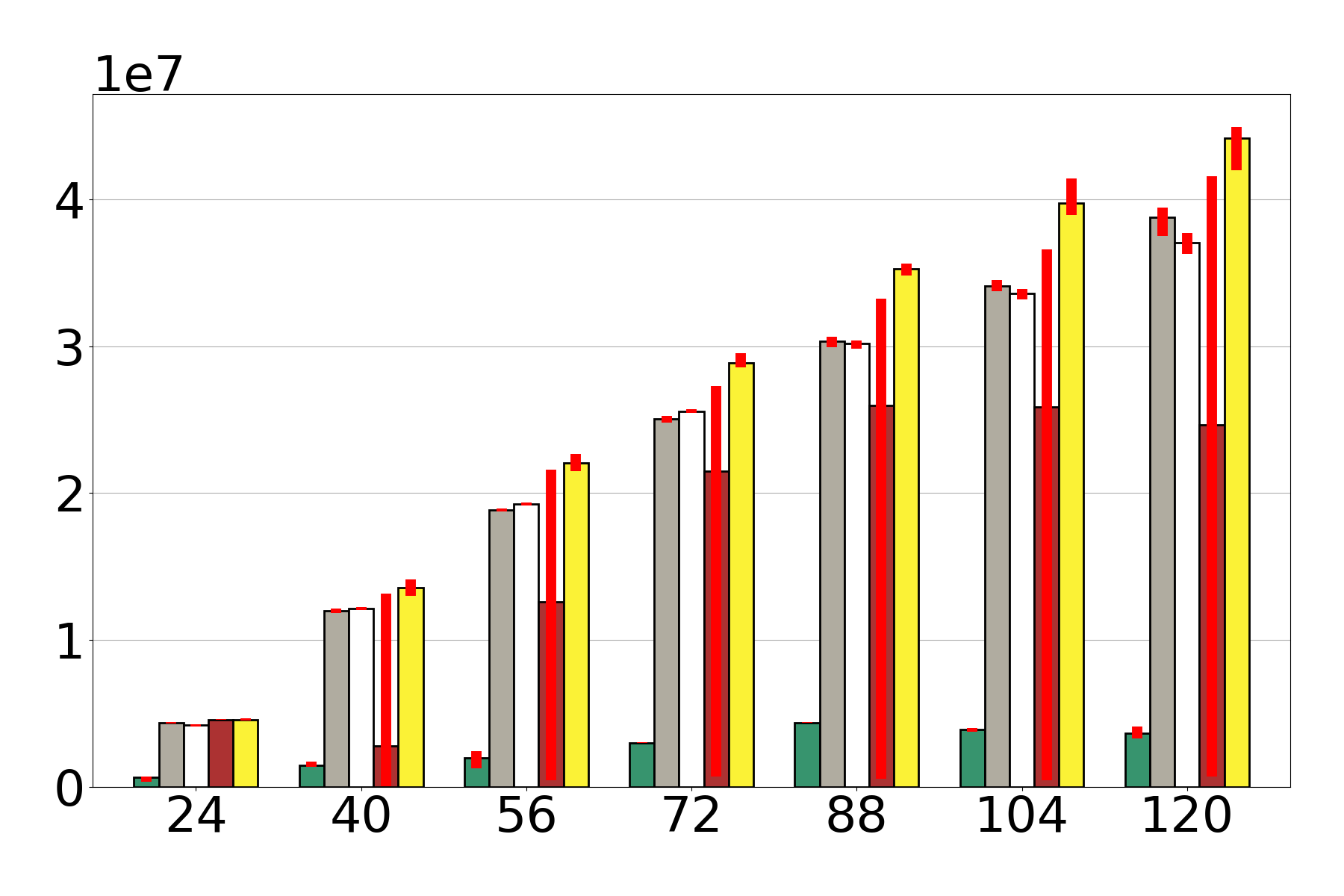}
        \end{subfigure} 
        \begin{subfigure}{0.32\linewidth}
            \centering        
            \includegraphics[width=1\linewidth]{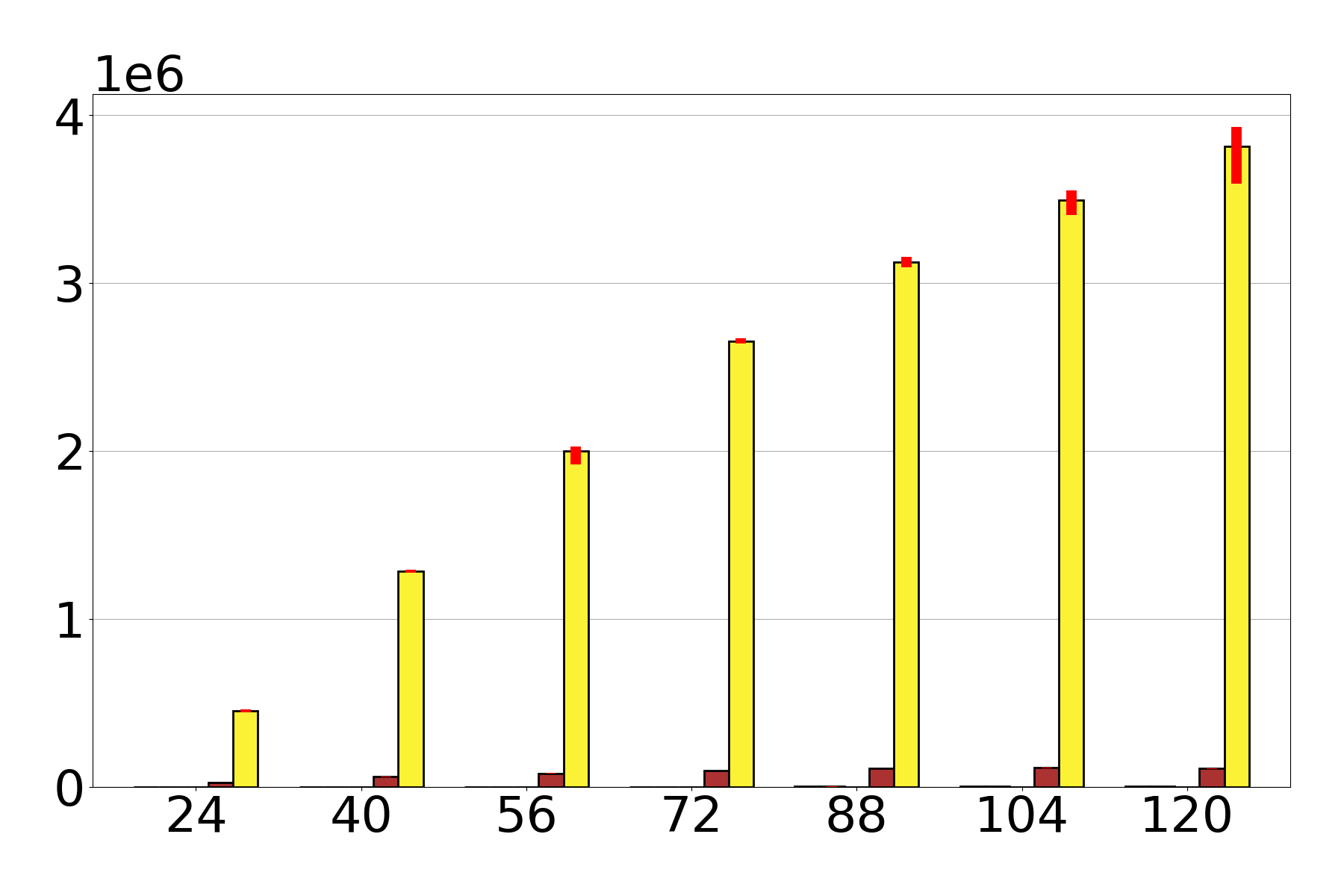}
        \end{subfigure}
        \begin{subfigure}{0.32\linewidth}
            \centering    
            \includegraphics[width=1\linewidth]{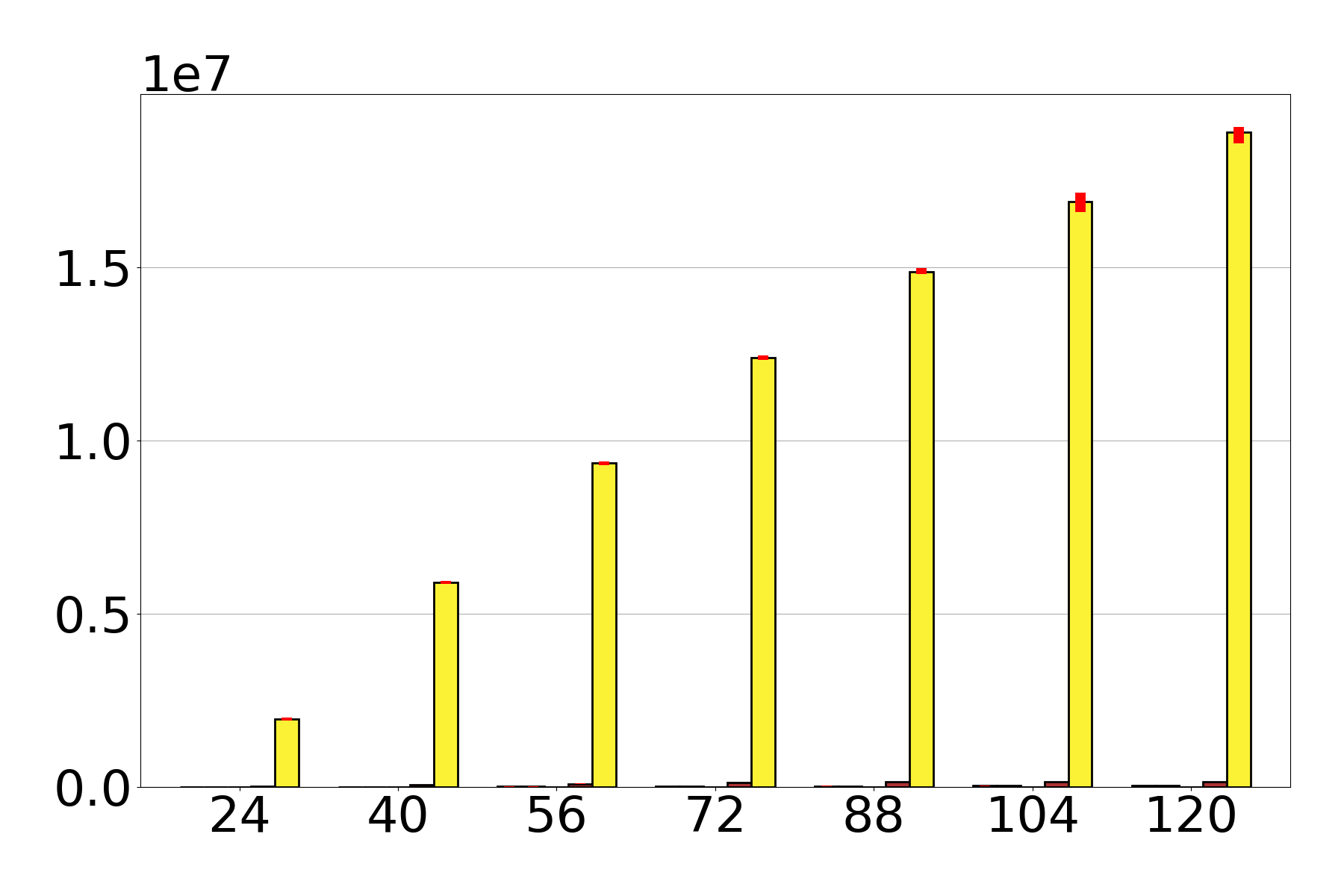}
        \end{subfigure}
    \end{subfigure}
    \begin{subfigure}{1.0\linewidth}
        \centering
        \includegraphics[width=0.4\linewidth]{plots/legend.png}
    \end{subfigure}     
    \vspace{-8mm}
    \caption{
    \centering Throughput for external binary search tree prefilled to 1 million keys using a uniform key access pattern. Y-axis is ops/sec. X-axis is number of threads. All workloads include 5\% insert and 5\% delete. RQ size is 10k (1\% of prefill size). Experiment ran on a single AMD EPYC 7662.
    }
    \Description{}
    \label{fig:throughput-bst-node0}
\end{figure*}

\begin{figure*}[t!]
    \begin{subfigure}{0.97\linewidth}
        \begin{subfigure}{0.24\linewidth}
            \centering
            90\% Search, 0\% SQ
            \includegraphics[width=1\linewidth]{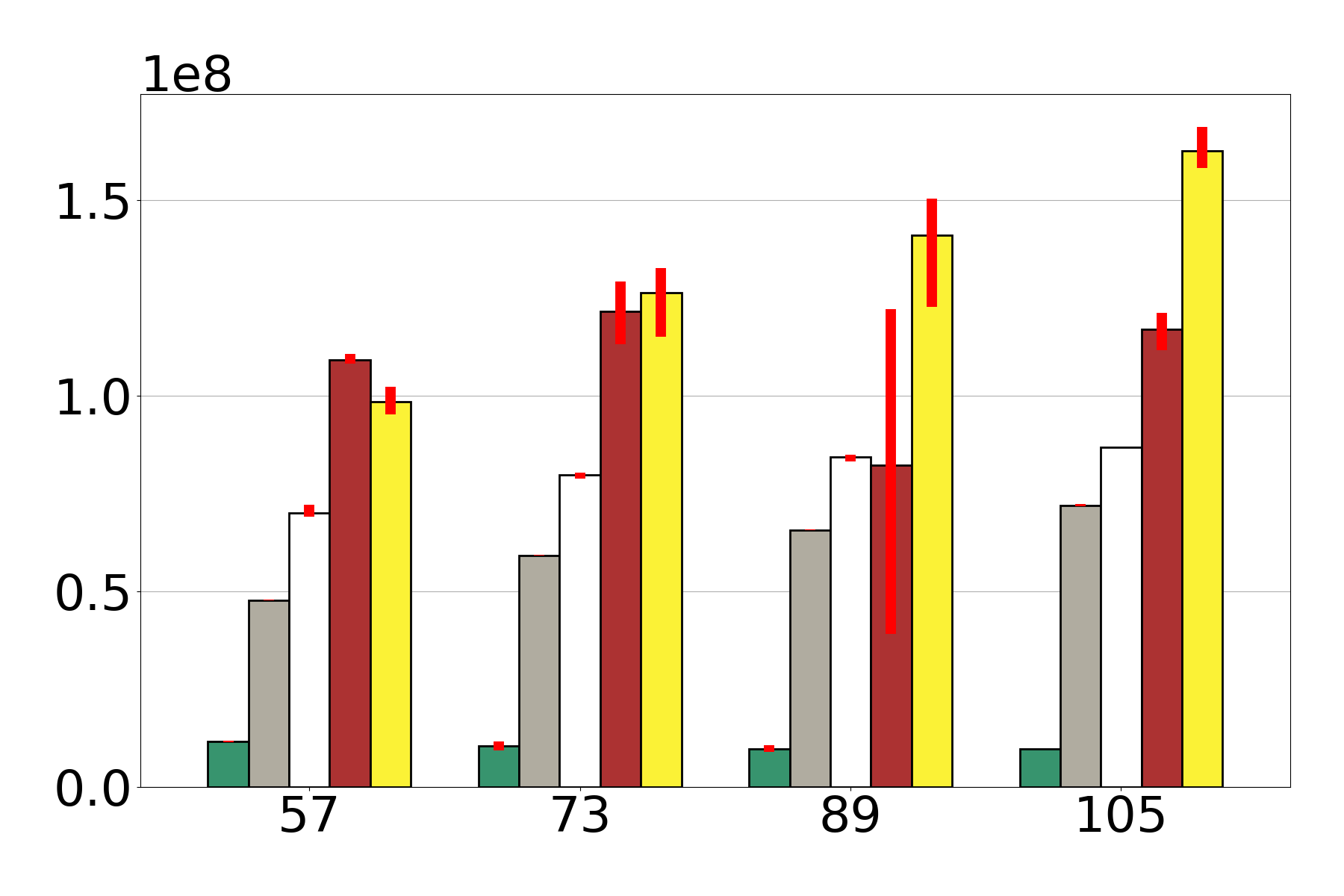}
        \end{subfigure} 
        \begin{subfigure}{0.24\linewidth}
            \centering        
            89.99\% Search, 0.01\% SQ
            \includegraphics[width=1\linewidth]{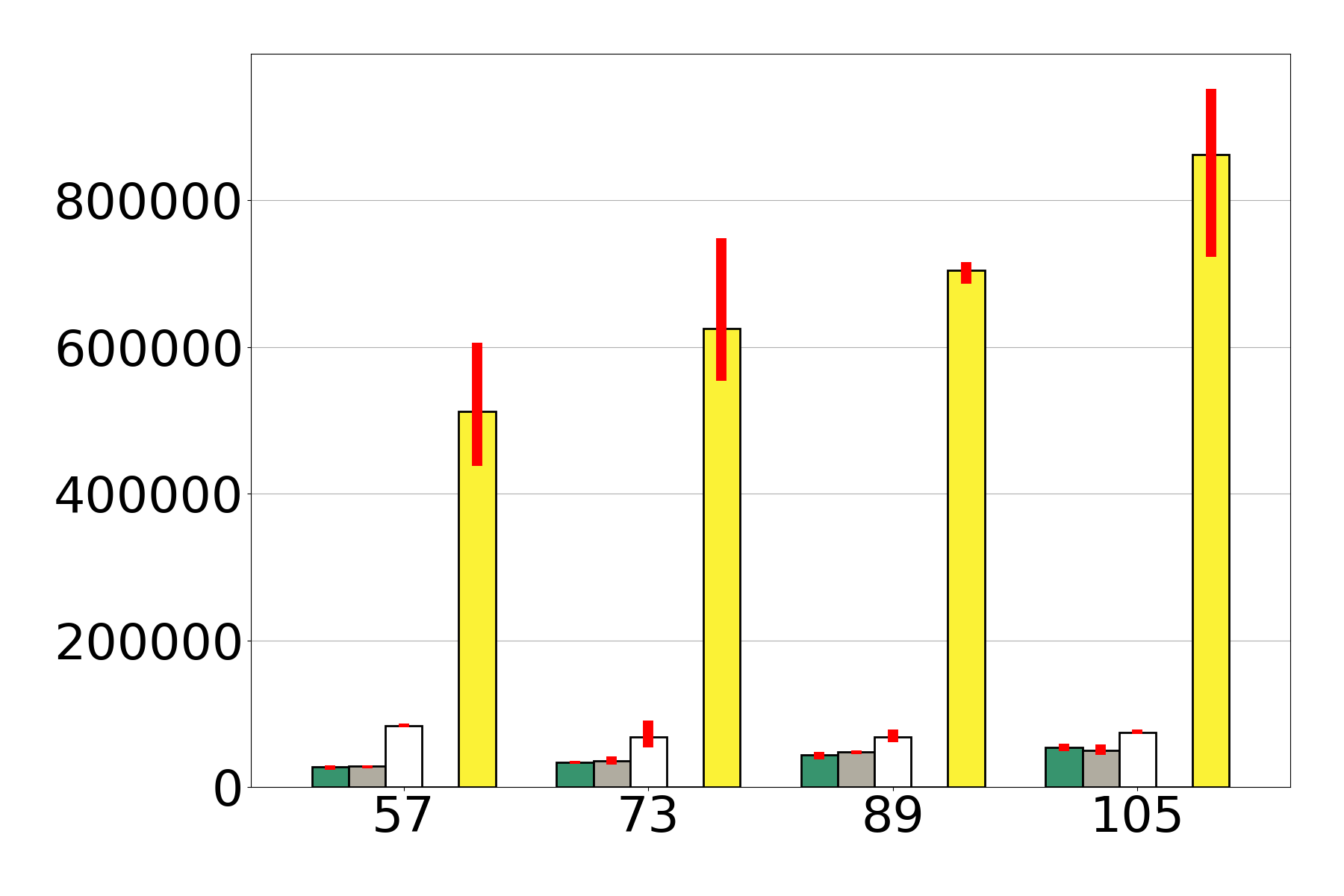}
        \end{subfigure}
        \begin{subfigure}{0.24\linewidth}
            \centering    
            90\% Search, 0\% Size SQ
            \includegraphics[width=1\linewidth]{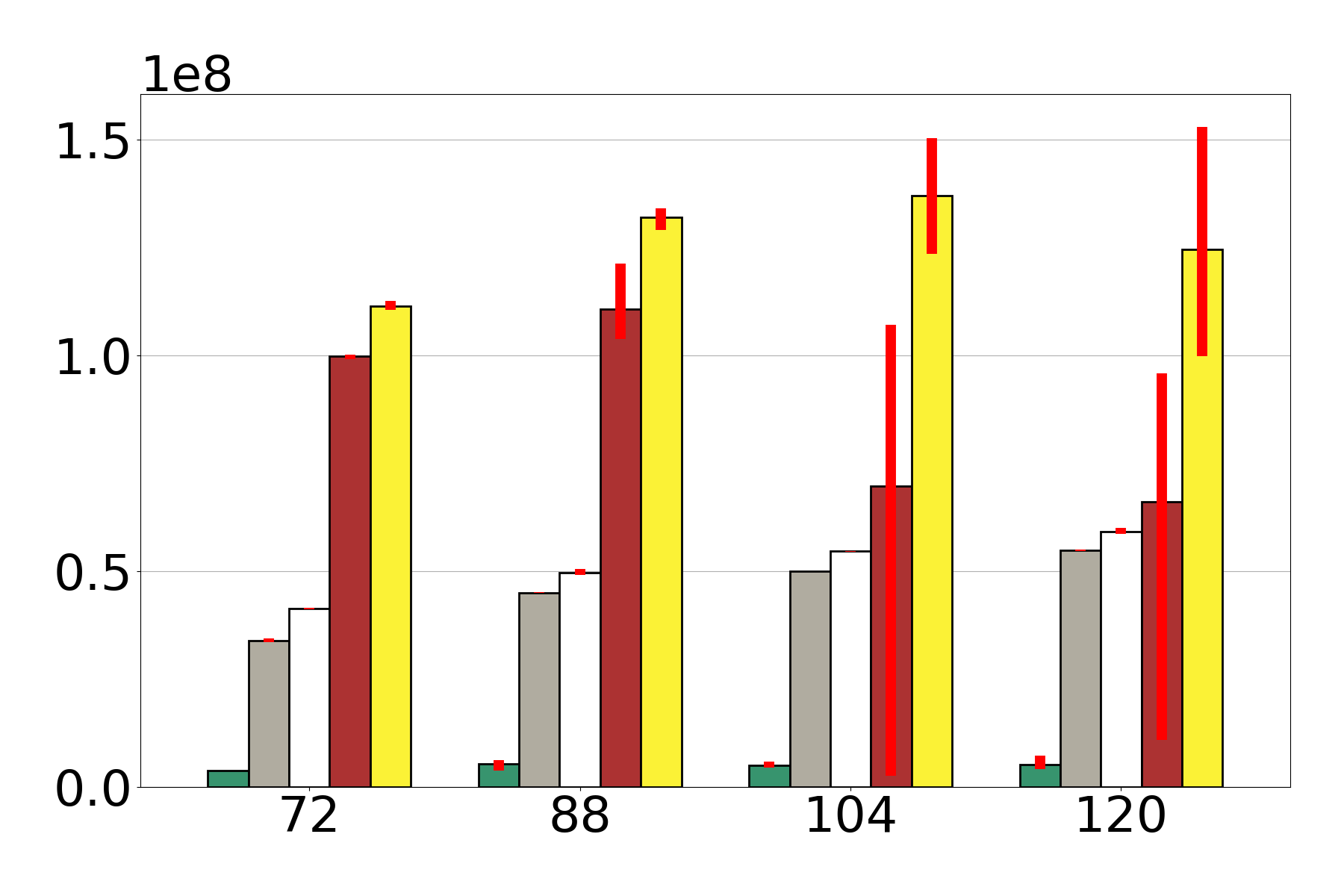}
        \end{subfigure}
        \begin{subfigure}{0.24\linewidth}
            \centering    
            89.99\% Search, 0.01\% SQ
            \includegraphics[width=1\linewidth]{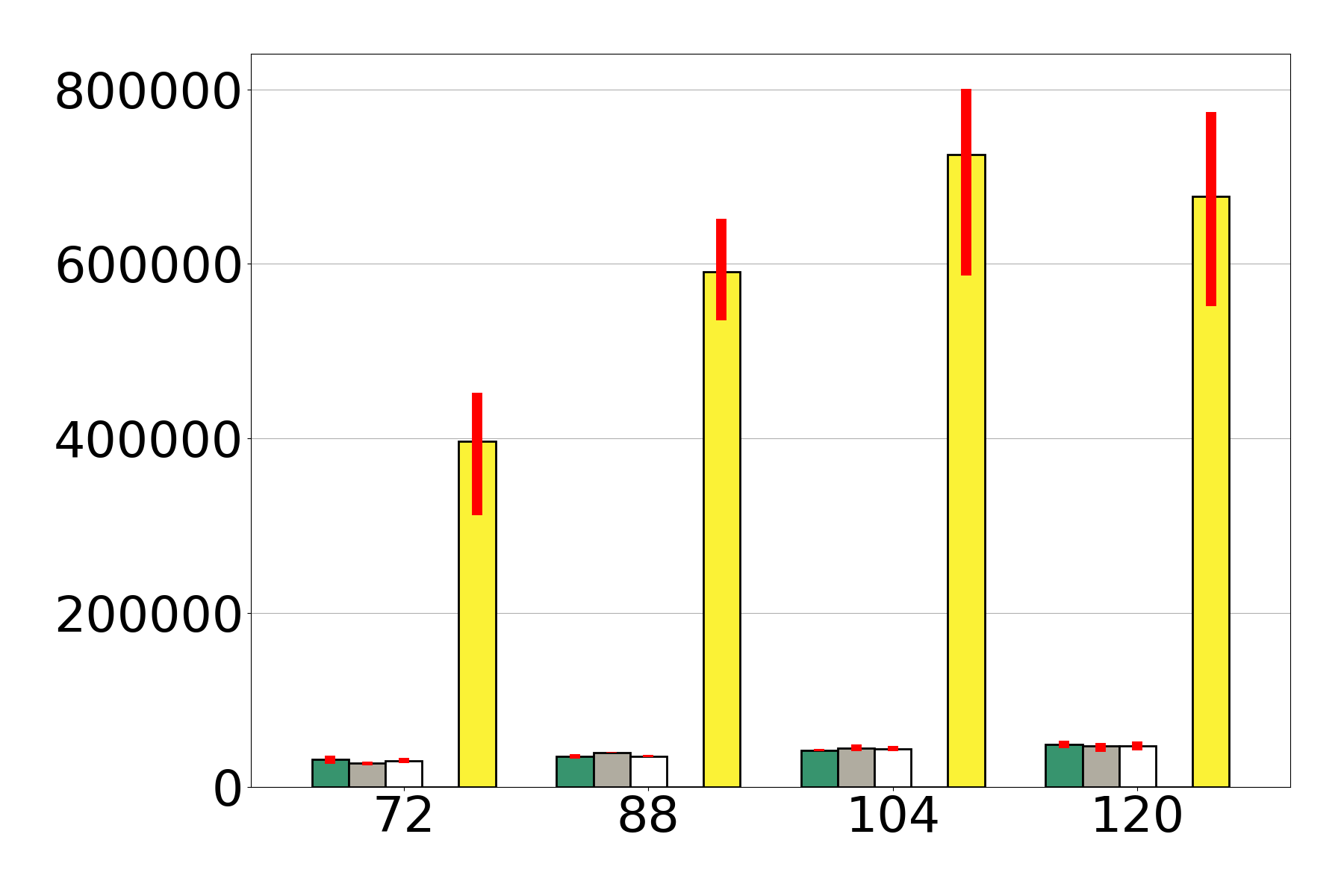}
        \end{subfigure}
    \end{subfigure}
    \begin{subfigure}{0.97\linewidth}
        \begin{subfigure}{0.49\linewidth}
            \caption{1 Updater}
        \end{subfigure} 
        \hfill        
        \begin{subfigure}{0.49\linewidth}
            \caption{16 Updaters}
        \end{subfigure}
    \end{subfigure}
    \begin{subfigure}{1.0\linewidth}
        \centering
        \includegraphics[width=0.4\linewidth]{plots/legend.png}
    \end{subfigure}     
    \vspace{-8mm}
    \caption{
    \centering Throughput for hashmap with 1 million buckets prefilled to 100k keys using a uniform key access pattern. Y-axis is ops/sec. X-axis is number of threads. All workloads include 5\% insert and 5\% delete. SQ (size queries) refers to atomic size operations. Experiment ran on a single AMD EPYC 7662.
    }
    \Description{}
    \label{fig:throughput-hashmap-node0}
\end{figure*}

\begin{figure*}[t!]
    \begin{subfigure}{0.02\linewidth}        
        \raisebox{0.3\height}{\rotatebox{90}{0 Updaters}}
    \end{subfigure}
    \begin{subfigure}{0.97\linewidth}
        \begin{subfigure}{0.24\linewidth}
            \centering
            89.9\% Search, 0.1\% RQ
            \includegraphics[width=1\linewidth]{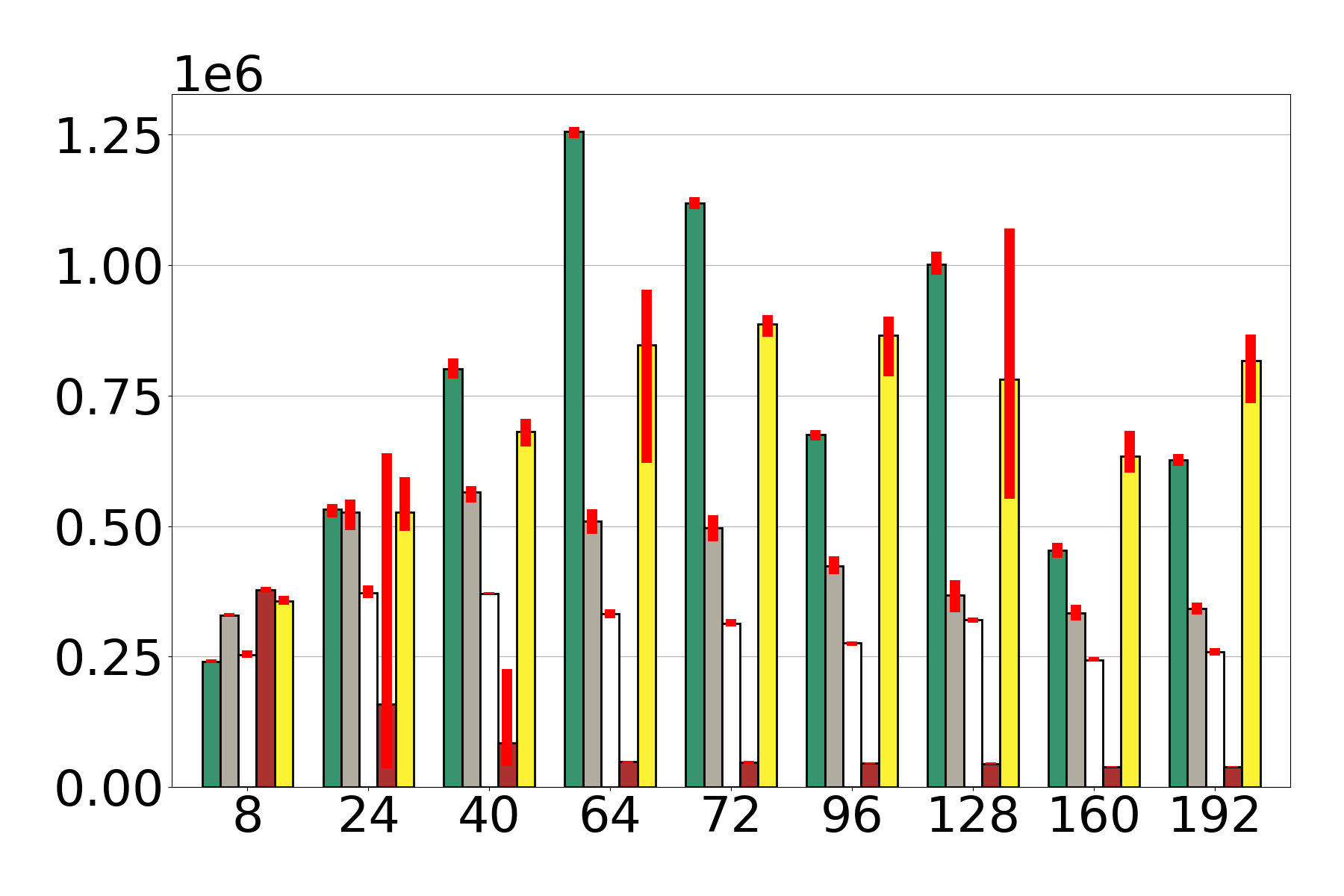}
        \end{subfigure} 
        \begin{subfigure}{0.24\linewidth}
            \centering        
            89.99\% Search, 0.01\% RQ
            \includegraphics[width=1\linewidth]{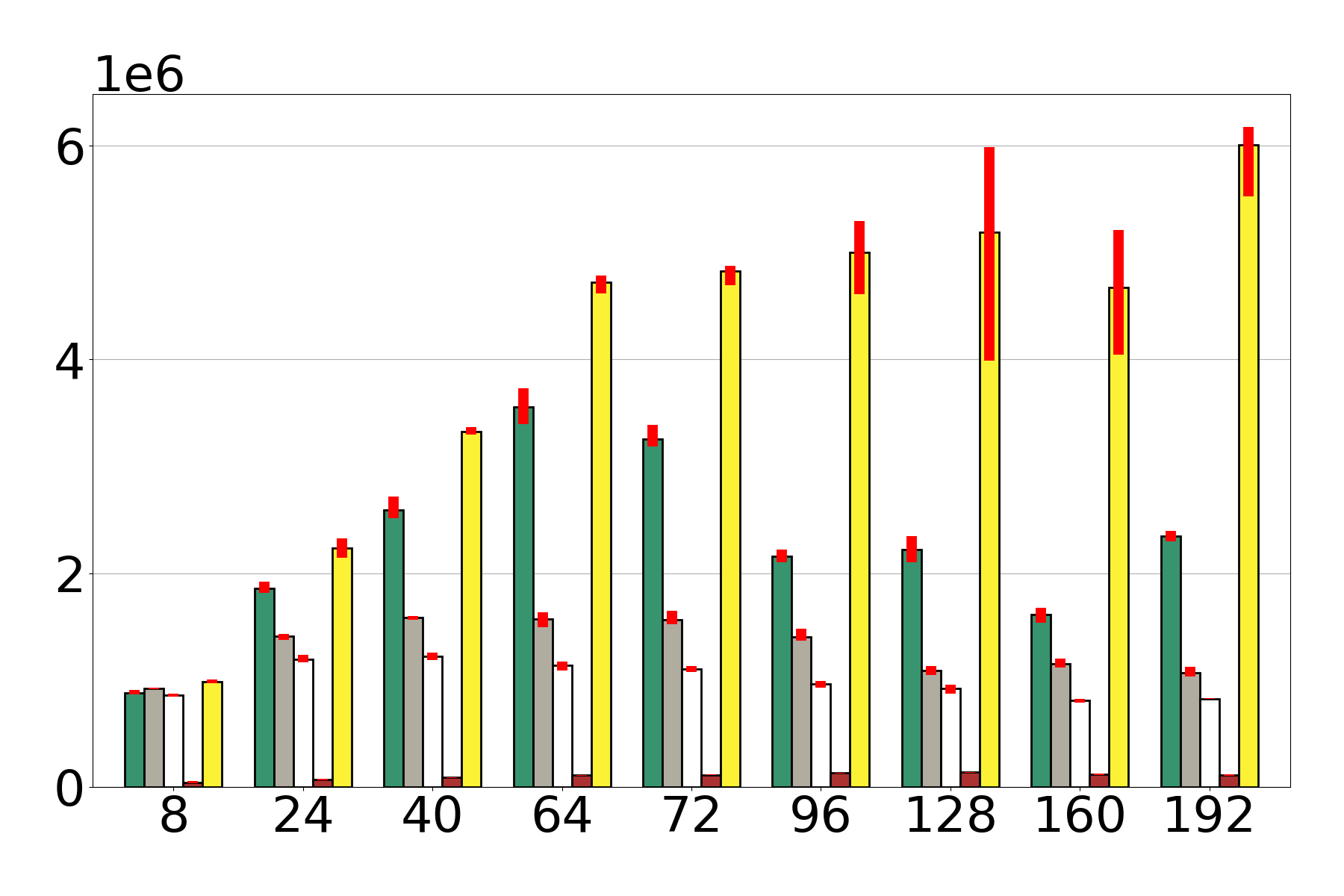}
        \end{subfigure}
        \begin{subfigure}{0.24\linewidth}
            \centering    
            59.9\% Search, 0.1\% RQ
            \includegraphics[width=1\linewidth]{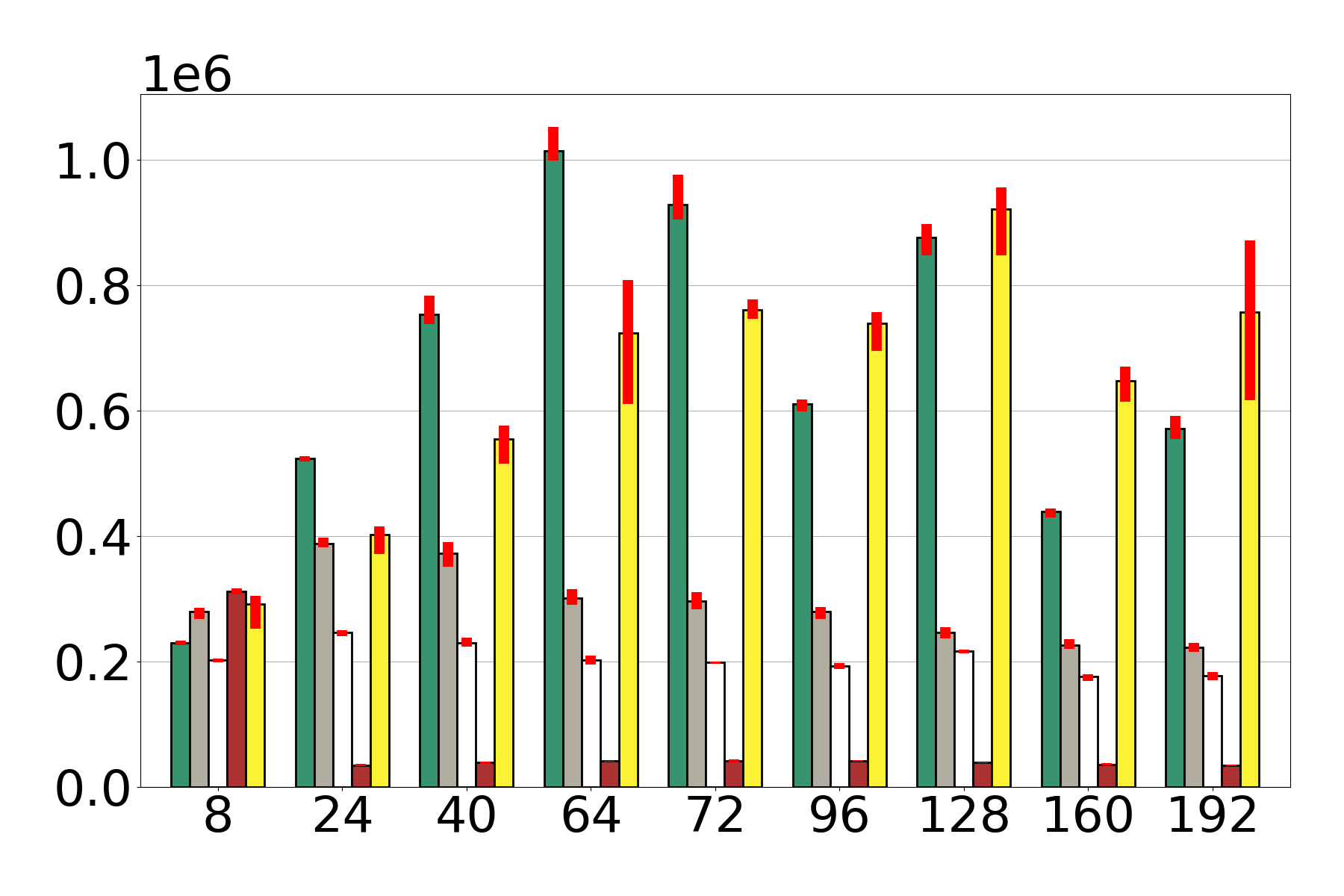}
        \end{subfigure}        
        \begin{subfigure}{0.24\linewidth}
            \centering 
            59.99\% Search, 0.01\% RQ
            \includegraphics[width=1\linewidth]{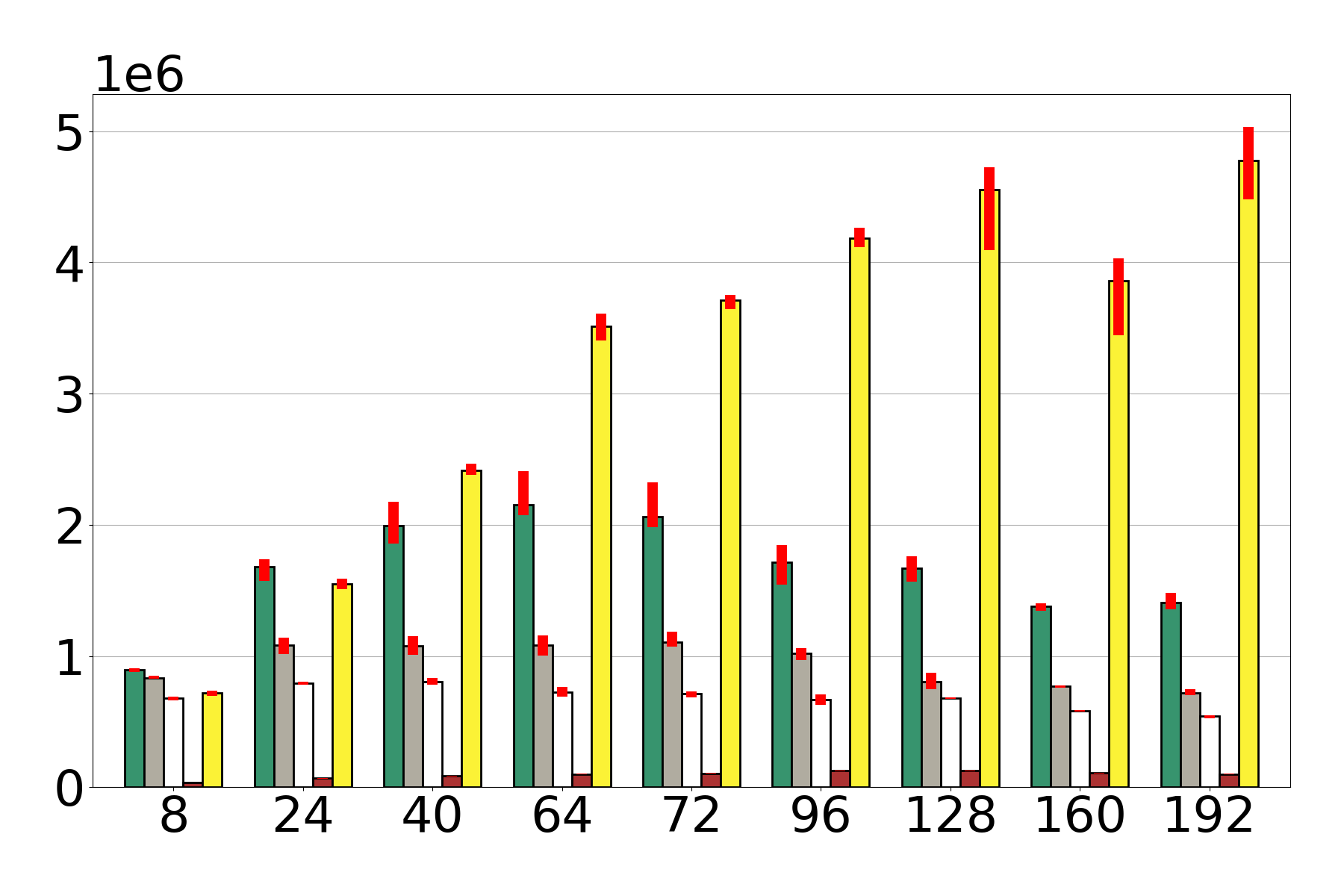}
        \end{subfigure}     
    \end{subfigure}
    \begin{subfigure}{0.02\linewidth}
        \raisebox{0.5\height}{\rotatebox{90}{16 Updaters}}
    \end{subfigure}
    \begin{subfigure}{0.97\linewidth}
        \begin{subfigure}{0.24\linewidth}
            \centering
            \includegraphics[width=1\linewidth]{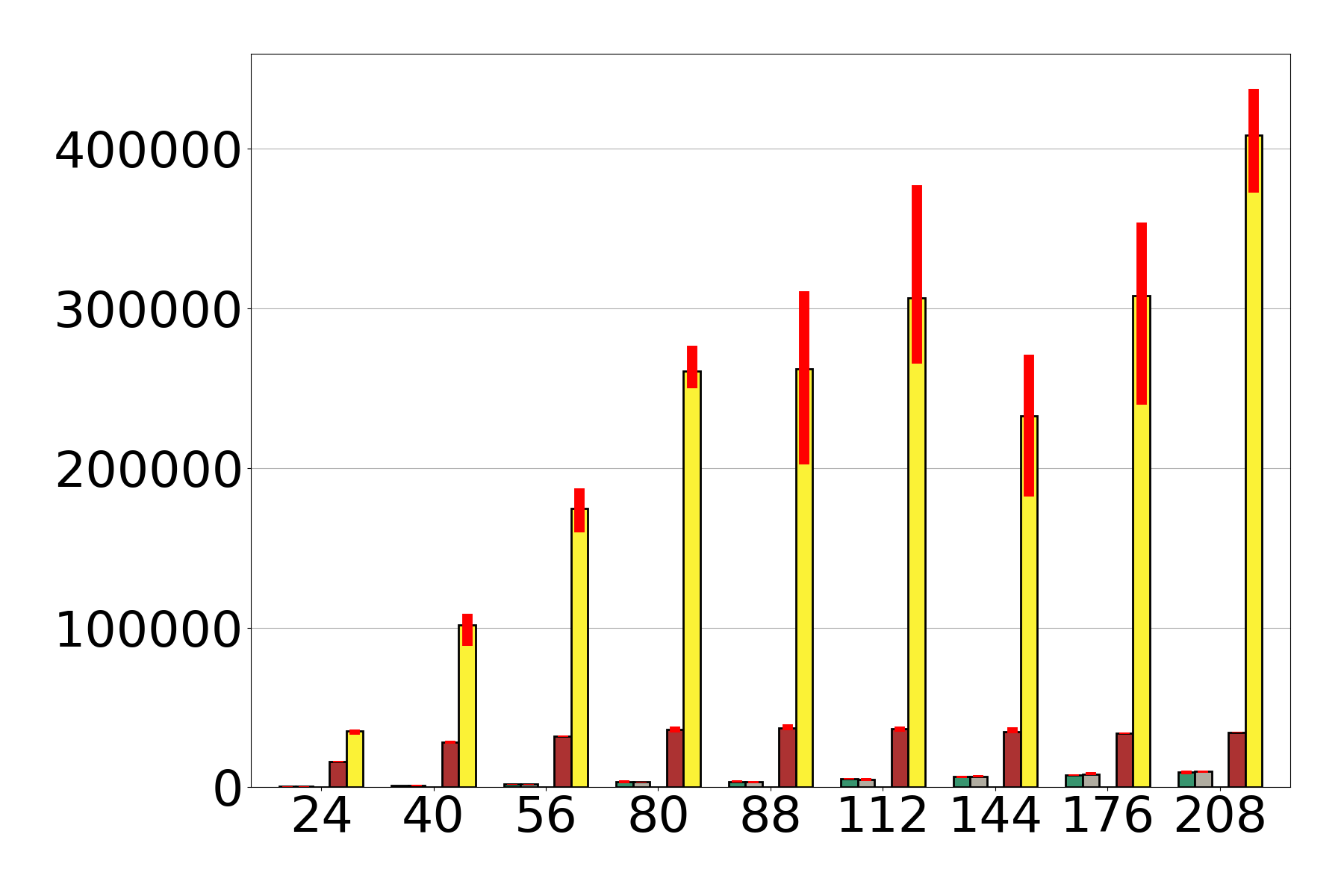}
        \end{subfigure} 
        \begin{subfigure}{0.24\linewidth}
            \centering        
            \includegraphics[width=1\linewidth]{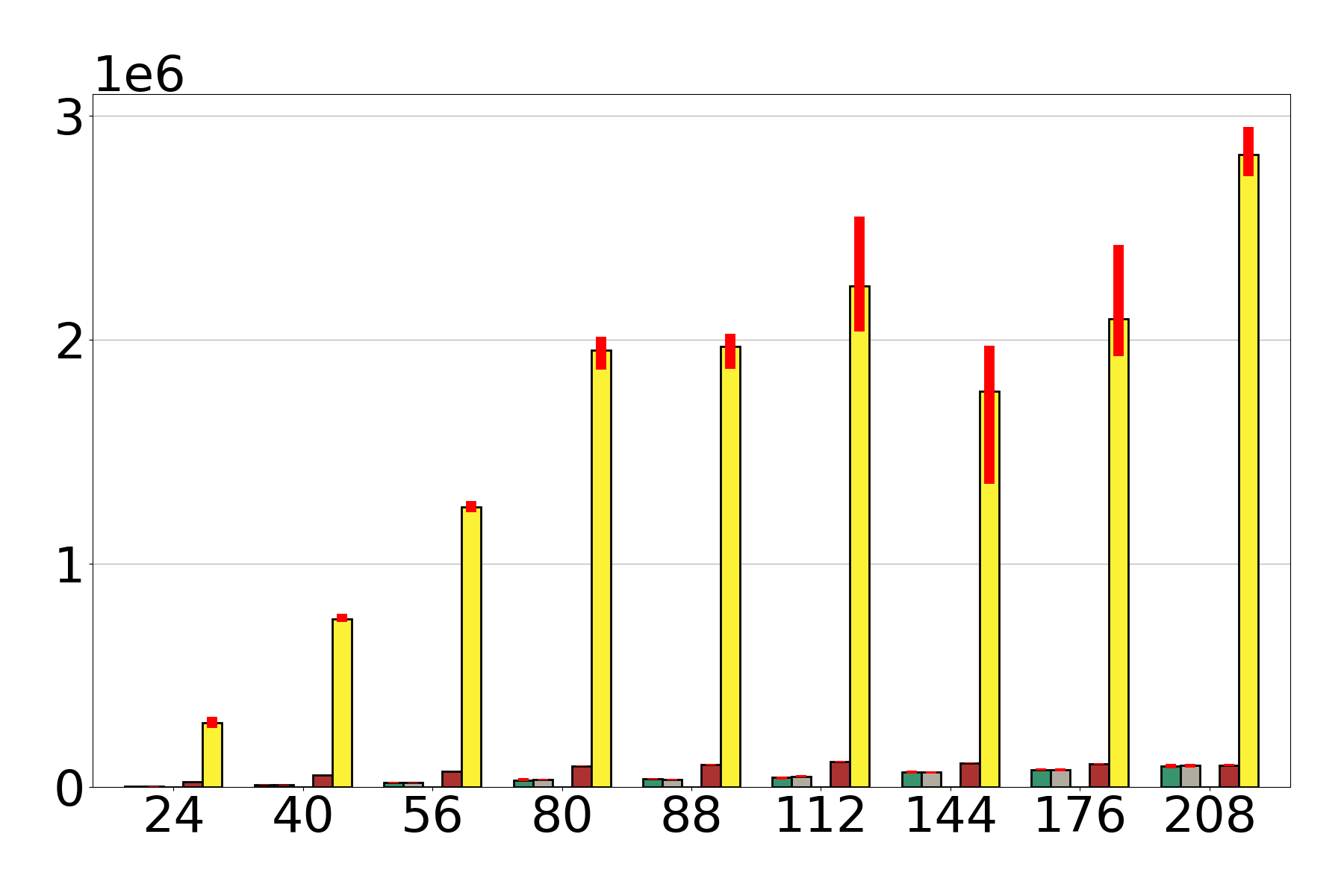}
        \end{subfigure}
        \begin{subfigure}{0.24\linewidth}
            \centering    
            \includegraphics[width=1\linewidth]{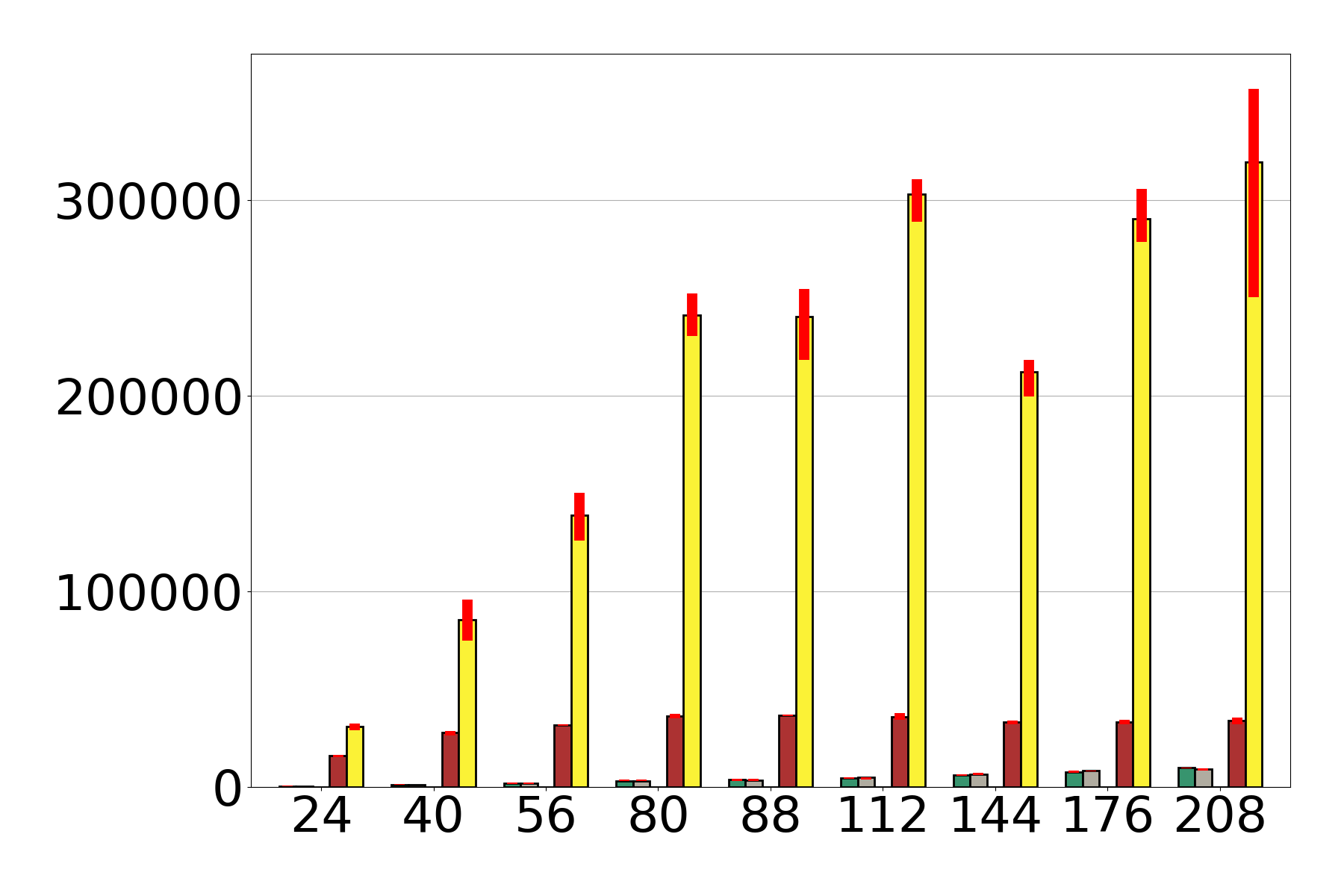}
        \end{subfigure}        
        \begin{subfigure}{0.24\linewidth}
            \centering 
            \includegraphics[width=1\linewidth]{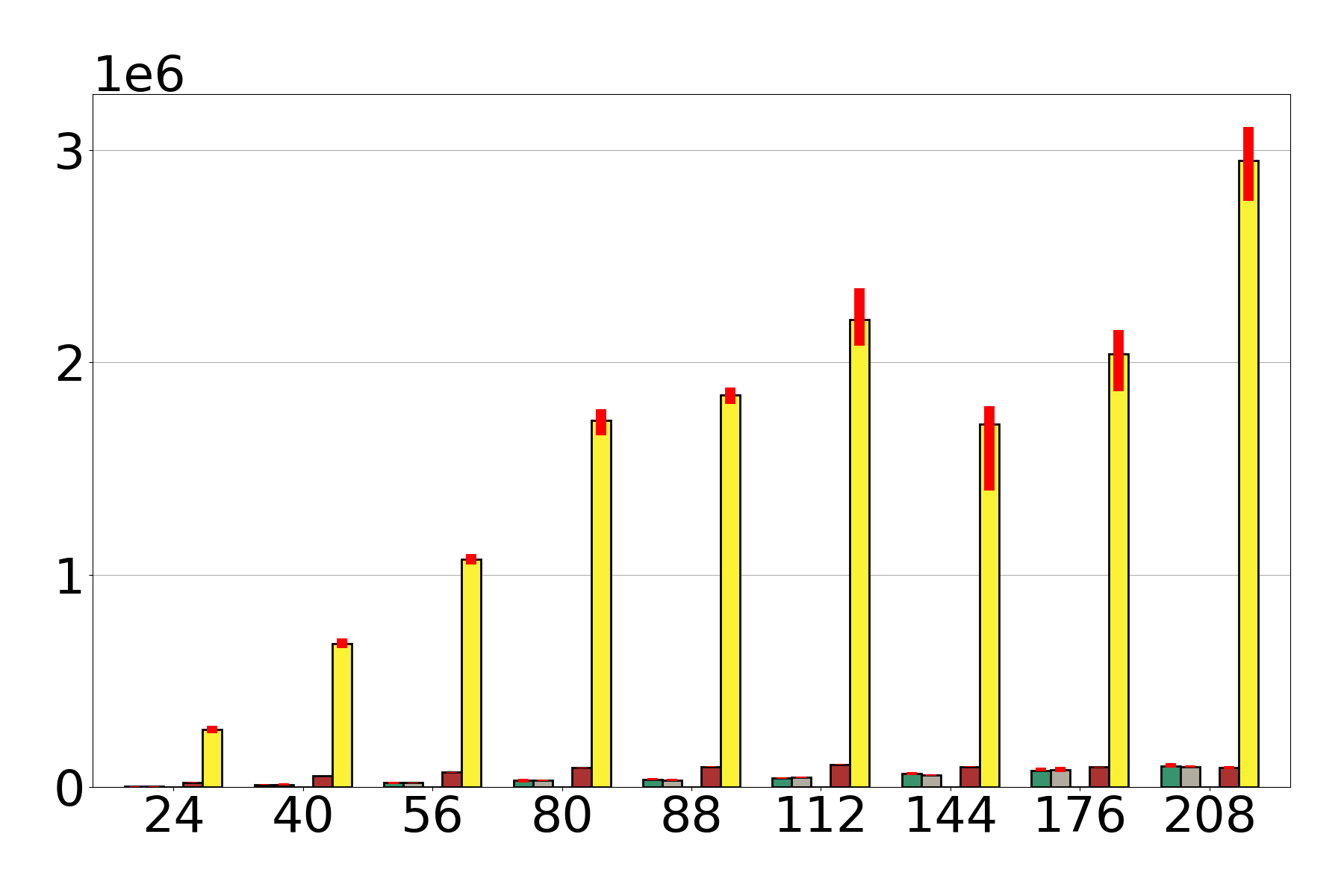}
        \end{subfigure}     
    \end{subfigure}
    \begin{subfigure}{1.0\linewidth}
        \centering
        \includegraphics[width=0.4\linewidth]{plots/legend.png}
    \end{subfigure}     
    \vspace{-8mm}
    \caption{\centering Throughput for AVL-tree prefilled to 1 million keys using a uniform key access pattern. Y-axis is ops/sec. X-axis is number of threads. All workloads include 5\% insert and 5\% delete. RQ size is 100k (10\% of prefill size).Experiment ran on dual AMD EPYC 7662.        
    }
    \Description{}
    \label{fig:throughput-avl}
\end{figure*}

\paragraph{Single AMD EPYC 7662 Experiments}
We include additional experiments that were run on an AMD EPYC 7662 processor.
For the internal AVL tree results see \Cref{fig:throughput-avl} and \Cref{fig:throughput-avl2}.
For the external binary search tree results see \Cref{fig:throughput-bst-node0}.
For the hashmap results see \Cref{fig:throughput-hashmap-node0}.
The experiments with these additional data structures are similar to what we have shown in the main paper.
Specifically, when there are no RQs \tmname\ achieves throughput comparable or better compared to the other STMs.
For workloads with RQs \tmname\ typically outperforms the other STMs, this is especially true when we introduce dedicated updaters.

\begin{figure*}[t!]
    \begin{subfigure}{0.02\linewidth}        
        \raisebox{0.5\height}{\rotatebox{90}{0 Updaters}}
    \end{subfigure}
    \begin{subfigure}{0.97\linewidth}
        \begin{subfigure}{0.32\linewidth}
            \centering
            90\% Search, 0\% RQ
            \includegraphics[width=1\linewidth]{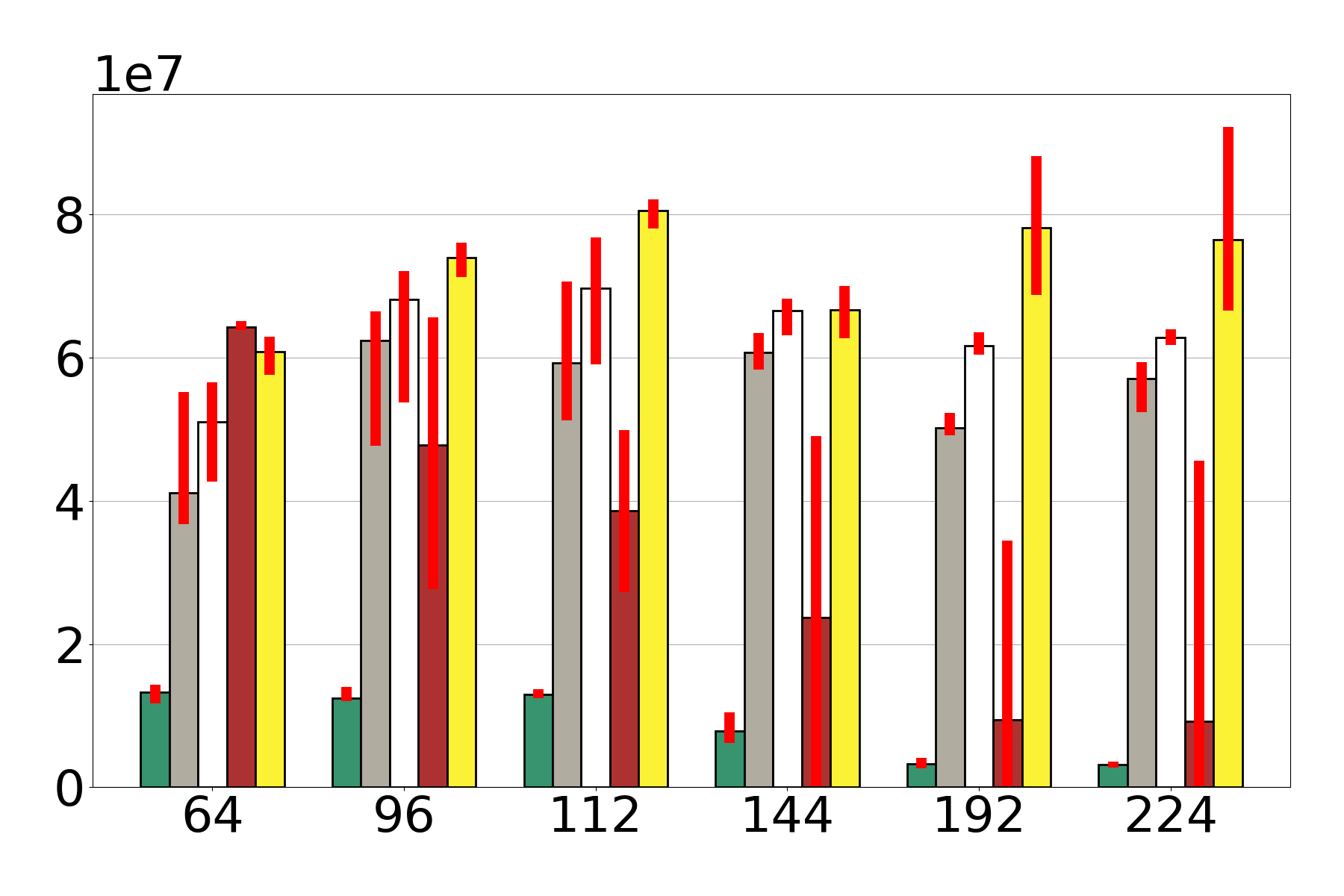}
        \end{subfigure} 
        \begin{subfigure}{0.32\linewidth}
            \centering        
            89.9\% Search, 0.1\% RQ
            \includegraphics[width=1\linewidth]{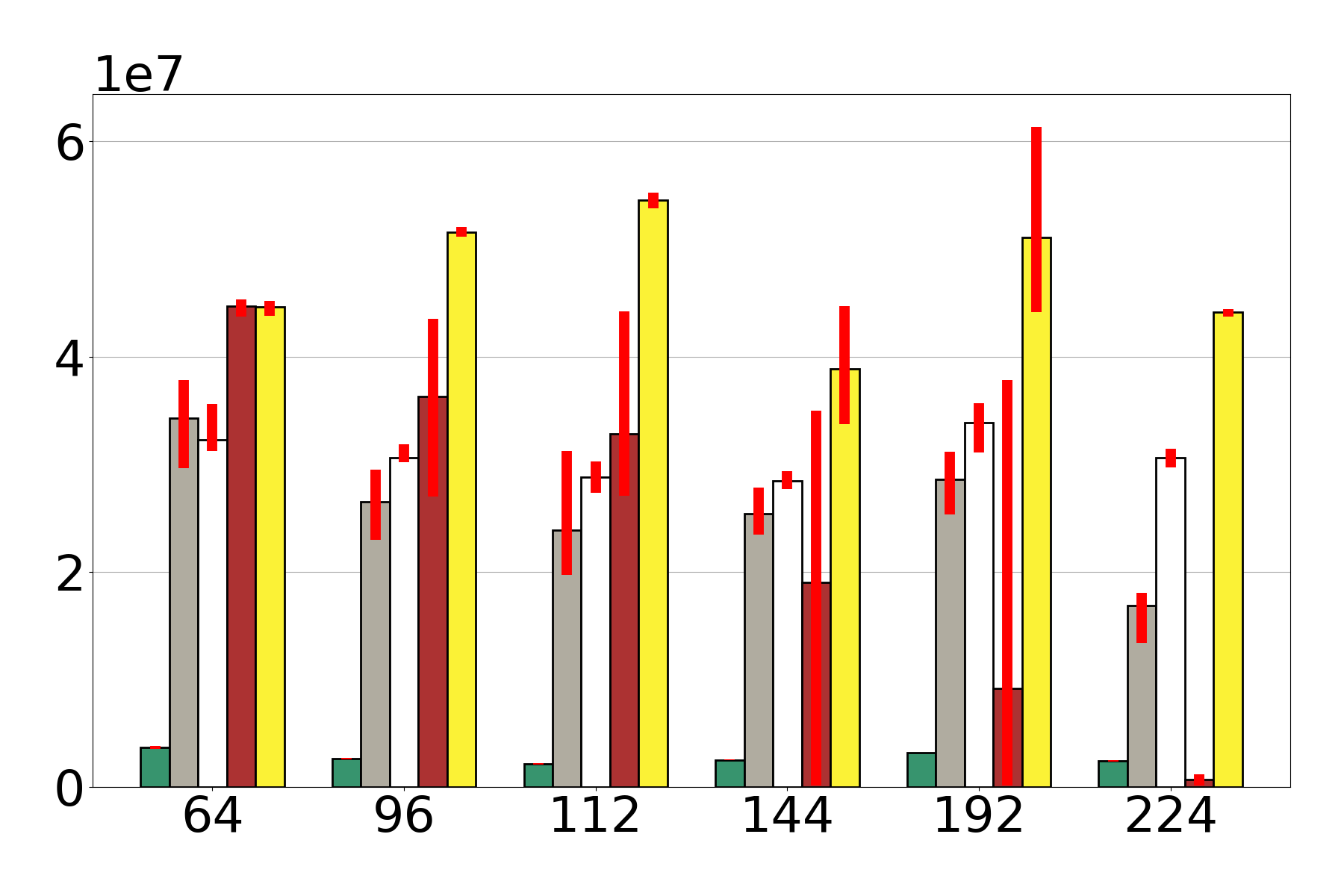}
        \end{subfigure}
        \begin{subfigure}{0.32\linewidth}
            \centering    
            89.99\% Search, 0.01\% RQ
            \includegraphics[width=1\linewidth]{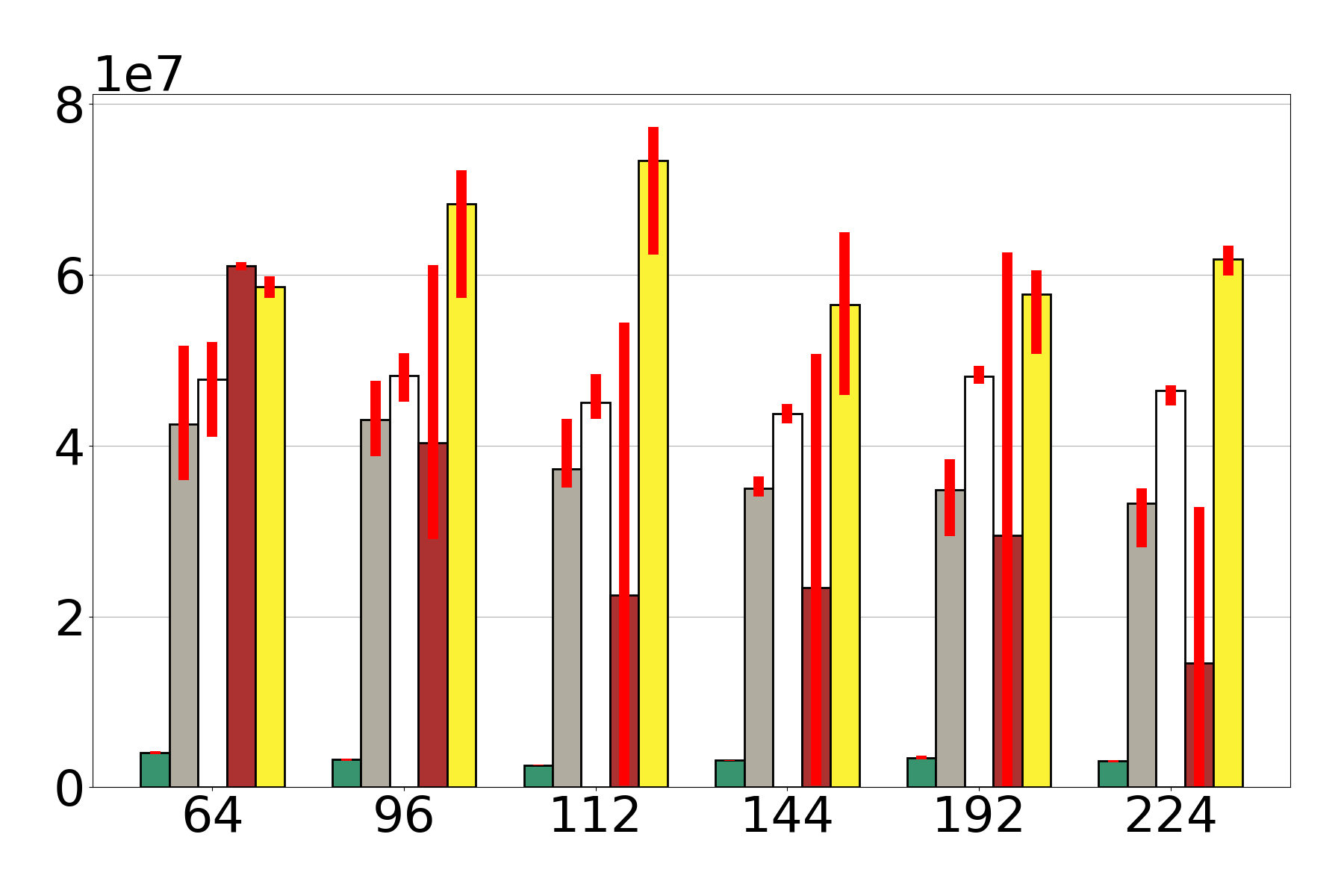}
        \end{subfigure}
    \end{subfigure}
    \begin{subfigure}{0.02\linewidth}
        \raisebox{0.5\height}{\rotatebox{90}{16 Updaters}}
    \end{subfigure}
    \begin{subfigure}{0.97\linewidth}
        \begin{subfigure}{0.32\linewidth}
            \centering
            \includegraphics[width=1\linewidth]{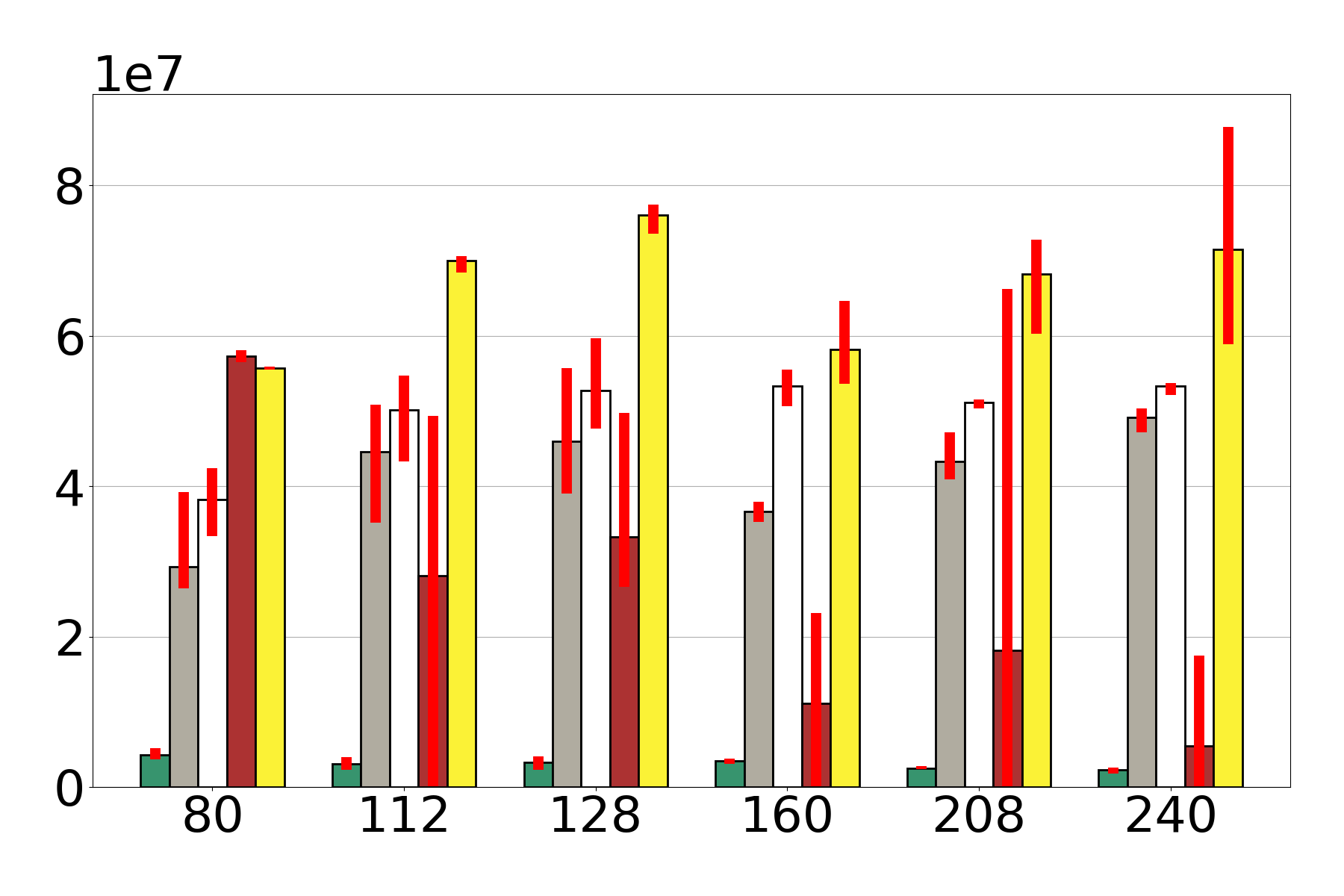}
        \end{subfigure} 
        \begin{subfigure}{0.32\linewidth}
            \centering        
            \includegraphics[width=1\linewidth]{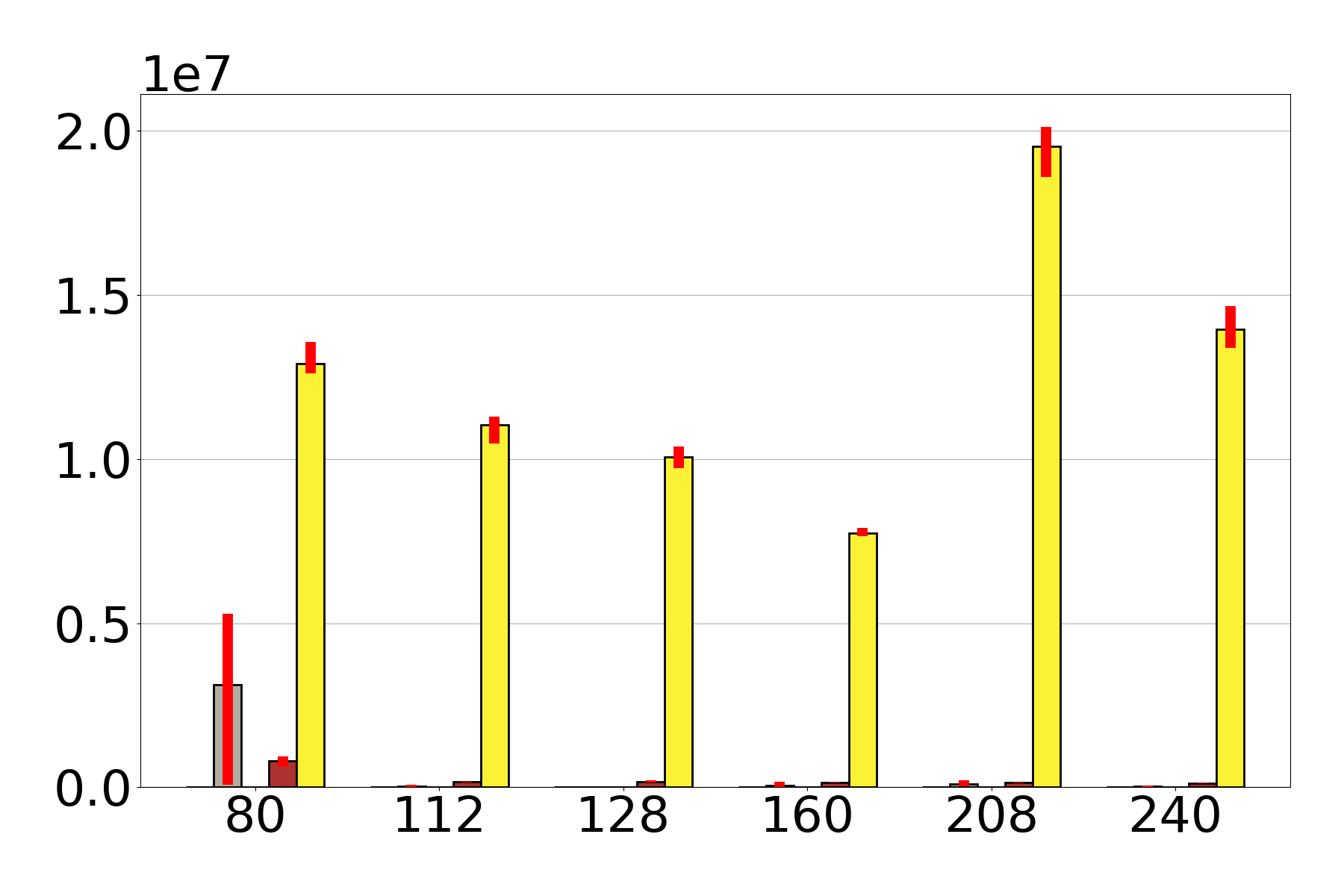}
        \end{subfigure}
        \begin{subfigure}{0.32\linewidth}
            \centering    
            \includegraphics[width=1\linewidth]{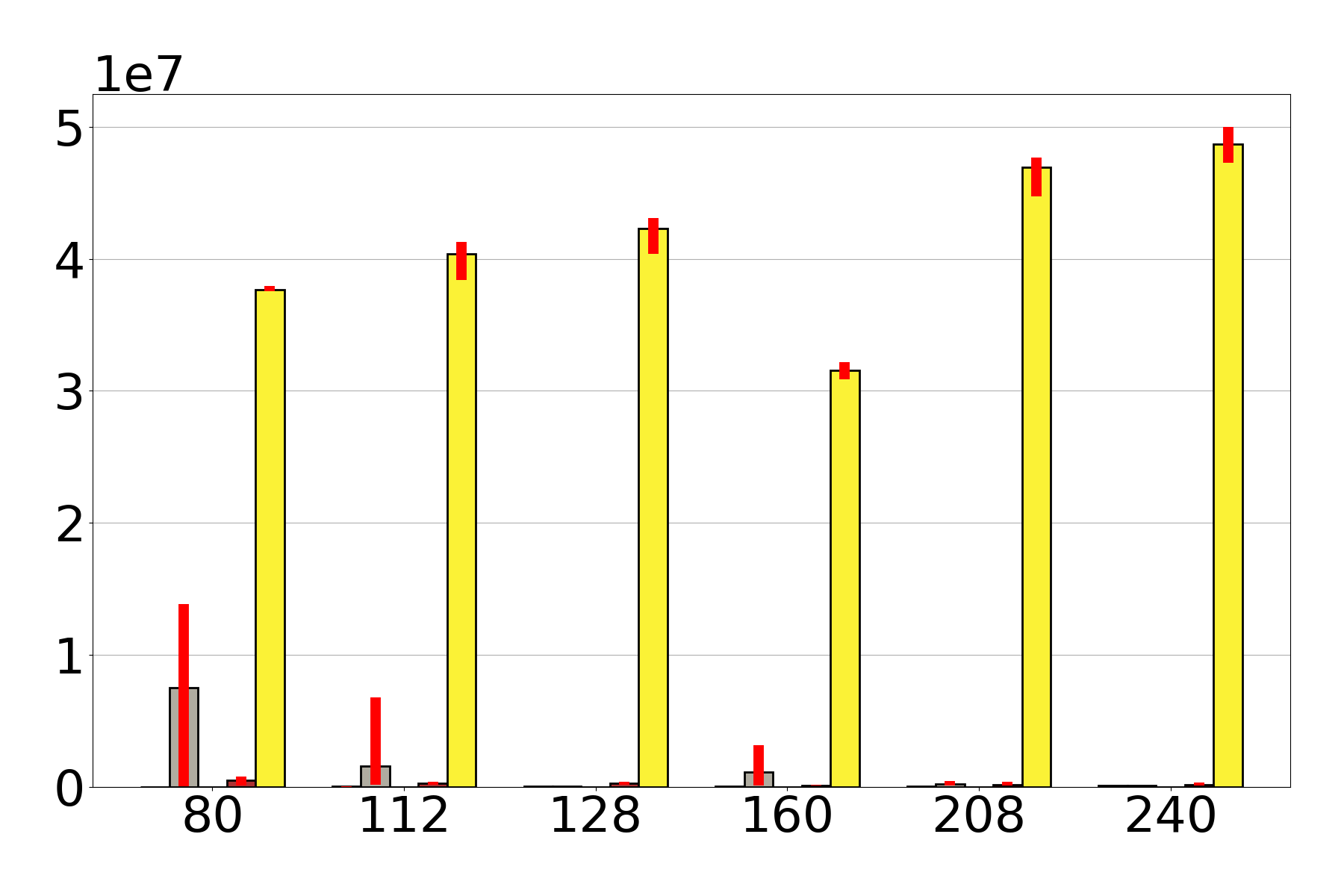}
        \end{subfigure}
    \end{subfigure}
    \begin{subfigure}{1.0\linewidth}
        \centering
        \includegraphics[width=0.4\linewidth]{plots/legend.png}
    \end{subfigure}     
    \vspace{-8mm}
    \caption{\centering Throughput for (a,b)-tree prefilled to 1 million keys using a uniform key access pattern. Y-axis is ops/sec. X-axis is number of threads. 
    All workloads include 5\% insert and 5\% delete. 
    RQ size is 10k (1\% of prefill size). Experiment ran on dual AMD EPYC 7662.    
    }
    \Description{}
    \label{fig:throughput-abtree-2socket}
\end{figure*}

\paragraph{Dual AMD EPYC 7662 Experiments}
We also ran experiments on dual AMD EPYC 7662 processors (see \Cref{fig:throughput-abtree-2socket} and \Cref{fig:throughput-avl}).
Unsurprisingly, we can observe some NUMA effects in the form of a slight throughput reduction for all algorithms when the thread count first reaches an amount that spans both NUMA nodes.

\begin{figure*}[t!]
    \begin{subfigure}{0.02\linewidth}        
        \raisebox{0.5\height}{\rotatebox{90}{0 Updaters}}
    \end{subfigure}
    \begin{subfigure}{0.97\linewidth}
        \begin{subfigure}{0.32\linewidth}
            \centering
            90\% Search, 0\% RQ
            \includegraphics[width=1\linewidth]{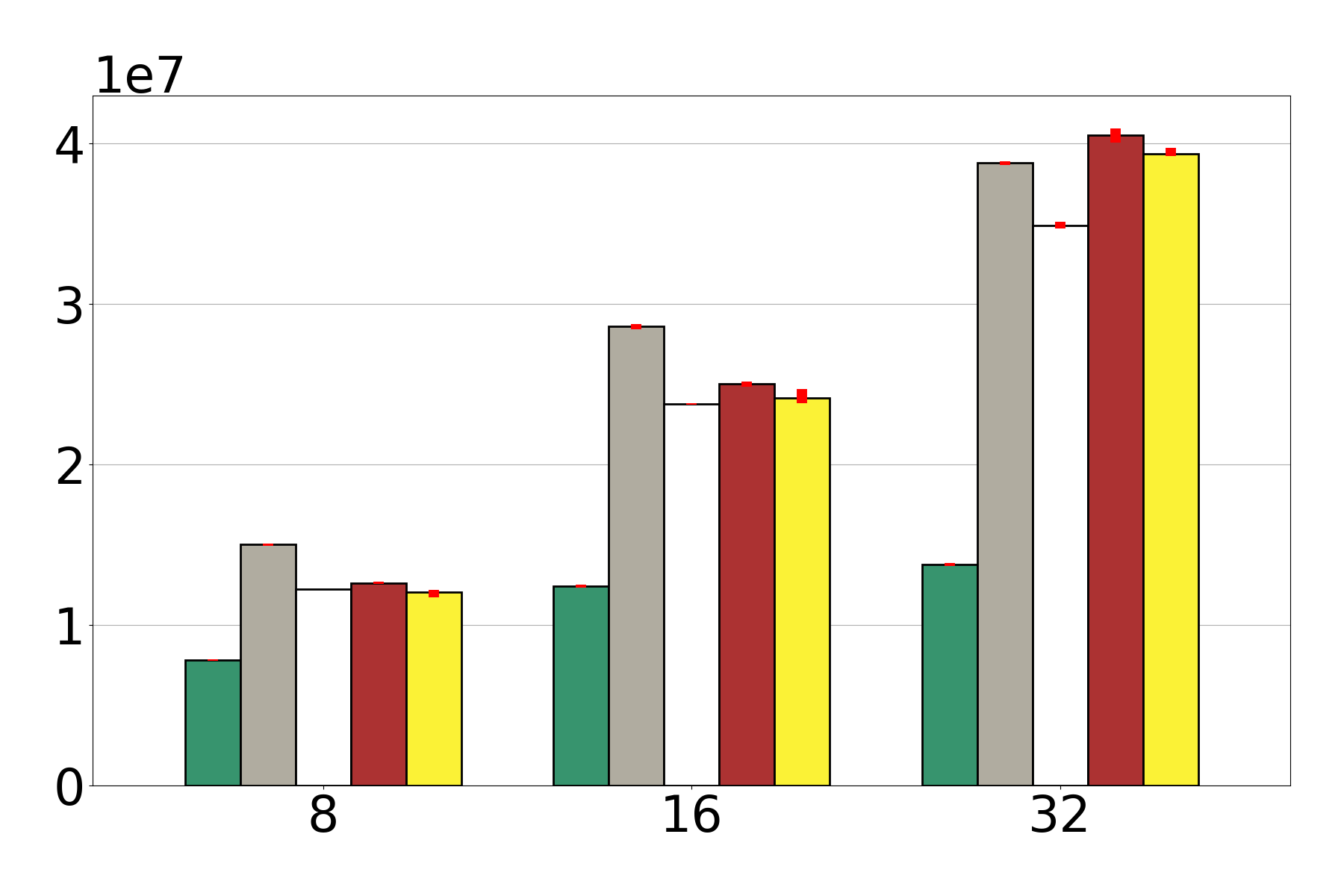}
        \end{subfigure} 
        \begin{subfigure}{0.32\linewidth}
            \centering        
            89.9\% Search, 0.1\% RQ
            \includegraphics[width=1\linewidth]{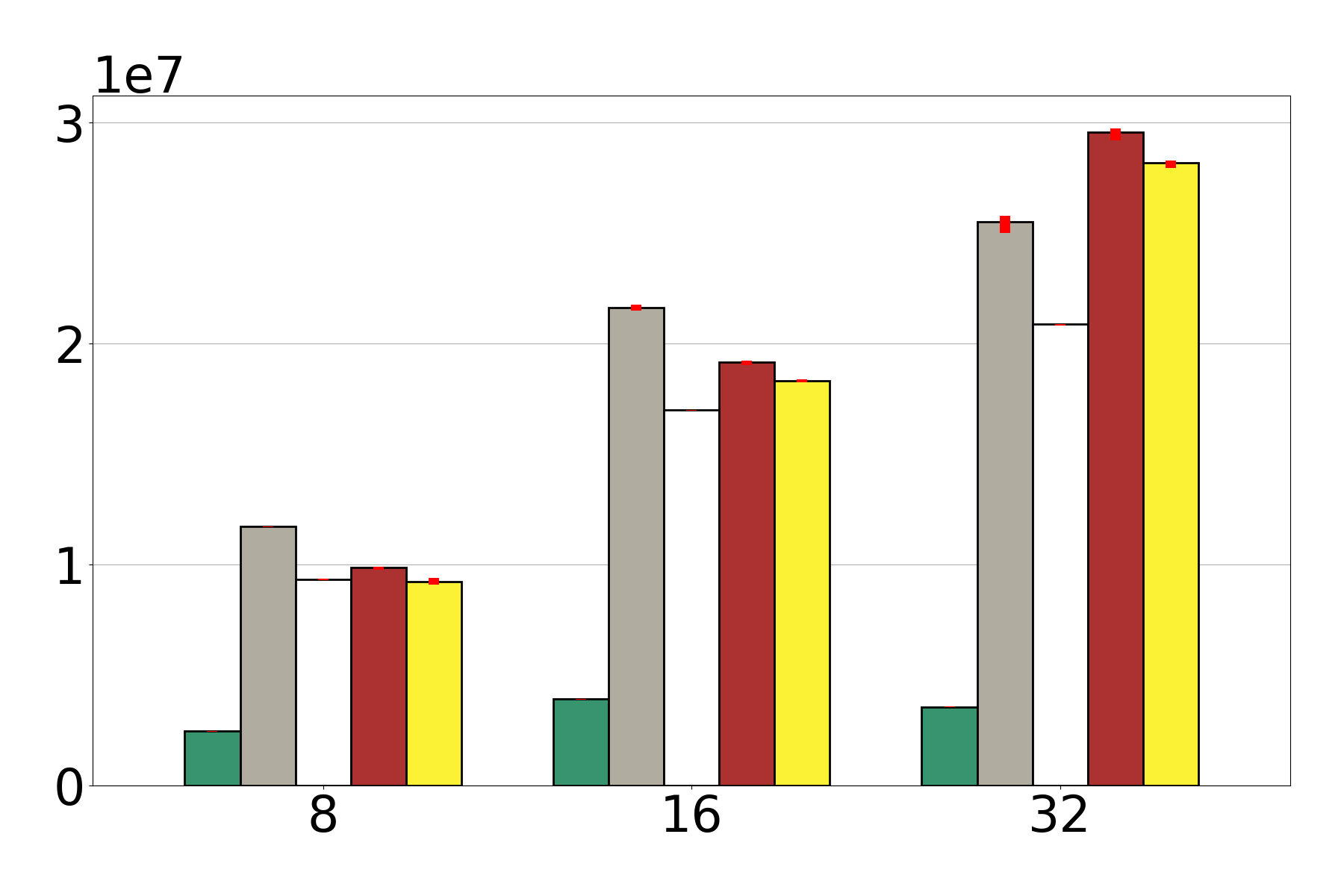}
        \end{subfigure}
        \begin{subfigure}{0.32\linewidth}
            \centering    
            89.99\% Search, 0.01\% RQ
            \includegraphics[width=1\linewidth]{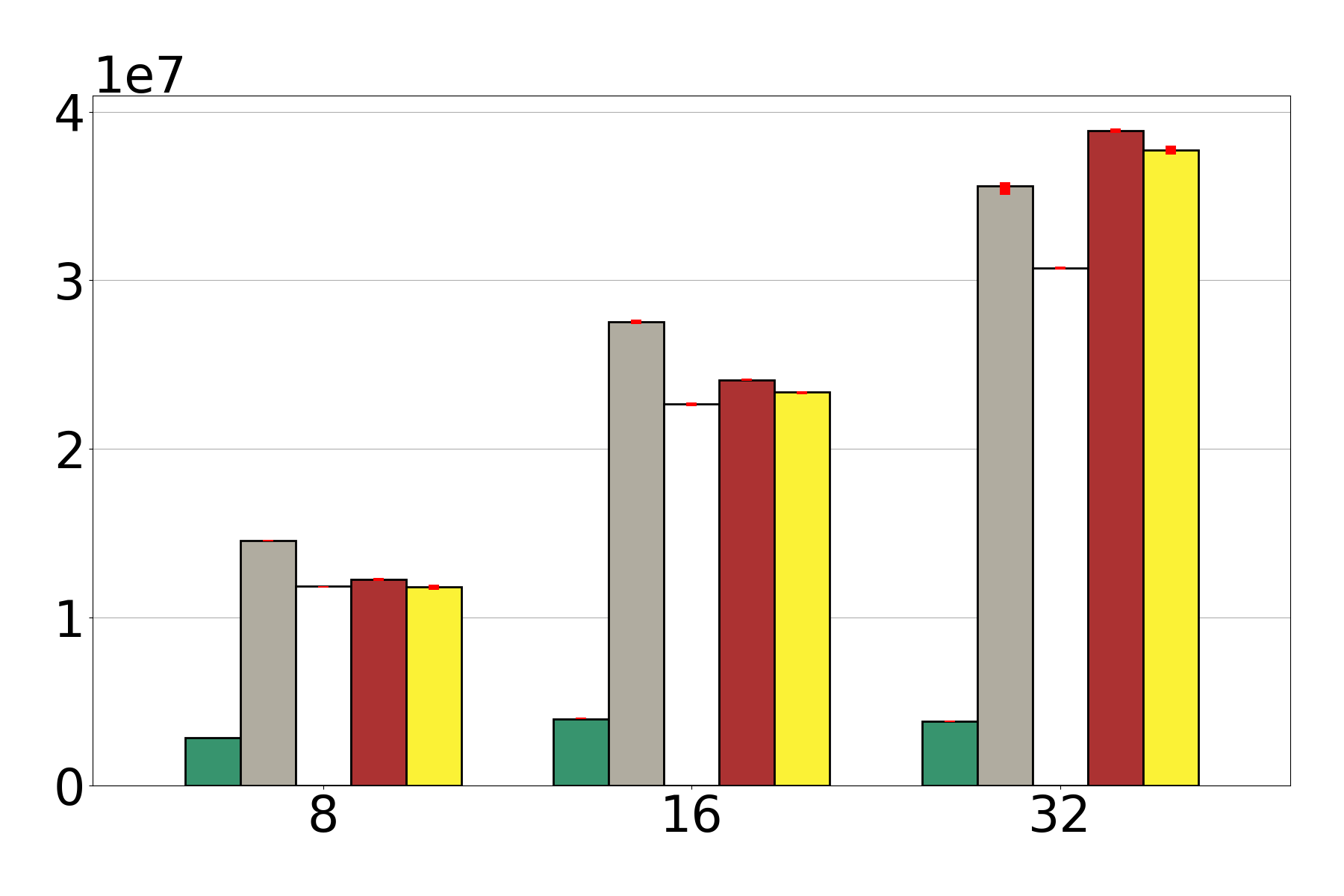}
        \end{subfigure}
    \end{subfigure}
    \begin{subfigure}{0.02\linewidth}
        \raisebox{0.5\height}{\rotatebox{90}{8 Updaters}}
    \end{subfigure}
    \begin{subfigure}{0.97\linewidth}
        \begin{subfigure}{0.32\linewidth}
            \centering
            \includegraphics[width=1\linewidth]{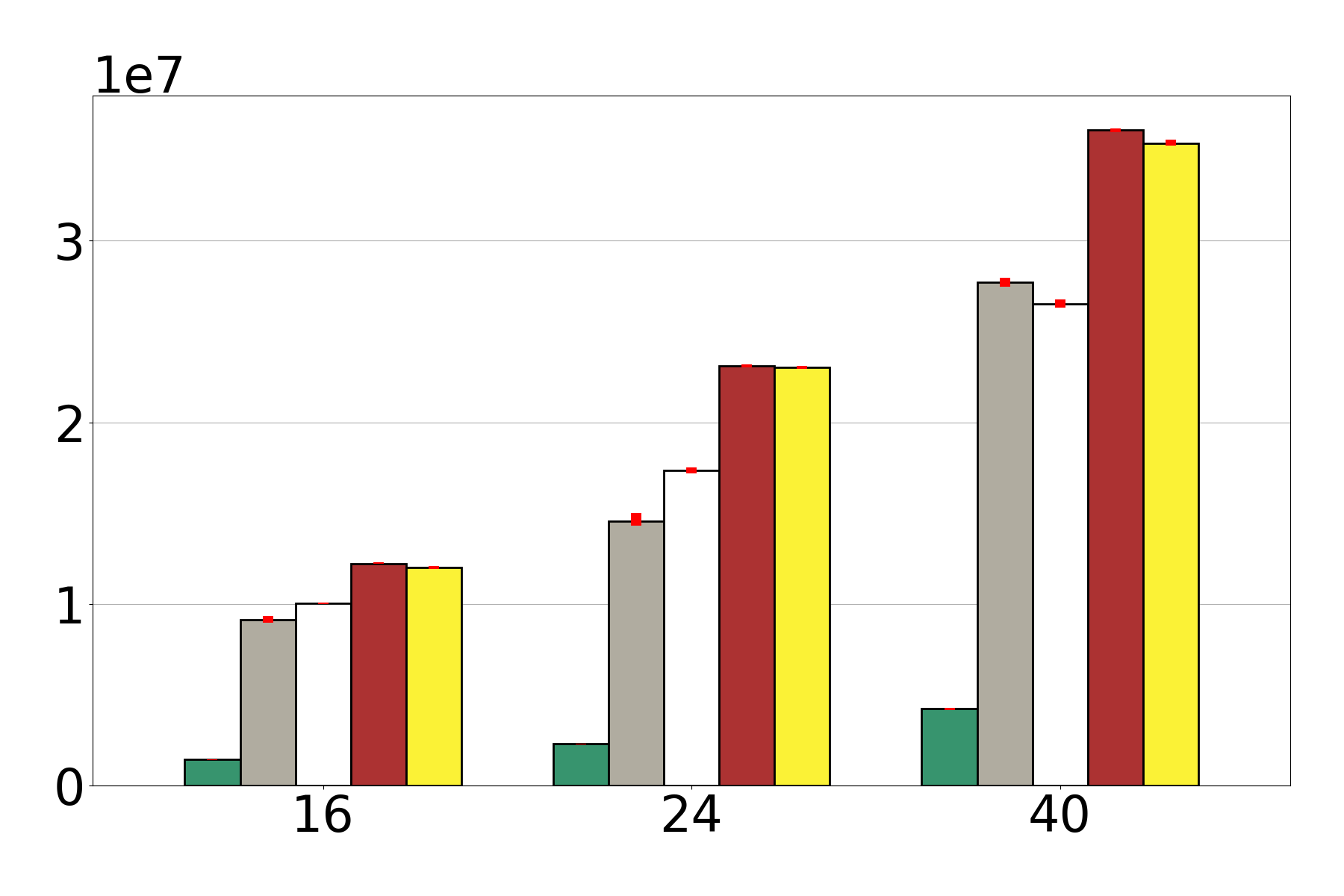}
        \end{subfigure} 
        \begin{subfigure}{0.32\linewidth}
            \centering        
            \includegraphics[width=1\linewidth]{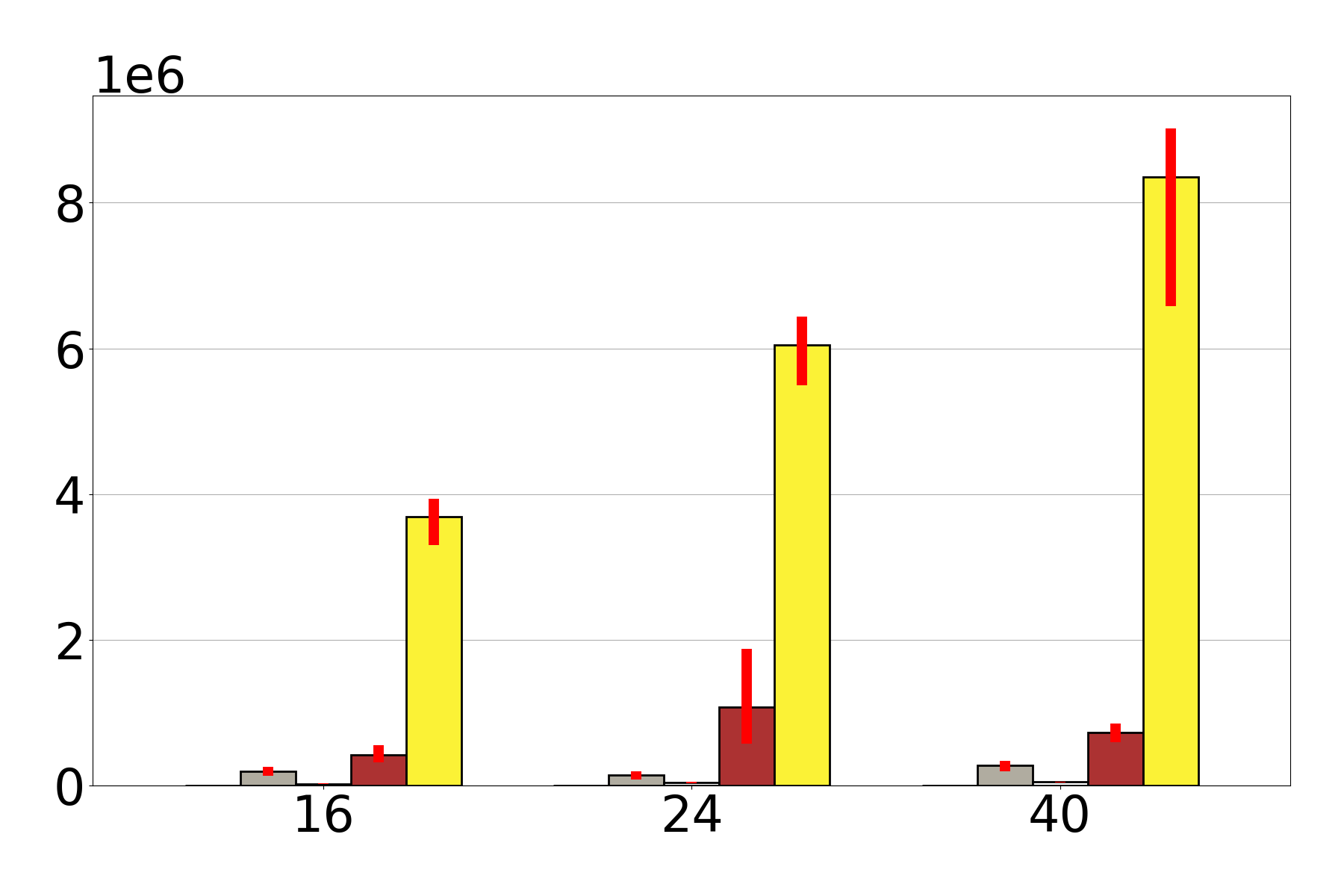}
        \end{subfigure}
        \begin{subfigure}{0.32\linewidth}
            \centering    
            \includegraphics[width=1\linewidth]{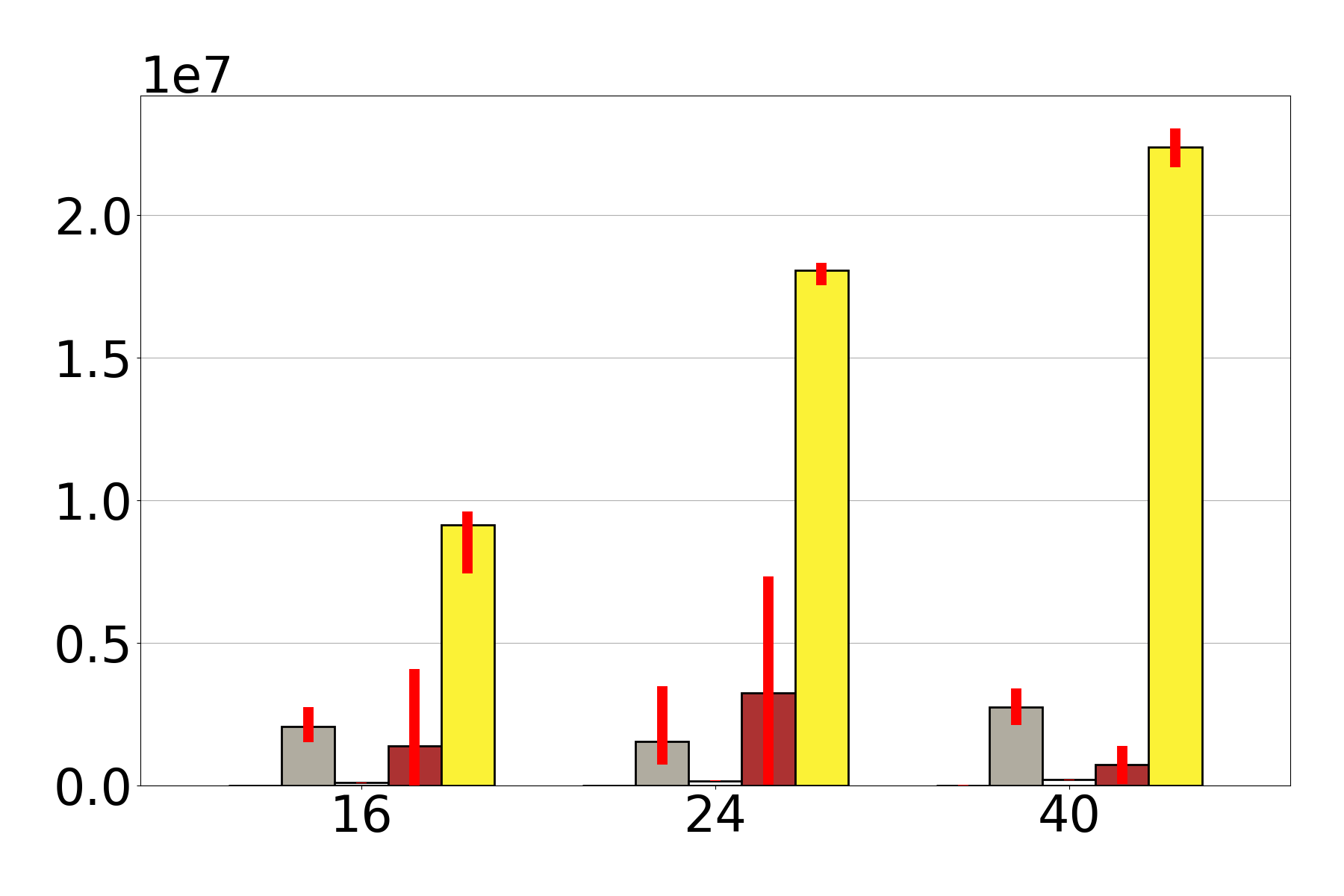}
        \end{subfigure}
    \end{subfigure} 
    \begin{subfigure}{0.02\linewidth}
        \raisebox{0.5\height}{\rotatebox{90}{16 Updaters}}
    \end{subfigure}
    \begin{subfigure}{0.97\linewidth}
        \begin{subfigure}{0.32\linewidth}
            \centering
            \includegraphics[width=1\linewidth]{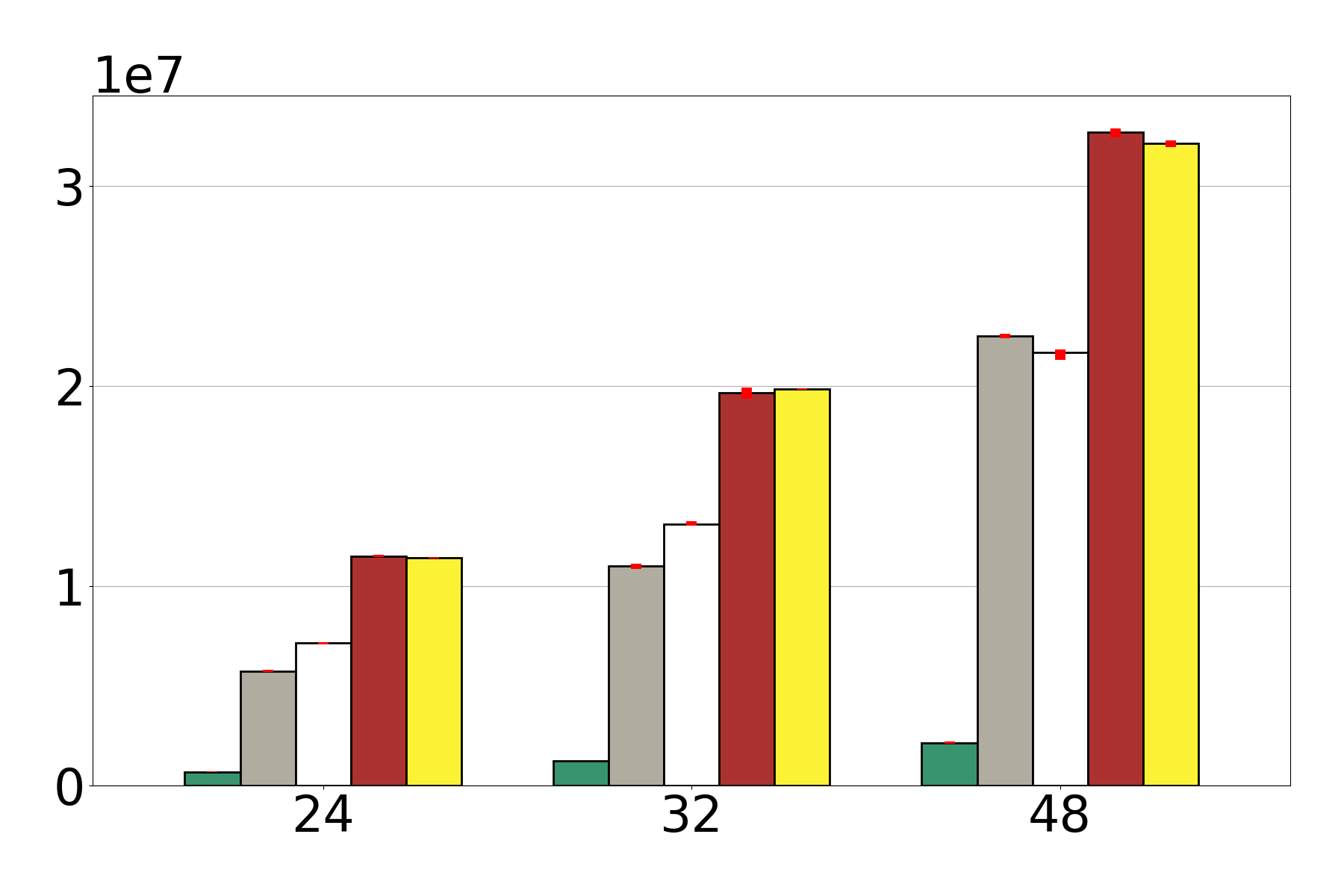}
        \end{subfigure} 
        \begin{subfigure}{0.32\linewidth}
            \centering        
            \includegraphics[width=1\linewidth]{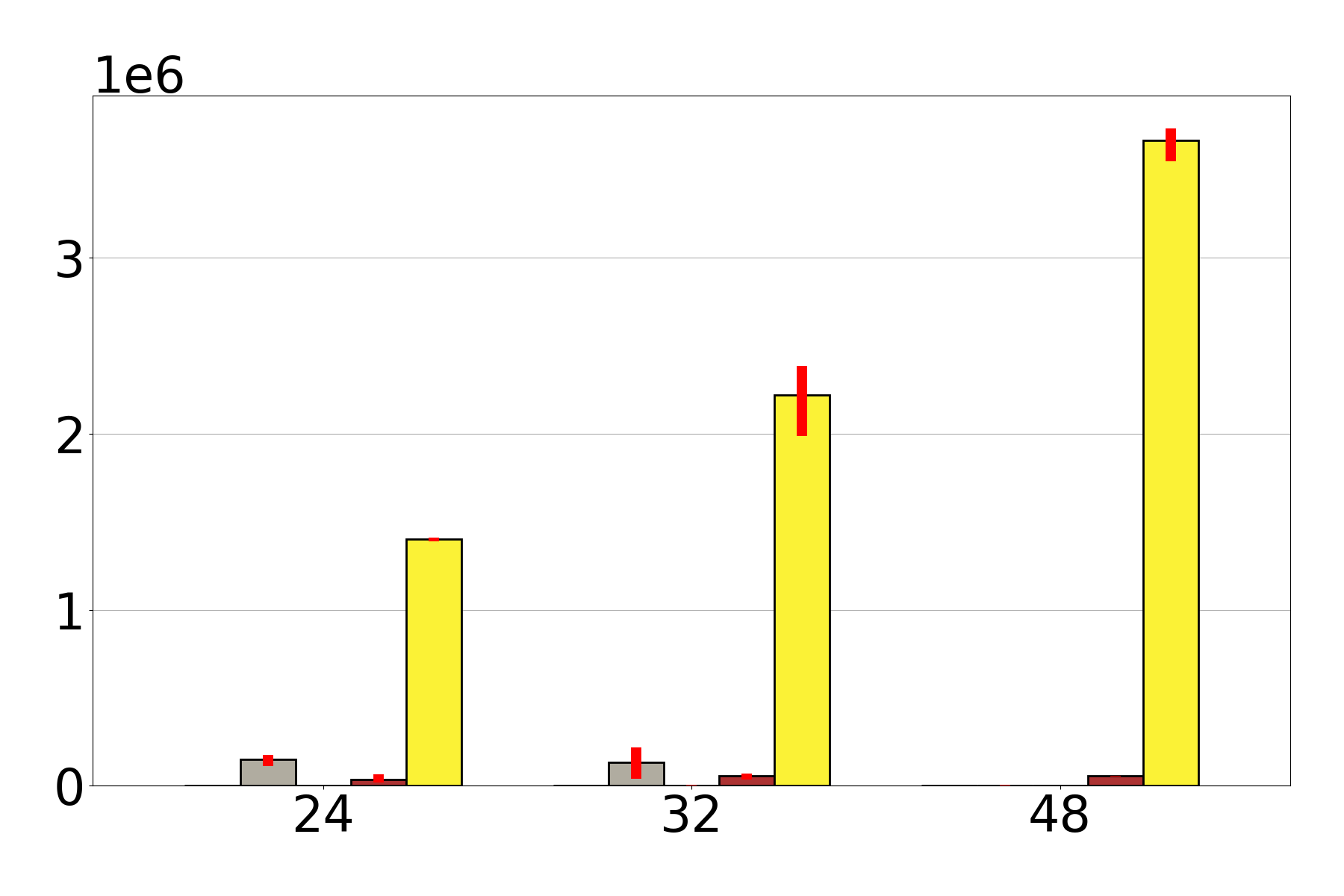}
        \end{subfigure}
        \begin{subfigure}{0.32\linewidth}
            \centering    
            \includegraphics[width=1\linewidth]{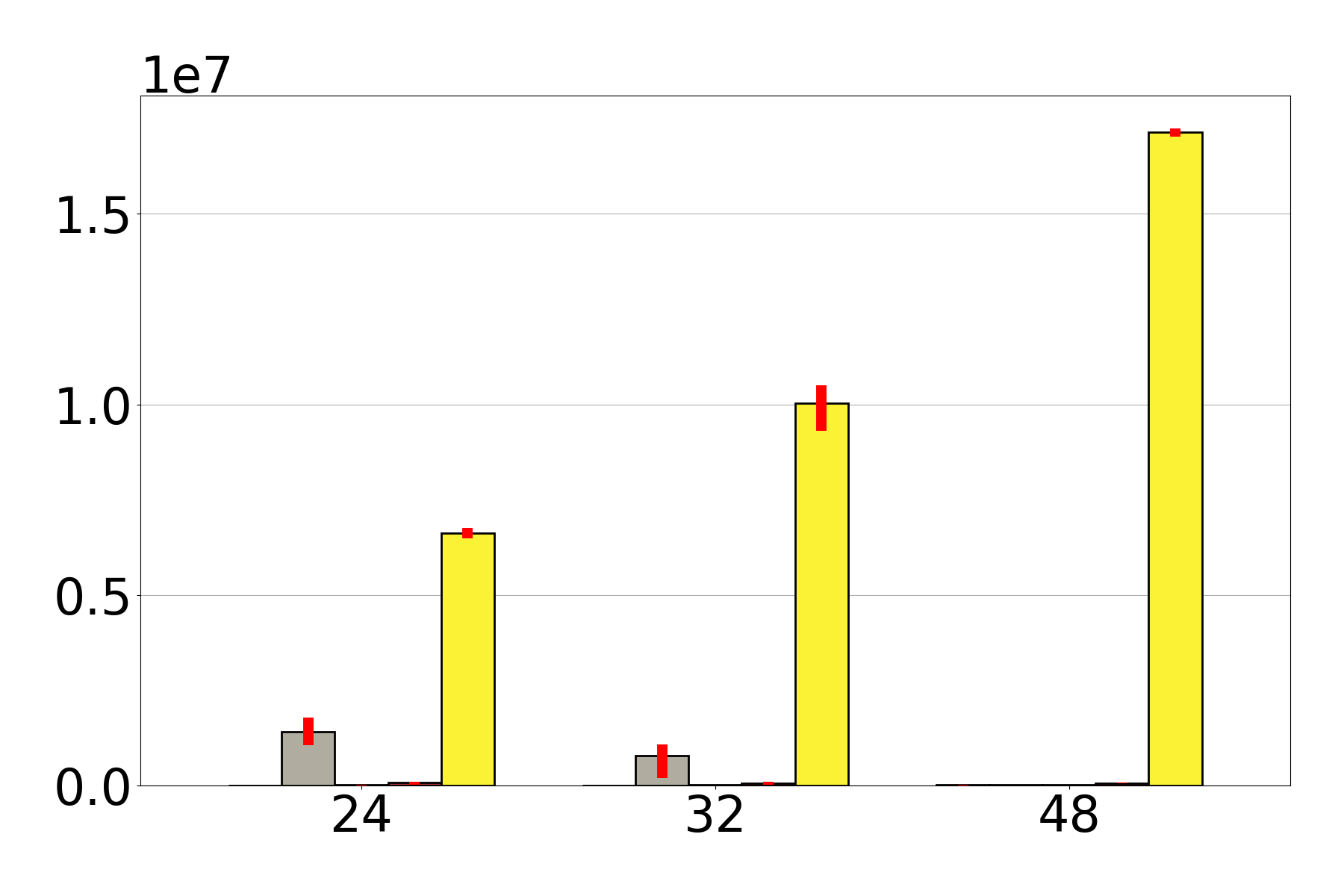}
        \end{subfigure}
    \end{subfigure}
    \begin{subfigure}{1.0\linewidth}
        \centering
        \includegraphics[width=0.4\linewidth]{plots/legend.png}
    \end{subfigure}     
    \vspace{-8mm}
    \caption{
    \centering Throughput for (a,b)-tree prefilled to 1 million keys using a uniform key access pattern. Y-axis is ops/sec. X-axis is number of threads. All workloads include 5\% insert and 5\% delete. RQ size is 10k (1\% of prefill size). Experiment ran on a single Intel Xeon Platinum 8160.
    }
    \Description{}
    \label{fig:jax-abtree-node0}
\end{figure*}

\begin{figure*}[t!]
    \begin{subfigure}{0.02\linewidth}        
        \raisebox{0.5\height}{\rotatebox{90}{0 Updaters}}
    \end{subfigure}
    \begin{subfigure}{0.97\linewidth}
        \begin{subfigure}{0.32\linewidth}
            \centering
            90\% Search, 0\% RQ
            \includegraphics[width=1\linewidth]{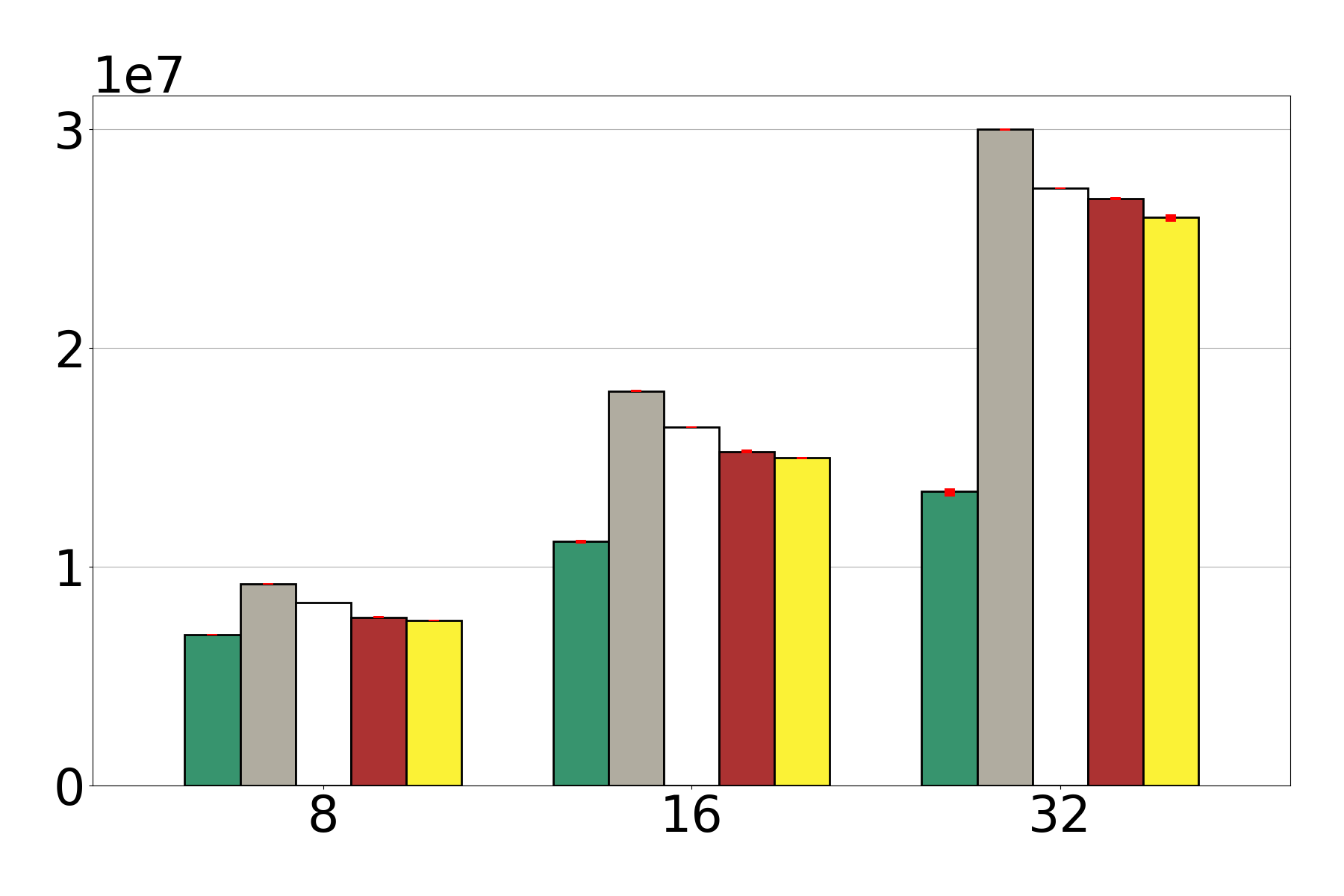}
        \end{subfigure} 
        \begin{subfigure}{0.32\linewidth}
            \centering        
            89.9\% Search, 0.1\% RQ
            \includegraphics[width=1\linewidth]{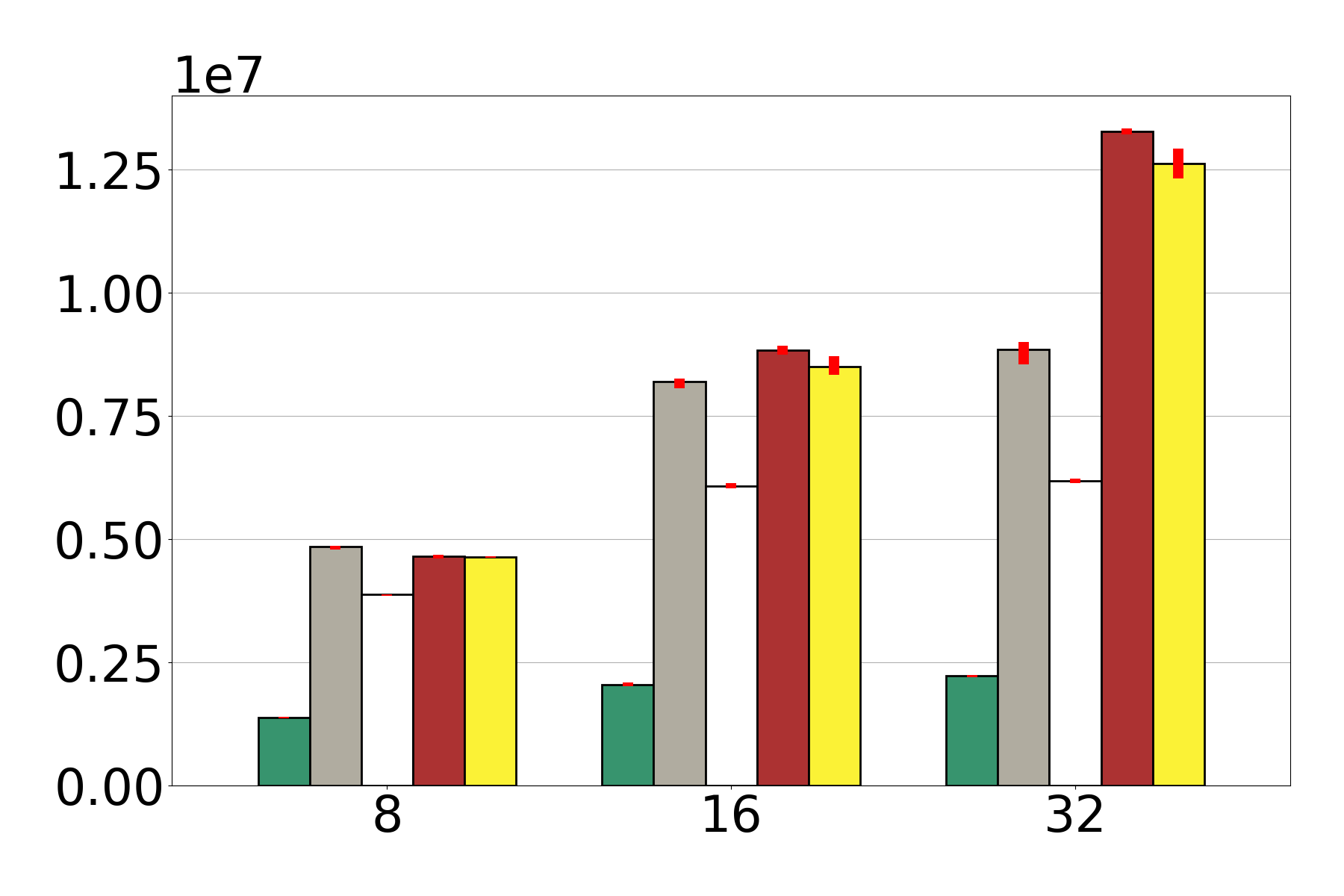}
        \end{subfigure}
        \begin{subfigure}{0.32\linewidth}
            \centering    
            89.99\% Search, 0.01\% RQ
            \includegraphics[width=1\linewidth]{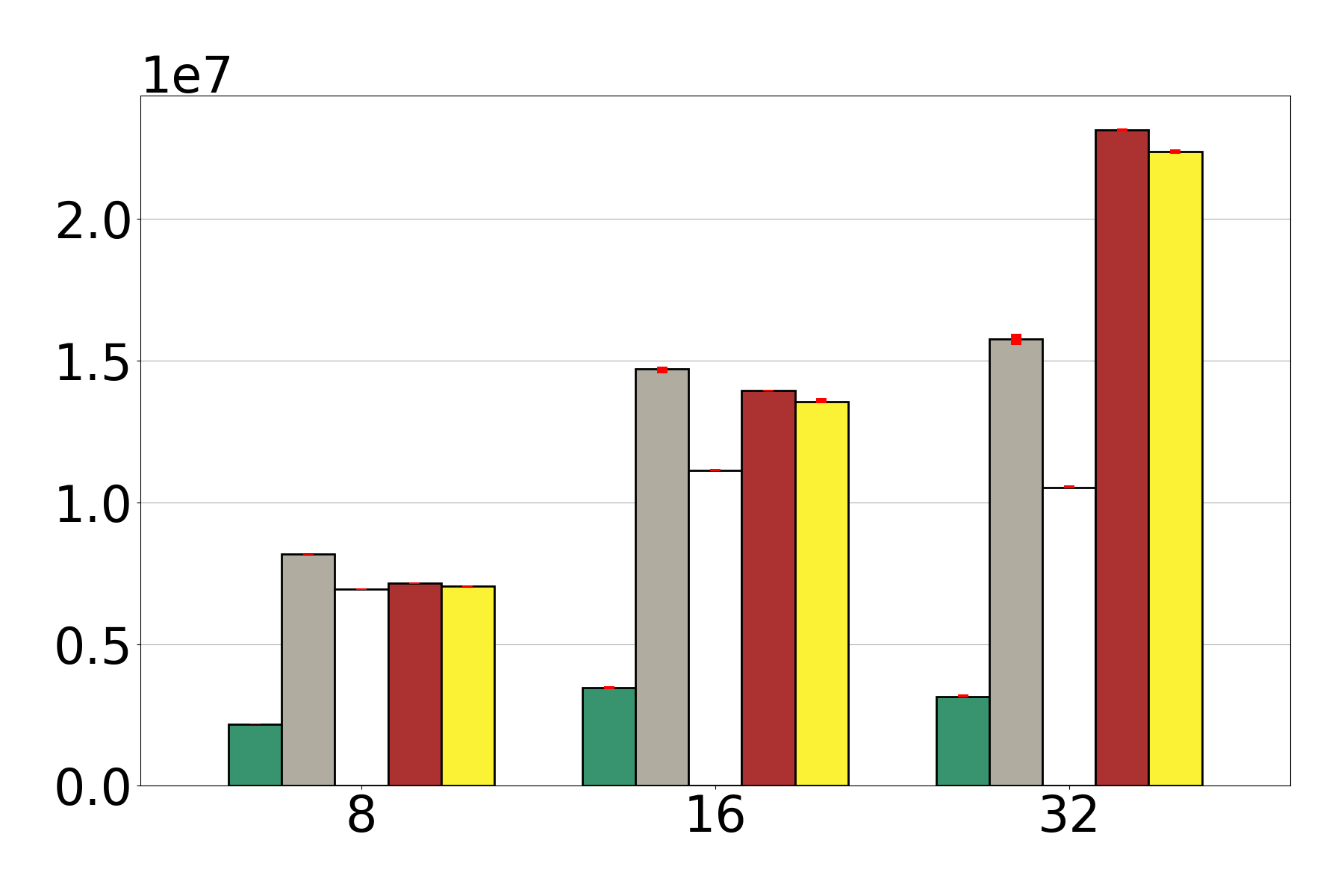}
        \end{subfigure}
    \end{subfigure}
    \begin{subfigure}{0.02\linewidth}
        \raisebox{0.5\height}{\rotatebox{90}{8 Updaters}}
    \end{subfigure}
    \begin{subfigure}{0.97\linewidth}
        \begin{subfigure}{0.32\linewidth}
            \centering
            \includegraphics[width=1\linewidth]{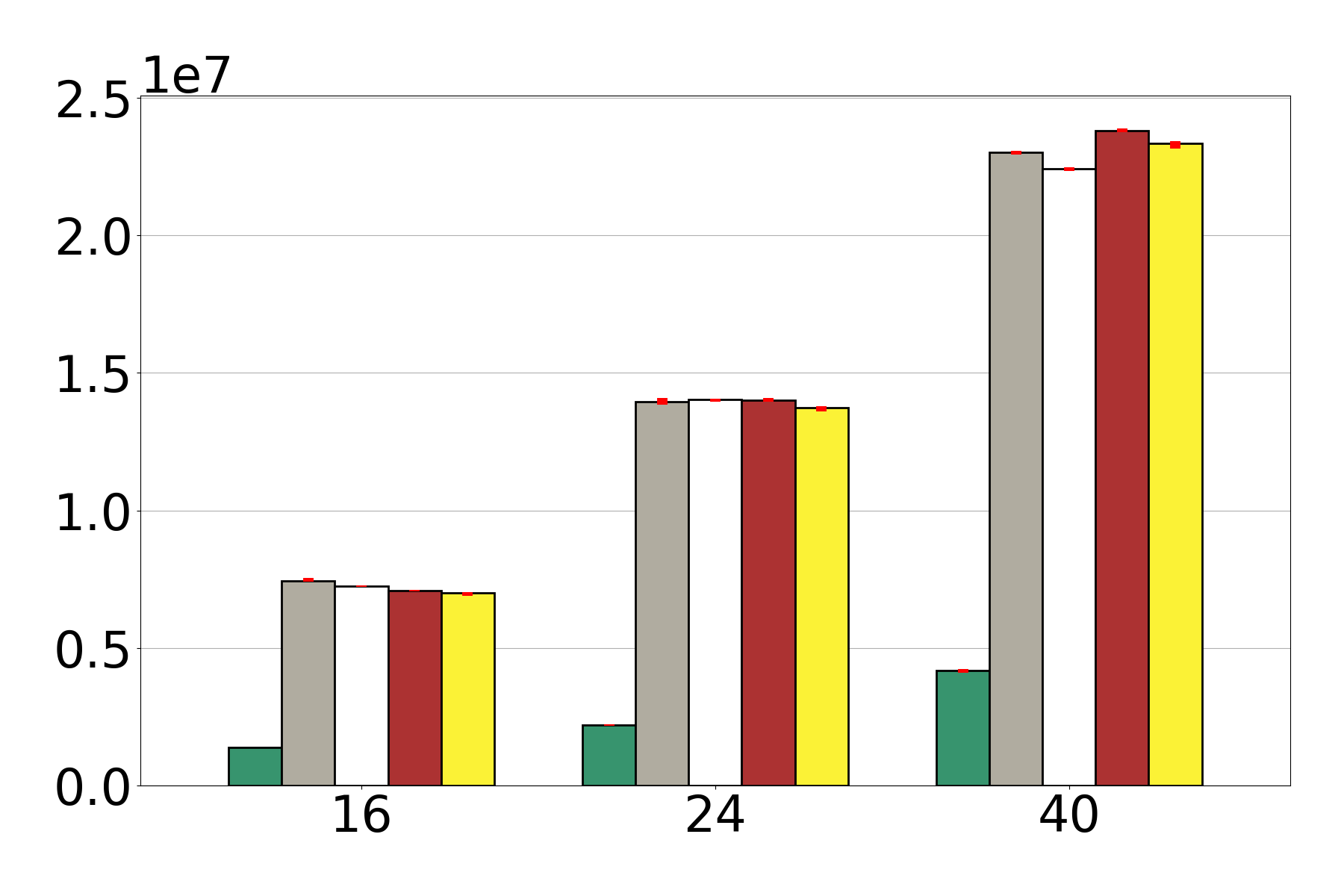}
        \end{subfigure} 
        \begin{subfigure}{0.32\linewidth}
            \centering        
            \includegraphics[width=1\linewidth]{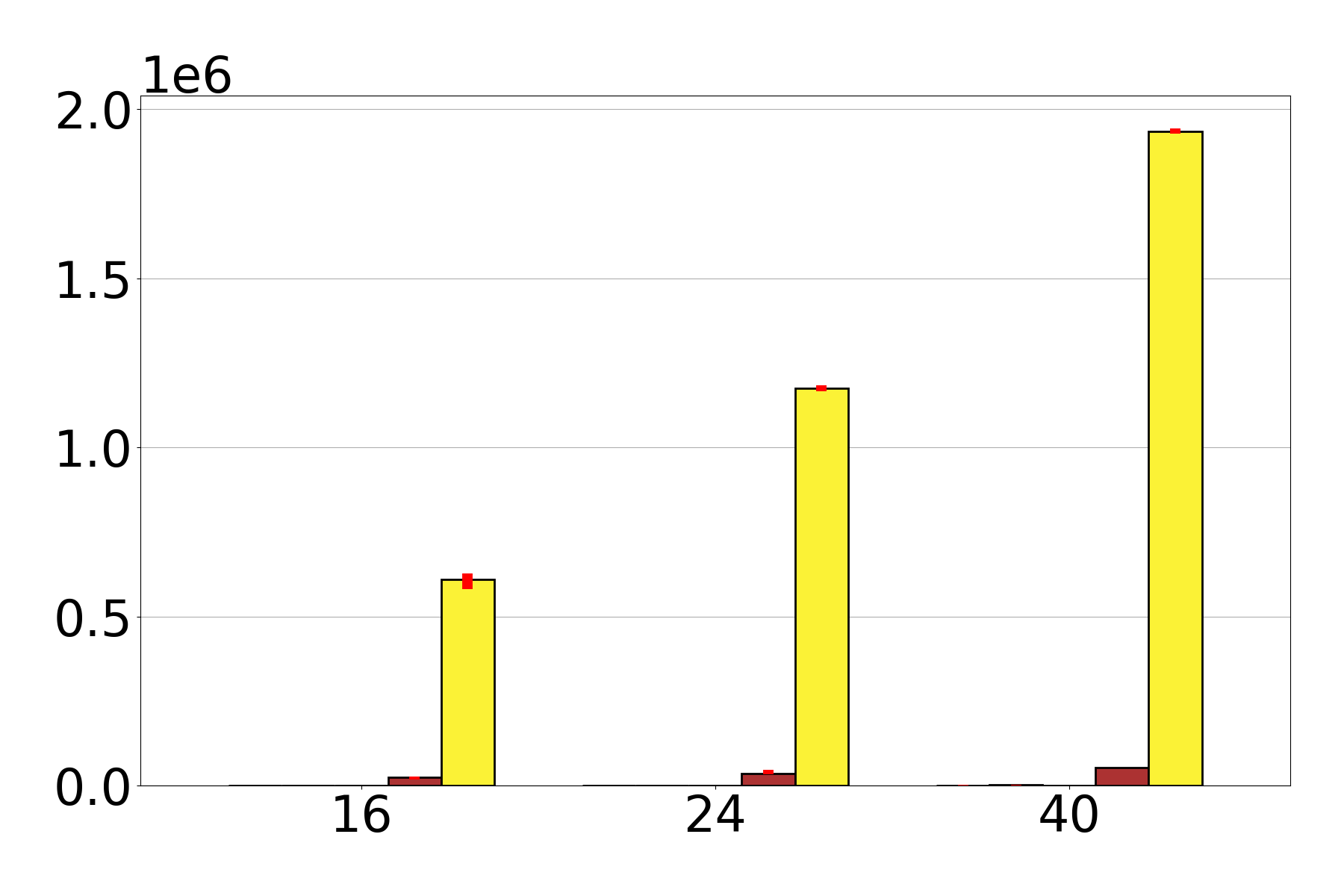}
        \end{subfigure}
        \begin{subfigure}{0.32\linewidth}
            \centering    
            \includegraphics[width=1\linewidth]{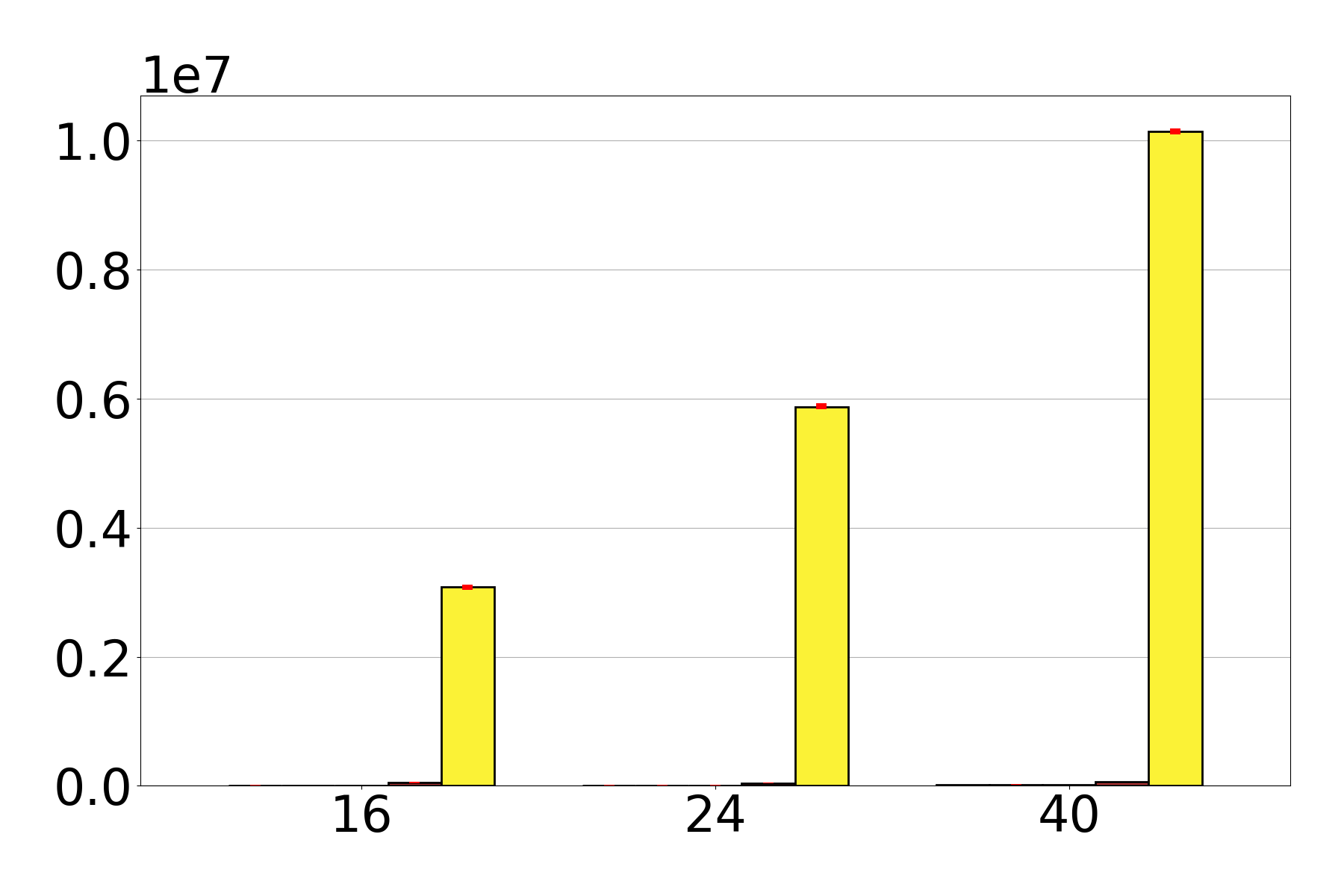}
        \end{subfigure}
    \end{subfigure} 
    \begin{subfigure}{0.02\linewidth}
        \raisebox{0.5\height}{\rotatebox{90}{16 Updaters}}
    \end{subfigure}
    \begin{subfigure}{0.97\linewidth}
        \begin{subfigure}{0.32\linewidth}
            \centering
            \includegraphics[width=1\linewidth]{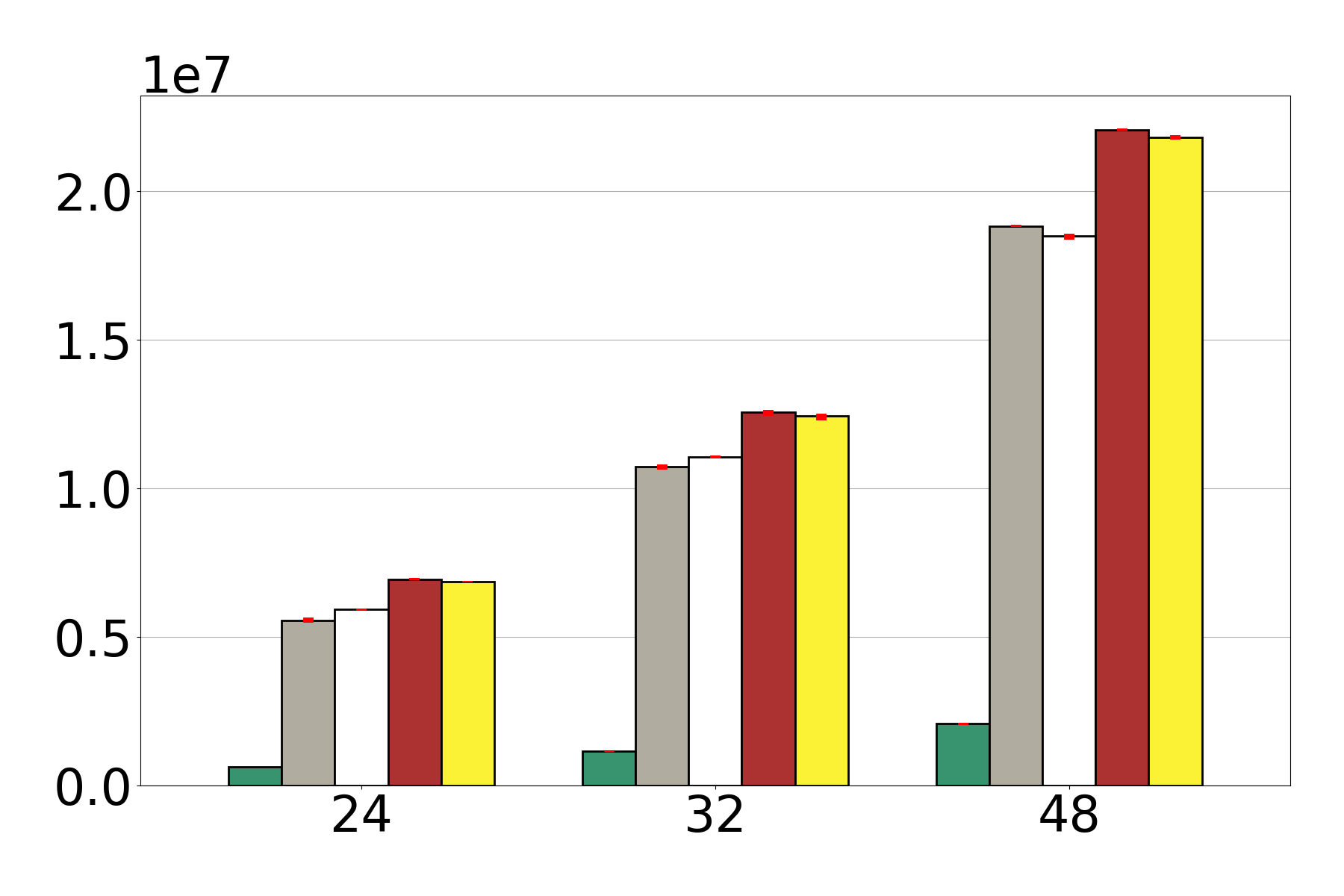}
        \end{subfigure} 
        \begin{subfigure}{0.32\linewidth}
            \centering        
            \includegraphics[width=1\linewidth]{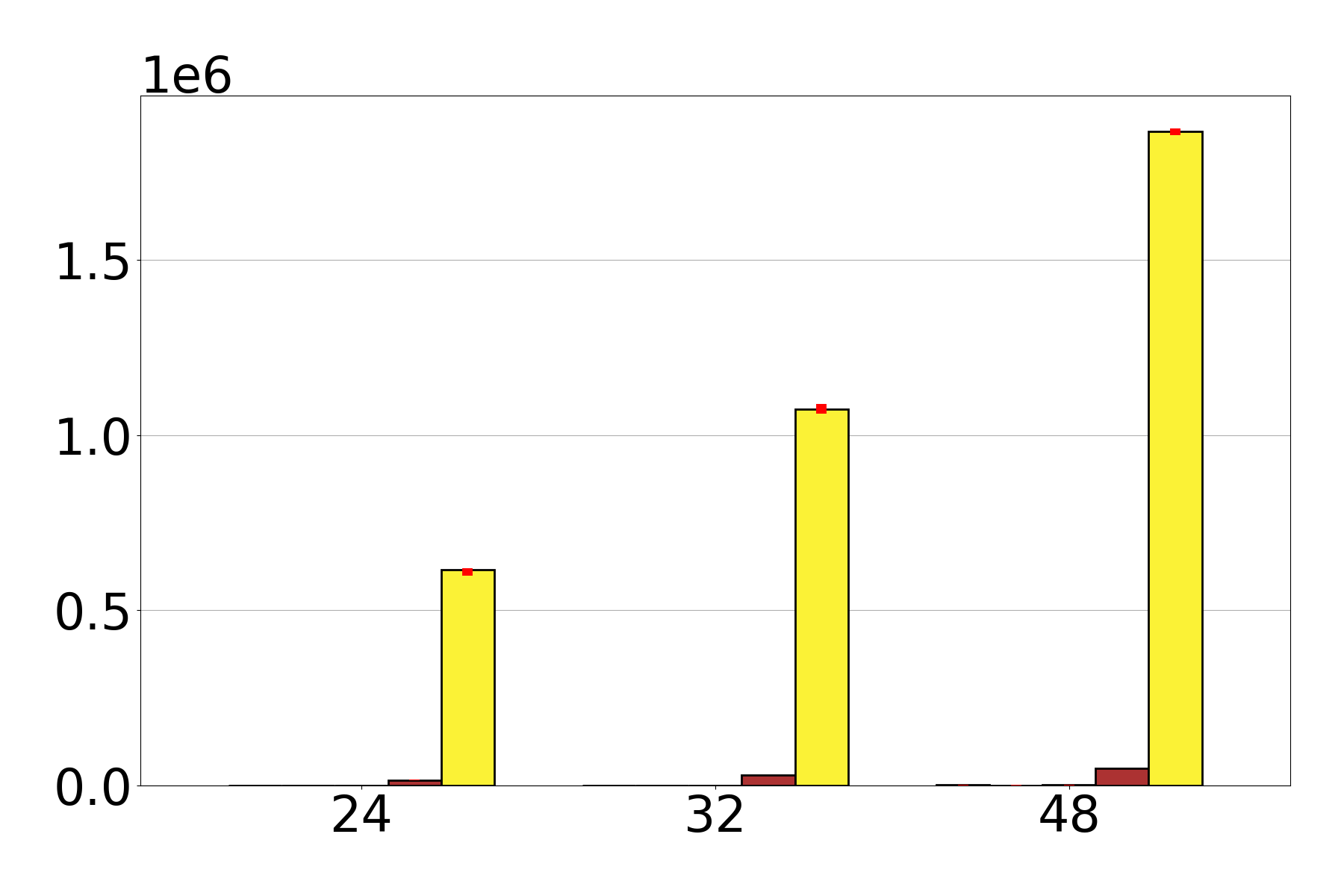}
        \end{subfigure}
        \begin{subfigure}{0.32\linewidth}
            \centering    
            \includegraphics[width=1\linewidth]{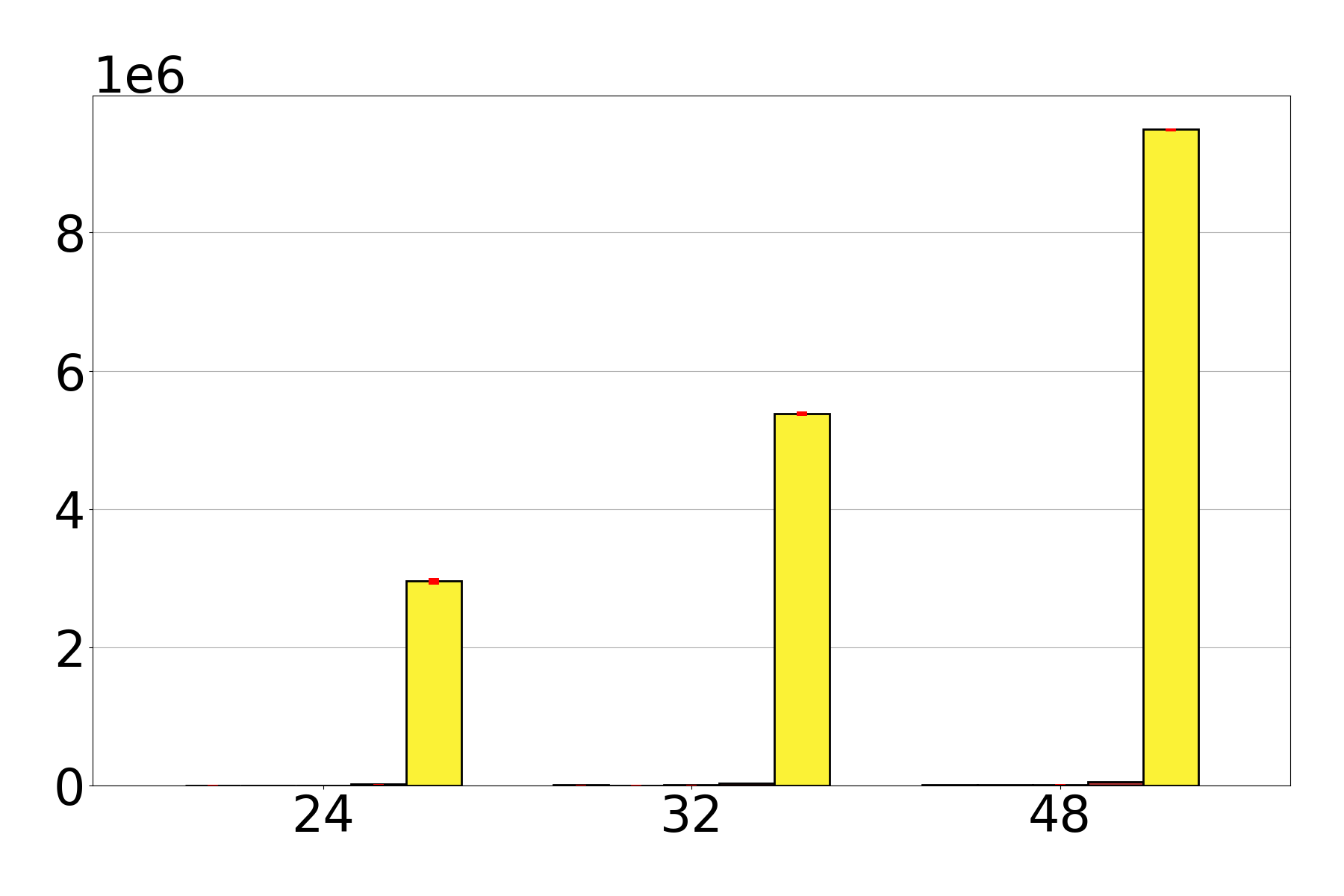}
        \end{subfigure}
    \end{subfigure}
    \begin{subfigure}{1.0\linewidth}
        \centering
        \includegraphics[width=0.4\linewidth]{plots/legend.png}
    \end{subfigure}     
    \vspace{-8mm}
    \caption{
    \centering Throughput for AVL-tree prefilled to 1 million keys using a uniform key access pattern. Y-axis is ops/sec. X-axis is number of threads. All workloads include 5\% insert and 5\% delete. RQ size is 10k (1\% of prefill size). Experiment ran on a single Intel Xeon Platinum 8160.
    }
    \Description{}
    \label{fig:jax-avl-node0}
\end{figure*}

\begin{figure*}[t!]
    \begin{subfigure}{0.02\linewidth}        
        \raisebox{0.5\height}{\rotatebox{90}{0 Updaters}}
    \end{subfigure}
    \begin{subfigure}{0.97\linewidth}
        \begin{subfigure}{0.32\linewidth}
            \centering
            90\% Search, 0\% RQ
            \includegraphics[width=1\linewidth]{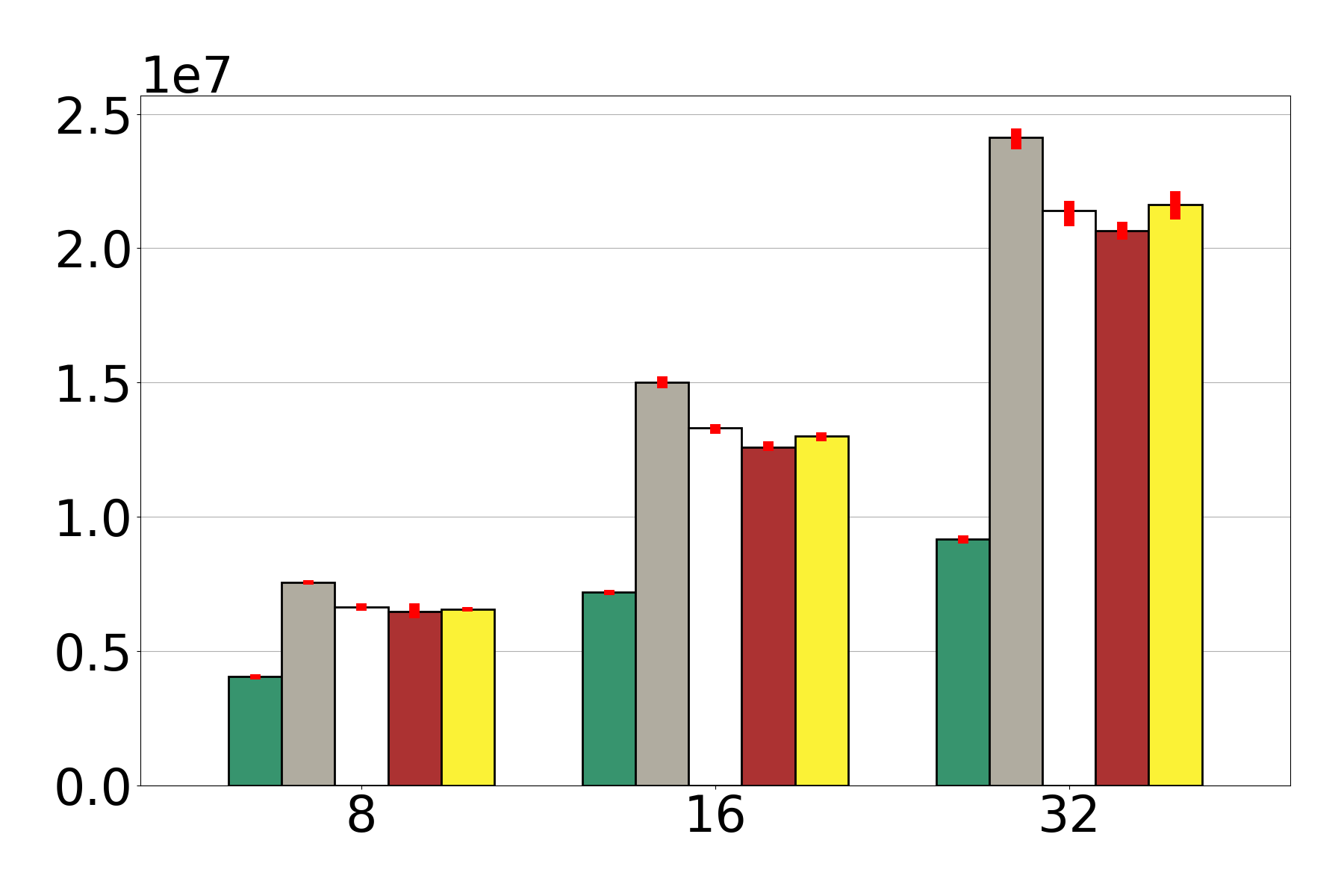}
        \end{subfigure} 
        \begin{subfigure}{0.32\linewidth}
            \centering        
            89.9\% Search, 0.1\% RQ
            \includegraphics[width=1\linewidth]{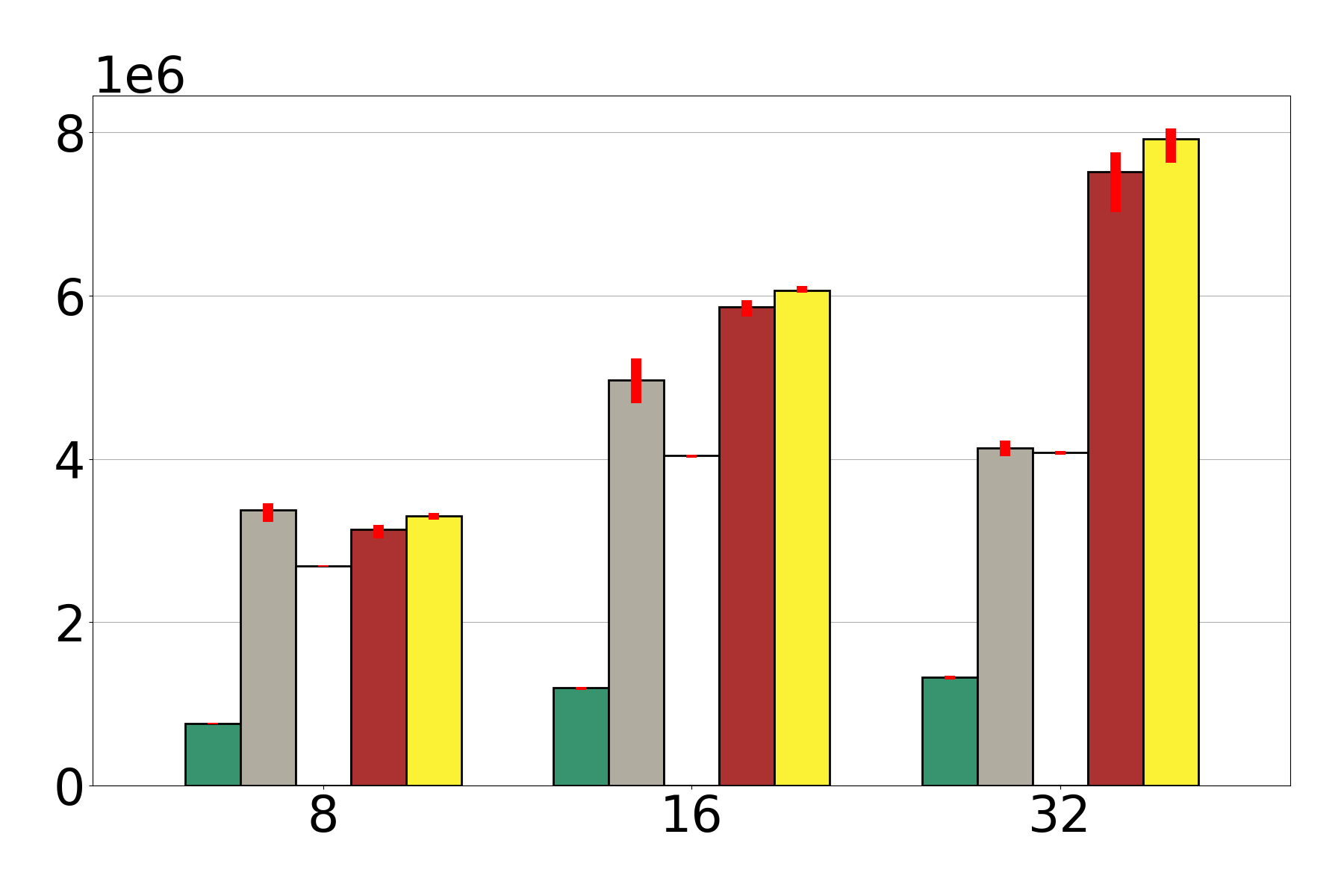}
        \end{subfigure}
        \begin{subfigure}{0.32\linewidth}
            \centering    
            89.99\% Search, 0.01\% RQ
            \includegraphics[width=1\linewidth]{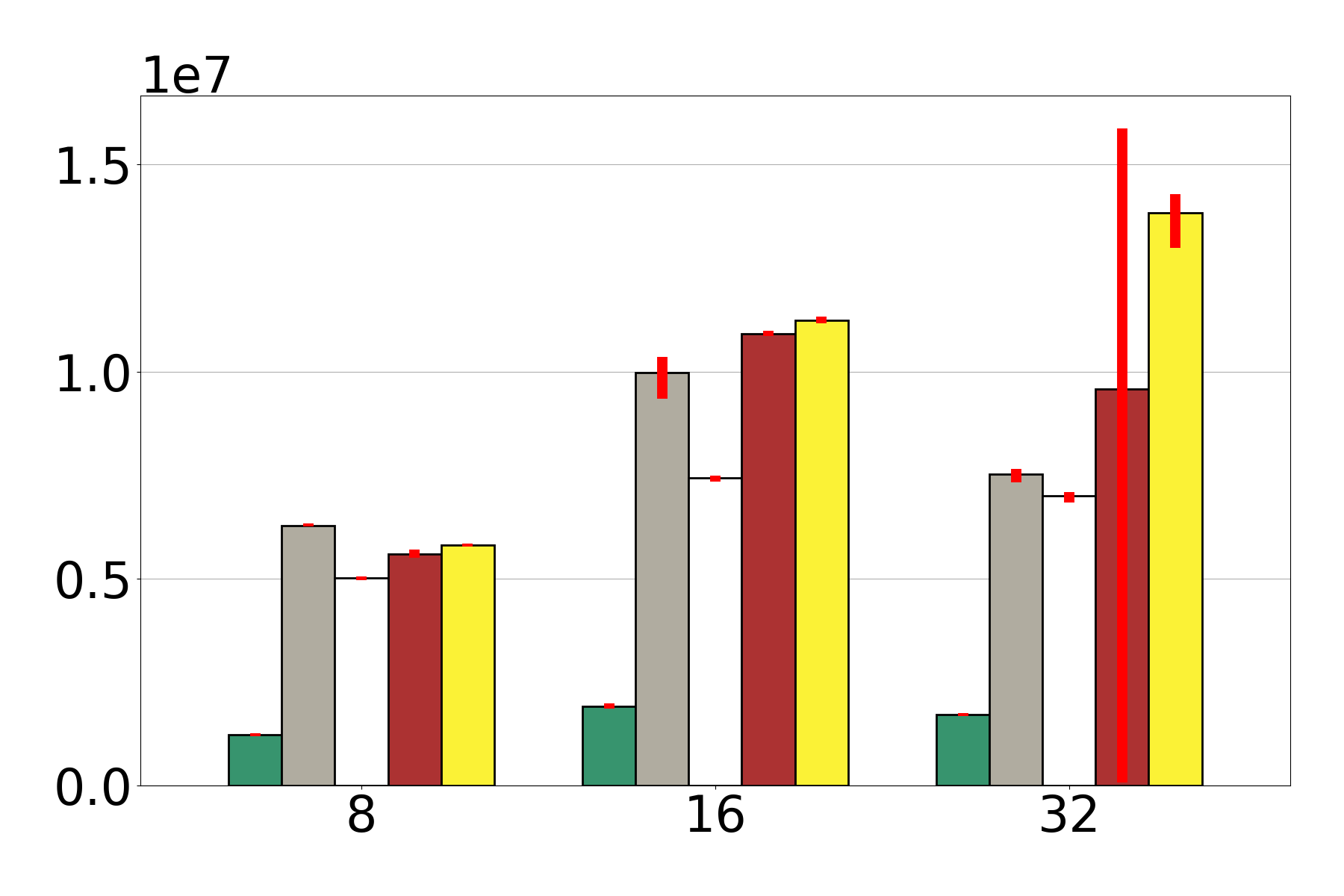}
        \end{subfigure}
    \end{subfigure}
    \begin{subfigure}{0.02\linewidth}
        \raisebox{0.5\height}{\rotatebox{90}{8 Updaters}}
    \end{subfigure}
    \begin{subfigure}{0.97\linewidth}
        \begin{subfigure}{0.32\linewidth}
            \centering
            \includegraphics[width=1\linewidth]{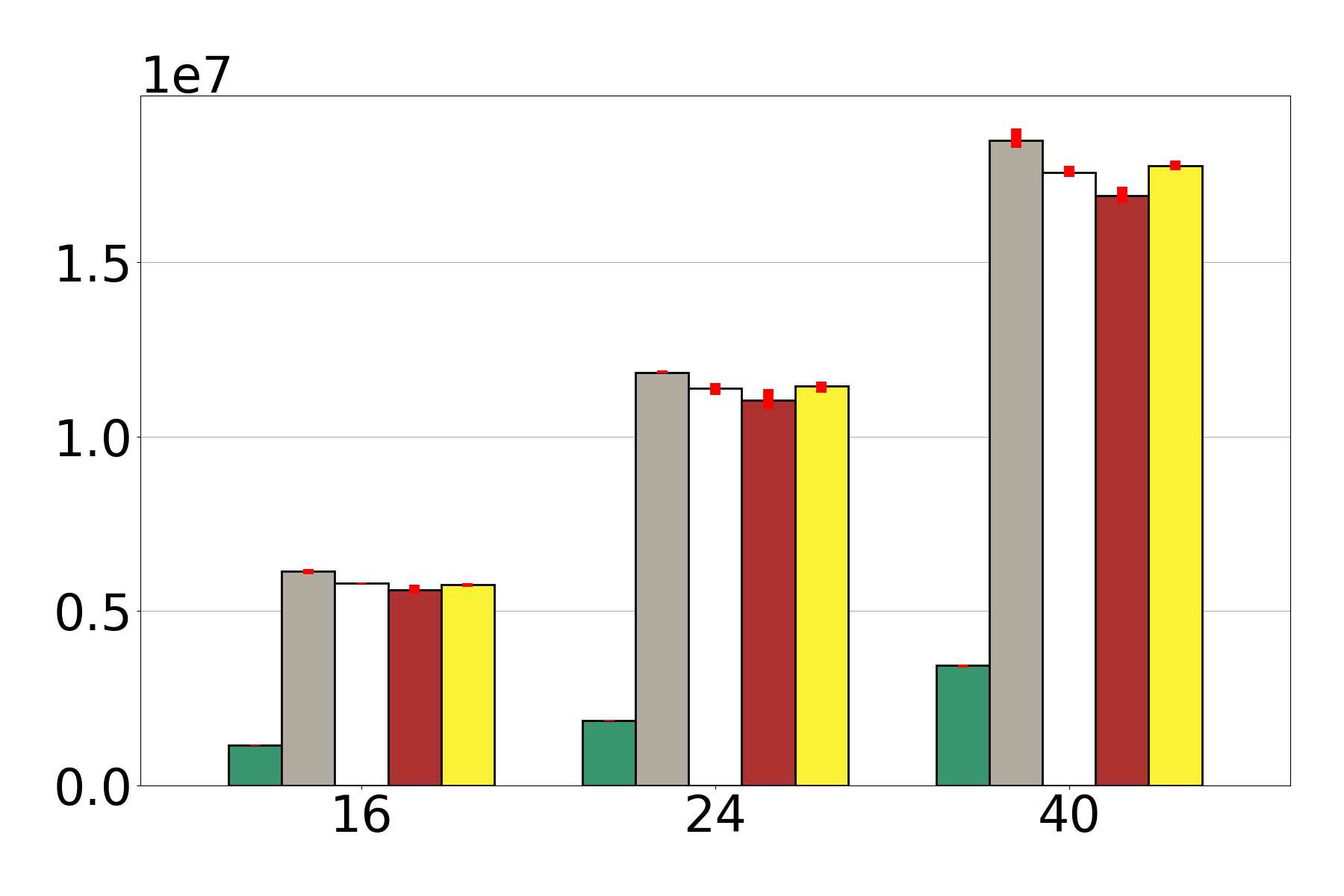}
        \end{subfigure} 
        \begin{subfigure}{0.32\linewidth}
            \centering        
            \includegraphics[width=1\linewidth]{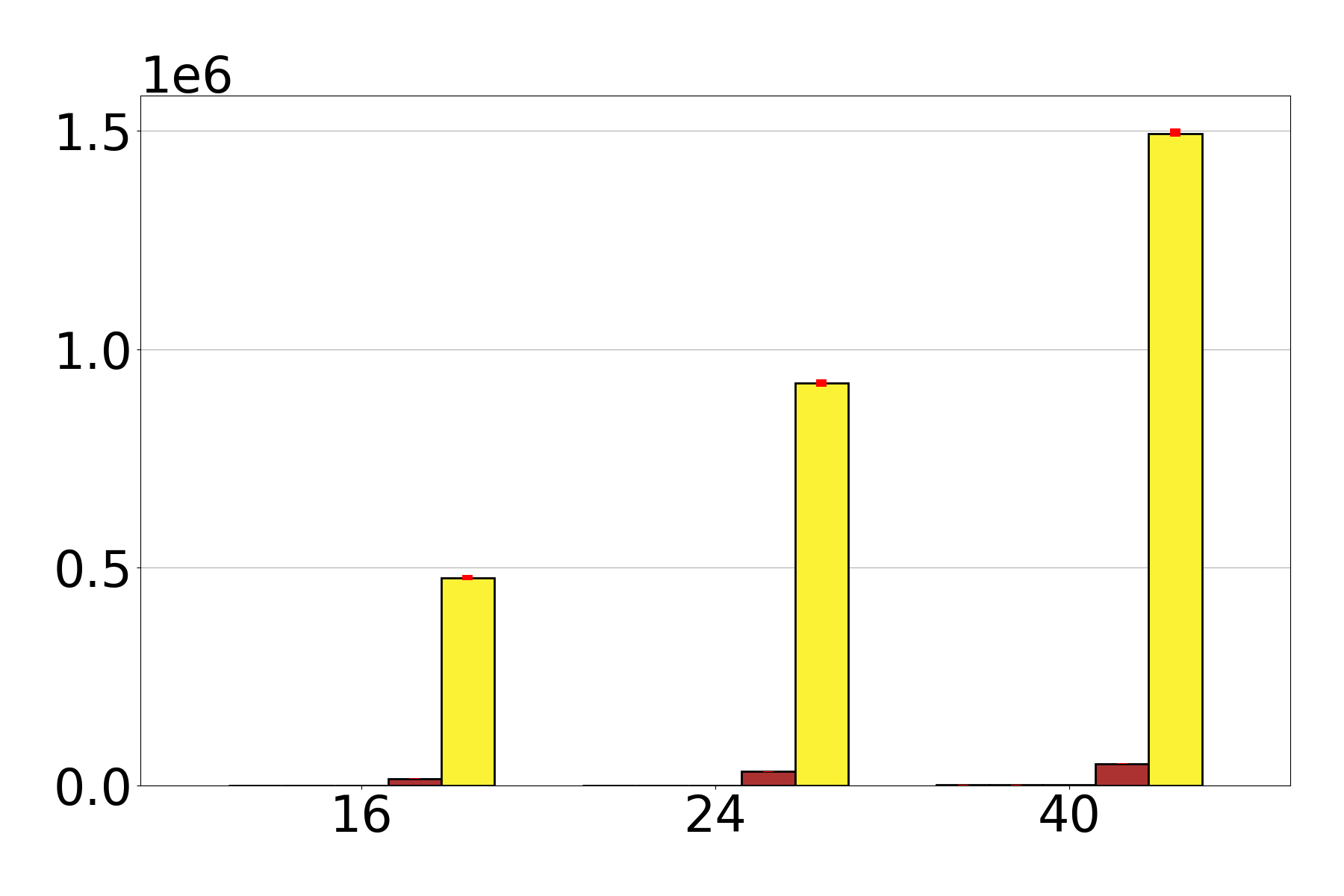}
        \end{subfigure}
        \begin{subfigure}{0.32\linewidth}
            \centering    
            \includegraphics[width=1\linewidth]{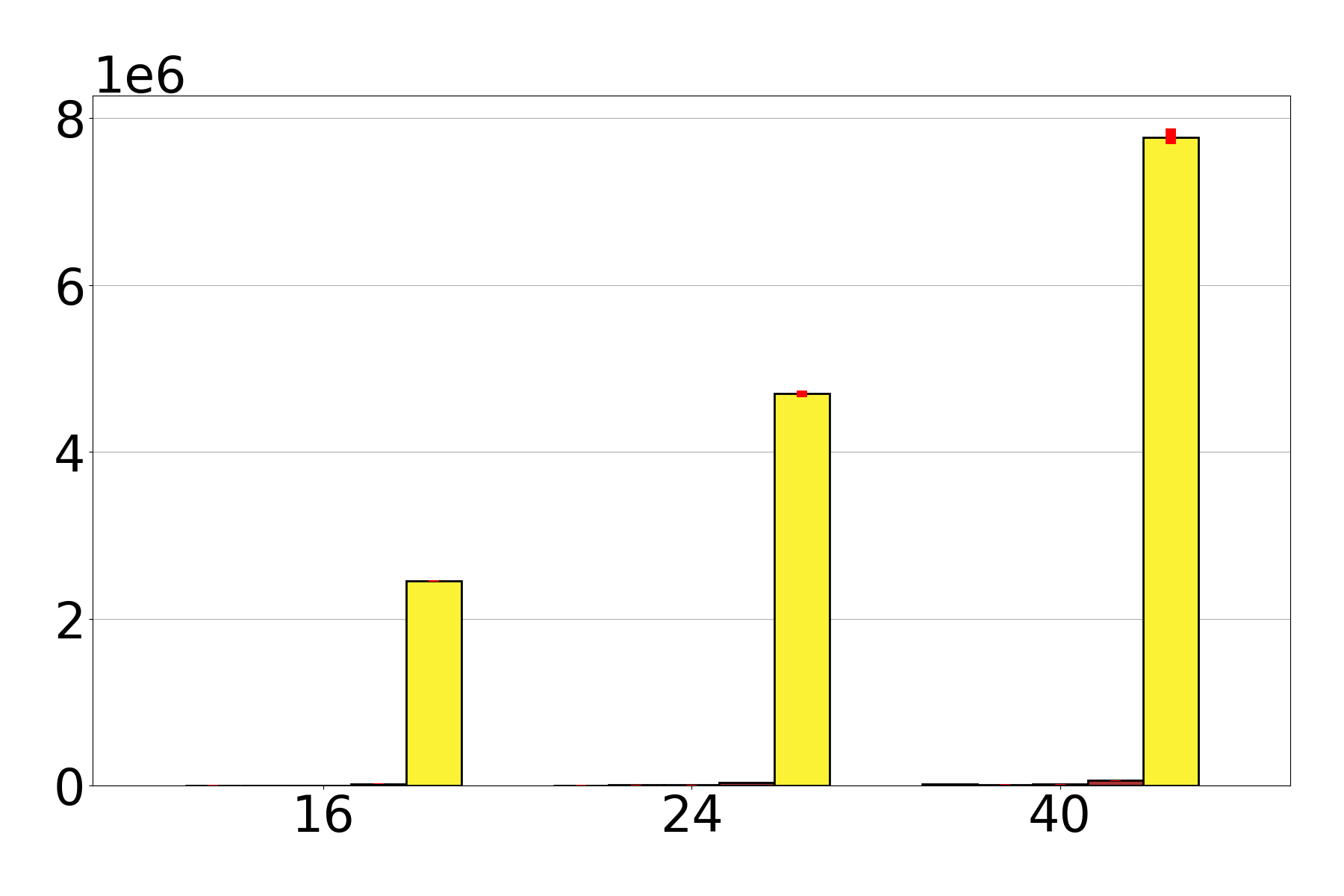}
        \end{subfigure}
    \end{subfigure} 
    \begin{subfigure}{0.02\linewidth}
        \raisebox{0.5\height}{\rotatebox{90}{16 Updaters}}
    \end{subfigure}
    \begin{subfigure}{0.97\linewidth}
        \begin{subfigure}{0.32\linewidth}
            \centering
            \includegraphics[width=1\linewidth]{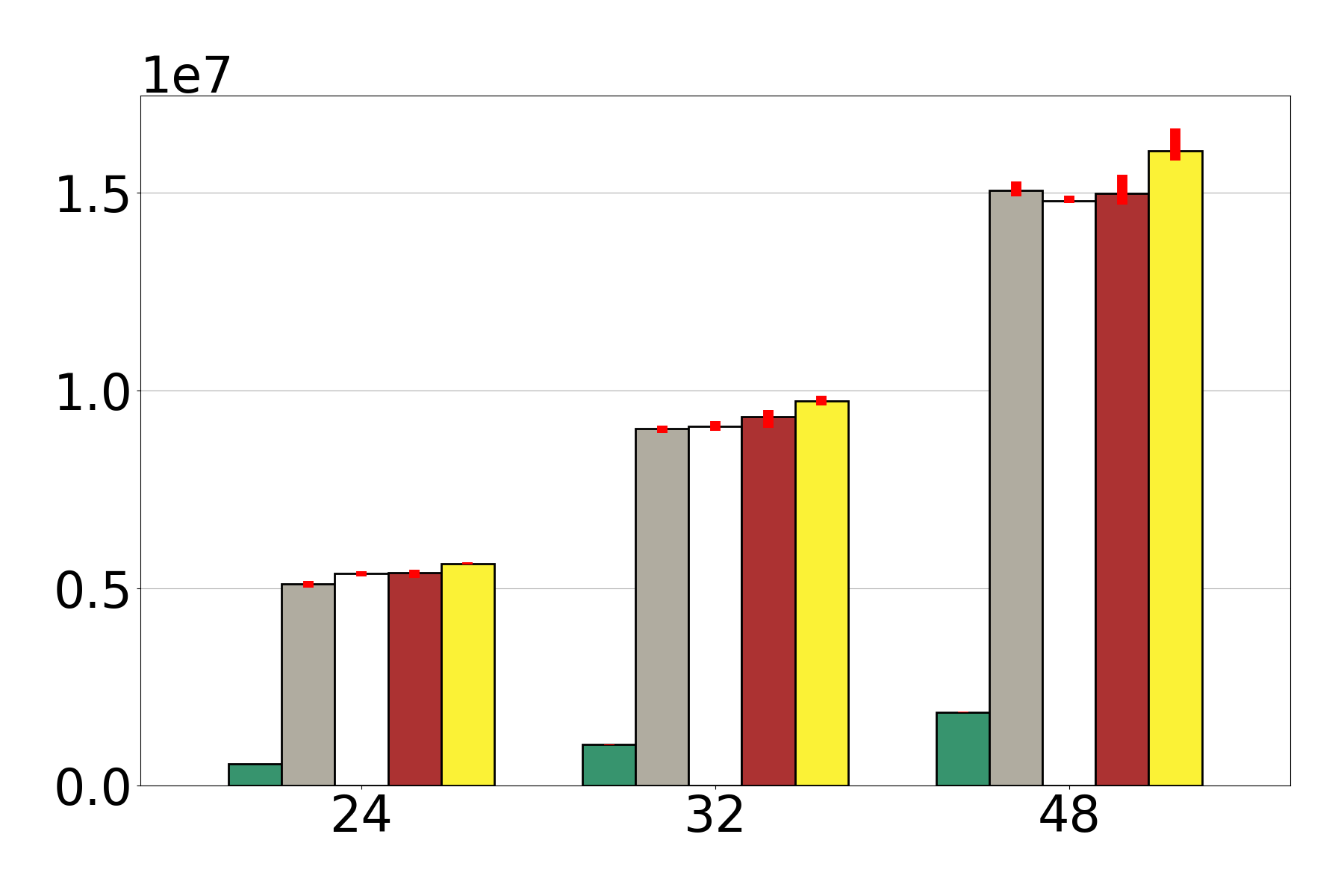}
        \end{subfigure} 
        \begin{subfigure}{0.32\linewidth}
            \centering        
            \includegraphics[width=1\linewidth]{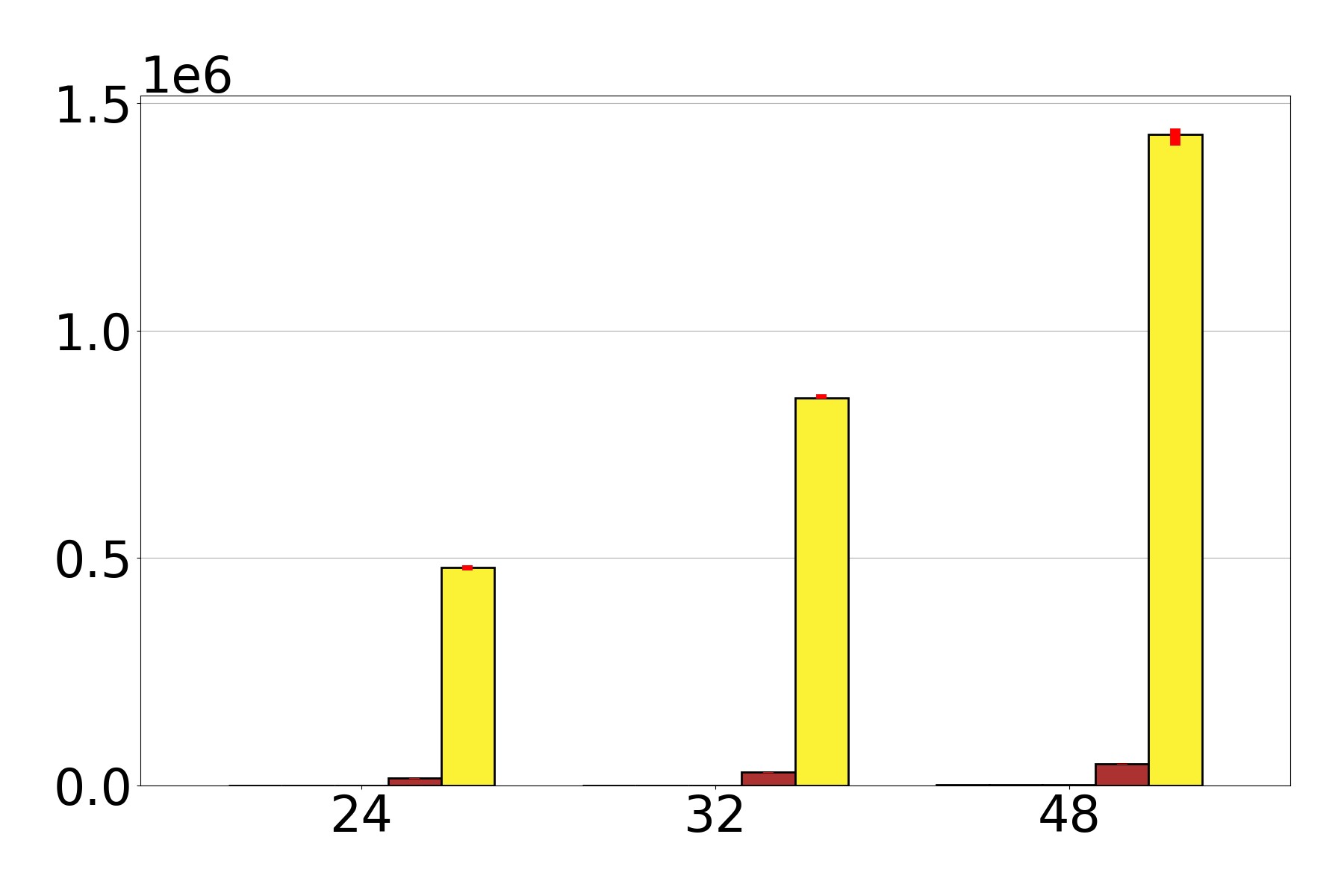}
        \end{subfigure}
        \begin{subfigure}{0.32\linewidth}
            \centering    
            \includegraphics[width=1\linewidth]{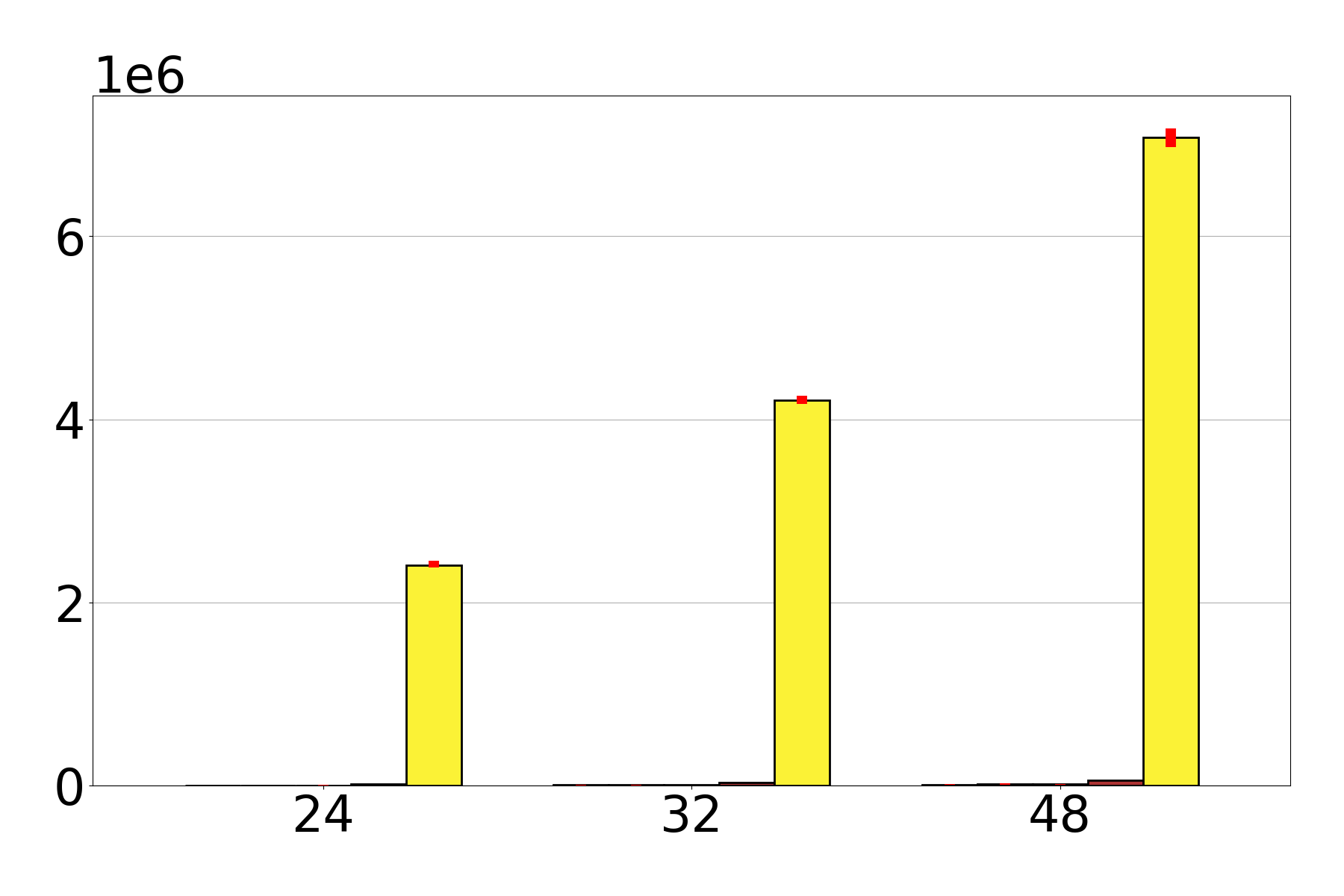}
        \end{subfigure}
    \end{subfigure}
    \begin{subfigure}{1.0\linewidth}
        \centering
        \includegraphics[width=0.4\linewidth]{plots/legend.png}
    \end{subfigure}     
    \vspace{-8mm}
    \caption{
    \centering Throughput for external binary search tree prefilled to 1 million keys using a uniform key access pattern. Y-axis is ops/sec. X-axis is number of threads. All workloads include 5\% insert and 5\% delete. RQ size is 10k (1\% of prefill size). Experiment ran on a single Intel Xeon Platinum 8160.
    }
    \Description{}
    \label{fig:jax-bst-node0}
\end{figure*}

\paragraph{Single Intel Xeon Platinum 8160 Experiments}
We include additional results that were run on a different system with a Intel Xeon Platinum 8160 which has 24 cores and 48 hardware threads.
Since the total hardware thread count is lower than our AMD machine we also include experiments with fewer dedicated updater threads (specifically 8).
As shown in \Cref{fig:jax-abtree-node0} through \Cref{fig:jax-bst-node0}, the results on our Intel machine are similar to the results of the experiments ran on our AMD machine.

\begin{figure*}[t!]
    \begin{subfigure}{0.02\linewidth}        
        \raisebox{0.5\height}{\rotatebox{90}{0 Updaters}}
    \end{subfigure}
    \begin{subfigure}{0.97\linewidth}
        \begin{subfigure}{0.32\linewidth}
            \centering
            90\% Search, 0\% RQ
            \includegraphics[width=1\linewidth]{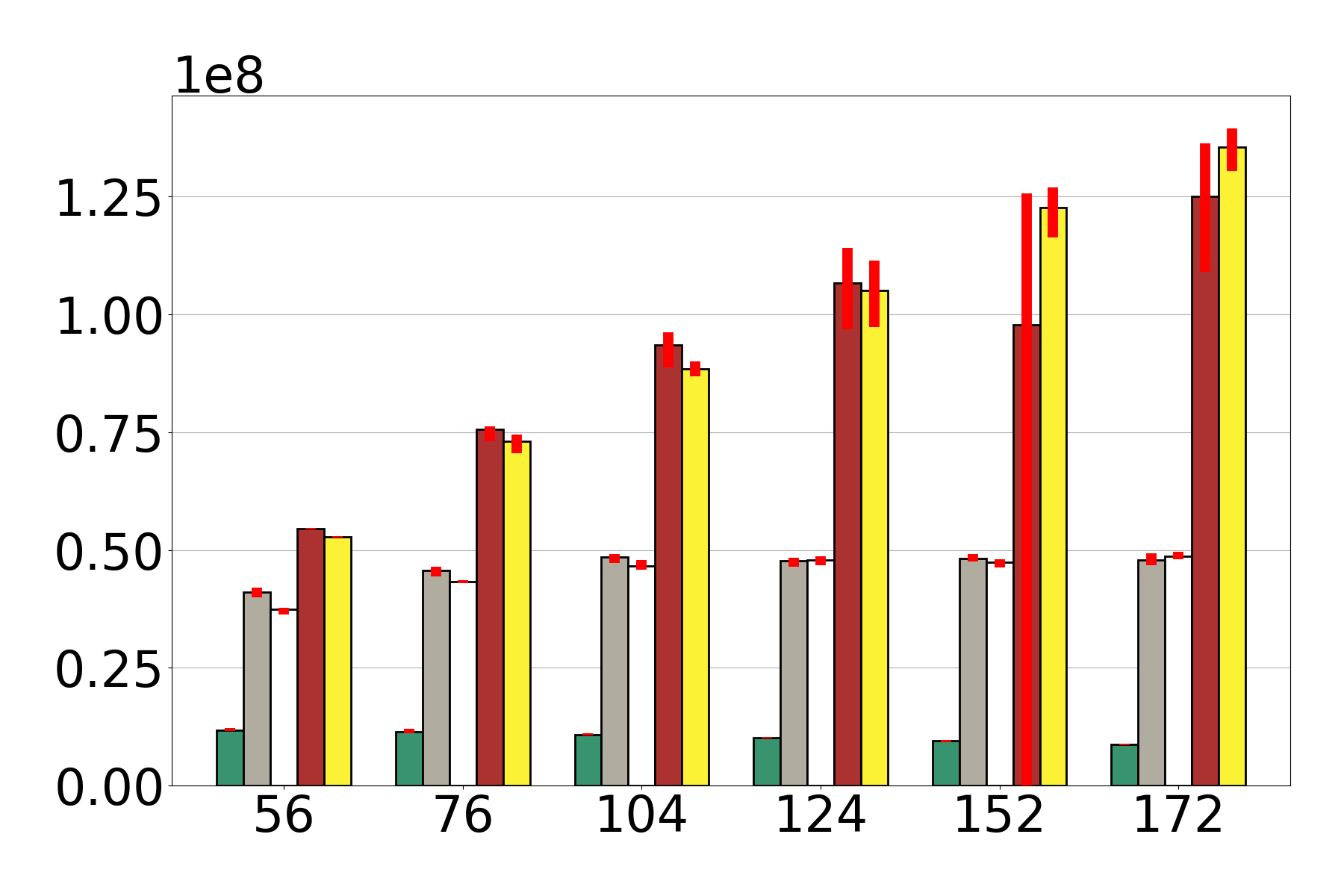}
        \end{subfigure} 
        \begin{subfigure}{0.32\linewidth}
            \centering        
            89.9\% Search, 0.1\% RQ
            \includegraphics[width=1\linewidth]{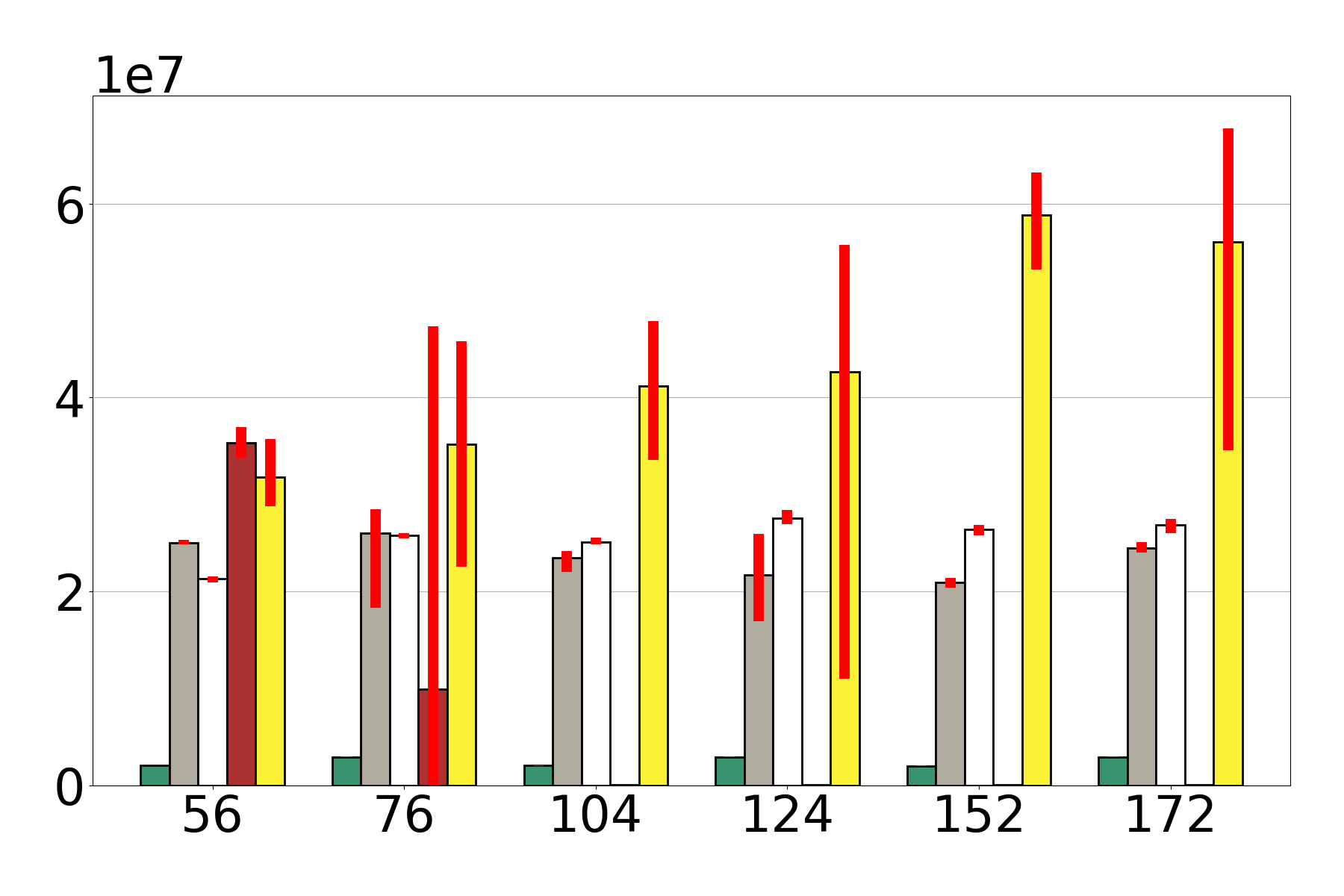}
        \end{subfigure}
        \begin{subfigure}{0.32\linewidth}
            \centering    
            89.99\% Search, 0.01\% RQ
            \includegraphics[width=1\linewidth]{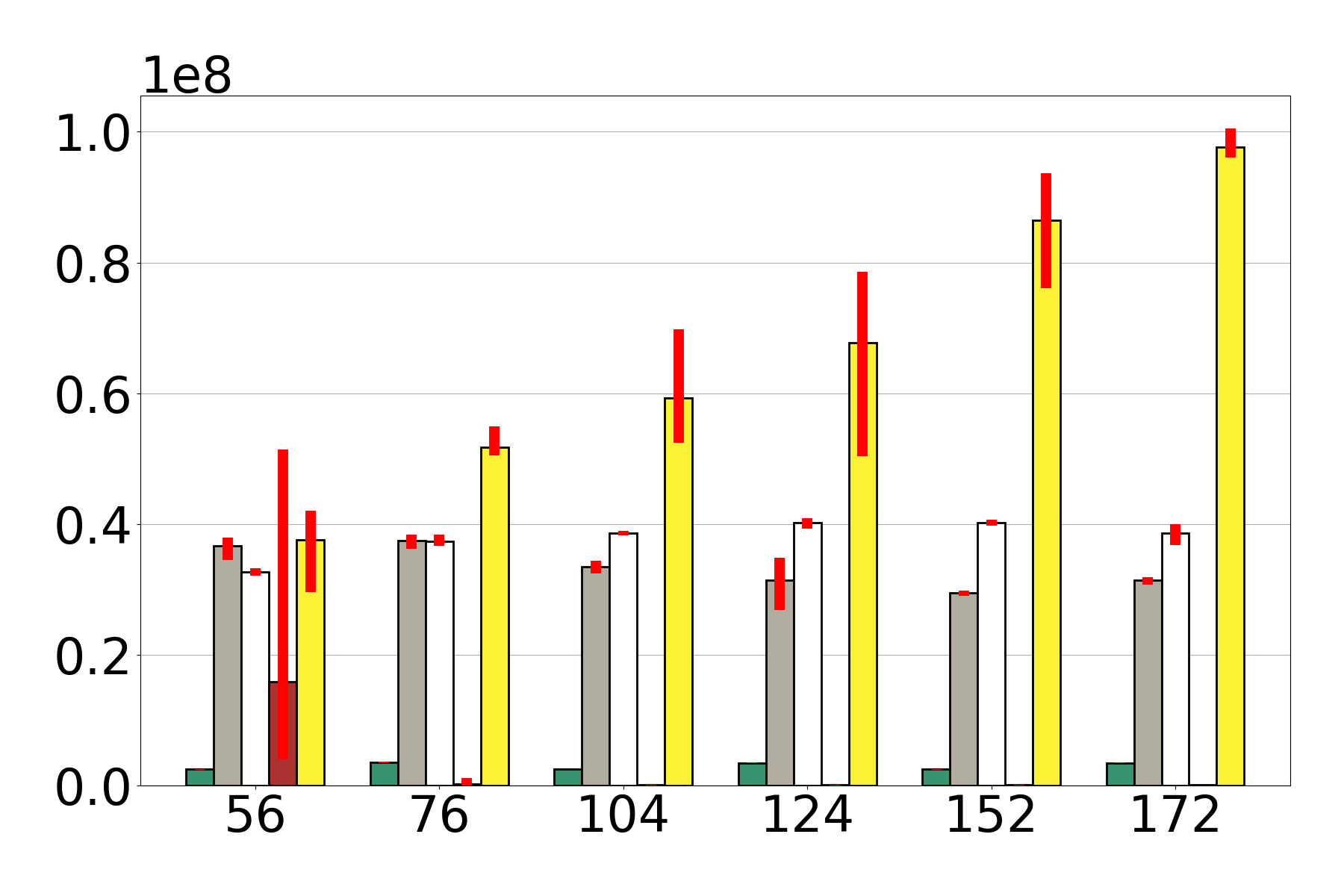}
        \end{subfigure}
    \end{subfigure}
    \begin{subfigure}{0.02\linewidth}
        \raisebox{0.5\height}{\rotatebox{90}{16 Updaters}}
    \end{subfigure}
    \begin{subfigure}{0.97\linewidth}
        \begin{subfigure}{0.32\linewidth}
            \centering
            \includegraphics[width=1\linewidth]{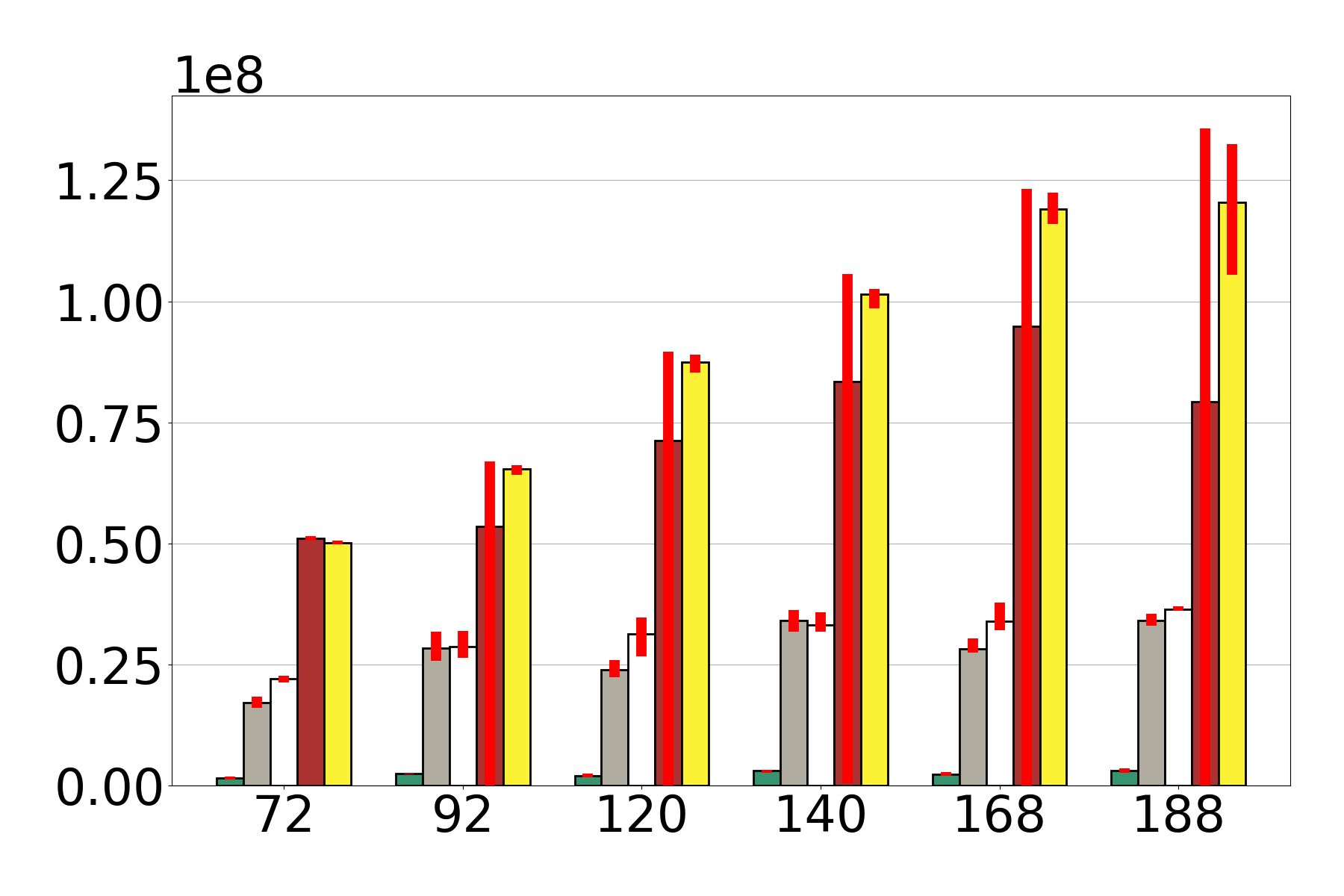}
        \end{subfigure} 
        \begin{subfigure}{0.32\linewidth}
            \centering        
            \includegraphics[width=1\linewidth]{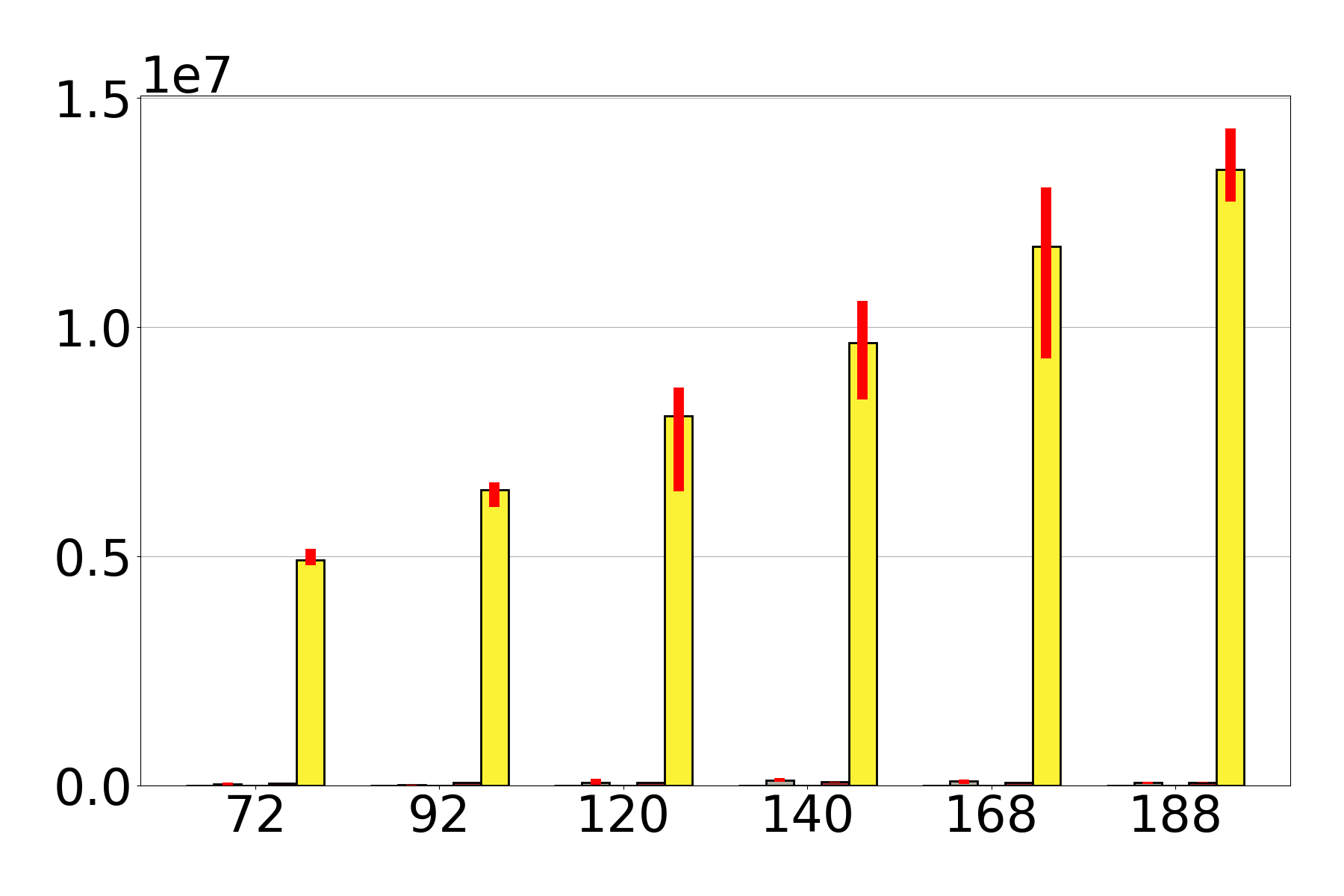}
        \end{subfigure}
        \begin{subfigure}{0.32\linewidth}
            \centering    
            \includegraphics[width=1\linewidth]{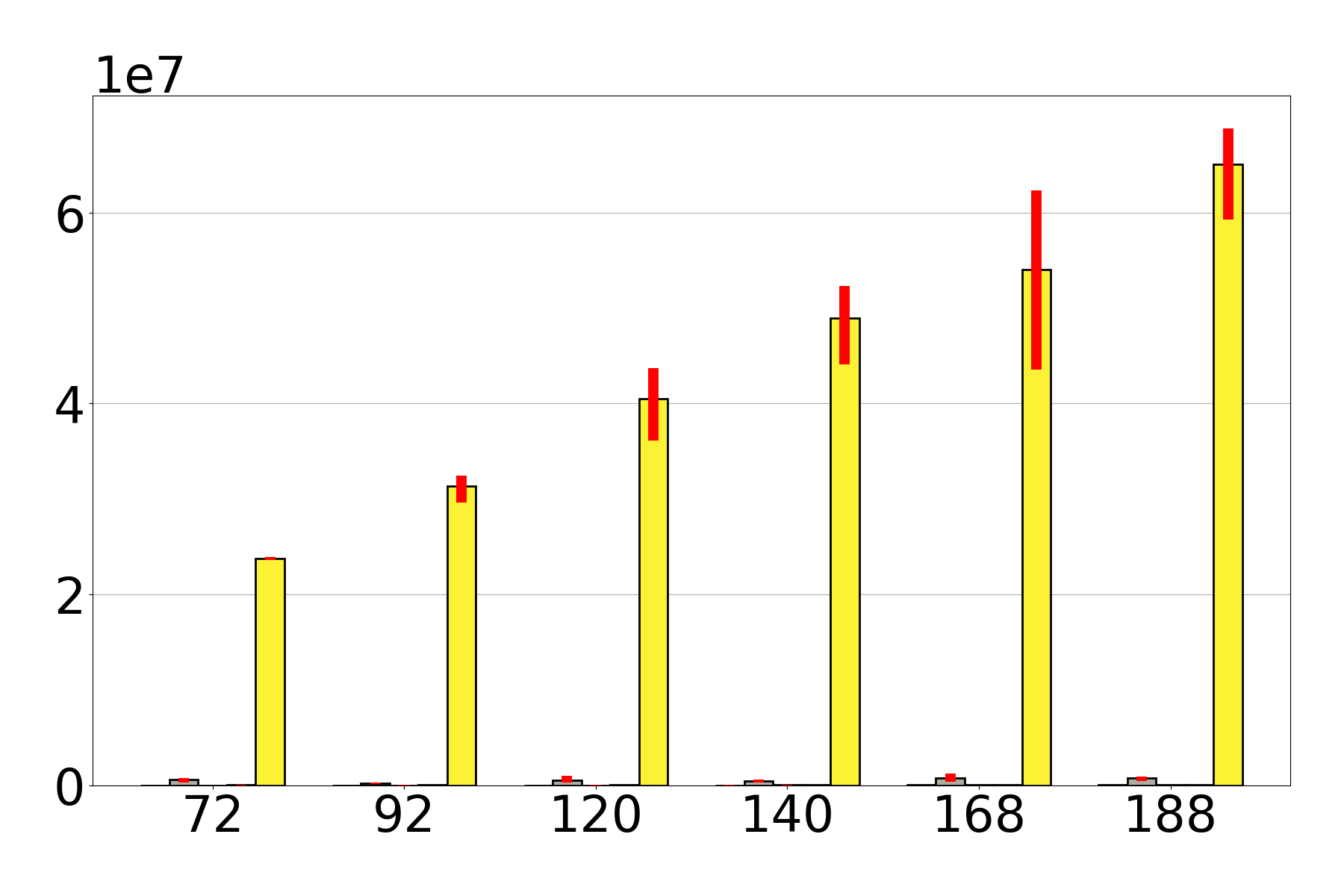}
        \end{subfigure}
    \end{subfigure}
    \begin{subfigure}{1.0\linewidth}
        \centering
        \includegraphics[width=0.4\linewidth]{plots/legend.png}
    \end{subfigure}     
    \vspace{-8mm}
    \caption{
    \centering Throughput for (a,b)-tree prefilled to 1 million keys using a uniform key access pattern. Y-axis is ops/sec. X-axis is number of threads. All workloads include 5\% insert and 5\% delete. RQ size is 10k (1\% of prefill size). Experiment ran on quad Intel Xeon Platinum 8160.
    }
    \Description{}
    \label{fig:jax-abtree-all_nodes}
\end{figure*}

\begin{figure*}[t!]
    \begin{subfigure}{0.02\linewidth}        
        \raisebox{0.5\height}{\rotatebox{90}{0 Updaters}}
    \end{subfigure}
    \begin{subfigure}{0.97\linewidth}
        \begin{subfigure}{0.32\linewidth}
            \centering
            90\% Search, 0\% RQ
            \includegraphics[width=1\linewidth]{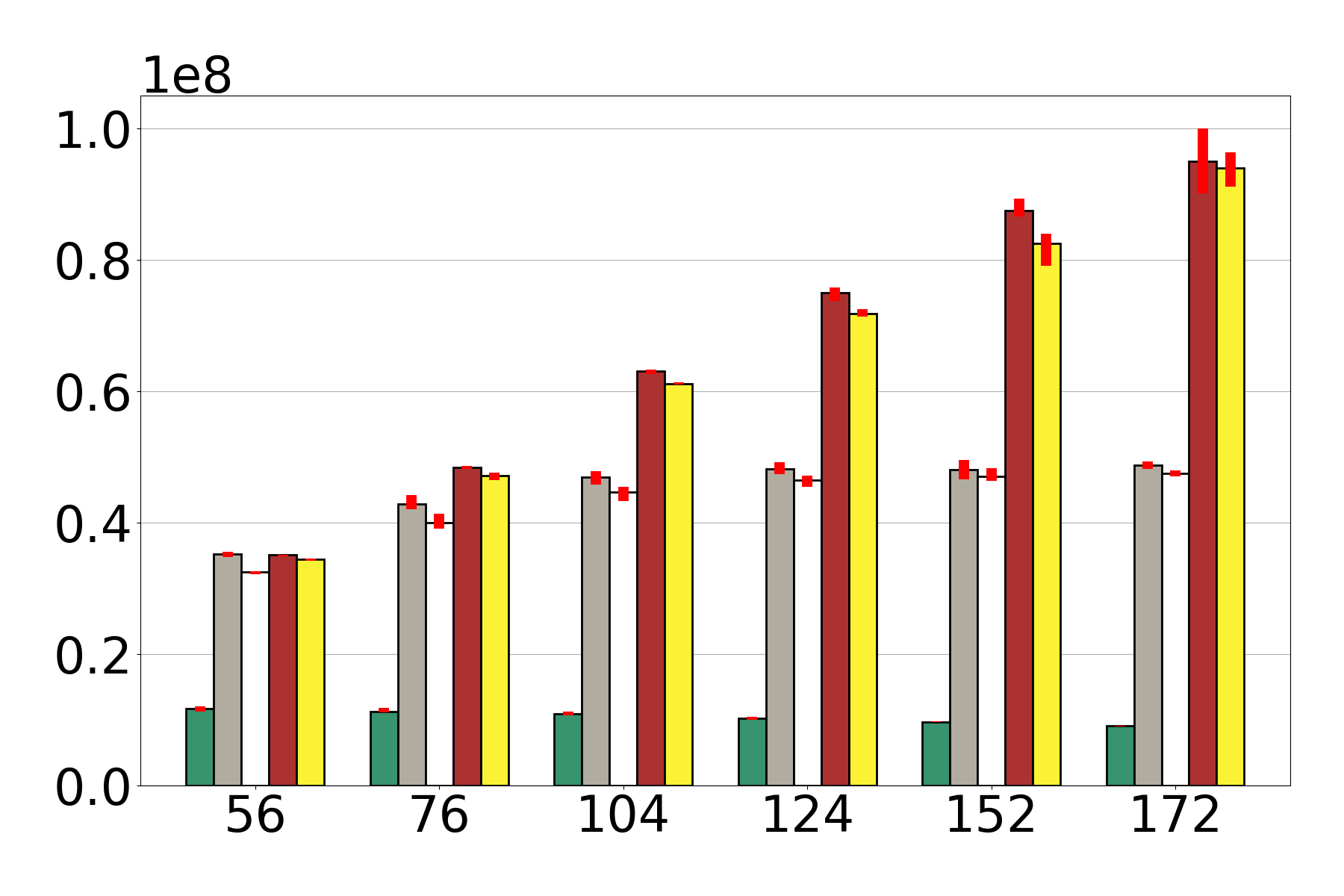}
        \end{subfigure} 
        \begin{subfigure}{0.32\linewidth}
            \centering        
            89.9\% Search, 0.1\% RQ
            \includegraphics[width=1\linewidth]{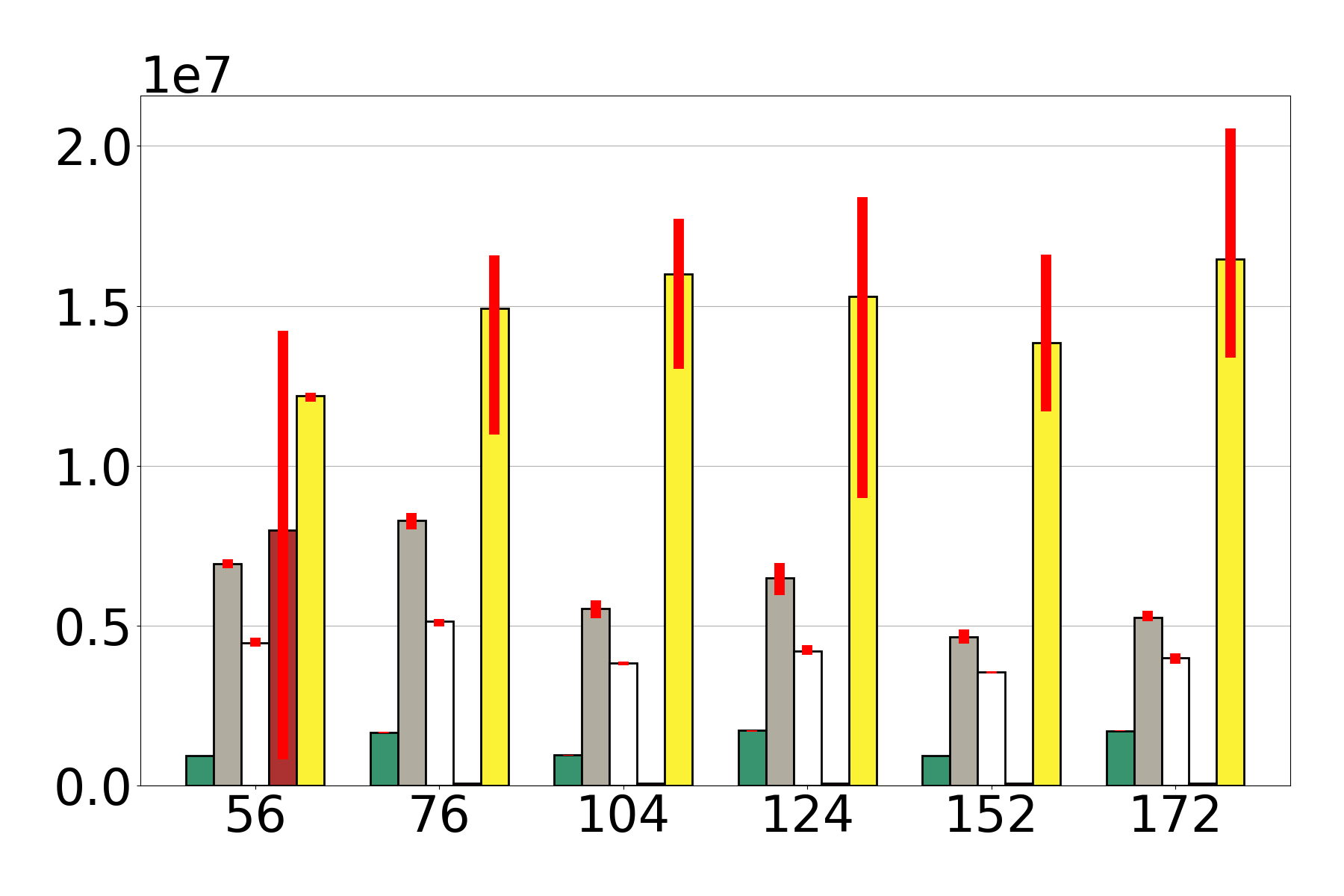}
        \end{subfigure}
        \begin{subfigure}{0.32\linewidth}
            \centering    
            89.99\% Search, 0.01\% RQ
            \includegraphics[width=1\linewidth]{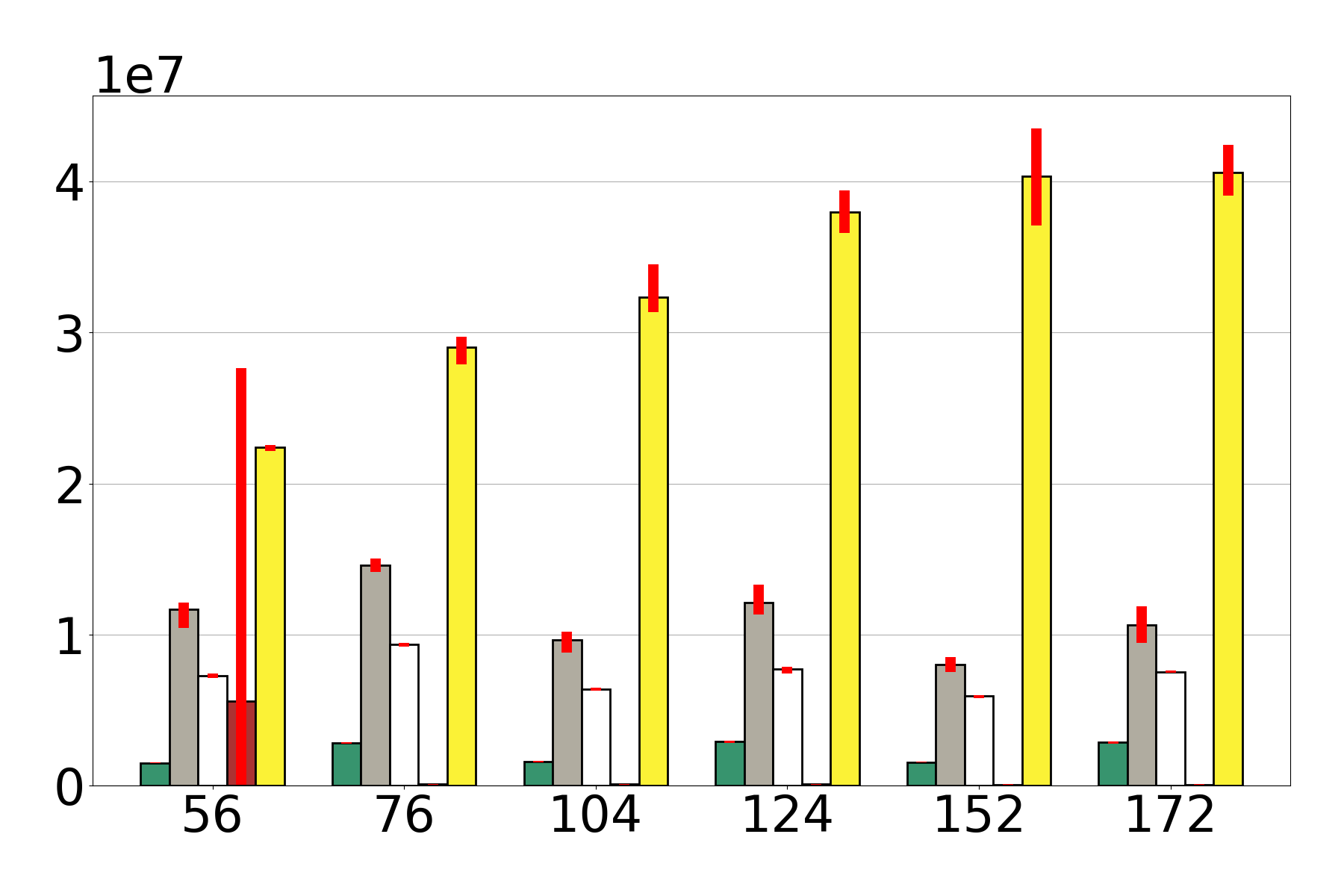}
        \end{subfigure}
    \end{subfigure}
    \begin{subfigure}{0.02\linewidth}
        \raisebox{0.5\height}{\rotatebox{90}{16 Updaters}}
    \end{subfigure}
    \begin{subfigure}{0.97\linewidth}
        \begin{subfigure}{0.32\linewidth}
            \centering
            \includegraphics[width=1\linewidth]{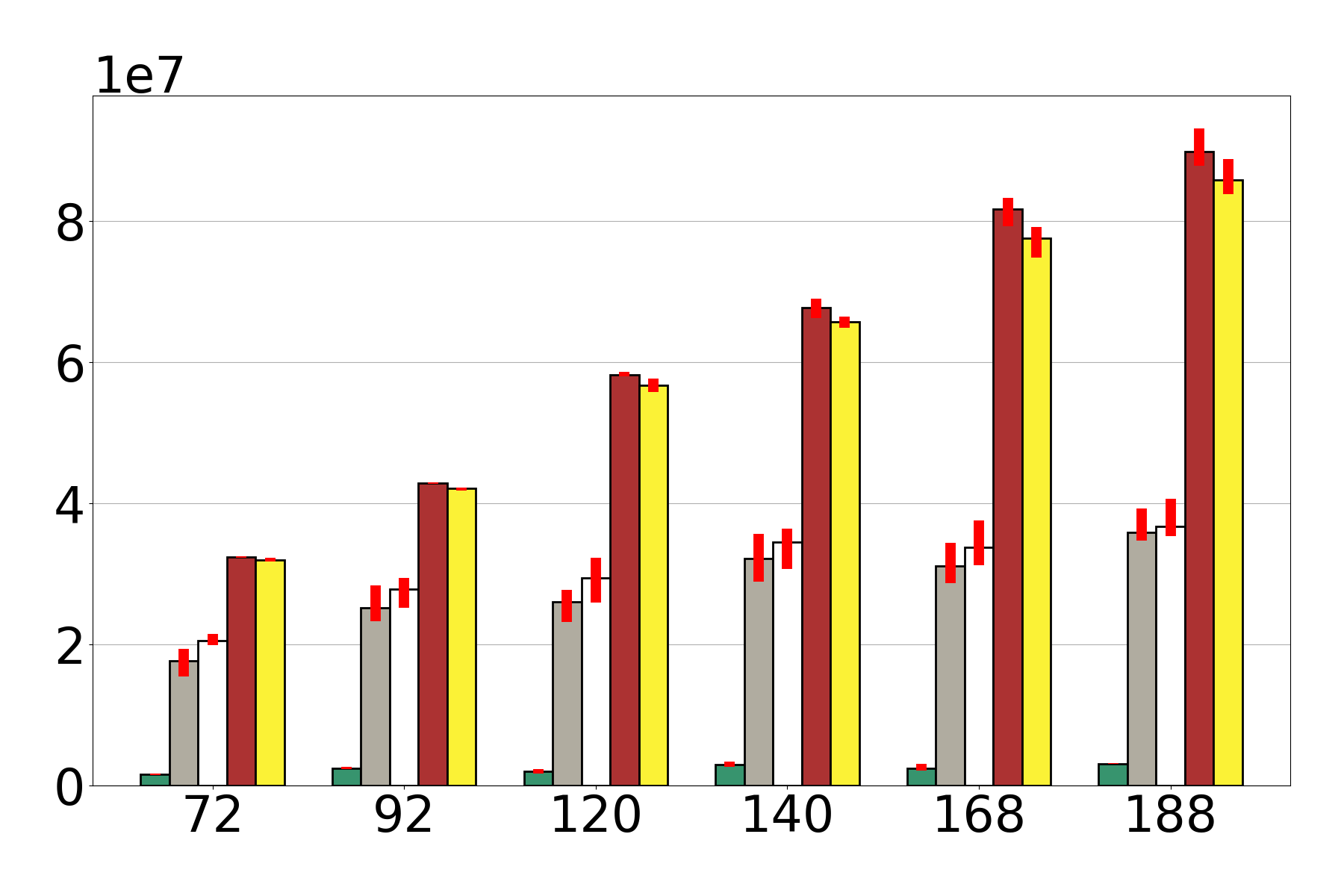}
        \end{subfigure} 
        \begin{subfigure}{0.32\linewidth}
            \centering        
            \includegraphics[width=1\linewidth]{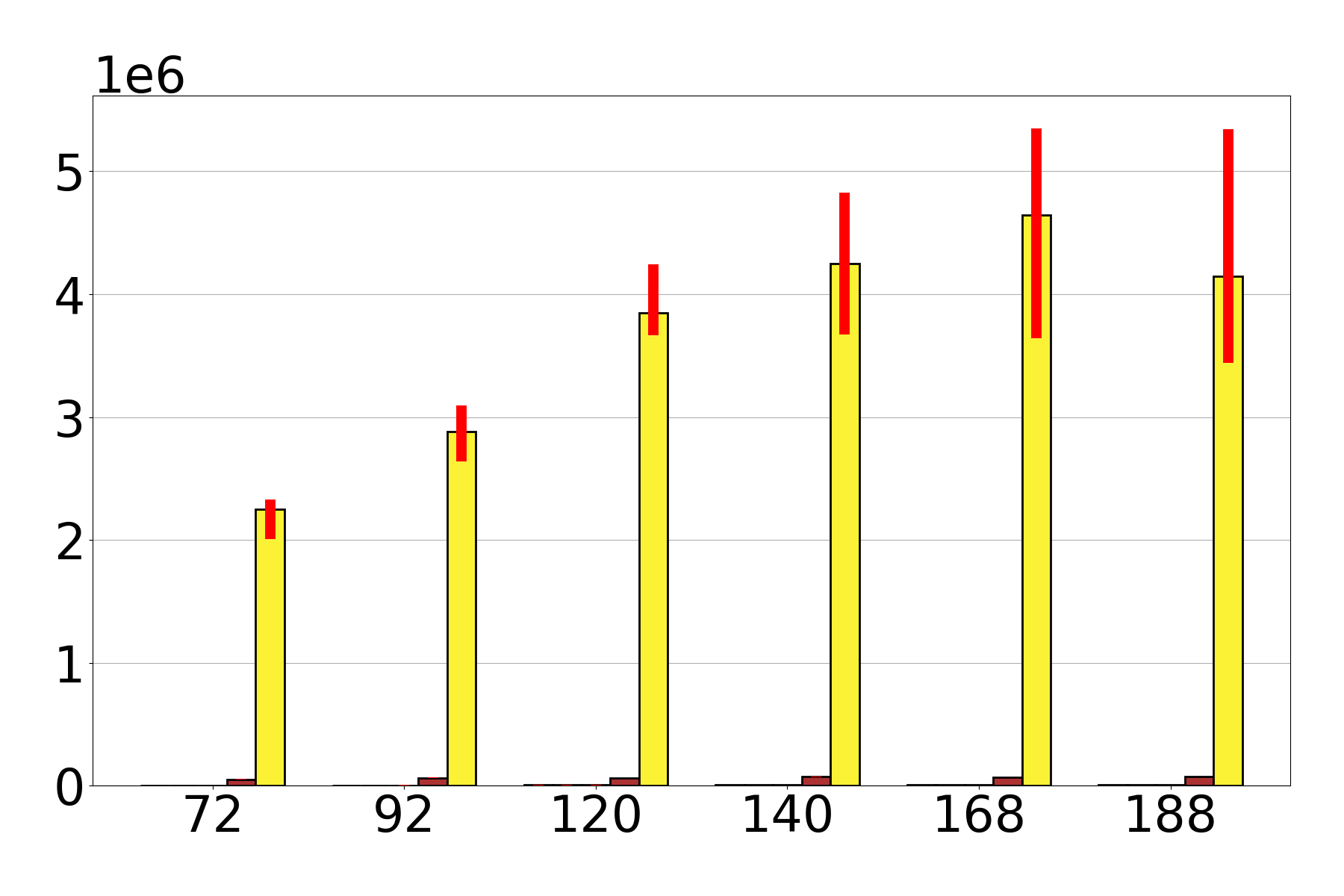}
        \end{subfigure}
        \begin{subfigure}{0.32\linewidth}
            \centering    
            \includegraphics[width=1\linewidth]{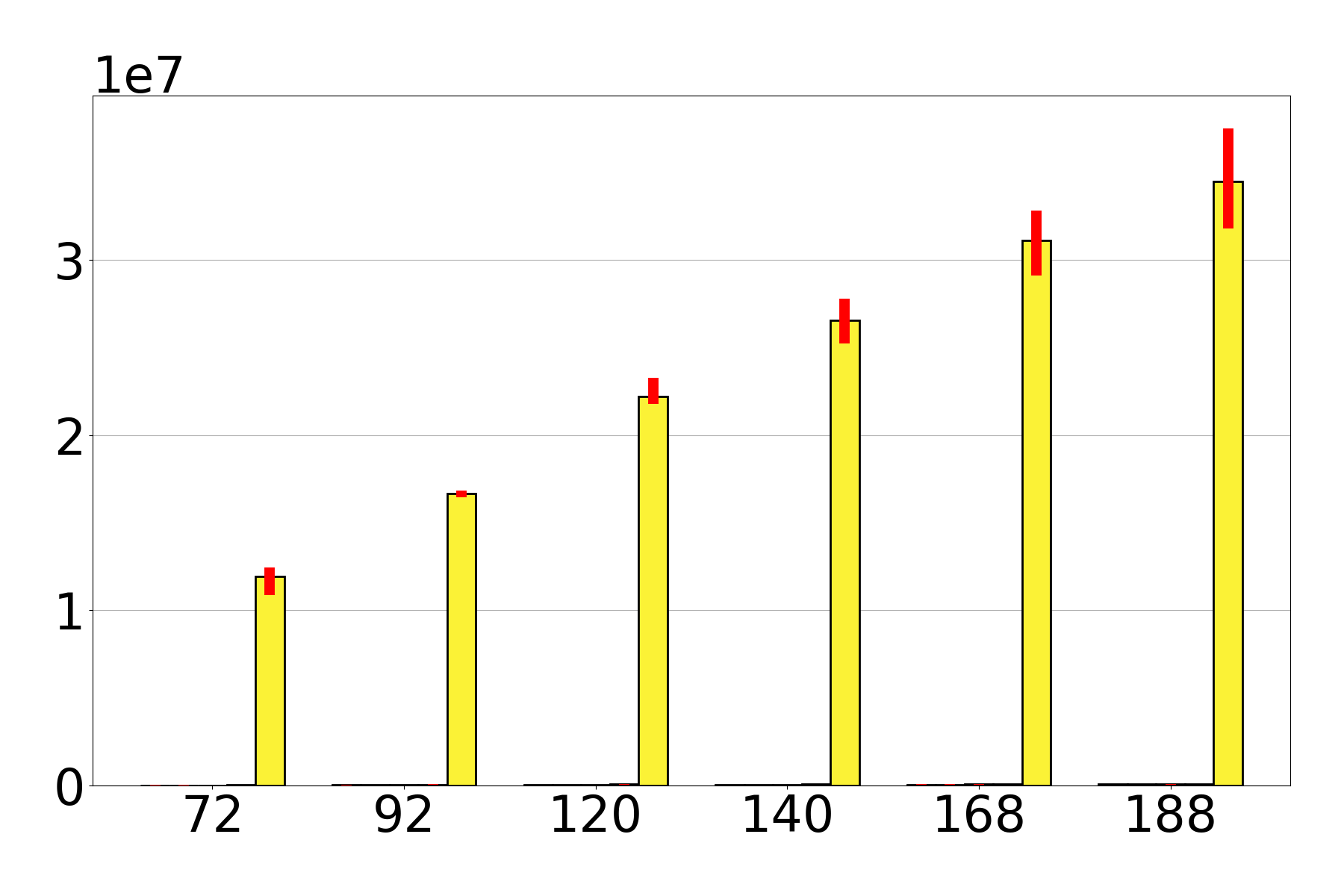}
        \end{subfigure}
    \end{subfigure}
    \begin{subfigure}{1.0\linewidth}
        \centering
        \includegraphics[width=0.4\linewidth]{plots/legend.png}
    \end{subfigure}     
    \vspace{-8mm}
    \caption{
    \centering Throughput for AVL-tree prefilled to 1 million keys using a uniform key access pattern. Y-axis is ops/sec. X-axis is number of threads. All workloads include 5\% insert and 5\% delete. RQ size is 10k (1\% of prefill size). Experiment ran on quad Intel Xeon Platinum 8160.
    }
    \Description{}
    \label{fig:jax-avl-all-nodes}
\end{figure*}

\begin{figure*}[t!]
    \begin{subfigure}{0.02\linewidth}        
        \raisebox{0.5\height}{\rotatebox{90}{0 Updaters}}
    \end{subfigure}
    \begin{subfigure}{0.97\linewidth}
        \begin{subfigure}{0.32\linewidth}
            \centering
            90\% Search, 0\% RQ
            \includegraphics[width=1\linewidth]{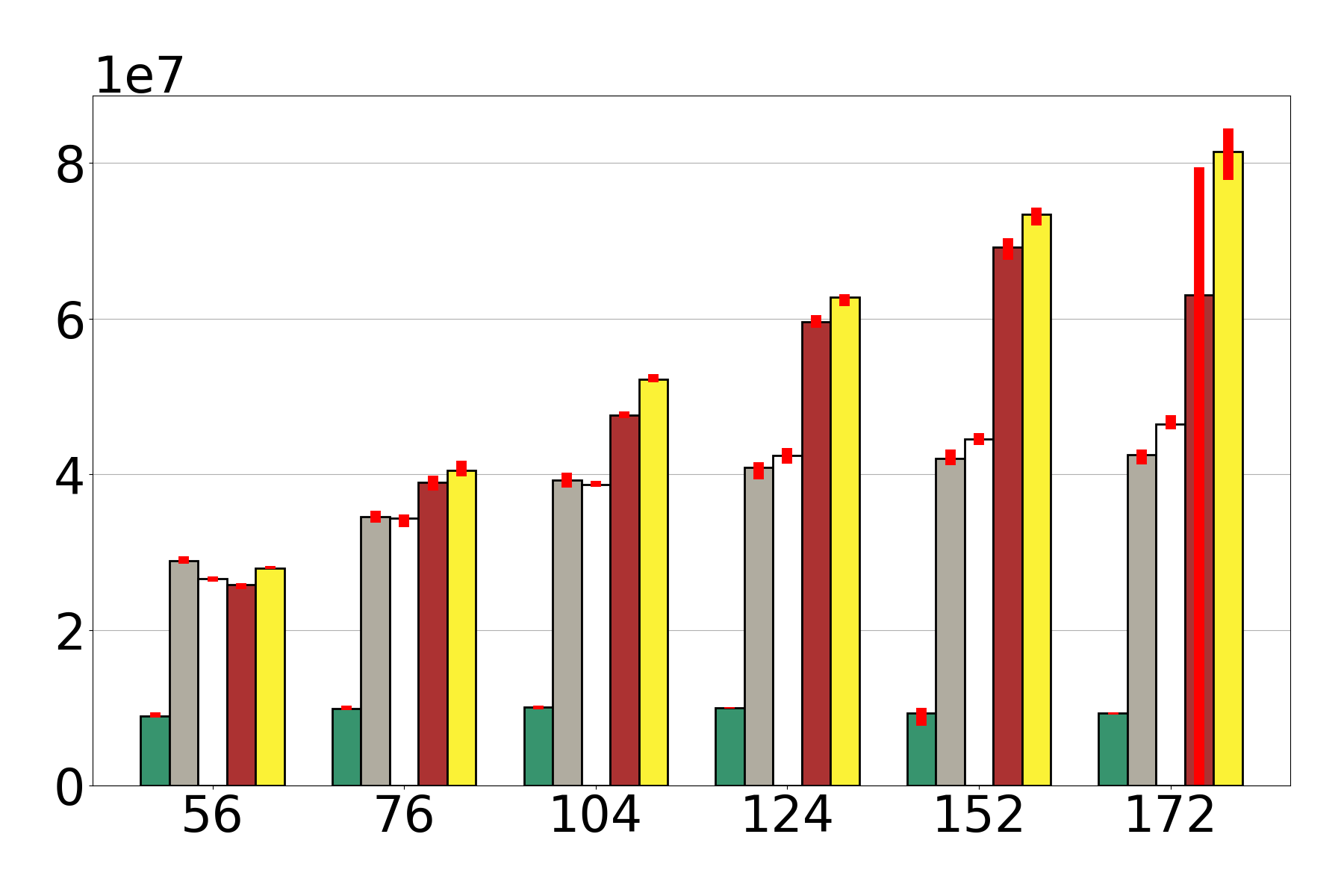}
        \end{subfigure} 
        \begin{subfigure}{0.32\linewidth}
            \centering        
            89.9\% Search, 0.1\% RQ
            \includegraphics[width=1\linewidth]{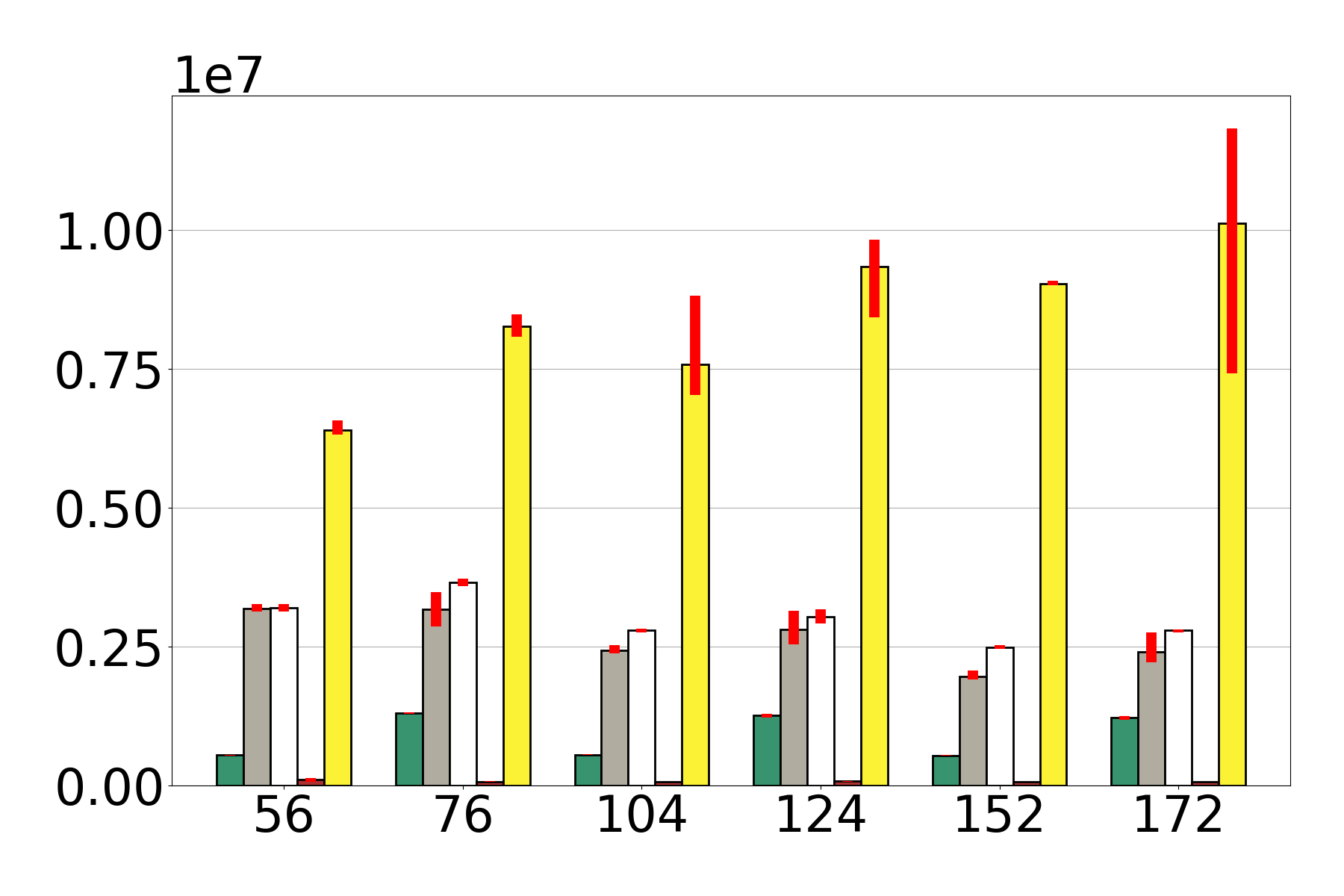}
        \end{subfigure}
        \begin{subfigure}{0.32\linewidth}
            \centering    
            89.99\% Search, 0.01\% RQ
            \includegraphics[width=1\linewidth]{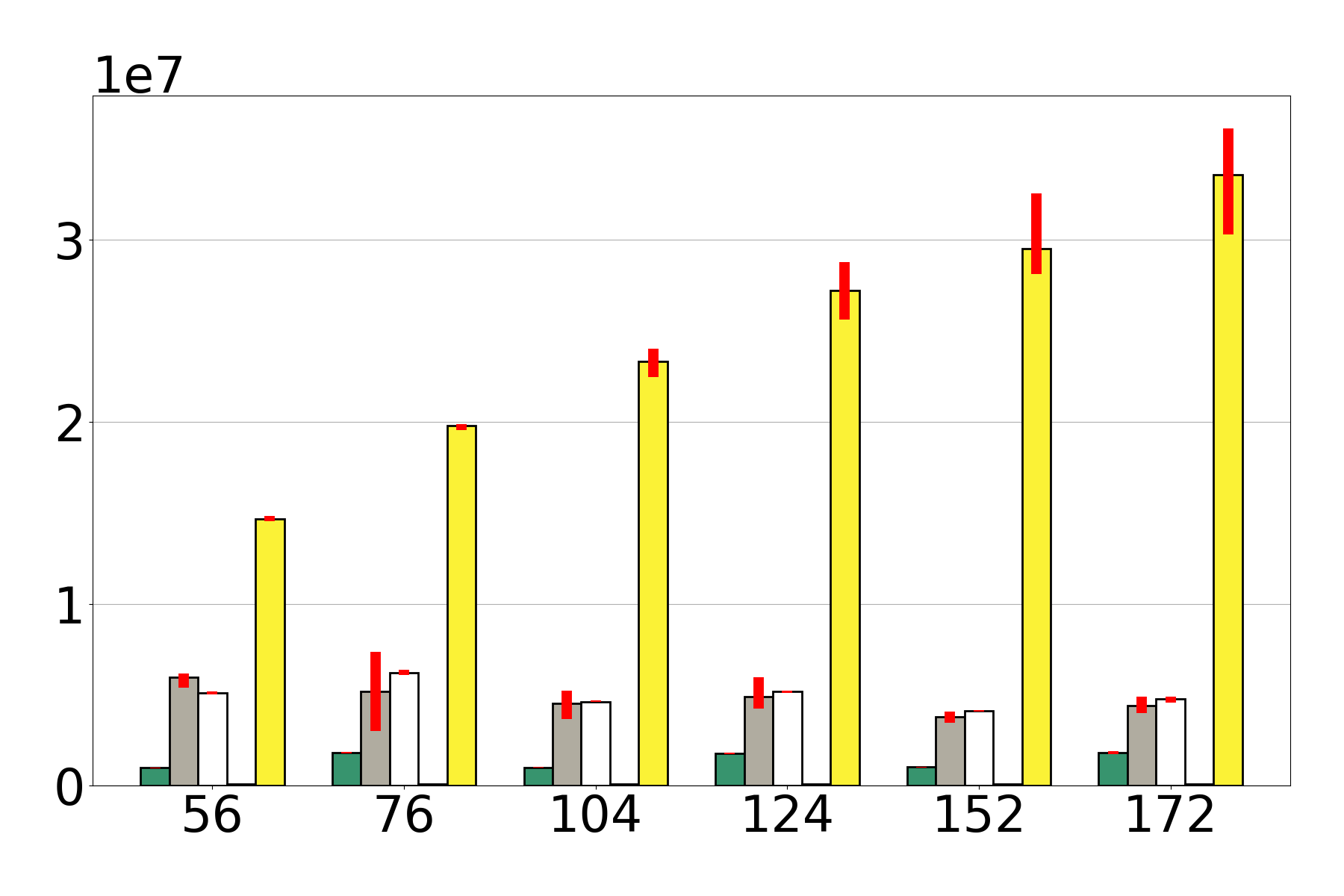}
        \end{subfigure}
    \end{subfigure}
    \begin{subfigure}{0.02\linewidth}
        \raisebox{0.5\height}{\rotatebox{90}{16 Updaters}}
    \end{subfigure}
    \begin{subfigure}{0.97\linewidth}
        \begin{subfigure}{0.32\linewidth}
            \centering
            \includegraphics[width=1\linewidth]{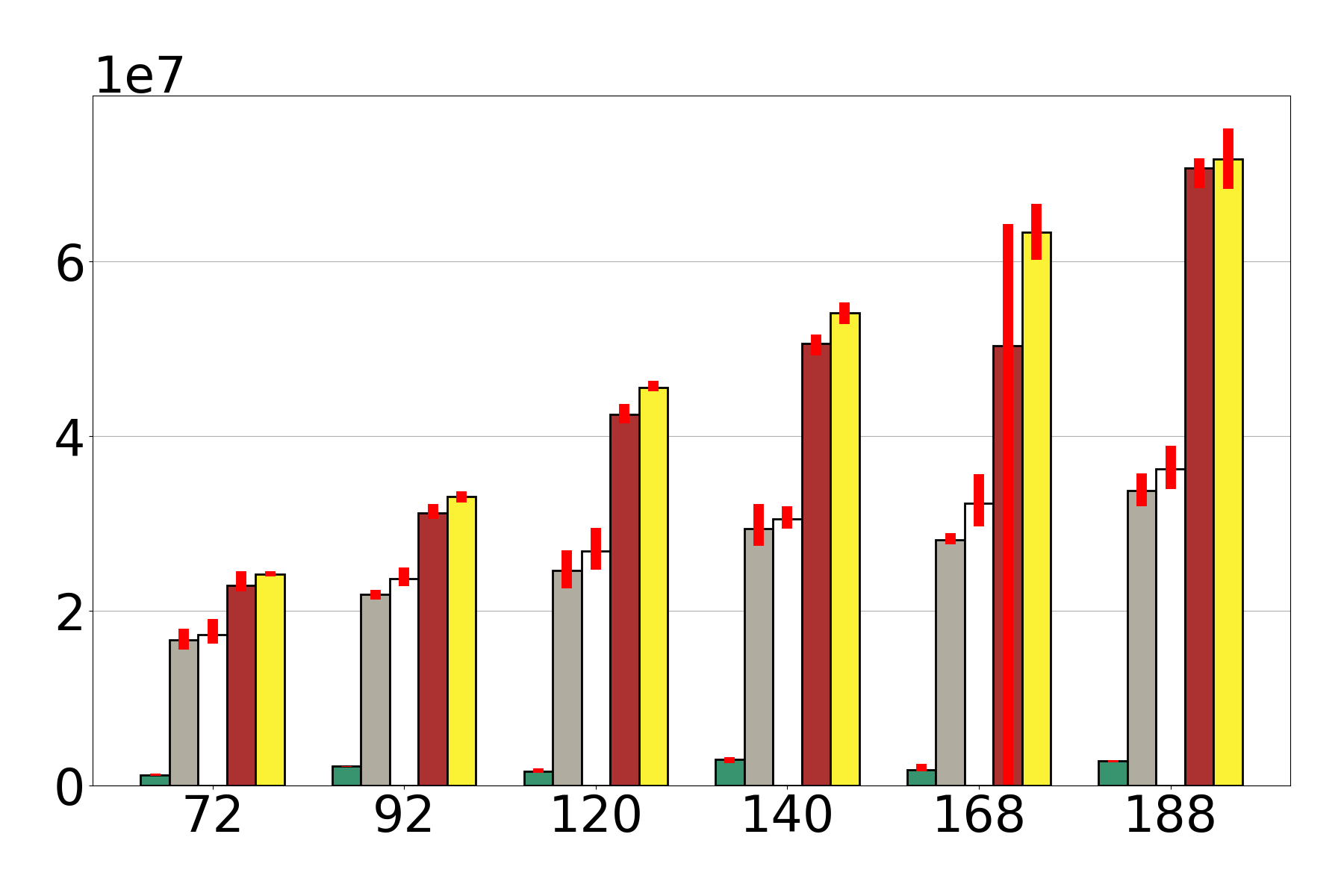}
        \end{subfigure} 
        \begin{subfigure}{0.32\linewidth}
            \centering        
            \includegraphics[width=1\linewidth]{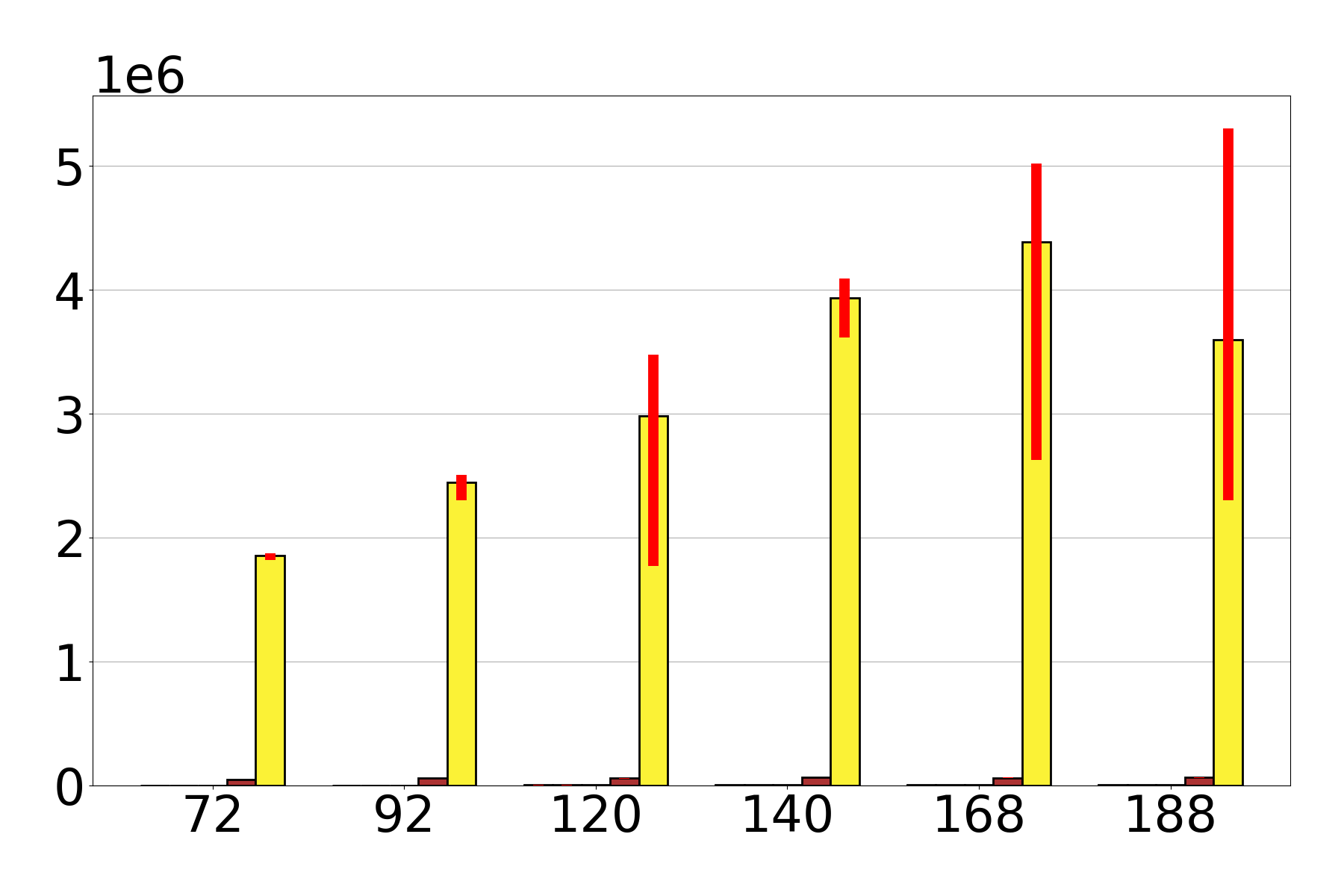}
        \end{subfigure}
        \begin{subfigure}{0.32\linewidth}
            \centering    
            \includegraphics[width=1\linewidth]{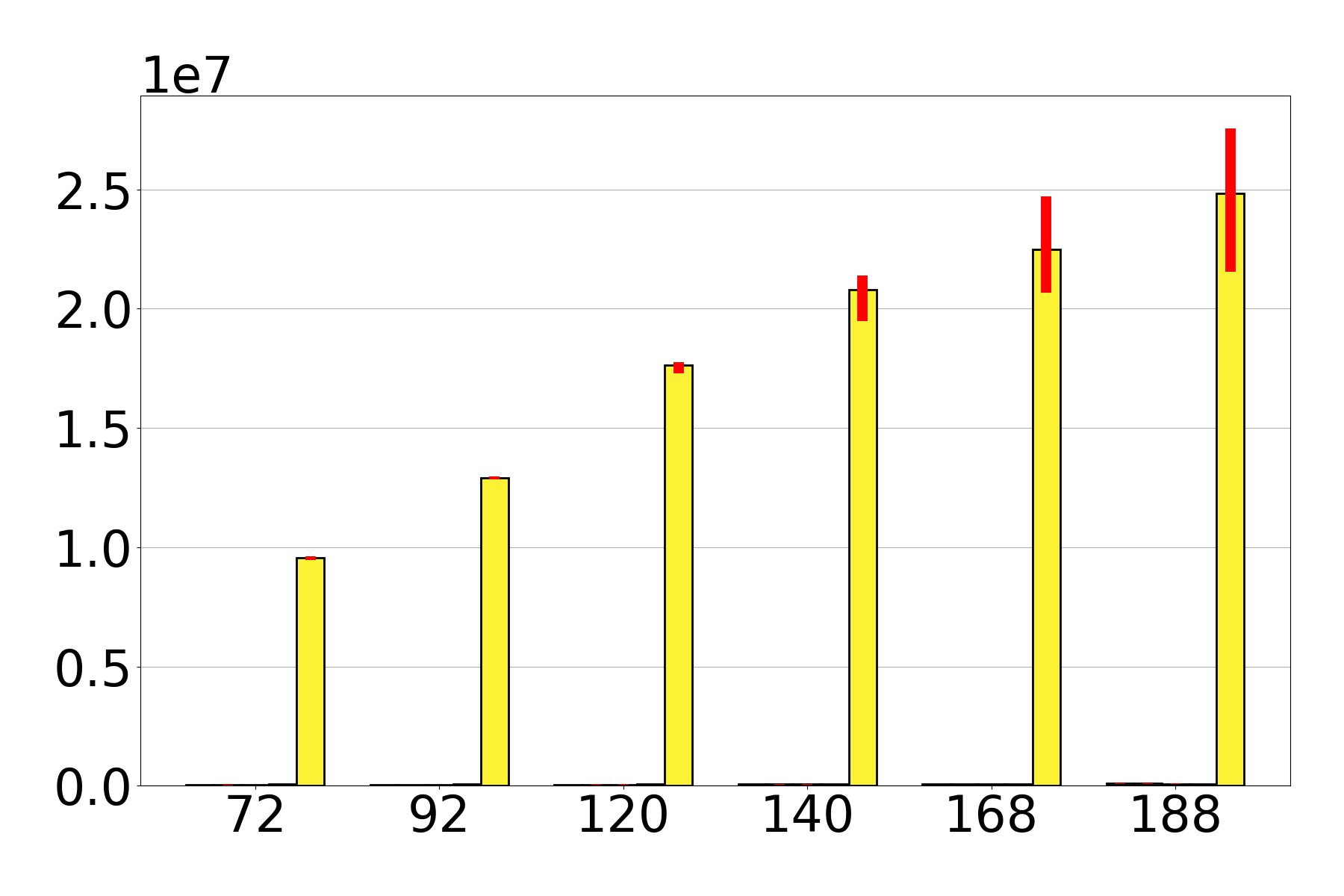}
        \end{subfigure}
    \end{subfigure}
    \begin{subfigure}{1.0\linewidth}
        \centering
        \includegraphics[width=0.4\linewidth]{plots/legend.png}
    \end{subfigure}     
    \vspace{-8mm}
    \caption{
    \centering Throughput for external binary search tree prefilled to 1 million keys using a uniform key access pattern. Y-axis is ops/sec. X-axis is number of threads. All workloads include 5\% insert and 5\% delete. RQ size is 10k (1\% of prefill size). Experiment ran on quad Intel Xeon Platinum 8160.
    }
    \Description{}
    \label{fig:jax-bst-all-nodes}
\end{figure*}

\paragraph{Quad Intel Xeon Platinum 8160 Experiments}
We also include results that were run on quad Intel Xeon Platinum 8160s.
See \Cref{fig:jax-abtree-all_nodes} through \Cref{fig:jax-bst-all-nodes}.

\end{document}